\documentclass[12pt]{elsarticle}

\usepackage{graphicx, import}
\usepackage{amsfonts,textgreek}
\usepackage{mathtools}
\usepackage{dsfont}
\usepackage{multicol}
\usepackage{bbm}
\usepackage{url}
\usepackage[dvipsnames]{xcolor}
\usepackage[colorlinks=true,allcolors=blue]{hyperref}
\usepackage[nameinlink,noabbrev,capitalize]{cleveref}
\usepackage{xfrac}
\usepackage{caption}
\usepackage{subcaption}
\usepackage{xspace}
\usepackage{stmaryrd}
\usepackage{enumerate}
\usepackage{soul}

\usepackage[utf8]{inputenc}
\usepackage{wrapfig,bm}
\usepackage{lineno}
\usepackage{siunitx} 
\usepackage{caption}
\usepackage{psfrag}
\usepackage{upgreek}
\usepackage{isomath}
\usepackage{latexsym}
\usepackage{array}
\usepackage{multirow}
\usepackage{booktabs}
\usepackage{colortbl}
\usepackage{verbatim}
\usepackage{graphicx}
\usepackage{epstopdf,overpic}
\usepackage{float}
\usepackage{soul}
\usepackage{color}
\usepackage{dirtytalk}
\usepackage{textcomp}
\usepackage{algorithm, algpseudocode, lmodern, mathtools}
\usepackage{tikz}
\usetikzlibrary{shapes.geometric, arrows}
\usepackage{amssymb}
\usepackage{geometry}
\usepackage{amsmath}
\usepackage{mathrsfs}  
\usepackage{graphicx}
\usepackage{titlesec}
\usepackage{indentfirst}
\usepackage{bigints}
\usepackage[export]{adjustbox}
\usepackage{pgfplots}
\usetikzlibrary{plotmarks,spy,calc}
\newlength{\figureheight}
\newlength{\figurewidth}

\usepackage{multirow}

\usepackage{soul}

\newcommand{\eref}[1]{Equation~(\ref{#1})}
\newcommand{\erefs}[1]{Equations~(\ref{#1})}
\newcommand{\fref}[1]{Figure~\ref{#1}}
\newcommand{\frefs}[1]{Figures~\ref{#1}}
\newcommand{\tref}[1]{Table~\ref{#1}}
\newcommand{\sref}[1]{Section~\ref{#1}}

\biboptions{sort&compress}

\pgfplotsset{compat=1.15}
\begin{document}

\begin{frontmatter}


\title{Design of metamaterial-based heat manipulators by isogeometric shape optimization}



\author[label1]{Chintan Jansari}
\author[label1,label2]{St\'ephane P.A. Bordas\corref{cor1}}
\ead{stephane.bordas@alum.northwestern.edu}
\author[label3]{Elena Atroshchenko}

\cortext[cor1]{Corresponding author}
\address[label1]{Institute of Computational Engineering, Faculty of Sciences, Technology and Medicine, University of Luxembourg, Luxembourg City, Luxembourg.}
\address[label2]{Clyde Visiting Fellow, Department of Mechanical Engineering, The University of Utah, Salt Lake City, Utah, United States.}
\address[label3]{School of Civil and Environmental Engineering, University of New South Wales, Sydney, Australia.}

\begin{abstract}
\par There has been a growing interest in controlled heat flux manipulation to increase the efficiency of thermal apparatus. Heat manipulators control and manipulate heat flow. A key to the effective performance of these heat manipulators is their thermal design. Such designs can be achieved by a periodic assembly of unit cells (known as metamaterials or meta-structure), whose geometry and material properties can be optimized for a specific objective. In this work, we focus on thermal metamaterial-based heat manipulators such as thermal concentrator (which concentrates the heat flux in a specified region of the domain). The main scope of the current work is to optimize the shape of the heat manipulators using Particle Swarm Optimization (PSO) method. The geometry is defined using NURBS basis functions due to the higher smoothness and continuity and the thermal boundary value problem is solved using Isogeometric Analysis (IGA). Often, nodes as design variables (as in Lagrange finite element method) generate the serrate shapes of boundaries which need to be smoothened later. For the NURBS-based boundary with the control points as design variables, the required smoothness can be predefined through knot vectors and smoothening in the post-processing can be avoided. The optimized shape generated by PSO is compared with the other shape exploited in the literature. The effects of the number of design variables, the thermal conductivity of the materials used, as well as some of the geometry parameters on the optimum shapes are also demonstrated.
\end{abstract}


\end{frontmatter}

\section{Introduction}
\label{Sec:Introduction}

\par The modern world of technological and engineering advancements is concerned about global warming and energy. Renewable energy is central to the advances. Even though much emphasis is given to producing affordable and clean energy, improving heat transfer is central to improving the energetical efficiency of engineering  systems. Consequently, new perspectives to enhance the efficiency and accuracy of thermal appliances by optimizing the heat transfer process are emerging.  One such perspective focuses on the development of heat manipulators, which are devices that control heat flow. However, the concept of controlled heat flow is not thoroughly investigated compared to other forms of energy transport such as electric and photonic currents. The ability to manage heat flow can lead to the development of thermal equivalents of devices such as electric transistors, resistors, rectifiers and diodes. 
\par On the other hand, the invention of artificial metamaterials lights up a new spark in material sciences and, subsequently, in the field of heat transfer. Because of their well-designed artificial structures, thermal metamaterials offer outstanding heat transfer capabilities that surpass those of natural materials. Recognizing these properties, thermal metamaterials were explored for creating several heat manipulators. The idea of a metamaterial-based thermal cloak was formulated~\cite{chen2008} and later experimentally investigated~\cite{Narayana2012}. A thermal cloak is a device that minimizes the temperature disturbance caused by an item and makes it unidentifiable through temperature measurements~\cite{Narayana2012,Guenneau2012,Schittny2013,Han2014,Han2014_2,Sklan2016,Li2019,FUJII2019}. Similarly, other heat manipulators such as thermal concentrator (that concentrates the heat flux in the specified region of the domain)~\cite{Narayana2012,Guenneau2012,Schittny2013,Shen2016,Li2019}, thermal camouflage (that creates multiple images of an item away from its actual position)~\cite{Han2014,Peng2020}, heat flux inverter (that inverts the direction of the heat flux in the specific region of the domain)~\cite{Narayana2012} have also been developed. All these heat manipulators work on macro-scale heat manipulation offered by conductive thermal metamaterials. 
\par As the name suggests, conductive thermal metamaterials guide the flux flow over a path of interest by engineering the requisite thermal conductivity. The required thermal conductivity is constrained by the spatial distribution of member materials. Hence, acknowledging the pivotal role played by the spatial configuration in the efficacy of a metamaterial-based heat manipulator, we present a structural optimization method to optimize the relative shape of the regions in a heat manipulator and their associated thermal conductivity.
\par Structural optimization aims at identifying a possible material distribution to achieve the defined objective. The objective could be reducing stress or concentrating heat flux that can be defined in the quantifiable form. Mainly two types of structural optimization, shape and topology optimization, are prevalent. In shape optimization, a structure cannot change its topology, but it can modify the shapes of topological features. Meanwhile, in topological topology optimization, the aim is to find the optimal material distribution in a design domain and optimize the topology.
\par Once the optimization problem is defined, we need an appropriate algorithm to find the solution. Optimization algorithms are divided into two types: gradient-based and gradient-free. Gradient-based algorithms can provide faster convergence than gradient-free algorithms. But they may get stuck in the local minima, which depends on the initial guess of the design variables. On the other hand, gradient-free algorithms, though comparatively slower, are easy to implement and do not require the optimization problem to be differentiable. We use one such algorithm called the particle swarm optimization (PSO). PSO was proposed by Kennedy and Eberhart~\cite{Kennedy1995}. The PSO algorithm works on a simple rule and that makes the process to search for a new solution pool faster. In comparison with another class of gradient-free algorithm - genetic algorithm (GA), PSO does not use crossover or mutation operations and works on real numbers instead of coding operators.
\par One more aspect of the present work is the application of isogeometric analysis (IGA)~\cite{HUGHES2005} to analyze the thermal boundary value problem and describe the boundary of the domains and subdomains. As we use IGA, both geometry and temperature fields are interpolated using non-uniform rational B-splines (NURBS) basis functions. The NURBS shape functions represent all conic geometries exactly. In addition to that, as CAD geometries are also based on NURBS shape functions, IGA creates a seamless integration of modelling and analysis and is suitable for a fully integrated design-analysis-optimization model. We can use the control points for the design variables instead of nodes as in Lagrange finite element methods (FEM). Often, the nodes as design variables generate the serrate shapes of boundary which needs to be smoothened later to avoid irregularities. However, we can predefine the required smoothness through knot vectors in the case of NURBS-based boundaries. In addition, we can easily provide higher continuity and smoothness across the elements in IGA compared to FEM. Only the geometries with predefined smoothness in the boundaries are explored during optimization. With this predefined higher smoothness through IGA, we can avoid serrate shapes and other irregularities in the boundaries. Several articles have been published that exploit the spline-based bases for optimization problems e.g topology optimization~\cite{Hassani2012,Shojaee2012,Dede2012,Tavakkoli2013,Tavakkoli2014,SEO2010,QIAN2013}, shape optimization~\cite{Espath2011,KOO2013,Park2013,Blanchard2014,KIENDL2014,Lian2016,LIAN2017}. 
\par Limited work has been published on optimization for heat manipulators~\cite{FUJII2019,Fujii2019_2,Fujii2020}. Existing work has focused on topology optimization with a stochastic evolution strategy. It is based on immersed-boundary level set method in combination with finite elements. However, to the best of our knowledge, there has not been any work done on shape optimization for heat manipulators. 
The salient features of this work are:
\begin{itemize}
    \item The shape optimization method, which requires design variables only on the boundary, is investigated as a tool to obtain better geometries for metamaterial-based heat manipulators. IGA is utilized for the thermal boundary value problem, while a gradient-free PSO algorithm is utilized for optimization. 
    \item The use of control points as design variables, as well as the NURBS approximation for geometry and solution fields, enables easy control of smoothness and provides accurate geometrical representation.
    \item In addition, the effect of several factors (such as geometric parameters, material conductivities, boundary conditions, etc.) on the optimized shapes is studied. 
\end{itemize}
\par The paper is organized as follows: \sref{sec:Boundary value problem} exhibits the governing equations, the weak formulation, NURBS approximation and matrix formulation of the thermal boundary problem using IGA. The optimization problem and PSO algorithm are presented in \sref{sec:Optimization}. In this work, we present two examples of heat manipulators, but the proposed method is generic and can be applied to other heat manipulators as well. \sref{sec:Thermal Concentrator} and \sref{sec:Thermal cloaked-concentrator} show detailed examples of a thermal concentrator and a thermal cloaked concentrator respectively. \sref{sec:Conclusions} presents the main conclusions of the proposed work.

\section{Boundary value problem}
\label{sec:Boundary value problem}
\begin{figure}[!htbp]
\centering
\includegraphics[width=4in]{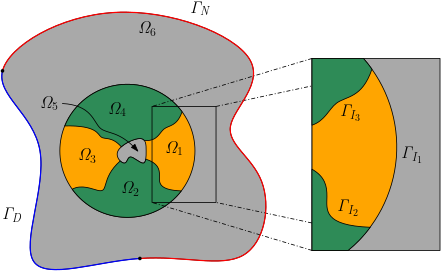}
\caption {Domain description of the boundary value problem} 
\label{fig:BVP domain}
\end{figure}

Consider a metamaterial-based heat manipulator embedded in a plate made of isotropic material as shown in \fref{fig:BVP domain}. The whole domain under consideration can be denoted as $\Omega=\bigcup\limits_{m} \Omega_m  \in\mathbb{R}^{2}$ bounded by $\Gamma$=$\partial\Omega$. The boundary $\Gamma$ is uniquely decomposed into two parts $\Gamma_D$ \& $\Gamma_N$, where Dirichlet and Neumann boundary conditions are applied, respectively. Furthermore, several internal boundaries in the form of material interfaces are present, which are collectively denoted by $\Gamma_I$, $\Gamma_I=\bigcup\limits_{k}\Gamma_{I_k}$. We assume that there is no internal heat generation and conduction is the only mode of heat transfer present. Let $T$ be the temperature field over $\Omega$, the steady state heat conduction equation can be written as,
\begin{equation}
    \nabla \cdot \left( \boldsymbol{\kappa}\nabla T\right) = 0 \quad \text{in} \quad \Omega,
    \label{eq:Laplace equation}
    \end{equation}
with the boundary conditions
\begin{subequations}
\begin{align} 
 T &= T_D \quad \textrm{on} \quad \Gamma_D,\\
  \nabla T\cdot \boldsymbol{n} &= Q_N \quad \textrm{on} \quad \Gamma_N,
\end{align}
\label{eq:Boundary conditions}
\end{subequations}
where $\boldsymbol{\kappa}$ be the thermal conductivity matrix (for isotropic material, $\boldsymbol{\kappa}=\kappa \mathrm{\mathbf{I}_2}$ with $\mathrm{\mathbf{I}_2}$ be an identity matrix of $\mathbb{R}^{2}$), $Q_N$ the externally applied flux on boundary $\Gamma_N$, $T_D$ the prescribed temperature on boundary $\Gamma_D$, $\boldsymbol{n}$ the unit normal on boundary and $\nabla= \small{\left({ \dfrac{\textstyle\partial}{\textstyle\partial x},\dfrac{\textstyle\partial}{\textstyle\partial y}}\right)}$. 
\par Following the standard Galerkin approach, the weak form of the given heat conduction boundary value problem (\erefs{eq:Laplace equation}-(\ref{eq:Boundary conditions})) is written as follows: Find $T^h \in \mathscr{T}^h$ such that $\forall S^h \in \mathscr{S}^h_0$,
\begin{equation}
a(T^h,S^h) = \ell(S^h),
\label{eq:weak_form}
\end{equation}
with $\mathscr{T}^h$ and $\mathscr{S}^h_0$ be the trial and test spaces, respectively,
\begin{subequations}
\begin{align} 
 &\mathscr{T}^h \subseteq \mathscr{T} = \left\lbrace T \in \mathbb{H}^1 ({\Omega}), T=T_D  \hspace{0.15cm} \textrm{on} \ {\Gamma}_D \right\rbrace,\\
  &\mathscr{S}^h_0 \subseteq \mathscr{S}_0 = \left\lbrace S \in \mathbb{H}^1 ({\Omega}), S=0 \hspace{0.15cm} \textrm{on} \ {\Gamma}_D \right\rbrace,
\end{align}
\label{eq:trail-test space}
\end{subequations}
\noindent and
\begin{equation}
a(T^h,S^h) = \int _{\Omega} (\nabla  S^h)^{\rm T} \boldsymbol{\kappa}\nabla T^h d\Omega,
\label{eq:weak_form_a}
\end{equation}
\begin{equation}
\ell{(S^h)} =  \int _{\Gamma_{N}} (S^h)^{\rm T} Q_N  d\Gamma.
\label{eq:weak_form_l}
\end{equation} 
\par In the present study, we assume that the temperature and normal flux is continuous along each material interface. If the patches connected at interface $\Gamma_I$ are denoted by 1 and 2, then the corresponding interface conditions are written as,
\begin{subequations}
\begin{align} 
  \left\llbracket T\right\rrbracket &=0 \quad &\textrm{on} \quad \Gamma_I,\\
  \boldsymbol{n}^1\cdot \boldsymbol{\kappa}^1\nabla T^1 &= -\boldsymbol{n}^2\cdot \boldsymbol{\kappa}^2\nabla T^2\quad &\textrm{on} \quad \Gamma_I,
\end{align}
\label{eq:Interface conditions}
\end{subequations}
where $\boldsymbol{n}^1$, $\boldsymbol{n}^2$; ${\kappa}^1$, ${\kappa}^2$ and  $T^1$, $T^2$ are unit normals, conductivity matrices, and temperatures associated with patch 1 and 2, respectively. The jump operator $\llbracket\cdot\rrbracket$ is defined in the next paragraph.  
\par The continuity conditions shown in \eref{eq:Interface conditions} are applied in the given boundary value problem by modifying the weak form using Nitsche's method. Nitsche's method is a method between the Lagrange multiplier method and the penalty method. In the weak formulation, it replaces the Lagrange multipliers by their physical representation, normal flux. In addition to that, it keeps the coercivity of the bilinear form intact and the variational form consistent. Nitsche's method has been successfully applied for patch coupling in~\cite{Nguyen2014,HU2018}. When Nitsche's method is applied to couple patches, the linear form (\eref{eq:weak_form_l}) on the right side remains the same, while the bilinear form (\eref{eq:weak_form_a}) is altered as follows,
\begin{multline}
a(T^h,S^h) = \int _{\Omega} (\nabla  S^h)^{\rm T} \boldsymbol{\kappa}\nabla T^h d\Omega - \int _{\Gamma_I} \left(\boldsymbol{n}\cdot \{\boldsymbol{\kappa}\nabla S^h\}\right)^{\rm T}\llbracket T^h \rrbracket~d\Gamma \\- \int _{\Gamma_I} \llbracket S^h  \rrbracket^{\rm T} \left(\boldsymbol{n}\cdot\{\boldsymbol{\kappa}\nabla T^h\}\right)  d\Gamma + \int _{\Gamma_I} \beta~\llbracket S^h  \rrbracket^{\rm T} \llbracket S^h  \rrbracket~d\Gamma, 
\label{eq:modified_weak_form_a}
 \end{multline}
where $\boldsymbol{n}$ is the normal at $\Gamma_I$ for any one patch from the patches connected at $\Gamma_I$ ($\boldsymbol{n}=\boldsymbol{n}^1= -~\boldsymbol{n}^2$), $\beta$ is the stabilization parameter. The jump operator $\llbracket\cdot\rrbracket$ described in \eref{eq:Interface conditions} and average operator $\{\cdot\}$ in \eref{eq:modified_weak_form_a} can be defined as,
\begin{equation}
    \llbracket\theta\rrbracket= \theta^1-\theta^2, \quad
    \{\theta\}=\gamma\theta^1 + (1-\gamma)\theta^2,
    \label{eq:average operator}
\end{equation}
where $\theta$ is a property of interest, superscript denotes the patch (with which property $\theta$ relates to) and $\gamma$ is the averaging parameter ($0<\gamma<1$). \begin{figure}[!htbp]
\centering
\includegraphics[width=3.8in]{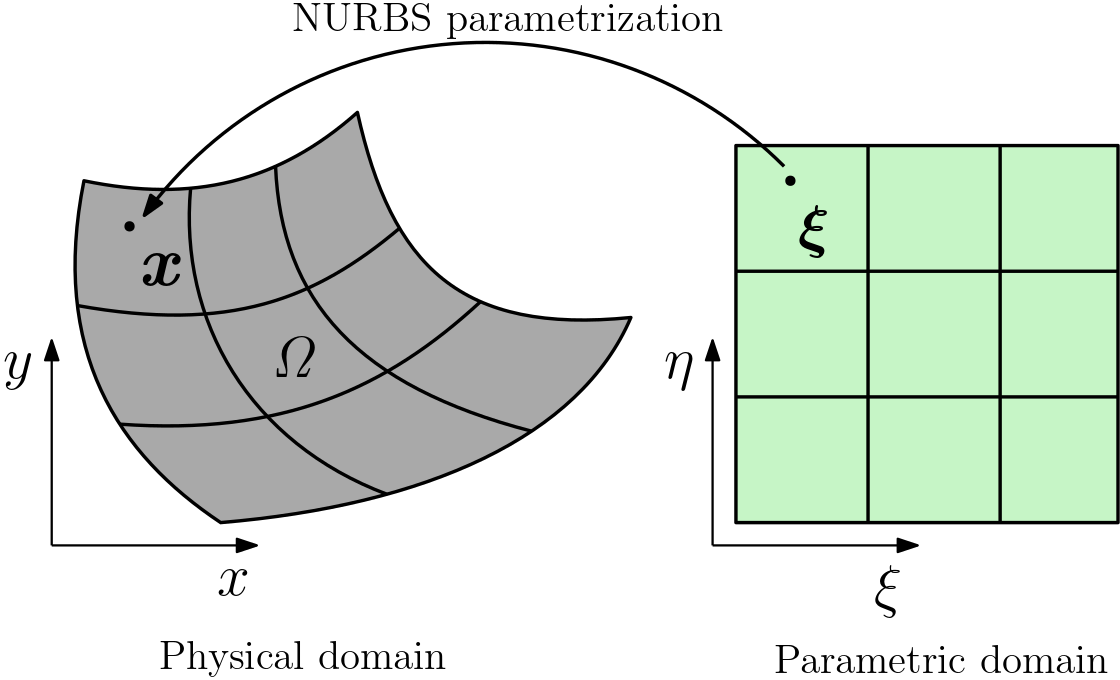}
\caption {Parametrization of a point from parametric domain to a point in physical domain using NURBS basis functions.} 
\label{fig:NURBS parameterization}
\end{figure} 
\par Let $\boldsymbol{x}\in \Omega$ and $\boldsymbol{\xi}$ be the corresponding knot value. We parametrize the domain using $n$ NURBS, $N_{i}$ (as shown in \fref{fig:NURBS parameterization})
\begin{equation}
    \boldsymbol{x} = \sum _{i=1}^{n}\mathbf{P}_{i}N_{i}(\boldsymbol{\xi}),
    \label{eq:NURBs_approx}
\end{equation}
\noindent where $\mathbf{P}_{i}$ is the $i^{th}$ control point.
\par As we are using isogeometric analysis with standard Galerkin approach, the test and trial function both are approximated with the same NURBS shape functions as geometry. The trial and test function approximation can be written as,
\begin{subequations}
\begin{align} 
  &T^h(\boldsymbol{\xi}) = \sum_{i=1}^{n}T_{i}N_{i}(\boldsymbol{\xi}),\\
 &S^h(\boldsymbol{\xi}) = \sum_{i=1}^{n}S_{i}N_{i}(\boldsymbol{\xi}),
\end{align}
\label{eq:Trail test approx}
\end{subequations}
where $T_i$, $S_{i}$ are the temperature and arbitrary temperature perturbation at $i^{th}$ control point.
\begin{figure}[!htbp]
\centering
\includegraphics[width=2.5in]{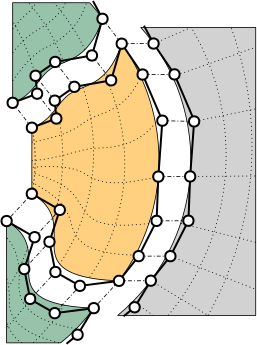}
\caption {The matching pairs of control points from different patches connected at interface by Nitsche method} \label{fig:Interface Nitsche}
\end{figure}
\par By substituting \eref{eq:Trail test approx} in \eref{eq:weak_form}, we obtain the linear system
\begin{equation}
\mathbf{K} \mathbf{T} = \mathbf{Q},
\end{equation}
where $\mathbf{T}$ is the vector of unknown temperature $T_{I}$ at all control points. The global stiffness matrix $\mathbf{K}$ and the global flux vector $\mathbf{F}$ are written as,
\begin{equation}
\mathbf{K}= \mathbf{K}^b + \mathbf{K}^n+ (\mathbf{K}^n)^{\rm T}+\mathbf{K}^s,
\end{equation}
\begin{equation}
\mathbf{F}= \int _{\Gamma_{N}} \mathbf{N}^{\textrm{T}} Q_N~d\Gamma,
\end{equation} 
where $\mathbf{K}^b$ is the bulk stiffness matrix defined as below,
\begin{equation}
\mathbf{K}^b =\sum_m \int _{\Omega_m} (\mathbf{B}^m)^{\textrm{T}}\boldsymbol{\kappa}^m\mathbf{B}^m~d\Omega. 
\label{eq:Kb}
\end{equation}
$\mathbf{K}^n$ and $\mathbf{K}^s$ are the interfacial stiffness matrices, as these matrices are used to couple the adjacent patches with conditions given in \eref{eq:Interface conditions}. A point to note, before defining $\mathbf{K}^n$ and $\mathbf{K}^s$, is that the connecting patches must have matching control points at the interface to apply Nitsche's method as shown in \fref{fig:Interface Nitsche}. Following the notation of \erefs{eq:Interface conditions}-(\ref{eq:average operator}), $\mathbf{K}^n$ and $\mathbf{K}^s$ are given by the following equations.
\renewcommand\arraystretch{2}
\begin{equation}
\mathbf{K}^n =
\begin{bmatrix} 
-\gamma\displaystyle\int_{\Gamma_I} (\mathbf{N}^1)^{\textrm{T}}\boldsymbol{n}\boldsymbol{\kappa}^1\mathbf{B}^1~d\Gamma 
& -(1-\gamma)\displaystyle\int_{\Gamma_I} (\mathbf{N}^1)^{\textrm{T}}\boldsymbol{n}\boldsymbol{\kappa}^2\mathbf{B}^2~d\Gamma    \\[0.1em]
\gamma\displaystyle\int_{\Gamma_I} (\mathbf{N}^2)^{\textrm{T}}\boldsymbol{n}\boldsymbol{\kappa}^1\mathbf{B}^1~d\Gamma  & (1-\gamma)\displaystyle\int_{\Gamma_I} (\mathbf{N}^2)^{\textrm{T}}\boldsymbol{n}\boldsymbol{\kappa}^2\mathbf{B}^2~d\Gamma 
\end{bmatrix}
\label{eq:Kn}
\end{equation}
\begin{equation}
\mathbf{K}^s =
\begin{bmatrix} 
\beta\displaystyle\int_{\Gamma_I} (\mathbf{N}^1)^{\textrm{T}}\mathbf{N}^1~d\Gamma 
& -\beta\displaystyle\int_{\Gamma_I} (\mathbf{N}^1)^{\textrm{T}}\mathbf{N}^2~d\Gamma  \\[0.1em]
-\beta\displaystyle\int_{\Gamma_I} (\mathbf{N}^2)^{\textrm{T}}\mathbf{N}^1~d\Gamma  & \beta\displaystyle\int_{\Gamma_I} (\mathbf{N}^2)^{\textrm{T}}\mathbf{N}^2~d\Gamma 
\end{bmatrix}
\label{eq:Ks}
\end{equation}
Matrices $\boldsymbol{\kappa}^m$, $\mathbf{B}^m$ and $\mathbf{N}^m$ in \eref{eq:Kb}-(\ref{eq:Ks}) are the conductivity matrix, shape function derivative matrix and shape function vector respectively (for a given patch $m$). The matrices $\mathbf{B}^m$ and $\mathbf{N}^m$ are given as below,
\begin{equation}
\mathbf{B}^m =
\begin{bmatrix} 
{N\strut_{1,x}^m} & {N\strut_{2,x}^m}&... & {N\strut_{I,x}^m} &...   \\[0.1em]
{N\strut_{1,y}^m} & {N\strut_{2,y}^m}&... & {N\strut_{I,y}^m} &... 
\end{bmatrix}, \quad
\mathbf{N}^m =
\begin{bmatrix} 
{N\strut_{1}^m} & {N\strut_{2}^m} & ... & {N\strut_{I}^m} & ... 
\end{bmatrix}.
 \end{equation}
 \begin{subequations}
 \renewcommand\arraystretch{1.5}
 \end{subequations}
 The verification is performed for the first example to check the accuracy of Nitsche's method (refer \sref{sec:Nitche method verification}). For the current work, the stabilization parameter $\beta=1\times 10^{12}$. With the large value of the stabilization parameter, Nitsche's method behaves close to the penalty method. Here the objective is to force the temperature continuity in a stricter sense and normal flux continuity in a weaker sense. In the literature~\cite{Nguyen2014,HU2018}, it is also reported that the large stabilization parameter might cause ill-conditioning of the system, but we did not face any conditioning issue for our boundary value problem. For the averaging parameter, a standard value $\gamma=0.5$ is taken, which gives equal weights to fluxes on both sides of the interface.

\section{Optimization}
\label{sec:Optimization}
\subsection{Optimization problem description}
\label{sec:Optimization problem description}
\par  In a standard shape optimization problem, the design variables are the parameters that control the geometry and try to optimize a function called objective function. In our case, suppose $\boldsymbol{x}$ is the vector of the $N_{\mathrm{var}}$ design variables and $f_{\mathrm{obj}}$ is the objective function, then the shape optimization problem for a heat manipulator can be defined in a mathematical form as,
\begin{equation}
\min_{\boldsymbol{x} \in  \mathbb{R}^{N_{\mathrm{var}}} }~f_{\mathrm{obj}}(\boldsymbol{x}),
\end{equation}
with
\begin{subequations}
\begin{alignat}{1}
&f_{\mathrm{obj}} : \mathbb{R}^{N_{\mathrm{var}}}  \rightarrow \mathbb{R},\\
&f_{\mathrm{obj}} : \boldsymbol{x} \mapsto f_{\mathrm{obj}}(\boldsymbol{x}),
\end{alignat}
\label{eq:obj fun}
\end{subequations}
\noindent\textrm{such that the following constraints are satisfied,}
\begin{align}
&\text{Equality constraints:}\quad &h_{i}(\boldsymbol{x})=~0 \quad
& i=1,2,...,n_h , \label{eq:equality constraints}\\
&\text{Inequality constraints:}\quad &g_{j}(\boldsymbol{x})\leq~0 \quad
& j=1,2,...,n_g , \label{eq:Inequality constraints}\\
&\text{Box constraints:}\quad &x_{k,\rm{min}}~\leq~x_{k}\leq~x_{k,\rm{max}} \quad &
k=1,2,...,N_{\mathrm{var}} ,  \label{eq:Box constraints}
\end{align} 
where $n_h$ and $n_g$ are the number of equality constraints and inequality constraints respectively. $x_{k,\rm{min}}$ and $x_{k,\rm{max}}$ are lower and upper bounds of the design variable $x_{k}$.

\subsection{Particle swarm optimization}
\label{sec:Particle swarm optimization}
Kennedy and Eberhart, in their article [59], proposed Particle Swarm Optimization (PSO), an optimization concept based on the swarming behaviour of birds flock and fish school. The PSO is a non-gradient based optimization method and can search very large spaces of candidate solutions for the optimization of continuous nonlinear functions. 
\par In this method, each candidate solution is considered as a particle. The algorithm starts with a swarm of initial particles, then advances towards the optimum solution by updating the positions and velocities of these particles according to their fitness to objective function. In our minimization-optimization problem, the position of a particle is a vector of design variable values, and the velocity is the correction applied to the position at the each iteration. Furthermore, the fitness measures how small the objective function is for a given particle. The position and velocity of a particle are updated based on its individual best position as well as the position of the best particle (the particle with the smallest objective function). 
\par Suppose, $\omega$ is a non-negative number which defines the contribution of old velocity on the updated velocity; $\mathrm{pbest}_k^i$ is the best position of the $k^{th}$ particle at the $i^{th}$-iteration, which has the smallest value of objective function among all positions utilized before by the $k^{th}$ particle; $\mathrm{gbest}^i$ is the position of the best particle that has the smallest objective function  up to the current iteration from the whole swarm; $\alpha_1$ and $\alpha_2$ are the acceleration coefficients and can be indicated as self adjustment weight and social adjustment weight respectively. The self adjustment weight defines the maximum correction in the direction of the individual best position, while social adjustment weight defines the maximum correction in the direction of the position of best particle; $\mathrm{rand}_1$ and $\mathrm{rand}_2$ are two random numbers between 0 and 1. Then, by using the velocity and position at the $i^{th}$-iteration, the equations to find the updated velocity and position of the $k^{th}$ particle at the $(i+1)^{th}$-iteration can be written as,
\begin{subequations}
\begin{equation}
 v_k^{i+1}= \omega v_k^{i} + \alpha_1~\mathrm{rand}_1 ~(\mathrm{pbest}_k^i -x_k^{i})\\+ \alpha_2~ \mathrm{rand}_2~(\mathrm{gbest}^i -x_k^{i}),
\end{equation}
\begin{equation}
 x_k^{i+1} = x_k^{i} + v_k^{i+1}.
\end{equation}
\label{eq:PSO_eqs}
\end{subequations}
\par
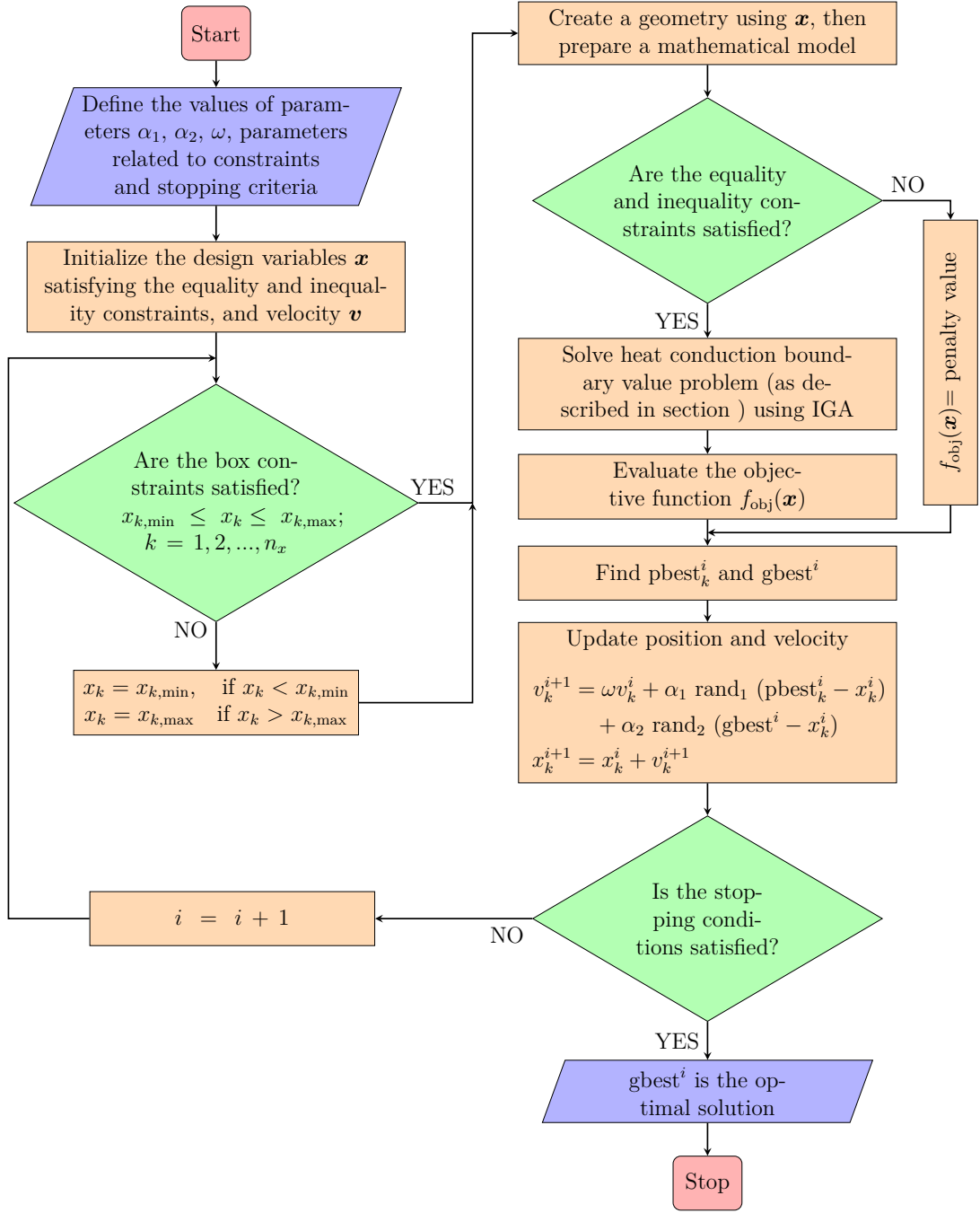
\begin{figure}[htbp!]
    \centering
   \scalebox{0.8}{\vspace{-0.5cm}

\tikzstyle{startstop} = [rectangle, rounded corners, minimum width=1cm,text width=1cm, minimum height=1cm,text centered, draw=black, fill=red!30]
\tikzstyle{io} = [trapezium, trapezium left angle=70, trapezium right angle=110, minimum width=4.5cm,text width=5cm, minimum height=1cm, text centered, draw=black, fill=blue!30,node distance=1.9cm,]
\tikzstyle{process} = [rectangle, minimum width=7cm, text width=6.5cm, minimum height=1cm, text centered, draw=black, fill=orange!30,node distance=1.9cm]
\tikzstyle{process2} = [rectangle, minimum width=0.1cm, text width=5cm, minimum height=1cm, text centered, draw=black, fill=orange!30]
\tikzstyle{decision} = [diamond, minimum width=0.3cm, minimum height=0.1cm, text centered,text width=3.5cm, draw=black, fill=green!30, aspect=1.7,node distance=4cm]
\tikzstyle{arrow} = [thick,->,>=stealth]
\begin{tikzpicture}[node distance=2.5cm]
\node (start) [startstop] {Start};
\node (in1) [io, below of=start,yshift=-0.2cm] {Define the values of parameters $\alpha_1$, $\alpha_2$, $\omega$, parameters related to constraints and stopping criteria};
\node (pro1) [process, below of=in1,yshift=-0.7cm] {Initialize the design variables $\boldsymbol{x}$ satisfying the equality and inequality constraints, and velocity $\boldsymbol{v}$};
\node (dec0) [decision, below of=pro1] {Are the box constraints satisfied? $x_{k,\rm{min}}~\leq~x_{k}\leq~x_{k,\rm{max}}$; $k=1,2,...,n_x$};
\node (pro3) [process, left of=start,xshift=11cm] {Create a geometry using $\boldsymbol{x}$, then prepare a mathematical model};
\node (dec02) [decision, below of=pro3,yshift=0.9cm] {Are the equality and inequality constraints satisfied?};
\node (pro32) [process, below of=dec02,yshift=-1.5cm] {Solve heat conduction boundary value problem (as described in section ) using IGA};
\node (pro4) [process, below of=pro32] {Evaluate the objective function $f_{\mathrm{obj}}(\boldsymbol{x})$};
\node (pro5) [process, below of=pro4,yshift=0.3cm] {Find $\mathrm{pbest}_k^i$ and $\mathrm{gbest}^i$};
\node (pro6) [process, below of=pro5,yshift=-0.5cm] {Update position and velocity
\begin{align*}
     v_k^{i+1} &= \omega v_k^{i} + \alpha_1~\mathrm{rand}_1 ~(\mathrm{pbest}_k^i -x_k^{i})\\&\quad +\alpha_2~ \mathrm{rand}_2~(\mathrm{gbest}^i -x_k^{i})\\
    x_k^{i+1} &= x_k^{i} + v_k^{i+1} 
    \end{align*}};
\node (dec1) [decision, below of=pro6] {Is the stopping conditions satisfied?};
\node (out1) [io, below of=dec1,yshift=-1.3cm] {$\mathrm{gbest}^i$ is the optimal solution};
\node (stop) [startstop, below of=out1,yshift=0.8cm] {Stop};
\node (sidePro1) [process2,left of=dec1,xshift=-6.3cm] {$i=i+1$};
\node (sidePro2) [process2,below of=dec0,yshift=-1.2cm] {$x_{k}=x_{k,\rm{min}},\quad \text{if}~x_{k}< x_{k,\rm{min}}$\\ $x_{k}=x_{k,\rm{max}}\quad \text{if}~ x_{k}>x_{k,\rm{max}}$};
\node (sidePro3) [process2,right of=dec02, rotate=90,xshift=-3cm,yshift=-2cm] {$f_{\mathrm{obj}}(\boldsymbol{x})$= penalty value};

\draw [arrow] (start) -- (in1);
\draw [arrow] (in1) -- (pro1);
\draw [arrow] (dec0.east) node[anchor=south west,xshift=-0.3cm] {YES}-| ++(1,0) |- coordinate[midway](m01)(pro3);
\coordinate (Below dec0) at ($(dec0.south) + (0,-1.0cm)$);
\coordinate (East dec0) at ($(dec0.east) + (1.0cm,0)$);
\draw [arrow] (dec0.south) -- node[anchor=south east] {NO} (sidePro2);
\draw [arrow] (sidePro2.east) -|(East dec0);
\draw [arrow] (pro3) -- (dec02);
\draw [arrow] (dec02) -- node[anchor=east] {YES} (pro32);
\draw [arrow] (pro32) -- (pro4);
\draw [arrow] (pro4) -- coordinate[midway](m02)(pro5);
\draw [arrow] (pro5) -- (pro6);
\draw [arrow] (pro6) -- (dec1);
\draw [arrow] (dec1) -- node[anchor=east] {YES}  (out1);
\draw [arrow] (pro1) -- coordinate[midway](m1)(dec0);
\draw [arrow] (dec1.west) node[anchor=north east] {NO} --  (sidePro1.east);

\coordinate (West sidePro1) at ($(sidePro1.west) + (-1.5cm,0)$);
\draw [arrow] (sidePro1.west) -| (West sidePro1)  |- (m1.west);
\draw [arrow] (dec02.east) node[anchor=south west] {NO} -| (sidePro3.east);
\draw [arrow] (sidePro3.west) |- (m02.east);
\draw [arrow] (out1) -- (stop);

\end{tikzpicture}
\vspace{-0.5cm}} \caption{Flow chart of particle swarm optimization algorithm}
\label{chart:PSO_flowchart}
\end{figure}

 The detailed methodology of the PSO is presented in the form of a flow chart in \fref{chart:PSO_flowchart}. At first, we need to define the values of the parameters utilized in the \eref{eq:PSO_eqs} such as $\alpha_1$, $\alpha_2$, and $\omega$. The next step is to initialize the position $\boldsymbol{x}$, such that it satisfies the equality and inequality constraints, and velocity $\boldsymbol{v}$. Then, the geometry and mathematical model are constructed by using $\boldsymbol{x}$. 
Moreover, all constraints are being checked. For box constraints, if any $x_k$ is out of a bound defined by \eref{eq:Box constraints}, it is set equal to the bound. However, if equality constraints or inequality constraints are not satisfied, then $f_{\mathrm{obj}}$ is set to a penalty number to avoid that specific particle as a solution.  
\par As we proceed further, using the geometry and mathematical model, the boundary value problem (In our case, heat conduction problem as described in section 3.1) is being solved and corresponding objective function $f_{\mathrm{obj}}$ is evaluated. Afterwards, $\mathrm{pbest}$ for each particle and $\mathrm{gbest}$ for the whole swarm are calculated based on $f_{\mathrm{obj}}$ values of all particles. At the end of each iteration, the stopping criteria are checked. If the criteria are not satisfied then the velocity and position of each particle are updated by \eref{eq:PSO_eqs} and the procedure will be repeated with the new values. This optimization loop continues until the stopping criteria are satisfied. Once, the stopping criteria are satisfied, the loop ends and $\mathrm{gbest}$ is taken as the optimum solution. 
\par In the present work, we utilize the inbuilt MATLAB function ``particleswarm" to  implement PSO algorithm. We consider tolerance of $1\times 10^{-6}$ in the change of objective function $f_{\mathrm{obj}}$ value and 15 consecutive stalled iterations (improvement in $f_{\mathrm{obj}}$ value is less than $1\times 10^{-6}$) as stopping criteria.  We take MATLAB default values of $\alpha_1$, $\alpha_1$=1.49 and $\alpha_2$, $\alpha_2$=1.49 for our case as well. Furthermore, the value of $\omega$ is adaptive between [0.1,1.1] as per the inbuilt scheme. The particular value of $\omega$ is decreased or increased based on stalled iteration count.
\section{Thermal Concentrator}
\label{sec:Thermal Concentrator}
\subsection{Problem definition}
\label{sec:Problem definition}
 \begin{figure}[htbp!]
    \centering
    \setlength\figureheight{1\textwidth}
    \setlength\figurewidth{1\textwidth}
    \begin{subfigure}[b]{0.30\textwidth}{\centering\includegraphics[width=1\textwidth]{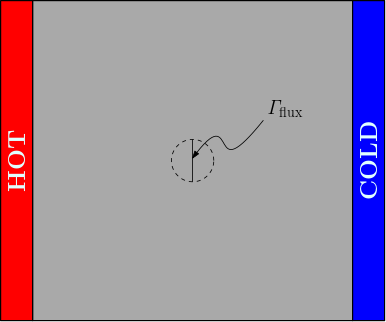}}
        \caption{A base material plate under application of constant flux.}
        \label{fig:Schematic flat plate}
    \end{subfigure}\quad
     \begin{subfigure}[b]{0.33\textwidth}{\centering\includegraphics[width=1\textwidth]{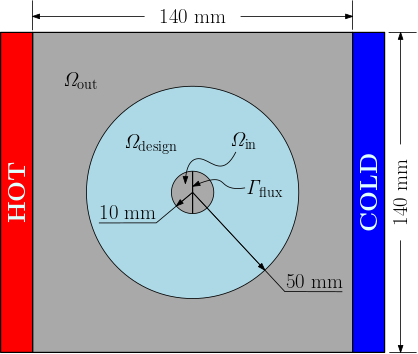}}
        \caption{A thermal concentrator embedded in the plate.}
        \label{fig:Schematic concentrator}
    \end{subfigure}\quad
    \begin{subfigure}[b]{0.30\textwidth}{\centering\includegraphics[width=1\textwidth]{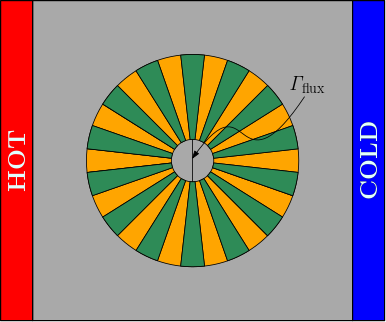}}
             \caption{A metamaterial based concentrator in the base material plate.}
             \label{fig:Schematic meta concentrator}
    \end{subfigure}
 \caption{Schematic design of (a) The base material domain $\Omega$ (nickel steel plate) under constant heat flux applied by high temperature source on the left side and low temperature sink on the right side; $\Omega_{\mathrm{in}}$ is the region of focus where we want to concentrate the flux, $\Gamma_{\mathrm{flux}}$ is the mid-section along which we measure the flux (b) A thermal concentrator embedded in the base material plate to increase the flux concentration in  $\Omega_{\mathrm{in}}$, $\Omega_{\mathrm{design}}$ is the area of the concentrator where we optimize the shape, $\Omega_{\mathrm{out}}$ is the outside area of remaining base material, $\Omega=\Omega_{\mathrm{in}} \cup \Omega_{\mathrm{design}}\cup \Omega_{\mathrm{out}}$ (c) An embedded matermaterial-based thermal concentrator created by putting alternative sectors of copper and PDMS.}
 \label{fig:Concentrator problem schematics}
\end{figure}

\begin{figure}[htbp!]
    \centering
   \scalebox{0.5}{\input{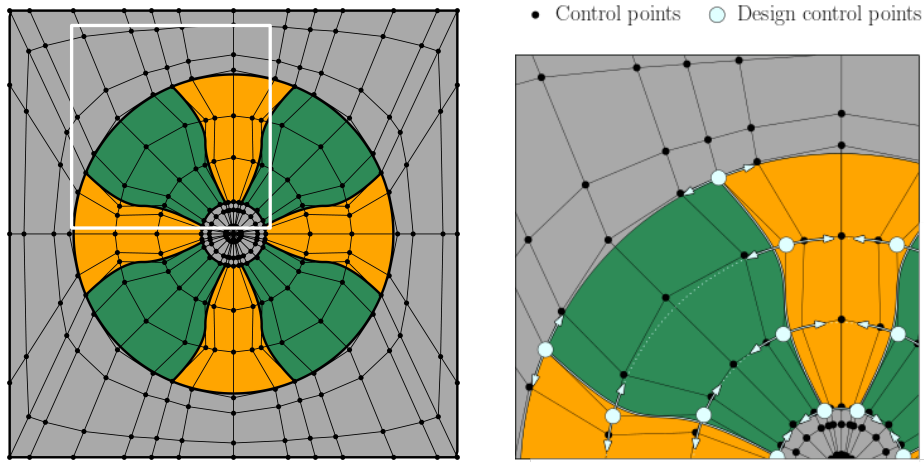}} \caption{NURBS parameterization of the domain $\Omega$. The detailed view of connectivity between adjacent patches as well as the design control points which are the means to manipulate the interfaces between copper and PDMS sectors, and eventually the shape of the concentrator.}
\label{fig:Concentrator problem NURBS approximation}
\end{figure}
\par In this example, we optimize a thermal concentrator, whose main objective is to concentrate the heat flux in the specified region of the domain. The schematics of the problem is given in \fref{fig:Concentrator problem schematics}. The geometry and problem details are referred from~\cite{Chen2015}. We consider a square base material-nickel steel plate ($\kappa_{\rm ni}=10 $ W/mK) with side length 140 mm. To increase the flux concentration in the circular core region of 10 mm radius at center, an annular sector-shaped concentrator of outer radius 50 mm is introduced around the core. The geometry is constructed by placing, side by side, annulus sectors made of two materials, one with higher conductivity and another with lower conductivity. The number of sectors employed in the geometry is denoted by $N_{\mathrm{sec}}$. To ascertain that each sector is adjacent to the sector of other material from both side, $N_{\mathrm{sec}}$ is always taken as an even number, $N_{\mathrm{sec}}=2\ell$ with $\ell=2,3,4,...$. Now, the sectors are arranged such that the center axis of one of sectors is overlaps with the $x$-axis. Moreover, the center axes of all sectors are uniformly distributed angle-wise on the annular region. Altogether, the concentrator appeared as a uniform series connection of two alternating materials in the azimuthal direction. Hence, the effective medium behaves as an anisotropic medium and alters the heat flux direction to follow a particular path of interest.  In our case, copper and polydimethylsiloxane (PDMS), with thermal conductivity $\kappa_{\rm copper}=398$ W/mK and $\kappa_{\rm PDMS}=0.27$ W/mK, are exploited to design the concentrator. Initially, the core material is taken as same as background base material nickel steel. The left and right side of the plate are fixed at $300$ K and $200$ K temperature respectively. 

\subsection{Objective function}
\label{sec:Cntr Objective function}
\par The objective of the concentrator is to concentrate the heat flux in inner core $\Omega_{\mathrm{in}}$. The concentrated flux is measured along the mid-section $\Gamma_{\mathrm{flux}}$ of $\Omega_{\mathrm{in}}$. Mathematically, the concentrator function can be written as,
\begin{equation}
    \Psi_{\mathrm{flux}}=\dfrac{1}{\overline{\Psi}_{\mathrm{flux}}} \int_{\Gamma_{\mathrm{flux}}} - \kappa_{\mathrm{in}} \nabla T \cdot \boldsymbol{n}~d\Gamma,
    \label{eq: value of psi flux}
    \end{equation}
where $\kappa_{\mathrm{in}}$ is the conductivity of core, and $\overline{\Psi}_{\mathrm{flux}}$ is the normalisation value given as,
\begin{equation}
    \overline{\Psi}_{\mathrm{flux}}= \int_{\Gamma_{\mathrm{flux}}} - \overline{\kappa}_{\mathrm{in}} \nabla \overline{T} \cdot \mathbf{n}~d\Gamma, 
    \label{eq:Normalised value of psi flux}
    \end{equation}
Here, all the properties with overline represent the reference case when whole domain is filled with base material. 
\par Now, since we are solving a minimization problem for optimization, we define the objective function as
\begin{equation}
    f_{\mathrm{obj}}= \left\vert\dfrac{1}{\Psi_{\mathrm{flux}}}\right\vert.  
    \label{eq:Cntr Obj Fn}
\end{equation}
\subsection{NURBS parameterization, design variables and constraints}
\label{sec:NURBS parameterization, design variables and constraints}
\par For the NURBS approximation of the geometry, each sector, inner core and outside base material are considered as a separate patch as shown in \fref{fig:Concentrator problem NURBS approximation}. As mentioned in \sref{sec:Boundary value problem}, Nitsche's method is implemented to provide continuity of temperature and normal flux across patch interfaces. All  external and internal boundaries are kept unchanged during optimization, except the interfaces between each pair of sectors (shown as white curves in the detailed view in \fref{fig:Concentrator problem NURBS approximation}). In fact, only some of the control points of these interfaces are used for shape manipulation, which are called design control points (shown as white dots in the detailed view in \fref{fig:Concentrator problem NURBS approximation}). The knot vectors and the weight of control points are kept fixed as well.
\par In the present work, the radial positions of design control points are fixed and distributed uniformly between the inner and outer radius. And the circumferential positions on the predefined circular paths by radial positions are interpreted as design variables. The number of design variables is denoted by $N_{\mathrm{var}}$.
To reduce $N_{\mathrm{var}}$ and subsequent computational burden on the optimization problem, symmetry along center axis is imposed for each sector. Therefore during optimization, the interface between only one pair of sectors is designed. All other interfaces will follow through the imposed symmetry. Now, due to overlap of the center axis of one sector on $x$-axis as mentioned earlier, the symmetry along $x$-axis is also preserved. The position of design control points (calculated using design variables), the predefined knot vector of the interface and the imposed symmetry will decide the overall shape of sector patches.
\par We consider the sector, whose center lies along the negative $x$-axis, as the first sector and assigned it the material with higher conductivity, copper. Then, each sector is assigned a material alternatively as shown in \fref{fig:Concentrator problem NURBS approximation}. Now, to simplify the explanation of the results in the next section, the possible configurations are divided into four types A,B,C and D according to $N_{\mathrm{sec}}$ as shown in \tref{table:Concentrator shape configuration}. The axes of symmetry present in each configuration are also mentioned in the table.
\newcolumntype{L}[1]{>{\raggedright\let\newline\\\arraybackslash\hspace{0pt}}m{#1}}
\newcolumntype{C}[1]{>{\centering\let\newline\\\arraybackslash\hspace{0pt}}m{#1}}
\newcolumntype{R}[1]{>{\raggedleft\let\newline\\\arraybackslash\hspace{0pt}}m{#1}}
\begin{table}[htbp!]
\renewcommand{\arraystretch}{1.7}   
\caption{Different possible configurations of the concentrator due to imposed symmetry and $N_{\mathrm{sec}}$.}  
\centering
\scalebox{0.9}{
\begin{tabular}{|C{6em}|C{7em} C{7em} C{7em} C{7em}|}
\hline		
  Type & A & B & C & D \\
\hline 
$N_{\mathrm{sec}}$ & $8\ell-4$ & $8\ell-2$ & $8\ell$ & $8\ell+2$
\\
Axes of symmetry & $x$-axis, $y$-axis and sector center axis  & $x$-axis and sector center axis & $x$-axis, $y$-axis and sector center axis & $x$-axis and sector center axis\\ 
Sector center-axes along $x$-axis & copper sectors  & one copper and one PDMS sector & copper sectors & one copper and one PDMS sector\\
Sector center-axes along $x$-axis & PDMS sectors  & - & copper sectors & -\\
Schematic & \includegraphics[width=0.18\textwidth]{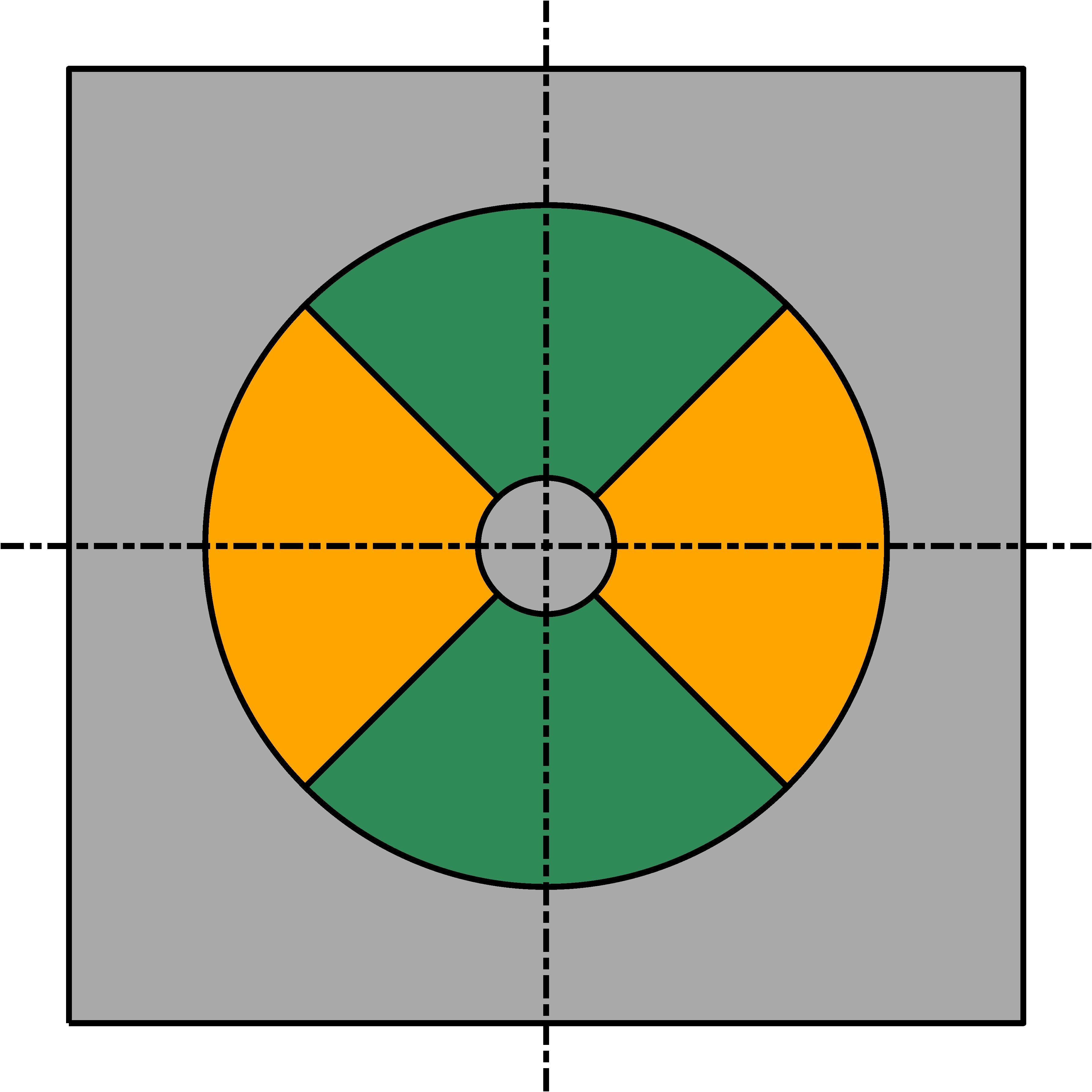} & \includegraphics[width=0.18\textwidth]{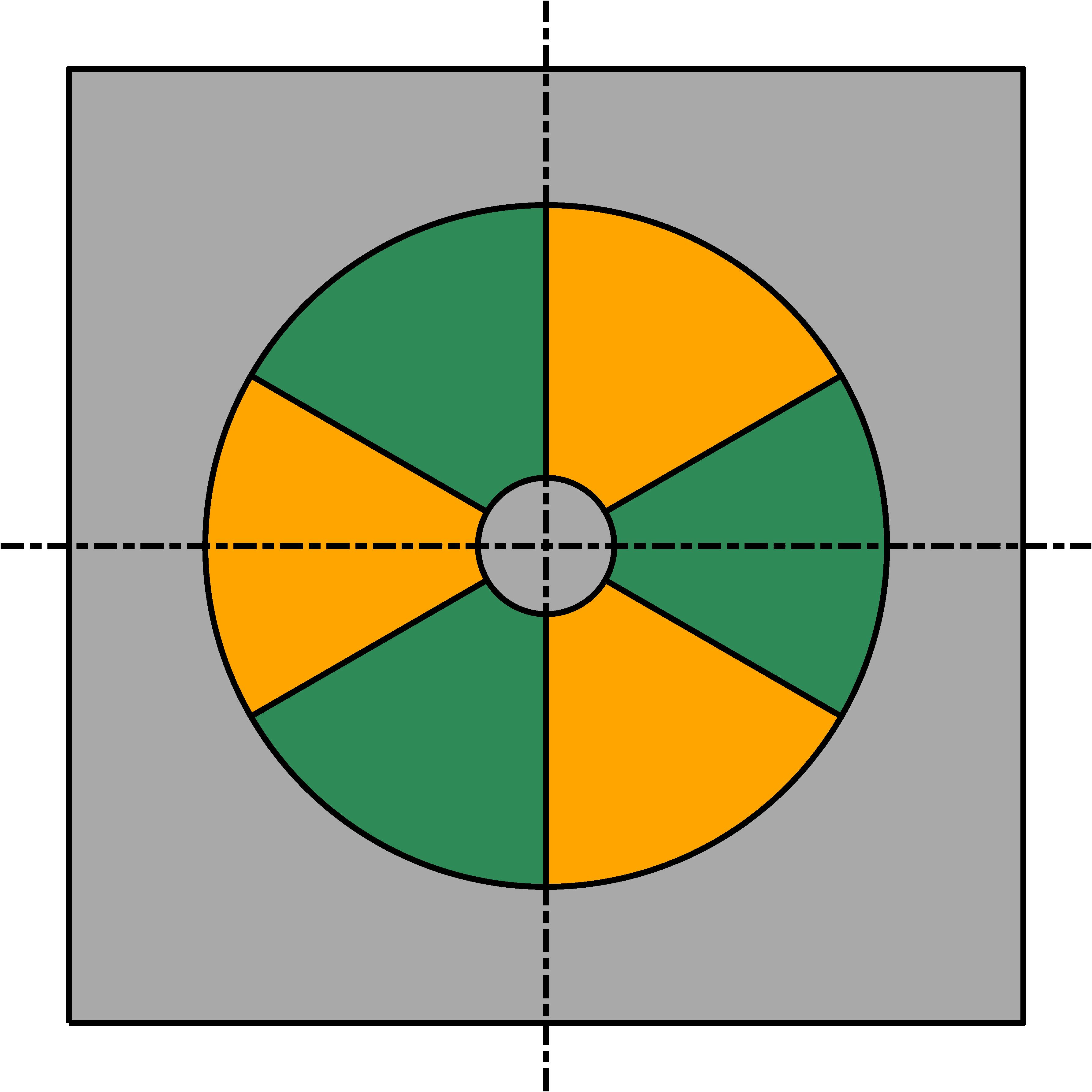} &
\includegraphics[width=0.18\textwidth]{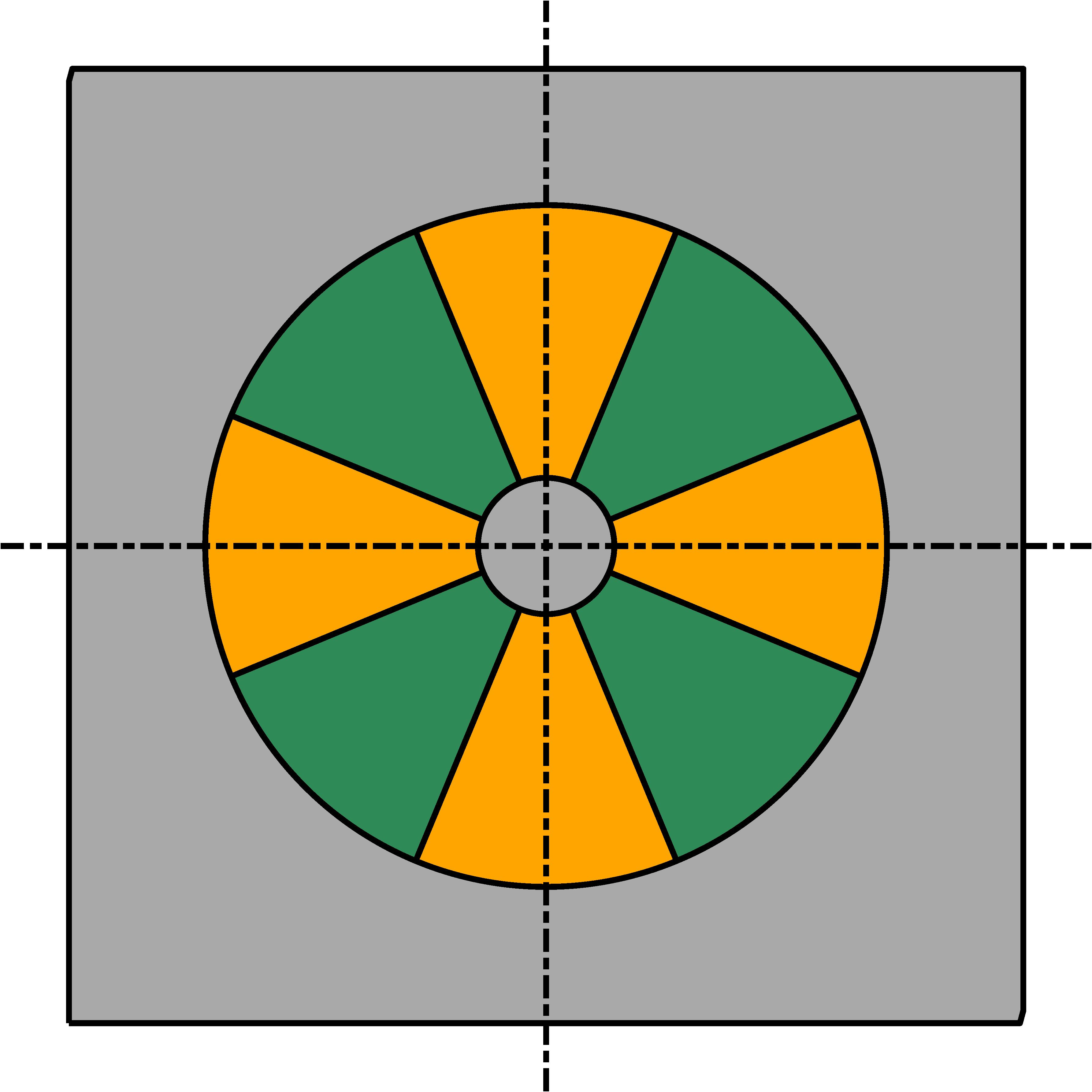} &
\includegraphics[width=0.18\textwidth]{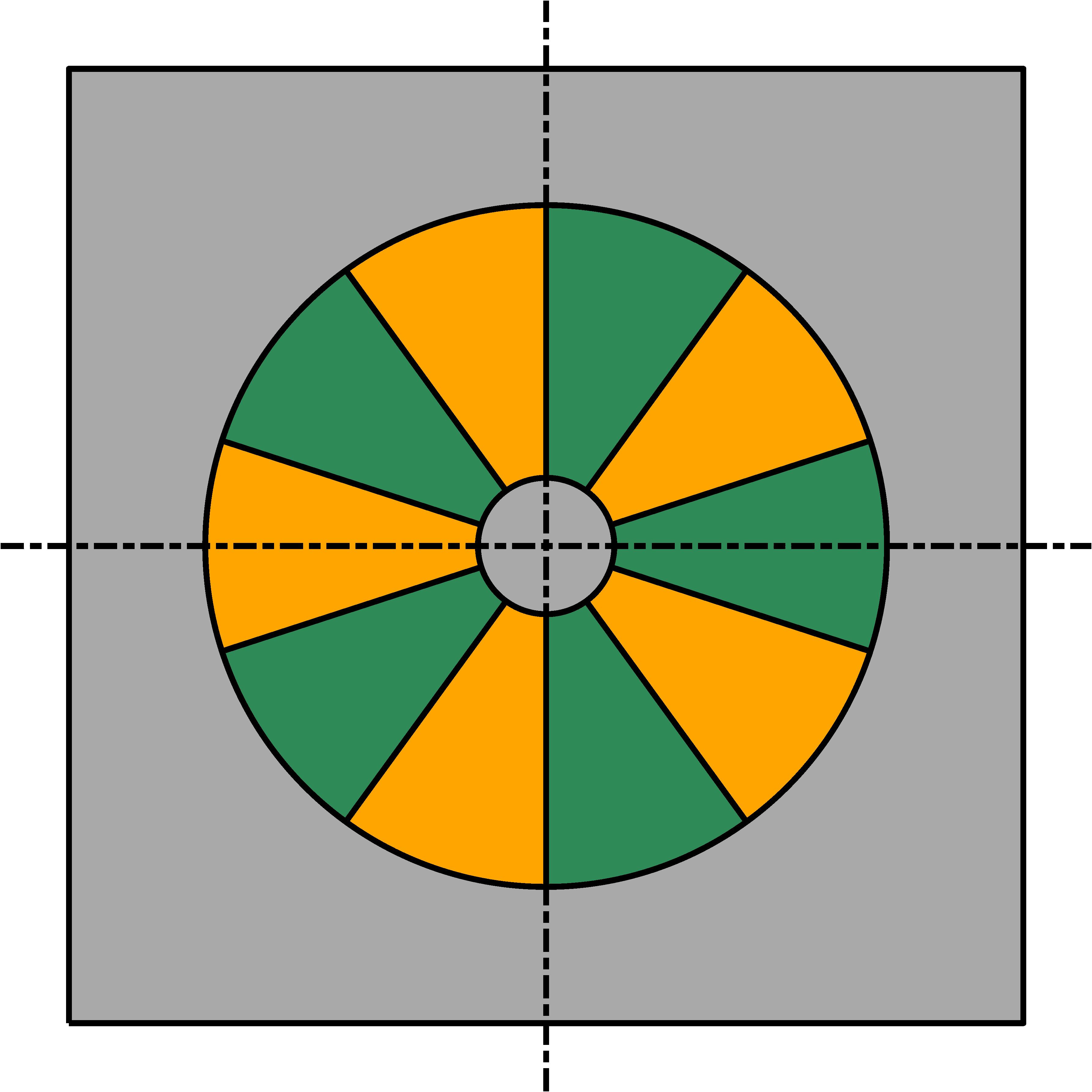}\\

\hline
\end{tabular}}
\label{table:Concentrator shape configuration} 
\end{table}
\par For the optimization, box constraints and inequality constraints are applied. The box constraints restrict the position of design control points between two center-axes, and avoid the singularities occurring due to the intersection of two interfaces. An inequality constraint ensuring the minimum area of each NURBS patch is imposed too. This constraint helps to evade very thin patches and the numerical instability created by them. In addition, another inequality constraint in terms of non-negativity of the Jacobain of NURBS parameterization is imposed, which guarantees only physically feasible geometry (avoid self-overlap or self-intersection).
 
\subsection{Results and discussion}
\label{sec:Results and discussion}
\subsubsection{Verification of Nitsche's method}
\label{sec:Nitche method verification}
\begin{figure}[htbp!]
\centering
\begin{subfigure}[b]{0.44\textwidth}{\centering\includegraphics[width=1\textwidth]{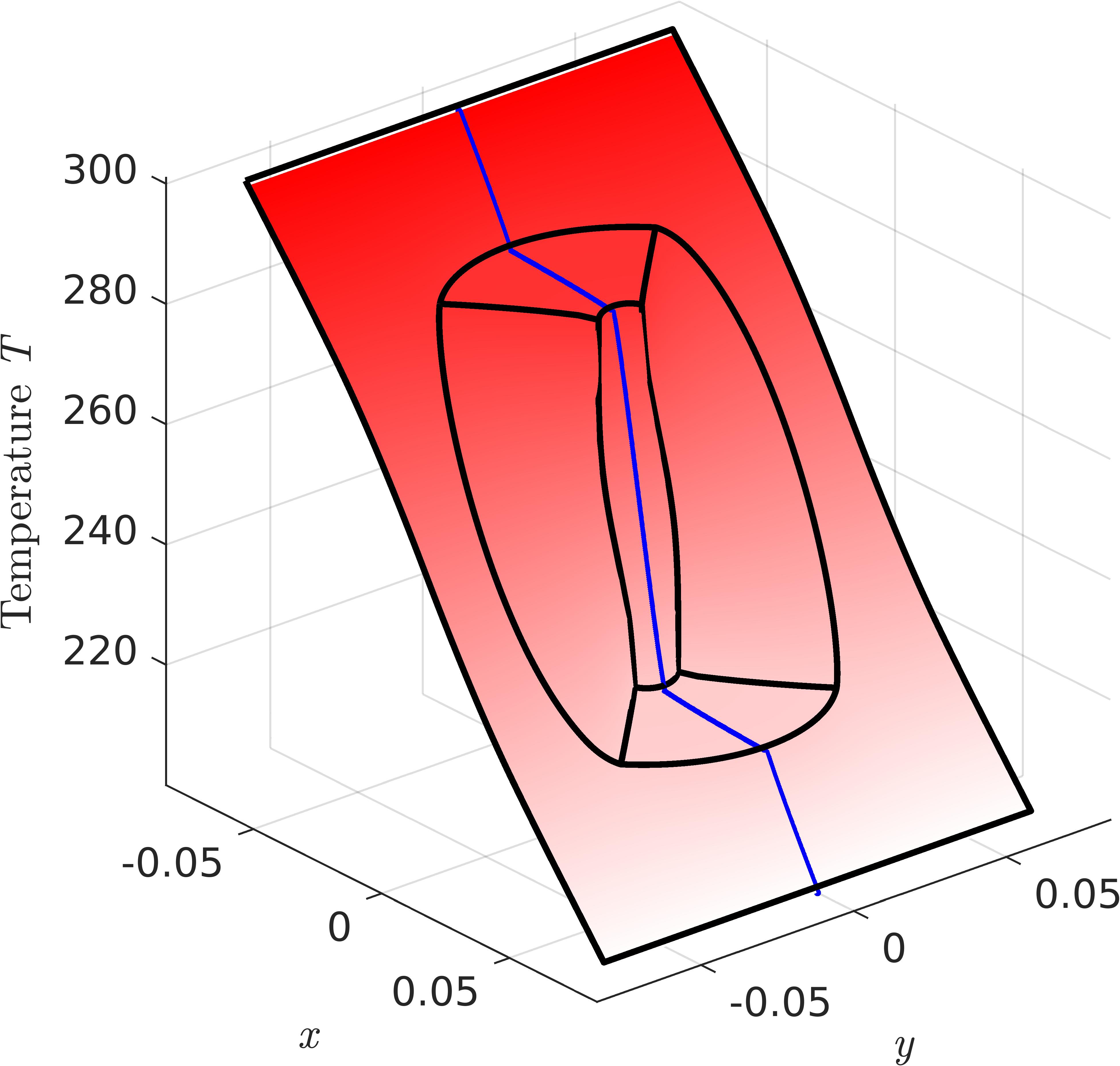}}
        \caption{\centering Temperature $T$ distribution with $\llbracket T \rrbracket=5.318 \times 10^{-10}$ at $\Gamma_{I}$}
        \label{fig:Cntr Nitsche verification a}
    \end{subfigure}\quad
    \begin{subfigure}[b]{0.36\textwidth}{\centering\includegraphics[width=1\textwidth]{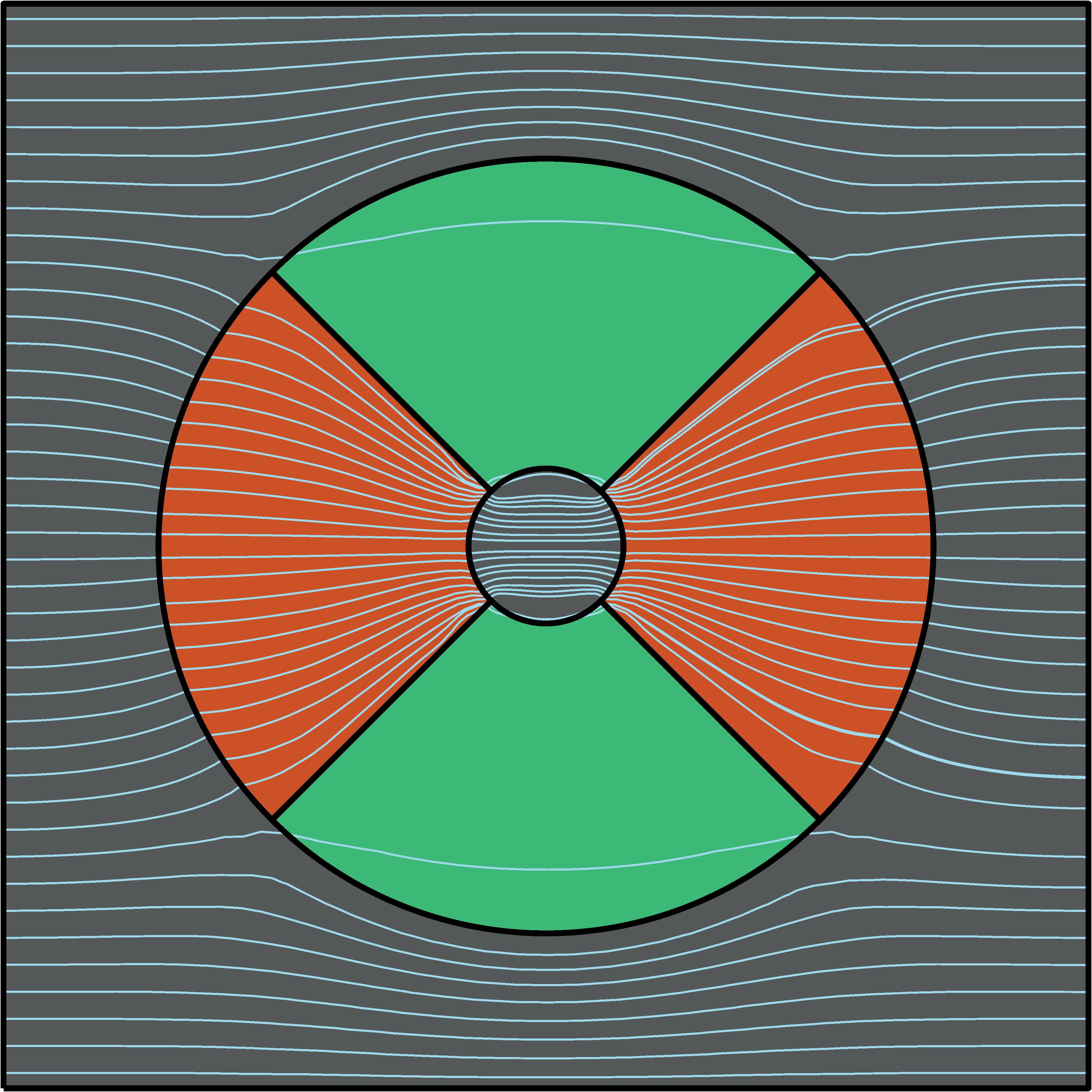}}
        \caption{\centering \centering Flux flow with $\dfrac{\boldsymbol{n} \cdot \llbracket \boldsymbol{\kappa} \nabla T \rrbracket}{|\boldsymbol{n} \cdot  \boldsymbol{\kappa}^{1} \nabla T^{1}|}=4.306 \times 10^{-3}$ at $\Gamma_{I}$}
        \label{fig:Cntr Nitsche verification b}
    \end{subfigure}
\caption{Verification of applied Nitsche's method through checking the continuity conditions  after boundary value problem solution of initial geometry with $N_{\mathrm{sec}}=4$.(a) Temperature distribution with total jump in temperature value $ \llbracket T \rrbracket = 5.318 \times 10^{-10}$ at $\Gamma_{I}$ (b) Flux flow with total jump in the relative normal flux value $ \dfrac{\boldsymbol{n} \cdot \llbracket \boldsymbol{\kappa} \nabla T \rrbracket}{| \boldsymbol{n} \cdot  \boldsymbol{\kappa}^{1} \nabla T^{1} |}=4.306 \times 10^{-3}$ at $\Gamma_{I}$.}
    \label{fig:Cntr Nitsche verification}
\end{figure}
\par In this section, we verified the Nitsch'e method utilized to apply the interface continuity condition. We consider the initial configuration (used for optimization) of the concentrator with straight radial edges of sectors with $N_{\mathrm{sec}}=4$. The boundary value problem is solved and the results are presented in~\fref{fig:Cntr Nitsche verification}. The temperature along the mid-section (along the $x$-axis) is shown in blue. From ~\fref{fig:Cntr Nitsche verification a}, it is evident that the temperature is continuous across all patch interfaces.  Now, from~\fref{fig:Cntr Nitsche verification b}, we can see that the flux streamlines flow from one patch to another without any jump. The numerical values of total jump of temperature and relative normal flux (considering all the interfaces) are measured $5.318 \times 10^{-10}$ and $4.306 \times 10^{-3}$. From these results, we can verify that the Nitsche's method impose the required continuity condition.
\subsubsection{A convergence study}
\label{sec:Mesh convergence}
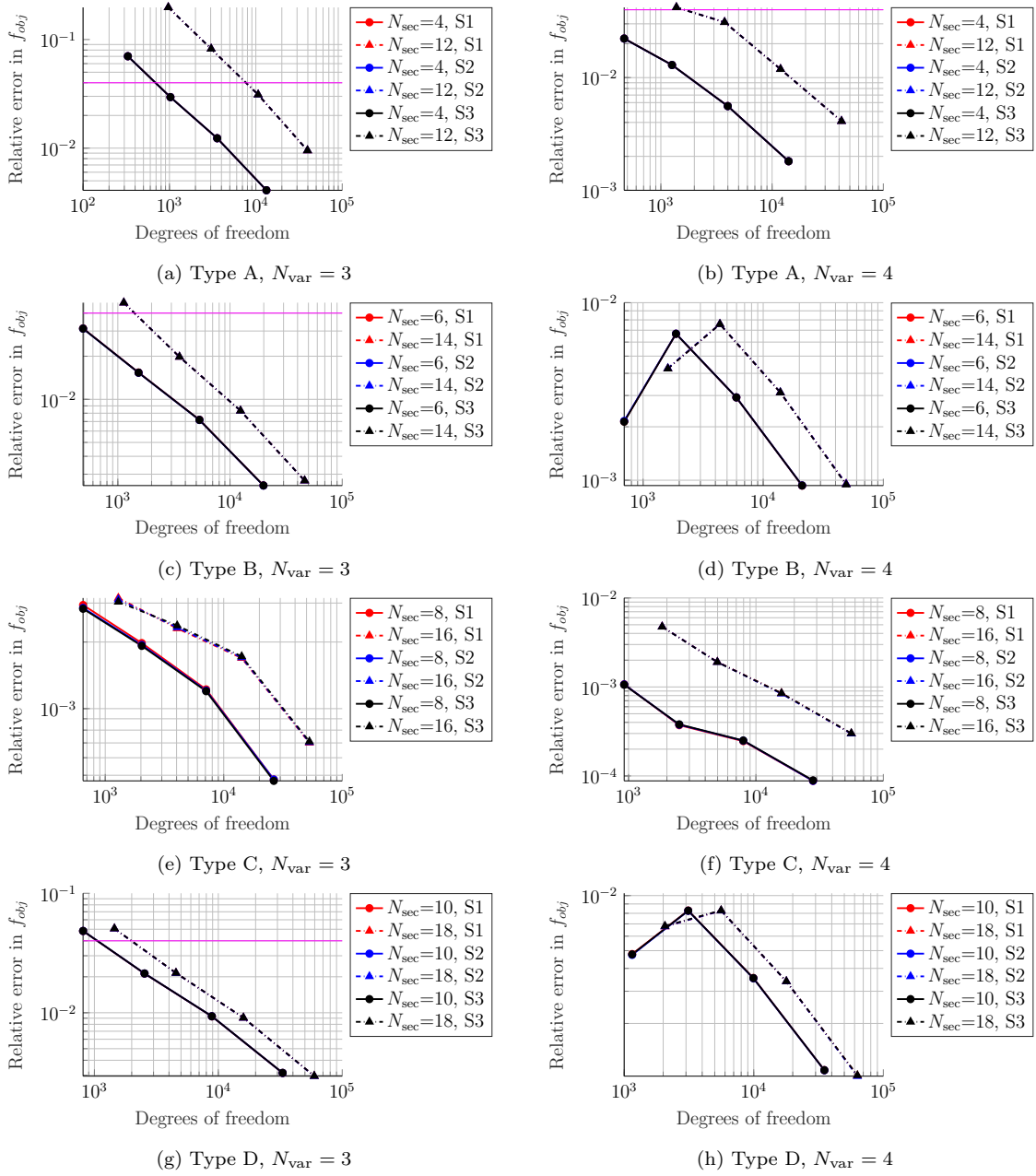
\begin{figure}[htbp!]
\centering
\begin{subfigure}[b]{0.48\textwidth}
\scalebox{0.65}[0.66]{
%
%
\definecolor{mycolor1}{rgb}{1.00000,0.00000,1.00000}%
\begin{tikzpicture}

\begin{axis}[%
width=5.75cm,
height=4cm,
at={(0cm,0cm)},
scale only axis,
xmode=log,
xmin=100,
xmax=100000,
xtick={   100,   1000,  10000, 100000},
xminorticks=true,
xlabel style={font=\color{white!15!black}},
xlabel={Degrees of freedom},
ymode=log,
ymin=0.00409281708618385,
ymax=0.197740580006442,
ytick={0.0001,  0.001,   0.01,    0.1},
yminorticks=true,
ylabel style={font=\color{white!15!black}},
ylabel={Relative error in $f_{obj}$},
axis background/.style={fill=white},
axis x line*=bottom,
axis y line*=left,
xminorgrids,
yminorgrids,
legend style={at={(1.03,1)}, anchor=north west, legend cell align=left, align=left, draw=white!15!black}
]
\addplot [color=red, line width=1.1pt, mark size=2.0pt, mark=*, mark options={solid, red}]
  table[row sep=crcr]{%
330	0.0702319684301543\\
1026	0.0294988525041586\\
3570	0.012341351997579\\
13266	0.00409628313403414\\
51090	0\\
};
\addlegendentry{$N_{\mathrm{sec}}$=4, S1}

\addplot [color=red, dashdotted, line width=1.1pt, mark size=2.3pt, mark=triangle*, mark options={solid, fill=red, red}]
  table[row sep=crcr]{%
970	0.197698087544243\\
3042	0.0821034768159768\\
10642	0.0310797076367626\\
39666	0.00951719128330362\\
153010	0\\
};
\addlegendentry{$N_{\mathrm{sec}}$=12, S1}

\addplot [color=blue, line width=1.1pt, mark size=2.0pt, mark=*, mark options={solid, blue}]
  table[row sep=crcr]{%
330	0.0702208540626088\\
1026	0.0294945417926209\\
3570	0.0123336109964315\\
13266	0.00409281708618385\\
51090	0\\
};
\addlegendentry{$N_{\mathrm{sec}}$=4, S2}

\addplot [color=blue, dashdotted, line width=1.1pt, mark size=2.3pt, mark=triangle*, mark options={solid, fill=blue, blue}]
  table[row sep=crcr]{%
970	0.197711888433461\\
3042	0.0821094767486691\\
10642	0.0310813298452942\\
39666	0.00951831205416662\\
153010	0\\
};
\addlegendentry{$N_{\mathrm{sec}}$=12, S2}

\addplot [color=black, line width=1.1pt, mark size=2.0pt, mark=*, mark options={solid, black}]
  table[row sep=crcr]{%
330	0.0702049817980004\\
1026	0.0294862894615747\\
3570	0.0123352695687278\\
13266	0.00409439769781389\\
51090	0\\
};
\addlegendentry{$N_{\mathrm{sec}}$=4, S3}

\addplot [color=black, dashdotted, line width=1.1pt, mark size=2.3pt, mark=triangle*, mark options={solid, fill=black, black}]
  table[row sep=crcr]{%
970	0.197740580006442\\
3042	0.0821232239211728\\
10642	0.0310900731808673\\
39666	0.00952107819101591\\
153010	0\\
};
\addlegendentry{$N_{\mathrm{sec}}$=12, S3}

\addplot [color=mycolor1]
  table[row sep=crcr]{%
100	0.04\\
100000	0.04\\
};

\end{axis}
\end{tikzpicture}%
}
        \caption{Type A, $N_{\mathrm{var}}=3$}
        \label{fig:Lshape_k_40_a}
    \end{subfigure}
    \quad
 \begin{subfigure}[b]{0.48\textwidth}
\scalebox{0.65}[0.66]{
%
%
\definecolor{mycolor1}{rgb}{1.00000,0.00000,1.00000}%
\begin{tikzpicture}

\begin{axis}[%
width=5.75cm,
height=4cm,
at={(0cm,0cm)},
scale only axis,
xmode=log,
xmin=468,
xmax=100000,
xtick={   100,   1000,  10000, 100000},
xminorticks=true,
xlabel style={font=\color{white!15!black}},
xlabel={Degrees of freedom},
ymode=log,
ymin=0.001,
ymax=0.0419922376437875,
ytick={0.0001,  0.001,   0.01,    0.1},
yminorticks=true,
ylabel style={font=\color{white!15!black}},
ylabel={Relative error in $f_{obj}$},
axis background/.style={fill=white},
axis x line*=bottom,
axis y line*=left,
xminorgrids,
yminorgrids,
legend style={at={(1.03,1)}, anchor=north west, legend cell align=left, align=left, draw=white!15!black}
]
\addplot [color=red, line width=1.1pt, mark size=2.0pt, mark=*, mark options={solid, red}]
  table[row sep=crcr]{%
468	0.0222250243435008\\
1260	0.0129435715662528\\
3996	0.0055890506988163\\
14076	0.00181137726816787\\
52668	0\\
};
\addlegendentry{$N_{\mathrm{sec}}$=4, S1}

\addplot [color=red, dashdotted, line width=1.1pt, mark size=2.3pt, mark=triangle*, mark options={solid, fill=red, red}]
  table[row sep=crcr]{%
1380	0.0419922376437875\\
3740	0.0310276001137955\\
11916	0.0118820208606598\\
42092	0.00411297762186166\\
157740	0\\
};
\addlegendentry{$N_{\mathrm{sec}}$=12, S1}

\addplot [color=blue, line width=1.1pt, mark size=2.0pt, mark=*, mark options={solid, blue}]
  table[row sep=crcr]{%
468	0.0221856552114792\\
1260	0.0129320607365657\\
3996	0.00558358331080169\\
14076	0.00181256241772779\\
52668	0\\
};
\addlegendentry{$N_{\mathrm{sec}}$=4, S2}

\addplot [color=blue, dashdotted, line width=1.1pt, mark size=2.3pt, mark=triangle*, mark options={solid, fill=blue, blue}]
  table[row sep=crcr]{%
1380	0.0419868915819424\\
3740	0.031026635233942\\
11916	0.0118802453238618\\
42092	0.00411238675206514\\
157740	0\\
};
\addlegendentry{$N_{\mathrm{sec}}$=12, S2}

\addplot [color=black, line width=1.1pt, mark size=2.0pt, mark=*, mark options={solid, black}]
  table[row sep=crcr]{%
468	0.022212397629999\\
1260	0.0129411248659837\\
3996	0.00558450436623394\\
14076	0.00180949015850605\\
52668	0\\
};
\addlegendentry{$N_{\mathrm{sec}}$=4, S3}

\addplot [color=black, dashdotted, line width=1.1pt, mark size=2.3pt, mark=triangle*, mark options={solid, fill=black, black}]
  table[row sep=crcr]{%
1380	0.0419899568198359\\
3740	0.0310206429926552\\
11916	0.0118741780744335\\
42092	0.00410737314732383\\
157740	0\\
};
\addlegendentry{$N_{\mathrm{sec}}$=12, S3}

\addplot [color=mycolor1]
  table[row sep=crcr]{%
468	0.04\\
100000	0.04\\
};

\end{axis}
\end{tikzpicture}%
}
        \caption{Type A, $N_{\mathrm{var}}=4$}
        \label{fig:Lshape_k_40_a}
    \end{subfigure}
\\
\begin{subfigure}[b]{0.48\textwidth}
\scalebox{0.65}[0.66]{
%
%
\definecolor{mycolor1}{rgb}{1.00000,0.00000,1.00000}%
\begin{tikzpicture}

\begin{axis}[%
width=5.75cm,
height=4cm,
at={(0cm,0cm)},
scale only axis,
xmode=log,
xmin=490,
xmax=100000,
xtick={   100,   1000,  10000, 100000},
xminorticks=true,
xlabel style={font=\color{white!15!black}},
xlabel={Degrees of freedom},
ymode=log,
ymin=0.00249879557437543,
ymax=0.0473057519752641,
ytick={0.0001,  0.001,   0.01,    0.1},
yminorticks=true,
ylabel style={font=\color{white!15!black}},
ylabel={Relative error in $f_{obj}$},
axis background/.style={fill=white},
axis x line*=bottom,
axis y line*=left,
xminorgrids,
yminorgrids,
legend style={at={(1.03,1)}, anchor=north west, legend cell align=left, align=left, draw=white!15!black}
]
\addplot [color=red, line width=1.1pt, mark size=2.0pt, mark=*, mark options={solid, red}]
  table[row sep=crcr]{%
490	0.0312763719675801\\
1530	0.0153923057208961\\
5338	0.00719462226657139\\
19866	0.00250476620991553\\
76570	0\\
};
\addlegendentry{$N_{\mathrm{sec}}$=6, S1}

\addplot [color=red, dashdotted, line width=1.1pt, mark size=2.3pt, mark=triangle*, mark options={solid, fill=red, red}]
  table[row sep=crcr]{%
1130	0.0473057519752641\\
3546	0.0197902021837341\\
12410	0.00832248553014166\\
46266	0.00270651295269345\\
178490	0\\
};
\addlegendentry{$N_{\mathrm{sec}}$=14, S1}

\addplot [color=blue, line width=1.1pt, mark size=2.0pt, mark=*, mark options={solid, blue}]
  table[row sep=crcr]{%
490	0.0311642796006298\\
1530	0.0153460976518807\\
5338	0.00717636346137703\\
19866	0.00249879557437543\\
76570	0\\
};
\addlegendentry{$N_{\mathrm{sec}}$=6, S2}

\addplot [color=blue, dashdotted, line width=1.1pt, mark size=2.3pt, mark=triangle*, mark options={solid, fill=blue, blue}]
  table[row sep=crcr]{%
1130	0.0472969411727332\\
3546	0.019788041717442\\
12410	0.00832089349993388\\
46266	0.00270534839394801\\
178490	0\\
};
\addlegendentry{$N_{\mathrm{sec}}$=14, S2}

\addplot [color=black, line width=1.1pt, mark size=2.0pt, mark=*, mark options={solid, black}]
  table[row sep=crcr]{%
490	0.0311906526465649\\
1530	0.0153567939476653\\
5338	0.00718427341795676\\
19866	0.00250232313385888\\
76570	0\\
};
\addlegendentry{$N_{\mathrm{sec}}$=6, S3}

\addplot [color=black, dashdotted, line width=1.1pt, mark size=2.3pt, mark=triangle*, mark options={solid, fill=black, black}]
  table[row sep=crcr]{%
1130	0.0472915663605874\\
3546	0.0197859814686966\\
12410	0.0083197656946314\\
46266	0.0027064046277051\\
178490	0\\
};
\addlegendentry{$N_{\mathrm{sec}}$=14, S3}

\addplot [color=mycolor1]
  table[row sep=crcr]{%
490	0.04\\
100000	0.04\\
};

\end{axis}
\end{tikzpicture}%
}
        \caption{Type B, $N_{\mathrm{var}}=3$}
        \label{fig:Lshape_k_40_a}
    \end{subfigure}
    \quad
 \begin{subfigure}[b]{0.48\textwidth}
\scalebox{0.65}[0.66]{
%
%
\begin{tikzpicture}

\begin{axis}[%
width=5.75cm,
height=4cm,
at={(0cm,0cm)},
scale only axis,
xmode=log,
xmin=696,
xmax=100000,
xtick={   100,   1000,  10000, 100000},
xminorticks=true,
xlabel style={font=\color{white!15!black}},
xlabel={Degrees of freedom},
ymode=log,
ymin=0.000932912770175528,
ymax=0.01,
ytick={0.0001,  0.001,   0.01,    0.1},
yminorticks=true,
ylabel style={font=\color{white!15!black}},
ylabel={Relative error in $f_{obj}$},
axis background/.style={fill=white},
axis x line*=bottom,
axis y line*=left,
xminorgrids,
yminorgrids,
legend style={at={(1.03,1)}, anchor=north west, legend cell align=left, align=left, draw=white!15!black}
]
\addplot [color=red, line width=1.1pt, mark size=2.0pt, mark=*, mark options={solid, red}]
  table[row sep=crcr]{%
696	0.0021415972929176\\
1880	0.00668335776596361\\
5976	0.00292697927581776\\
21080	0.000932912770175528\\
78936	0\\
};
\addlegendentry{$N_{\mathrm{sec}}$=6, S1}

\addplot [color=red, dashdotted, line width=1.1pt, mark size=2.3pt, mark=triangle*, mark options={solid, fill=red, red}]
  table[row sep=crcr]{%
1608	0.00426008614152823\\
4360	0.00757192991602243\\
13896	0.0031177142109315\\
49096	0.000952672014626242\\
184008	0\\
};
\addlegendentry{$N_{\mathrm{sec}}$=14, S1}

\addplot [color=blue, line width=1.1pt, mark size=2.0pt, mark=*, mark options={solid, blue}]
  table[row sep=crcr]{%
696	0.00215600506064142\\
1880	0.00669120788139822\\
5976	0.00292926783048358\\
21080	0.000934638707008657\\
78936	0\\
};
\addlegendentry{$N_{\mathrm{sec}}$=6, S2}

\addplot [color=blue, dashdotted, line width=1.1pt, mark size=2.3pt, mark=triangle*, mark options={solid, fill=blue, blue}]
  table[row sep=crcr]{%
1608	0.00427009818643911\\
4360	0.00757640200823181\\
13896	0.00312461410005968\\
49096	0.000951643379764812\\
184008	0\\
};
\addlegendentry{$N_{\mathrm{sec}}$=14, S2}

\addplot [color=black, line width=1.1pt, mark size=2.0pt, mark=*, mark options={solid, black}]
  table[row sep=crcr]{%
696	0.00213602139860054\\
1880	0.00667960606012053\\
5976	0.00292662607165088\\
21080	0.000933527517868013\\
78936	0\\
};
\addlegendentry{$N_{\mathrm{sec}}$=6, S3}

\addplot [color=black, dashdotted, line width=1.1pt, mark size=2.3pt, mark=triangle*, mark options={solid, fill=black, black}]
  table[row sep=crcr]{%
1608	0.00424751332006757\\
4360	0.00756638098530162\\
13896	0.00311804496138128\\
49096	0.00094403264614735\\
184008	0\\
};
\addlegendentry{$N_{\mathrm{sec}}$=14, S3}

\end{axis}
\end{tikzpicture}%
}
        \caption{Type B, $N_{\mathrm{var}}=4$}
        \label{fig:Lshape_k_40_a}
    \end{subfigure}
\\
\begin{subfigure}[b]{0.48\textwidth}
\scalebox{0.65}[0.66]{
%
%
\begin{tikzpicture}

\begin{axis}[%
width=5.75cm,
height=4cm,
at={(0cm,0cm)},
scale only axis,
xmode=log,
xmin=650,
xmax=100000,
xtick={   100,   1000,  10000, 100000},
xminorticks=true,
xlabel style={font=\color{white!15!black}},
xlabel={Degrees of freedom},
ymode=log,
ymin=0.000471675801563683,
ymax=0.00316240680453421,
ytick={0.0001,  0.001,   0.01,    0.1},
yminorticks=true,
ylabel style={font=\color{white!15!black}},
ylabel={Relative error in $f_{obj}$},
axis background/.style={fill=white},
axis x line*=bottom,
axis y line*=left,
xminorgrids,
yminorgrids,
legend style={at={(1.03,1)}, anchor=north west, legend cell align=left, align=left, draw=white!15!black}
]
\addplot [color=red, line width=1.1pt, mark size=2.0pt, mark=*, mark options={solid, red}]
  table[row sep=crcr]{%
650	0.00293962014018106\\
2034	0.00197848994131503\\
7106	0.0012216179315912\\
26466	0.000478675151071081\\
102050	0\\
};
\addlegendentry{$N_{\mathrm{sec}}$=8, S1}

\addplot [color=red, dashdotted, line width=1.1pt, mark size=2.3pt, mark=triangle*, mark options={solid, fill=red, red}]
  table[row sep=crcr]{%
1290	0.00316240680453421\\
4050	0.00230405870926647\\
14178	0.00169274316745731\\
52866	0.000700807864896365\\
203970	0\\
};
\addlegendentry{$N_{\mathrm{sec}}$=16, S1}

\addplot [color=blue, line width=1.1pt, mark size=2.0pt, mark=*, mark options={solid, blue}]
  table[row sep=crcr]{%
650	0.00285258257279554\\
2034	0.00194038960929289\\
7106	0.0012073643793498\\
26466	0.000478356336590213\\
102050	0\\
};
\addlegendentry{$N_{\mathrm{sec}}$=8, S2}

\addplot [color=blue, dashdotted, line width=1.1pt, mark size=2.3pt, mark=triangle*, mark options={solid, fill=blue, blue}]
  table[row sep=crcr]{%
1290	0.00311122134306149\\
4050	0.00232794205997723\\
14178	0.00170335900487344\\
52866	0.000704548715646211\\
203970	0\\
};
\addlegendentry{$N_{\mathrm{sec}}$=16, S2}

\addplot [color=black, line width=1.1pt, mark size=2.0pt, mark=*, mark options={solid, black}]
  table[row sep=crcr]{%
650	0.0028269287701966\\
2034	0.00192145636976146\\
7106	0.00119919743801196\\
26466	0.000471675801563683\\
102050	0\\
};
\addlegendentry{$N_{\mathrm{sec}}$=8, S3}

\addplot [color=black, dashdotted, line width=1.1pt, mark size=2.3pt, mark=triangle*, mark options={solid, fill=black, black}]
  table[row sep=crcr]{%
1290	0.00303704790908645\\
4050	0.00237351048752022\\
14178	0.00172222178824185\\
52866	0.000710038100807467\\
203970	0\\
};
\addlegendentry{$N_{\mathrm{sec}}$=16, S3}

\end{axis}
\end{tikzpicture}%
}
        \caption{Type C, $N_{\mathrm{var}}=3$}
        \label{fig:Lshape_k_40_a}
    \end{subfigure}
    \quad
 \begin{subfigure}[b]{0.48\textwidth}
\scalebox{0.65}[0.66]{
%
%
\begin{tikzpicture}

\begin{axis}[%
width=5.75cm,
height=4cm,
at={(0cm,0cm)},
scale only axis,
xmode=log,
xmin=924,
xmax=100000,
xtick={   100,   1000,  10000, 100000},
xminorticks=true,
xlabel style={font=\color{white!15!black}},
xlabel={Degrees of freedom},
ymode=log,
ymin=8.80561205327572e-05,
ymax=0.01,
ytick={0.0001,  0.001,   0.01,    0.1},
yminorticks=true,
ylabel style={font=\color{white!15!black}},
ylabel={Relative error in $f_{obj}$},
axis background/.style={fill=white},
axis x line*=bottom,
axis y line*=left,
xminorgrids,
yminorgrids,
legend style={at={(1.03,1)}, anchor=north west, legend cell align=left, align=left, draw=white!15!black}
]
\addplot [color=red, line width=1.1pt, mark size=2.0pt, mark=*, mark options={solid, red}]
  table[row sep=crcr]{%
924	0.00107068780750549\\
2500	0.00037206493922476\\
7956	0.000246023591054027\\
28084	8.80889007503936e-05\\
105204	0\\
};
\addlegendentry{$N_{\mathrm{sec}}$=8, S1}

\addplot [color=red, dashdotted, line width=1.1pt, mark size=2.3pt, mark=triangle*, mark options={solid, fill=red, red}]
  table[row sep=crcr]{%
1836	0.00476258261600476\\
4980	0.00191123692976607\\
15876	0.000849803884160903\\
56100	0.000298856359282815\\
210276	0\\
};
\addlegendentry{$N_{\mathrm{sec}}$=16, S1}

\addplot [color=blue, line width=1.1pt, mark size=2.0pt, mark=*, mark options={solid, blue}]
  table[row sep=crcr]{%
924	0.00106713571654515\\
2500	0.000376716545191346\\
7956	0.000249065271131335\\
28084	8.80561205327572e-05\\
105204	0\\
};
\addlegendentry{$N_{\mathrm{sec}}$=8, S2}

\addplot [color=blue, dashdotted, line width=1.1pt, mark size=2.3pt, mark=triangle*, mark options={solid, fill=blue, blue}]
  table[row sep=crcr]{%
1836	0.00475408563606065\\
4980	0.00190488044182251\\
15876	0.000841514629524586\\
56100	0.000298468501275516\\
210276	0\\
};
\addlegendentry{$N_{\mathrm{sec}}$=16, S2}

\addplot [color=black, line width=1.1pt, mark size=2.0pt, mark=*, mark options={solid, black}]
  table[row sep=crcr]{%
924	0.00105856241317993\\
2500	0.00037977368798816\\
7956	0.000250034559820846\\
28084	8.91231644928064e-05\\
105204	0\\
};
\addlegendentry{$N_{\mathrm{sec}}$=8, S3}

\addplot [color=black, dashdotted, line width=1.1pt, mark size=2.3pt, mark=triangle*, mark options={solid, fill=black, black}]
  table[row sep=crcr]{%
1836	0.00474163700456495\\
4980	0.00190173646669288\\
15876	0.000850177024508214\\
56100	0.000301051735088219\\
210276	0\\
};
\addlegendentry{$N_{\mathrm{sec}}$=16, S3}

\end{axis}
\end{tikzpicture}%
}
        \caption{Type C, $N_{\mathrm{var}}=4$}
        \label{fig:Lshape_k_40_a}
    \end{subfigure}
\\
\begin{subfigure}[b]{0.48\textwidth}
\scalebox{0.65}[0.66]{
%
%
\definecolor{mycolor1}{rgb}{1.00000,0.00000,1.00000}%
\begin{tikzpicture}

\begin{axis}[%
width=5.75cm,
height=4cm,
at={(0cm,0cm)},
scale only axis,
xmode=log,
xmin=810,
xmax=100000,
xtick={   100,   1000,  10000, 100000},
xminorticks=true,
xlabel style={font=\color{white!15!black}},
xlabel={Degrees of freedom},
ymode=log,
ymin=0.0029446636310576,
ymax=0.1,
ytick={0.0001,  0.001,   0.01,    0.1},
yminorticks=true,
ylabel style={font=\color{white!15!black}},
ylabel={Relative error in $f_{obj}$},
axis background/.style={fill=white},
axis x line*=bottom,
axis y line*=left,
xminorgrids,
yminorgrids,
legend style={at={(1.03,1)}, anchor=north west, legend cell align=left, align=left, draw=white!15!black}
]
\addplot [color=red, line width=1.1pt, mark size=2.0pt, mark=*, mark options={solid, red}]
  table[row sep=crcr]{%
810	0.0482949257566168\\
2538	0.0212744161660041\\
8874	0.0093233077438704\\
33066	0.00313098576575534\\
127530	0\\
};
\addlegendentry{$N_{\mathrm{sec}}$=10, S1}

\addplot [color=red, dashdotted, line width=1.1pt, mark size=2.3pt, mark=triangle*, mark options={solid, fill=red, red}]
  table[row sep=crcr]{%
1450	0.0506012465570902\\
4554	0.0215016578568031\\
15946	0.00906963851323951\\
59466	0.0029522157506571\\
229450	0\\
};
\addlegendentry{$N_{\mathrm{sec}}$=18, S1}

\addplot [color=blue, line width=1.1pt, mark size=2.0pt, mark=*, mark options={solid, blue}]
  table[row sep=crcr]{%
810	0.0483769403125358\\
2538	0.0213067801022279\\
8874	0.00933680014783454\\
33066	0.00313167391514725\\
127530	0\\
};
\addlegendentry{$N_{\mathrm{sec}}$=10, S2}

\addplot [color=blue, dashdotted, line width=1.1pt, mark size=2.3pt, mark=triangle*, mark options={solid, fill=blue, blue}]
  table[row sep=crcr]{%
1450	0.0505516138456861\\
4554	0.0214783308890052\\
15946	0.00905808078009914\\
59466	0.0029446636310576\\
229450	0\\
};
\addlegendentry{$N_{\mathrm{sec}}$=18, S2}

\addplot [color=black, line width=1.1pt, mark size=2.0pt, mark=*, mark options={solid, black}]
  table[row sep=crcr]{%
810	0.0483855114993335\\
2538	0.0213127625503666\\
8874	0.00933957948555138\\
33066	0.00313494913003604\\
127530	0\\
};
\addlegendentry{$N_{\mathrm{sec}}$=10, S3}

\addplot [color=black, dashdotted, line width=1.1pt, mark size=2.3pt, mark=triangle*, mark options={solid, fill=black, black}]
  table[row sep=crcr]{%
1450	0.0506599223364431\\
4554	0.0215327239959148\\
15946	0.00907956157967048\\
59466	0.00295427652905452\\
229450	0\\
};
\addlegendentry{$N_{\mathrm{sec}}$=18, S3}

\addplot [color=mycolor1]
  table[row sep=crcr]{%
810	0.04\\
100000	0.04\\
};

\end{axis}
\end{tikzpicture}%
}
        \caption{Type D, $N_{\mathrm{var}}=3$}
        \label{fig:Lshape_k_40_a}
    \end{subfigure}
    \quad
 \begin{subfigure}[b]{0.48\textwidth}
\scalebox{0.65}[0.66]{
%
%
\begin{tikzpicture}

\begin{axis}[%
width=5.75cm,
height=4cm,
at={(0cm,0cm)},
scale only axis,
xmode=log,
xmin=1000,
xmax=100000,
xtick={   100,   1000,  10000, 100000},
xminorticks=true,
xlabel style={font=\color{white!15!black}},
xlabel={Degrees of freedom},
ymode=log,
ymin=0.00103001854134146,
ymax=0.01029381221897737,
ytick={0.0001,  0.001,   0.01,    0.1},
yminorticks=true,
ylabel style={font=\color{white!15!black}},
ylabel={Relative error in $f_{obj}$},
axis background/.style={fill=white},
axis x line*=bottom,
axis y line*=left,
xminorgrids,
yminorgrids,
legend style={at={(1.03,1)}, anchor=north west, legend cell align=left, align=left, draw=white!15!black}
]
\addplot [color=red, line width=1.1pt, mark size=2.0pt, mark=*, mark options={solid, red}]
  table[row sep=crcr]{%
1152	0.00478911871302194\\
3120	0.00828178918495719\\
9936	0.00354297240488264\\
35088	0.00111108729872306\\
131472	0\\
};
\addlegendentry{$N_{\mathrm{sec}}$=10, S1}

\addplot [color=red, dashdotted, line width=1.1pt, mark size=2.3pt, mark=triangle*, mark options={solid, fill=red, red}]
  table[row sep=crcr]{%
2064	0.00677417027640607\\
5600	0.00827369697114146\\
17856	0.00339378101208072\\
63104	0.00103722344719448\\
236544	0\\
};
\addlegendentry{$N_{\mathrm{sec}}$=18, S1}

\addplot [color=blue, line width=1.1pt, mark size=2.0pt, mark=*, mark options={solid, blue}]
  table[row sep=crcr]{%
1152	0.00474147240508017\\
3120	0.00825741045815109\\
9936	0.0035324207741536\\
35088	0.00110975182749636\\
131472	0\\
};
\addlegendentry{$N_{\mathrm{sec}}$=10, S2}

\addplot [color=blue, dashdotted, line width=1.1pt, mark size=2.3pt, mark=triangle*, mark options={solid, fill=blue, blue}]
  table[row sep=crcr]{%
2064	0.00677830525626121\\
5600	0.00827052289141637\\
17856	0.00339628635432762\\
63104	0.00103001854134146\\
236544	0\\
};
\addlegendentry{$N_{\mathrm{sec}}$=18, S2}

\addplot [color=black, line width=1.1pt, mark size=2.0pt, mark=*, mark options={solid, black}]
  table[row sep=crcr]{%
1152	0.00477094043198439\\
3120	0.00827177056774247\\
9936	0.00353881840335868\\
35088	0.0011107606854012\\
131472	0\\
};
\addlegendentry{$N_{\mathrm{sec}}$=10, S3}

\addplot [color=black, dashdotted, line width=1.1pt, mark size=2.3pt, mark=triangle*, mark options={solid, fill=black, black}]
  table[row sep=crcr]{%
2064	0.00681732234409404\\
5600	0.00829381221897737\\
17856	0.00339819527482877\\
63104	0.00104042294719834\\
236544	0\\
};
\addlegendentry{$N_{\mathrm{sec}}$=18, S3}

\end{axis}
\end{tikzpicture}%
}
        \caption{Type D, $N_{\mathrm{var}}=4$}
        \label{fig:Lshape_k_40_a}
    \end{subfigure}
\caption{Convergence of the objective function $f_{\mathrm{obj}}$ for thermal concentrator problem for Type-A ($N_{\mathrm{sec}}=$ 4, 12), Type-B ($N_{\mathrm{sec}}=$ 6, 14), Type-C ($N_{\mathrm{sec}}=$ 8, 16), Type-D ($N_{\mathrm{sec}}=$ 10, 18) configurations, and $N_{\mathrm{var}}=$ 3, 4. $4\%$ relative error (shown by pink horizontal line) is taken as the tolerance value to select the minimum mesh size for optimization. Samples 1, 2, and 3 are denoted by S1, S2 and S3. Type-A requires a finer mesh compared to Type-B and Type-D. Type-C needs a comparatively coarse mesh.}
    \label{fig:Cntr convergence 1}
\end{figure}

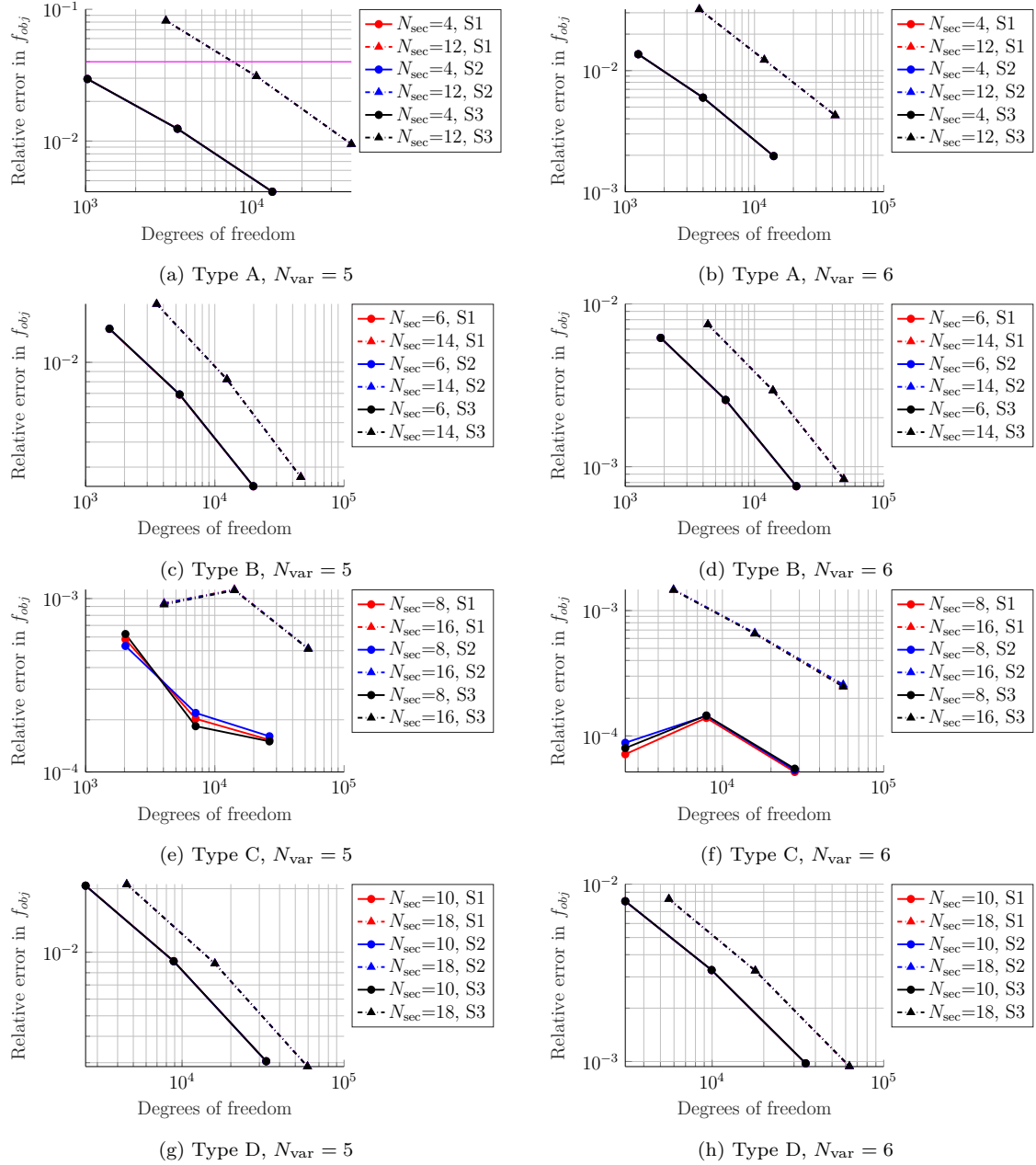
\begin{figure}[htbp!]
\centering
\begin{subfigure}[b]{0.48\textwidth}
\scalebox{0.65}[0.66]{
%
%
\definecolor{mycolor1}{rgb}{1.00000,0.00000,1.00000}%
\begin{tikzpicture}

\begin{axis}[%
width=5.913cm,
height=4cm,
at={(0cm,0cm)},
scale only axis,
xmode=log,
xmin=1000,
xmax=39666,
xtick={   100,   1000,  10000, 100000},
xminorticks=true,
xlabel style={font=\color{white!15!black}},
xlabel={Degrees of freedom},
ymode=log,
ymin=0.00412416299025193,
ymax=0.1,
ytick={0.0001,  0.001,   0.01,    0.1},
yminorticks=true,
ylabel style={font=\color{white!15!black}},
ylabel={Relative error in $f_{obj}$},
axis background/.style={fill=white},
axis x line*=bottom,
axis y line*=left,
xminorgrids,
yminorgrids,
legend style={at={(1.03,1)}, anchor=north west, legend cell align=left, align=left, draw=white!15!black}
]
\addplot [color=red, line width=1.1pt, mark size=2.0pt, mark=*, mark options={solid, red}]
  table[row sep=crcr]{%
1026	0.0296264182177653\\
3570	0.0124151179165845\\
13266	0.00412904711005713\\
51090	0\\
};
\addlegendentry{$N_{\mathrm{sec}}$=4, S1}

\addplot [color=red, dashdotted, line width=1.1pt, mark size=2.3pt, mark=triangle*, mark options={solid, fill=red, red}]
  table[row sep=crcr]{%
3042	0.0820584448713664\\
10642	0.0310204599307202\\
39666	0.00948719235737246\\
153010	0\\
};
\addlegendentry{$N_{\mathrm{sec}}$=12, S1}

\addplot [color=blue, line width=1.1pt, mark size=2.0pt, mark=*, mark options={solid, blue}]
  table[row sep=crcr]{%
1026	0.0295860172392532\\
3570	0.0123927192657078\\
13266	0.00412416299025193\\
51090	0\\
};
\addlegendentry{$N_{\mathrm{sec}}$=4, S2}

\addplot [color=blue, dashdotted, line width=1.1pt, mark size=2.3pt, mark=triangle*, mark options={solid, fill=blue, blue}]
  table[row sep=crcr]{%
3042	0.0820668499787832\\
10642	0.031028745539308\\
39666	0.00948976127516523\\
153010	0\\
};
\addlegendentry{$N_{\mathrm{sec}}$=12, S2}

\addplot [color=black, line width=1.1pt, mark size=2.0pt, mark=*, mark options={solid, black}]
  table[row sep=crcr]{%
1026	0.0295955507812331\\
3570	0.0124002793594313\\
13266	0.00412837060115336\\
51090	0\\
};
\addlegendentry{$N_{\mathrm{sec}}$=4, S3}

\addplot [color=black, dashdotted, line width=1.1pt, mark size=2.3pt, mark=triangle*, mark options={solid, fill=black, black}]
  table[row sep=crcr]{%
3042	0.0820424487920138\\
10642	0.0310144929177049\\
39666	0.00948652454264718\\
153010	0\\
};
\addlegendentry{$N_{\mathrm{sec}}$=12, S3}

\addplot [color=mycolor1]
  table[row sep=crcr]{%
1000	0.04\\
39666	0.04\\
};

\end{axis}
\end{tikzpicture}%
}
        \caption{Type A, $N_{\mathrm{var}}=5$}
        \label{fig:Lshape_k_40_a}
    \end{subfigure}
    \quad
 \begin{subfigure}[b]{0.48\textwidth}
\scalebox{0.65}[0.66]{
%
%
\begin{tikzpicture}

\begin{axis}[%
width=5.75cm,
height=4cm,
at={(0cm,0cm)},
scale only axis,
xmode=log,
xmin=1000,
xmax=100000,
xtick={   100,   1000,  10000, 100000},
xminorticks=true,
xlabel style={font=\color{white!15!black}},
xlabel={Degrees of freedom},
ymode=log,
ymin=0.001,
ymax=0.0320470614820234,
ytick={0.0001,  0.001,   0.01,    0.1},
yminorticks=true,
ylabel style={font=\color{white!15!black}},
ylabel={Relative error in $f_{obj}$},
axis background/.style={fill=white},
axis x line*=bottom,
axis y line*=left,
xminorgrids,
yminorgrids,
legend style={at={(1.03,1)}, anchor=north west, legend cell align=left, align=left, draw=white!15!black}
]
\addplot [color=red, line width=1.1pt, mark size=2.0pt, mark=*, mark options={solid, red}]
  table[row sep=crcr]{%
1260	0.013657517530964\\
3996	0.00598639297493611\\
14076	0.00197213617890294\\
52668	0\\
};
\addlegendentry{$N_{\mathrm{sec}}$=4, S1}

\addplot [color=red, dashdotted, line width=1.1pt, mark size=2.3pt, mark=triangle*, mark options={solid, fill=red, red}]
  table[row sep=crcr]{%
3740	0.0320470614820234\\
11916	0.0122599701408773\\
42092	0.00427380420808583\\
157740	0\\
};
\addlegendentry{$N_{\mathrm{sec}}$=12, S1}

\addplot [color=blue, line width=1.1pt, mark size=2.0pt, mark=*, mark options={solid, blue}]
  table[row sep=crcr]{%
1260	0.0136506572716699\\
3996	0.00598919600190898\\
14076	0.0019688506311615\\
52668	0\\
};
\addlegendentry{$N_{\mathrm{sec}}$=4, S2}

\addplot [color=blue, dashdotted, line width=1.1pt, mark size=2.3pt, mark=triangle*, mark options={solid, fill=blue, blue}]
  table[row sep=crcr]{%
3740	0.0320443294043125\\
11916	0.0122602633708168\\
42092	0.00427420949612515\\
157740	0\\
};
\addlegendentry{$N_{\mathrm{sec}}$=12, S2}

\addplot [color=black, line width=1.1pt, mark size=2.0pt, mark=*, mark options={solid, black}]
  table[row sep=crcr]{%
1260	0.013646102125234\\
3996	0.00598579251662207\\
14076	0.00197152646748546\\
52668	0\\
};
\addlegendentry{$N_{\mathrm{sec}}$=4, S3}

\addplot [color=black, dashdotted, line width=1.1pt, mark size=2.3pt, mark=triangle*, mark options={solid, fill=black, black}]
  table[row sep=crcr]{%
3740	0.0320466621476727\\
11916	0.0122595244821278\\
42092	0.00427381720688131\\
157740	0\\
};
\addlegendentry{$N_{\mathrm{sec}}$=12, S3}

\end{axis}
\end{tikzpicture}%
}
        \caption{Type A, $N_{\mathrm{var}}=6$}
        \label{fig:Lshape_k_40_a}
    \end{subfigure}
\\
\begin{subfigure}[b]{0.48\textwidth}
\scalebox{0.65}[0.66]{
%
%
\begin{tikzpicture}

\begin{axis}[%
width=5.75cm,
height=4cm,
at={(0cm,0cm)},
scale only axis,
xmode=log,
xmin=1000,
xmax=100000,
xtick={   100,   1000,  10000, 100000},
xminorticks=true,
xlabel style={font=\color{white!15!black}},
xlabel={Degrees of freedom},
ymode=log,
ymin=0.00240428434090558,
ymax=0.019569153641313,
ytick={0.0001,  0.001,   0.01,    0.1},
yminorticks=true,
ylabel style={font=\color{white!15!black}},
ylabel={Relative error in $f_{obj}$},
axis background/.style={fill=white},
axis x line*=bottom,
axis y line*=left,
xminorgrids,
yminorgrids,
legend style={at={(1.03,1)}, anchor=north west, legend cell align=left, align=left, draw=white!15!black}
]
\addplot [color=red, line width=1.1pt, mark size=2.0pt, mark=*, mark options={solid, red}]
  table[row sep=crcr]{%
1530	0.0146679688099396\\
5338	0.0068987169512456\\
19866	0.00240428434090558\\
76570	0\\
};
\addlegendentry{$N_{\mathrm{sec}}$=6, S1}

\addplot [color=red, dashdotted, line width=1.1pt, mark size=2.3pt, mark=triangle*, mark options={solid, fill=red, red}]
  table[row sep=crcr]{%
3546	0.019569153641313\\
12410	0.00822699888373983\\
46266	0.00267887447931643\\
178490	0\\
};
\addlegendentry{$N_{\mathrm{sec}}$=14, S1}

\addplot [color=blue, line width=1.1pt, mark size=2.0pt, mark=*, mark options={solid, blue}]
  table[row sep=crcr]{%
1530	0.0147152942568647\\
5338	0.00691448483720711\\
19866	0.00240787821169986\\
76570	0\\
};
\addlegendentry{$N_{\mathrm{sec}}$=6, S2}

\addplot [color=blue, dashdotted, line width=1.1pt, mark size=2.3pt, mark=triangle*, mark options={solid, fill=blue, blue}]
  table[row sep=crcr]{%
3546	0.0195466019742877\\
12410	0.00821661976341727\\
46266	0.00267219932996871\\
178490	0\\
};
\addlegendentry{$N_{\mathrm{sec}}$=14, S2}

\addplot [color=black, line width=1.1pt, mark size=2.0pt, mark=*, mark options={solid, black}]
  table[row sep=crcr]{%
1530	0.0147202475983666\\
5338	0.00691714812387597\\
19866	0.00241178343997321\\
76570	0\\
};
\addlegendentry{$N_{\mathrm{sec}}$=6, S3}

\addplot [color=black, dashdotted, line width=1.1pt, mark size=2.3pt, mark=triangle*, mark options={solid, fill=black, black}]
  table[row sep=crcr]{%
3546	0.019547133012757\\
12410	0.00822090518775192\\
46266	0.00267359418526232\\
178490	0\\
};
\addlegendentry{$N_{\mathrm{sec}}$=14, S3}

\end{axis}
\end{tikzpicture}%
}
        \caption{Type B, $N_{\mathrm{var}}=5$}
        \label{fig:Lshape_k_40_a}
    \end{subfigure}
    \quad
 \begin{subfigure}[b]{0.48\textwidth}
\scalebox{0.65}[0.66]{
%
%
\begin{tikzpicture}

\begin{axis}[%
width=5.75cm,
height=4cm,
at={(0cm,0cm)},
scale only axis,
xmode=log,
xmin=1000,
xmax=100000,
xtick={   100,   1000,  10000, 100000},
xminorticks=true,
xlabel style={font=\color{white!15!black}},
xlabel={Degrees of freedom},
ymode=log,
ymin=0.000757263576628771,
ymax=0.01,
ytick={0.0001,  0.001,   0.01,    0.1},
yminorticks=true,
ylabel style={font=\color{white!15!black}},
ylabel={Relative error in $f_{obj}$},
axis background/.style={fill=white},
axis x line*=bottom,
axis y line*=left,
xminorgrids,
yminorgrids,
legend style={at={(1.03,1)}, anchor=north west, legend cell align=left, align=left, draw=white!15!black}
]
\addplot [color=red, line width=1.1pt, mark size=2.0pt, mark=*, mark options={solid, red}]
  table[row sep=crcr]{%
1880	0.00618192774986671\\
5976	0.00256622507044575\\
21080	0.000757263576628771\\
78936	0\\
};
\addlegendentry{$N_{\mathrm{sec}}$=6, S1}

\addplot [color=red, dashdotted, line width=1.1pt, mark size=2.3pt, mark=triangle*, mark options={solid, fill=red, red}]
  table[row sep=crcr]{%
4360	0.00748723618051816\\
13896	0.00294147484392572\\
49096	0.000843612047559233\\
184008	0\\
};
\addlegendentry{$N_{\mathrm{sec}}$=14, S1}

\addplot [color=blue, line width=1.1pt, mark size=2.0pt, mark=*, mark options={solid, blue}]
  table[row sep=crcr]{%
1880	0.00618595275709613\\
5976	0.0025727336145096\\
21080	0.000758829444784675\\
78936	0\\
};
\addlegendentry{$N_{\mathrm{sec}}$=6, S2}

\addplot [color=blue, dashdotted, line width=1.1pt, mark size=2.3pt, mark=triangle*, mark options={solid, fill=blue, blue}]
  table[row sep=crcr]{%
4360	0.00748555578814301\\
13896	0.00293974625136464\\
49096	0.000838134943441321\\
184008	0\\
};
\addlegendentry{$N_{\mathrm{sec}}$=14, S2}

\addplot [color=black, line width=1.1pt, mark size=2.0pt, mark=*, mark options={solid, black}]
  table[row sep=crcr]{%
1880	0.00619165970464249\\
5976	0.00257582288406063\\
21080	0.000760266190702037\\
78936	0\\
};
\addlegendentry{$N_{\mathrm{sec}}$=6, S3}

\addplot [color=black, dashdotted, line width=1.1pt, mark size=2.3pt, mark=triangle*, mark options={solid, fill=black, black}]
  table[row sep=crcr]{%
4360	0.00747994377101091\\
13896	0.00293632936878771\\
49096	0.000837348477458317\\
184008	0\\
};
\addlegendentry{$N_{\mathrm{sec}}$=14, S3}

\end{axis}
\end{tikzpicture}%
}
        \caption{Type B, $N_{\mathrm{var}}=6$}
        \label{fig:Lshape_k_40_a}
    \end{subfigure}
\\
\begin{subfigure}[b]{0.48\textwidth}
\scalebox{0.65}[0.66]{
%
%
\begin{tikzpicture}

\begin{axis}[%
width=5.75cm,
height=4cm,
at={(0cm,0cm)},
scale only axis,
xmode=log,
xmin=1000,
xmax=100000,
xtick={   100,   1000,  10000, 100000},
xminorticks=true,
xlabel style={font=\color{white!15!black}},
xlabel={Degrees of freedom},
ymode=log,
ymin=0.0001,
ymax=0.00112723373912567,
ytick={0.0001,  0.001,   0.01,    0.1},
yminorticks=true,
ylabel style={font=\color{white!15!black}},
ylabel={Relative error in $f_{obj}$},
axis background/.style={fill=white},
axis x line*=bottom,
axis y line*=left,
xminorgrids,
yminorgrids,
legend style={at={(1.03,1)}, anchor=north west, legend cell align=left, align=left, draw=white!15!black}
]
\addplot [color=red, line width=1.1pt, mark size=2.0pt, mark=*, mark options={solid, red}]
  table[row sep=crcr]{%
2034	0.000581874146276525\\
7106	0.000202672836833919\\
26466	0.000153080318667047\\
102050	0\\
};
\addlegendentry{$N_{\mathrm{sec}}$=8, S1}

\addplot [color=red, dashdotted, line width=1.1pt, mark size=2.3pt, mark=triangle*, mark options={solid, fill=red, red}]
  table[row sep=crcr]{%
4050	0.000939667276868754\\
14178	0.00112723373912567\\
52866	0.000513031581345051\\
203970	0\\
};
\addlegendentry{$N_{\mathrm{sec}}$=16, S1}

\addplot [color=blue, line width=1.1pt, mark size=2.0pt, mark=*, mark options={solid, blue}]
  table[row sep=crcr]{%
2034	0.000532870417972792\\
7106	0.000218930457700817\\
26466	0.000160619913419409\\
102050	0\\
};
\addlegendentry{$N_{\mathrm{sec}}$=8, S2}

\addplot [color=blue, dashdotted, line width=1.1pt, mark size=2.3pt, mark=triangle*, mark options={solid, fill=blue, blue}]
  table[row sep=crcr]{%
4050	0.000937886711743163\\
14178	0.00112422353186268\\
52866	0.000515715249344567\\
203970	0\\
};
\addlegendentry{$N_{\mathrm{sec}}$=16, S2}

\addplot [color=black, line width=1.1pt, mark size=2.0pt, mark=*, mark options={solid, black}]
  table[row sep=crcr]{%
2034	0.000624306320819001\\
7106	0.000183705741954648\\
26466	0.000150178177289965\\
102050	0\\
};
\addlegendentry{$N_{\mathrm{sec}}$=8, S3}

\addplot [color=black, dashdotted, line width=1.1pt, mark size=2.3pt, mark=triangle*, mark options={solid, fill=black, black}]
  table[row sep=crcr]{%
4050	0.0009225180986947\\
14178	0.00111429510973255\\
52866	0.000511398394594157\\
203970	0\\
};
\addlegendentry{$N_{\mathrm{sec}}$=16, S3}

\end{axis}
\end{tikzpicture}%
}
        \caption{Type C, $N_{\mathrm{var}}=5$}
        \label{fig:Lshape_k_40_a}
    \end{subfigure}
    \quad
 \begin{subfigure}[b]{0.48\textwidth}
\scalebox{0.65}[0.66]{
%
%
\begin{tikzpicture}

\begin{axis}[%
width=5.75cm,
height=4cm,
at={(0cm,0cm)},
scale only axis,
xmode=log,
xmin=2500,
xmax=100000,
xtick={   100,   1000,  10000, 100000},
xminorticks=true,
xlabel style={font=\color{white!15!black}},
xlabel={Degrees of freedom},
ymode=log,
ymin=5.16055573832421e-05,
ymax=0.00147320421101962,
ytick={0.0001,  0.001,   0.01,    0.1},
yminorticks=true,
ylabel style={font=\color{white!15!black}},
ylabel={Relative error in $f_{obj}$},
axis background/.style={fill=white},
axis x line*=bottom,
axis y line*=left,
xminorgrids,
yminorgrids,
legend style={at={(1.03,1)}, anchor=north west, legend cell align=left, align=left, draw=white!15!black}
]
\addplot [color=red, line width=1.1pt, mark size=2.0pt, mark=*, mark options={solid, red}]
  table[row sep=crcr]{%
2500	7.14440839977682e-05\\
7956	0.000139245284942886\\
28084	5.16055573832421e-05\\
105204	0\\
};
\addlegendentry{$N_{\mathrm{sec}}$=8, S1}

\addplot [color=red, dashdotted, line width=1.1pt, mark size=2.3pt, mark=triangle*, mark options={solid, fill=red, red}]
  table[row sep=crcr]{%
4980	0.0014700277876022\\
15876	0.000657337920642253\\
56100	0.00025116182080125\\
210276	0\\
};
\addlegendentry{$N_{\mathrm{sec}}$=16, S1}

\addplot [color=blue, line width=1.1pt, mark size=2.0pt, mark=*, mark options={solid, blue}]
  table[row sep=crcr]{%
2500	8.81997949843414e-05\\
7956	0.000144539547001984\\
28084	5.32754828787204e-05\\
105204	0\\
};
\addlegendentry{$N_{\mathrm{sec}}$=8, S2}

\addplot [color=blue, dashdotted, line width=1.1pt, mark size=2.3pt, mark=triangle*, mark options={solid, fill=blue, blue}]
  table[row sep=crcr]{%
4980	0.00147320421101962\\
15876	0.000663345373164586\\
56100	0.0002570317172186\\
210276	0\\
};
\addlegendentry{$N_{\mathrm{sec}}$=16, S2}

\addplot [color=black, line width=1.1pt, mark size=2.0pt, mark=*, mark options={solid, black}]
  table[row sep=crcr]{%
2500	8.00078130187661e-05\\
7956	0.00014569733609313\\
28084	5.48438975802033e-05\\
105204	0\\
};
\addlegendentry{$N_{\mathrm{sec}}$=8, S3}

\addplot [color=black, dashdotted, line width=1.1pt, mark size=2.3pt, mark=triangle*, mark options={solid, fill=black, black}]
  table[row sep=crcr]{%
4980	0.00145659525601702\\
15876	0.000652813333300081\\
56100	0.000246655176255888\\
210276	0\\
};
\addlegendentry{$N_{\mathrm{sec}}$=16, S3}

\end{axis}
\end{tikzpicture}%
}
        \caption{Type C, $N_{\mathrm{var}}=6$}
        \label{fig:Lshape_k_40_a}
    \end{subfigure}
\\
\begin{subfigure}[b]{0.48\textwidth}
\scalebox{0.65}[0.66]{
%
%
\begin{tikzpicture}

\begin{axis}[%
width=5.75cm,
height=4cm,
at={(0cm,0cm)},
scale only axis,
xmode=log,
xmin=2538,
xmax=100000,
xtick={   100,   1000,  10000, 100000},
xminorticks=true,
xlabel style={font=\color{white!15!black}},
xlabel={Degrees of freedom},
ymode=log,
ymin=0.00287490032633513,
ymax=0.0210146362870987,
ytick={0.0001,  0.001,   0.01,    0.1},
yminorticks=true,
ylabel style={font=\color{white!15!black}},
ylabel={Relative error in $f_{obj}$},
axis background/.style={fill=white},
axis x line*=bottom,
axis y line*=left,
xminorgrids,
yminorgrids,
legend style={at={(1.03,1)}, anchor=north west, legend cell align=left, align=left, draw=white!15!black}
]
\addplot [color=red, line width=1.1pt, mark size=2.0pt, mark=*, mark options={solid, red}]
  table[row sep=crcr]{%
2538	0.0206758643958986\\
8874	0.00907587639757489\\
33066	0.00304880709604744\\
127530	0\\
};
\addlegendentry{$N_{\mathrm{sec}}$=10, S1}

\addplot [color=red, dashdotted, line width=1.1pt, mark size=2.3pt, mark=triangle*, mark options={solid, fill=red, red}]
  table[row sep=crcr]{%
4554	0.0209788023889824\\
15946	0.00884580123003661\\
59466	0.00287923439934592\\
229450	0\\
};
\addlegendentry{$N_{\mathrm{sec}}$=18, S1}

\addplot [color=blue, line width=1.1pt, mark size=2.0pt, mark=*, mark options={solid, blue}]
  table[row sep=crcr]{%
2538	0.0206441586379833\\
8874	0.00906236488248371\\
33066	0.0030466646082823\\
127530	0\\
};
\addlegendentry{$N_{\mathrm{sec}}$=10, S2}

\addplot [color=blue, dashdotted, line width=1.1pt, mark size=2.3pt, mark=triangle*, mark options={solid, fill=blue, blue}]
  table[row sep=crcr]{%
4554	0.0210146362870987\\
15946	0.00885664058099768\\
59466	0.00287889685402949\\
229450	0\\
};
\addlegendentry{$N_{\mathrm{sec}}$=18, S2}

\addplot [color=black, line width=1.1pt, mark size=2.0pt, mark=*, mark options={solid, black}]
  table[row sep=crcr]{%
2538	0.0206627573209798\\
8874	0.00907093598716402\\
33066	0.00304723884866654\\
127530	0\\
};
\addlegendentry{$N_{\mathrm{sec}}$=10, S3}

\addplot [color=black, dashdotted, line width=1.1pt, mark size=2.3pt, mark=triangle*, mark options={solid, fill=black, black}]
  table[row sep=crcr]{%
4554	0.0209493385849431\\
15946	0.00883293277931162\\
59466	0.00287490032633513\\
229450	0\\
};
\addlegendentry{$N_{\mathrm{sec}}$=18, S3}

\end{axis}
\end{tikzpicture}%
}
        \caption{Type D, $N_{\mathrm{var}}=5$}
        \label{fig:Lshape_k_40_a}
    \end{subfigure}
    \quad
 \begin{subfigure}[b]{0.48\textwidth}
\scalebox{0.65}[0.66]{
%
%
\begin{tikzpicture}

\begin{axis}[%
width=5.75cm,
height=4cm,
at={(0cm,0cm)},
scale only axis,
xmode=log,
xmin=3120,
xmax=100000,
xtick={   100,   1000,  10000, 100000},
xminorticks=true,
xlabel style={font=\color{white!15!black}},
xlabel={Degrees of freedom},
ymode=log,
ymin=0.00093640804976279,
ymax=0.01,
ytick={0.0001,  0.001,   0.01,    0.1},
yminorticks=true,
ylabel style={font=\color{white!15!black}},
ylabel={Relative error in $f_{obj}$},
axis background/.style={fill=white},
axis x line*=bottom,
axis y line*=left,
xminorgrids,
yminorgrids,
legend style={at={(1.03,1)}, anchor=north west, legend cell align=left, align=left, draw=white!15!black}
]
\addplot [color=red, line width=1.1pt, mark size=2.0pt, mark=*, mark options={solid, red}]
  table[row sep=crcr]{%
3120	0.008004904529089\\
9936	0.0032744424711304\\
35088	0.00097457397126266\\
131472	0\\
};
\addlegendentry{$N_{\mathrm{sec}}$=10, S1}

\addplot [color=red, dashdotted, line width=1.1pt, mark size=2.3pt, mark=triangle*, mark options={solid, fill=red, red}]
  table[row sep=crcr]{%
5600	0.00823539688300802\\
17856	0.00325017608266855\\
63104	0.00093640804976279\\
236544	0\\
};
\addlegendentry{$N_{\mathrm{sec}}$=18, S1}

\addplot [color=blue, line width=1.1pt, mark size=2.0pt, mark=*, mark options={solid, blue}]
  table[row sep=crcr]{%
3120	0.00800570985188765\\
9936	0.00327577187369242\\
35088	0.000975898925904175\\
131472	0\\
};
\addlegendentry{$N_{\mathrm{sec}}$=10, S2}

\addplot [color=blue, dashdotted, line width=1.1pt, mark size=2.3pt, mark=triangle*, mark options={solid, fill=blue, blue}]
  table[row sep=crcr]{%
5600	0.00823824179346581\\
17856	0.00325023120287408\\
63104	0.000940053782893205\\
236544	0\\
};
\addlegendentry{$N_{\mathrm{sec}}$=18, S2}

\addplot [color=black, line width=1.1pt, mark size=2.0pt, mark=*, mark options={solid, black}]
  table[row sep=crcr]{%
3120	0.0080165961781533\\
9936	0.00328001350817352\\
35088	0.000977619754526353\\
131472	0\\
};
\addlegendentry{$N_{\mathrm{sec}}$=10, S3}

\addplot [color=black, dashdotted, line width=1.1pt, mark size=2.3pt, mark=triangle*, mark options={solid, fill=black, black}]
  table[row sep=crcr]{%
5600	0.00825370737470759\\
17856	0.00325488816178774\\
63104	0.000942295780767392\\
236544	0\\
};
\addlegendentry{$N_{\mathrm{sec}}$=18, S3}

\end{axis}
\end{tikzpicture}%
}
        \caption{Type D, $N_{\mathrm{var}}=6$}
        \label{fig:Lshape_k_40_a}
    \end{subfigure}
\caption{Convergence of the objective function $f_{\mathrm{obj}}$ for thermal concentrator problem for Type-A ($N_{\mathrm{sec}}=$ 4, 12), Type-B ($N_{\mathrm{sec}}=$ 6, 14), Type-C ($N_{\mathrm{sec}}=$ 8, 16), Type-D ($N_{\mathrm{sec}}=$ 10, 18) configurations, and $N_{\mathrm{var}}=$ 5, 6. $4\%$ relative error (shown by pink horizontal line) is taken as the tolerance value to select the minimum mesh size for optimization. Samples 1, 2, and 3 are denoted by S1, S2 and S3. Type-A requires a finer mesh compared to Type-B and Type-D. Type-C needs a comparatively coarse mesh.
}
    \label{fig:Cntr convergence 2}
\end{figure}
\par At first, we study the convergence of the objective function with refinement of mesh. In this study, the optimization (with chosen $N_{\mathrm{sec}}$ and $N_{\mathrm{var}}$) is run with a coarse mesh to find the optimized shape. Then, the mesh is refined without changing the shape to investigate the effect of refinement on the objective function value. This study assists us in finding a minimum mesh size that can ensure sufficient accuracy for the objective function during optimization.  In \frefs{fig:Cntr convergence 1}-\ref{fig:Cntr convergence 2}, the relative error in the objective function value with respect to the degrees of freedom is plotted for several numbers of sectors as well as for numbers of design variables. The relative error is measured with reference to the most refined mesh for each case. $N_{\mathrm{sec}}$= $4,6,8,10,12,14,16$ and 18, while $N_{\mathrm{var}}$= $3,4,5$ and 6. For $N_{\mathrm{var}}=3,5$, $C^{0}$ continuity is applied between the interface NURBS elements, while for $N_{\mathrm{var}}=4,6$, $C^{1}$ continuity is applied.  We assume that a mesh with a relative error within $4\%$ is  sufficiently fine and will subsequently use this discretization for optimization. 
\par From \frefs{fig:Cntr convergence 1}-\ref{fig:Cntr convergence 2}, it is apparent that the Type-A configuration requires a finer mesh for convergence compared to Type-B and Type-D. On the other hand, Type-C needs a comparatively coarse mesh for convergence. Overall, the relative error tolerance can be reached by a mesh with less than $10^4$ degrees of freedom.
\subsubsection{Shape optimization}
\label{sec:Shape optimization}
\begin{figure}[htpb!]
    \centering
    \setlength\figureheight{1\textwidth}
    \setlength\figurewidth{1\textwidth}
    \begin{subfigure}[b]{0.48\textwidth}{\centering\includegraphics[width=1\textwidth]{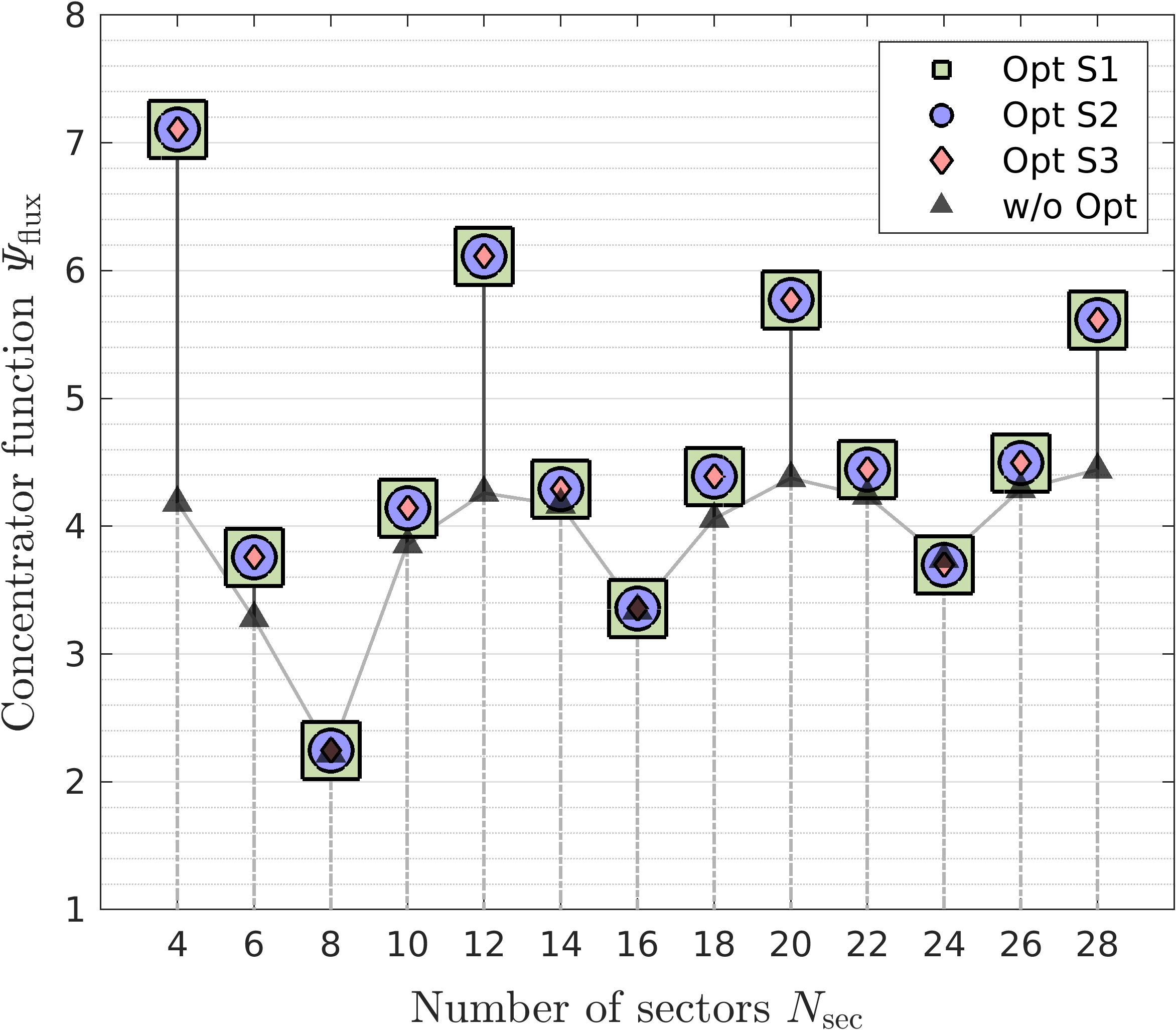}}
        \caption{$N_{\mathrm{var}}=3$}
        \label{figure:star_conv_plot_diff_p}
    \end{subfigure}
    \begin{subfigure}[b]{0.48\textwidth}{\centering\includegraphics[width=1\textwidth]{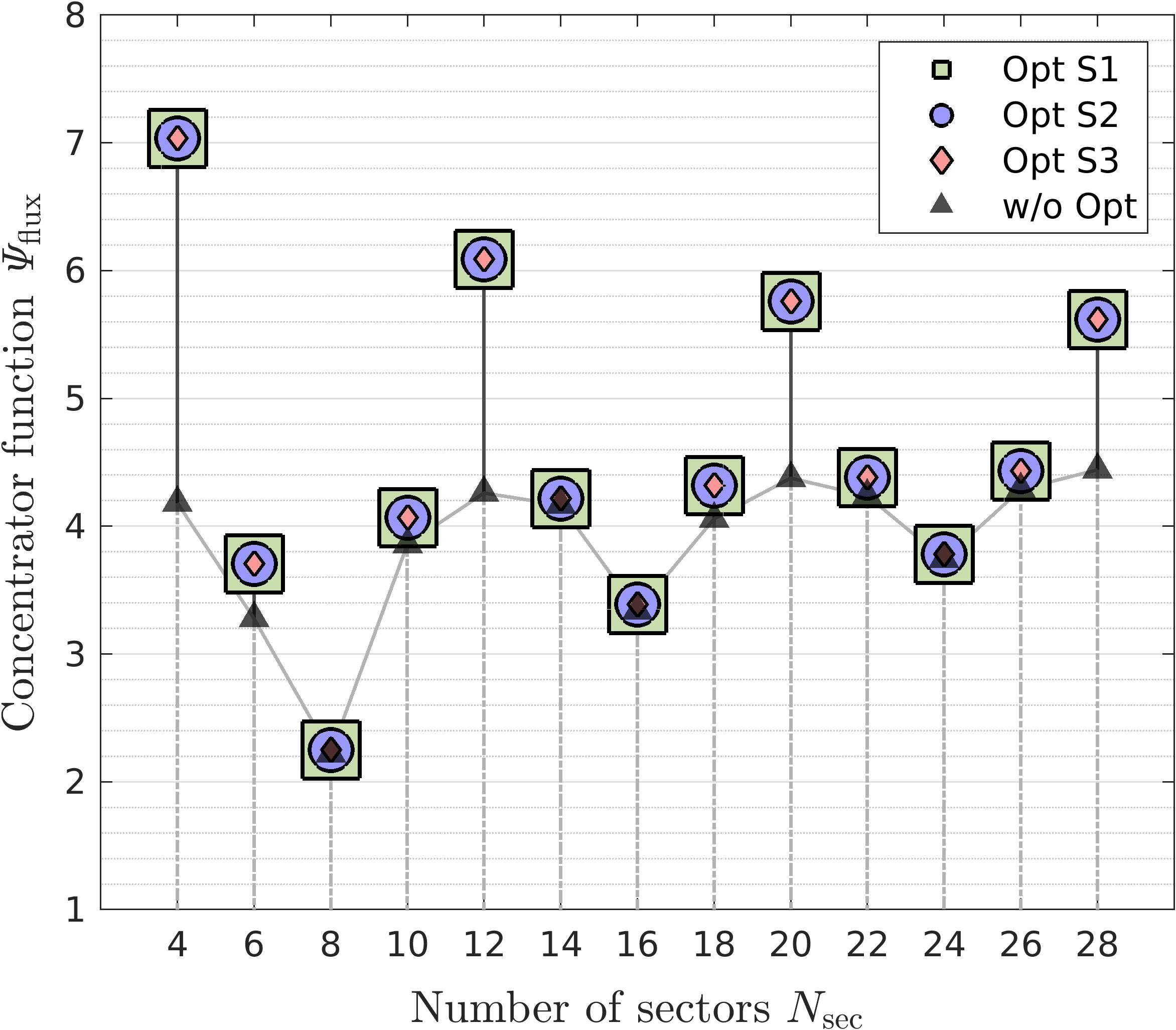}}
        \caption{$N_{\mathrm{var}}=4$}
        \label{figure:star_conv_plot_diff_p}
    \end{subfigure}\\
    \begin{subfigure}[b]{0.48\textwidth}{\centering\includegraphics[width=1\textwidth]{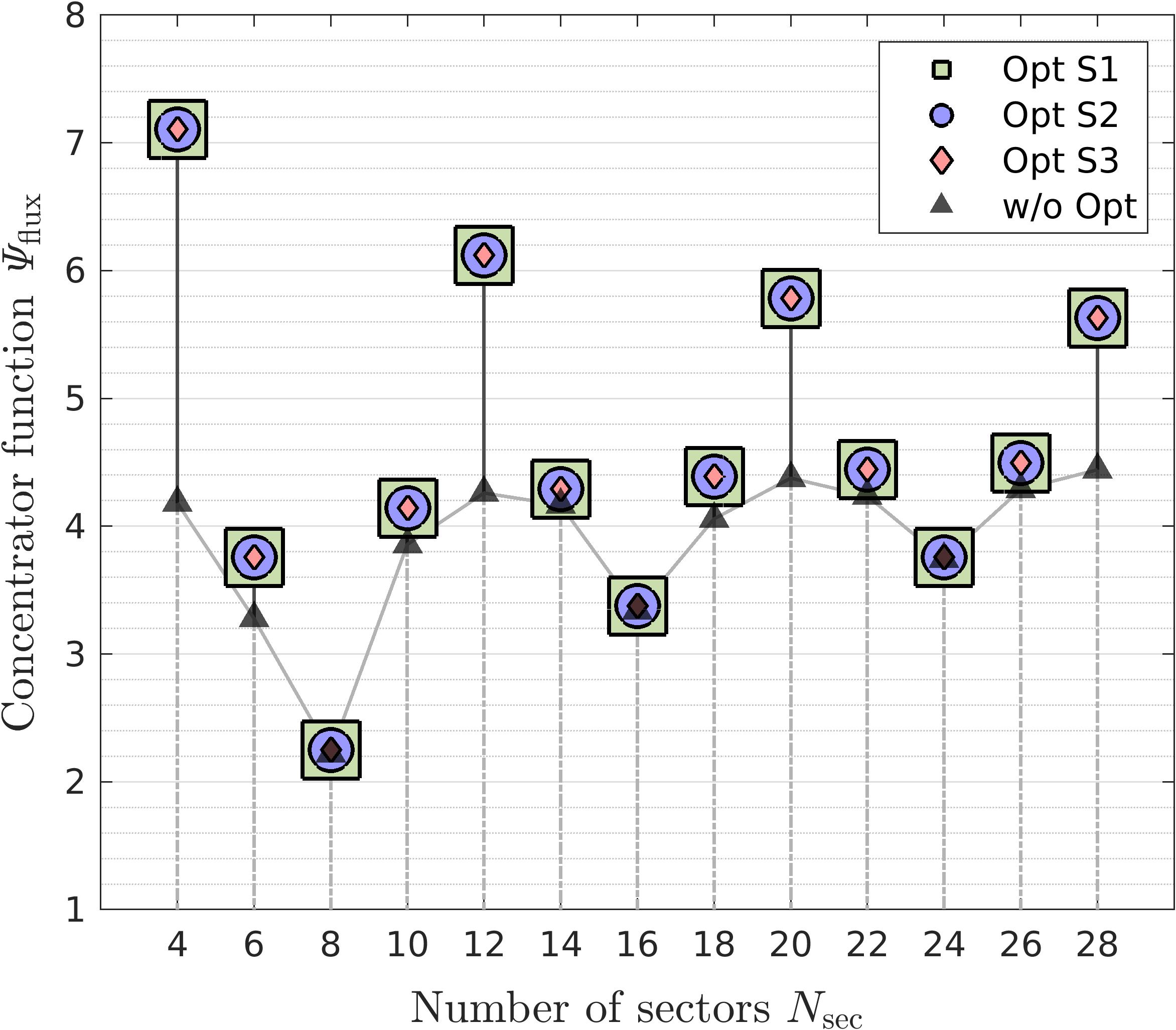}}
        \caption{$N_{\mathrm{var}}=5$}
        \label{figure:star_conv_plot_diff_p}
    \end{subfigure}
    \begin{subfigure}[b]{0.48\textwidth}{\centering\includegraphics[width=1\textwidth]{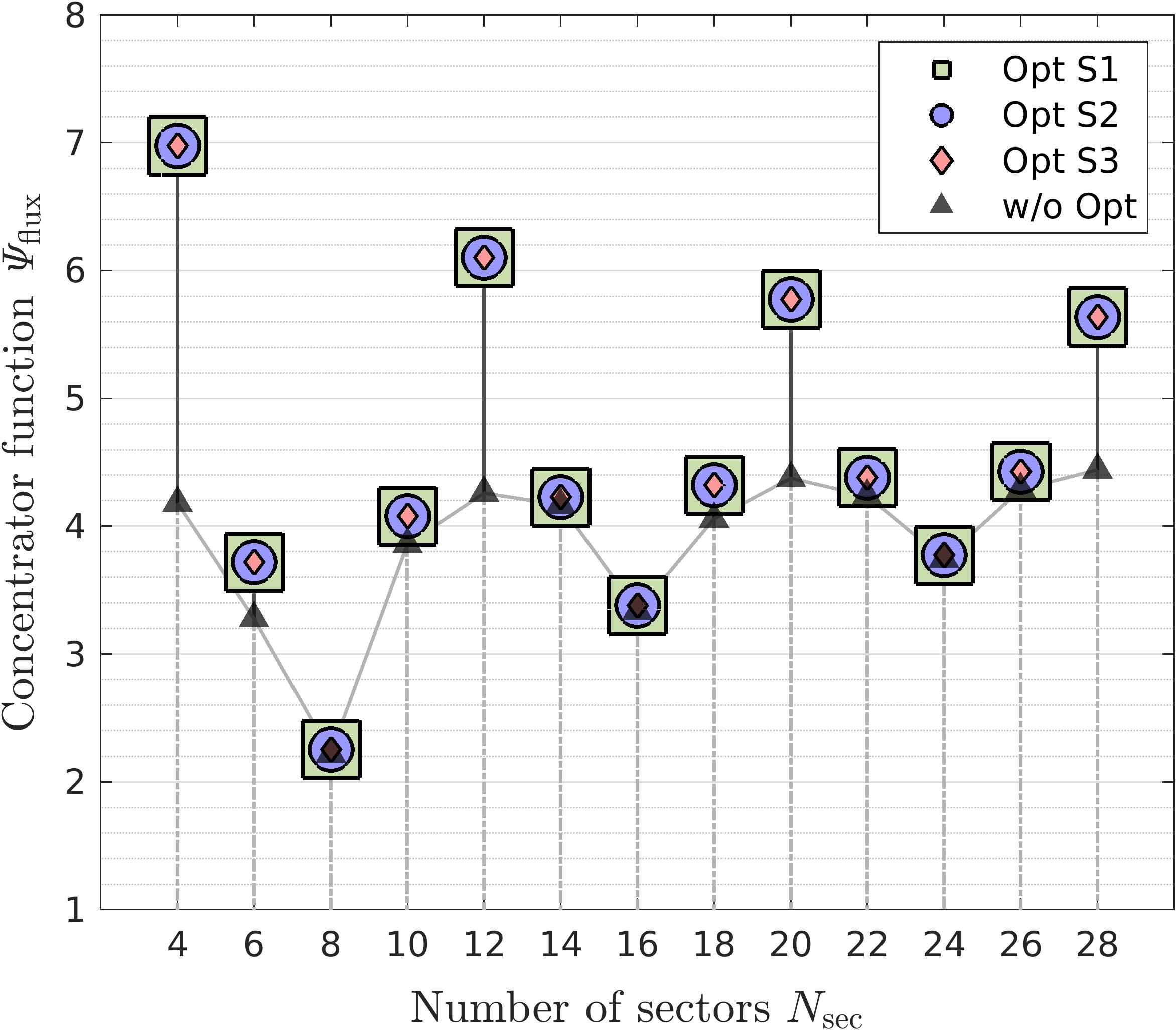}}
        \caption{$N_{\mathrm{var}}=6$}
        \label{figure:star_conv_plot_diff_p}
    \end{subfigure}
 \caption{Variation of concentrator function $\Psi_{\mathrm{flux}}$ of the optimized shape with respect to number of sectors for different number of design variables. $N_{\mathrm{var}}=$3, 4, 5, 6. Samples 1, 2, and 3 are denoted by S1, S2 and S3. The un-optimized geometry concentrates around 3 times more flux than a base material flat plate, while the optimized geometries concentrates up to 6 times more flux. Type-A configuration concentrates more flux compared to other configurations. There is no significant gain by using larger $N_{\mathrm{var}}$.}
 \label{fig:Cntr obj funtion variation}
\end{figure}

\begin{figure}[htpb!]
    \centering
    \setlength\figureheight{1\textwidth}
    \setlength\figurewidth{1\textwidth}
    \begin{subfigure}[b]{0.23\textwidth}{\centering\includegraphics[width=1\textwidth]{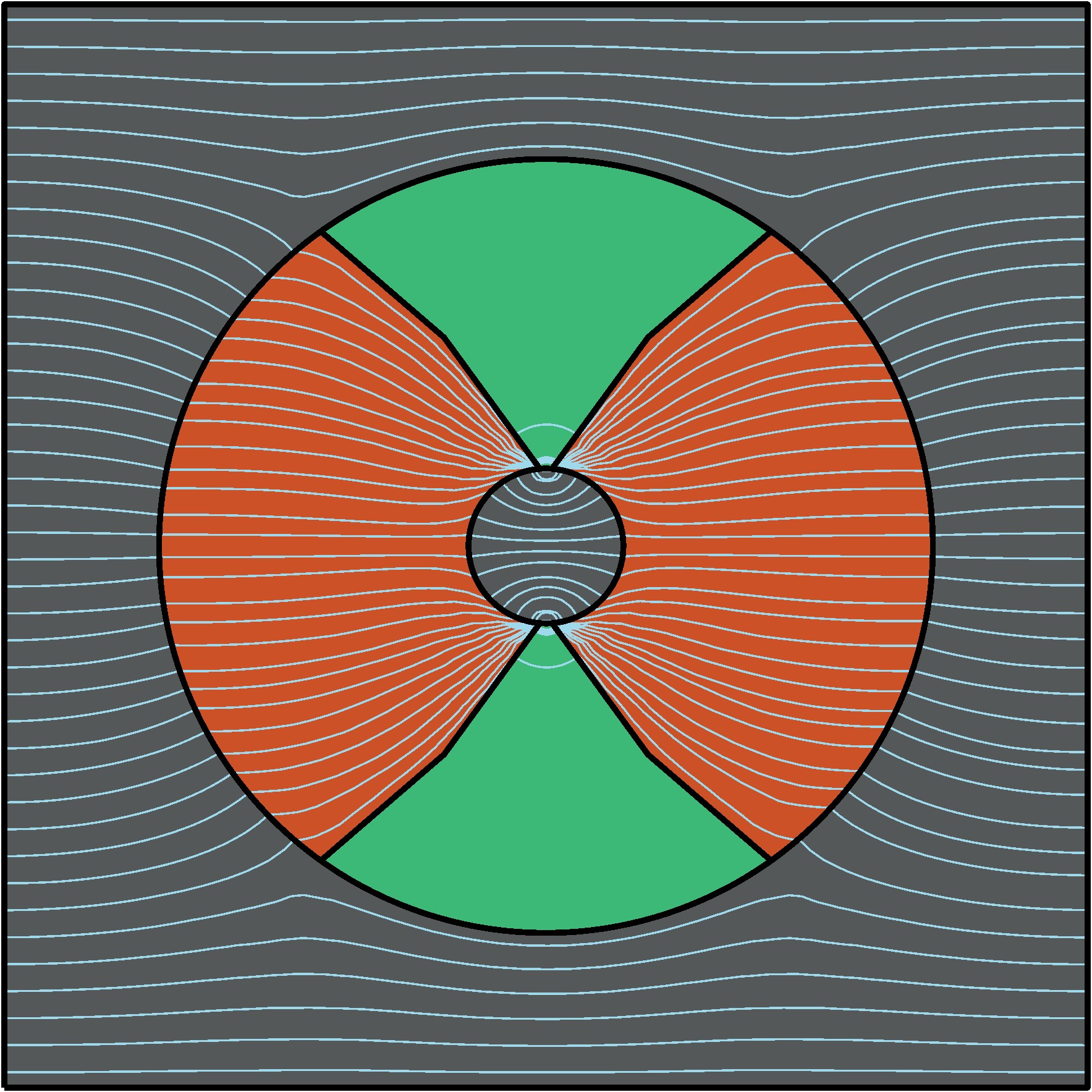}}
        \caption{\centering Type-A, $N_{\mathrm{sec}}=4$, $N_{\mathrm{var}}=3$}
        \label{figure:star_conv_plot_diff_p}
    \end{subfigure}
    \begin{subfigure}[b]{0.23\textwidth}{\centering\includegraphics[width=1\textwidth]{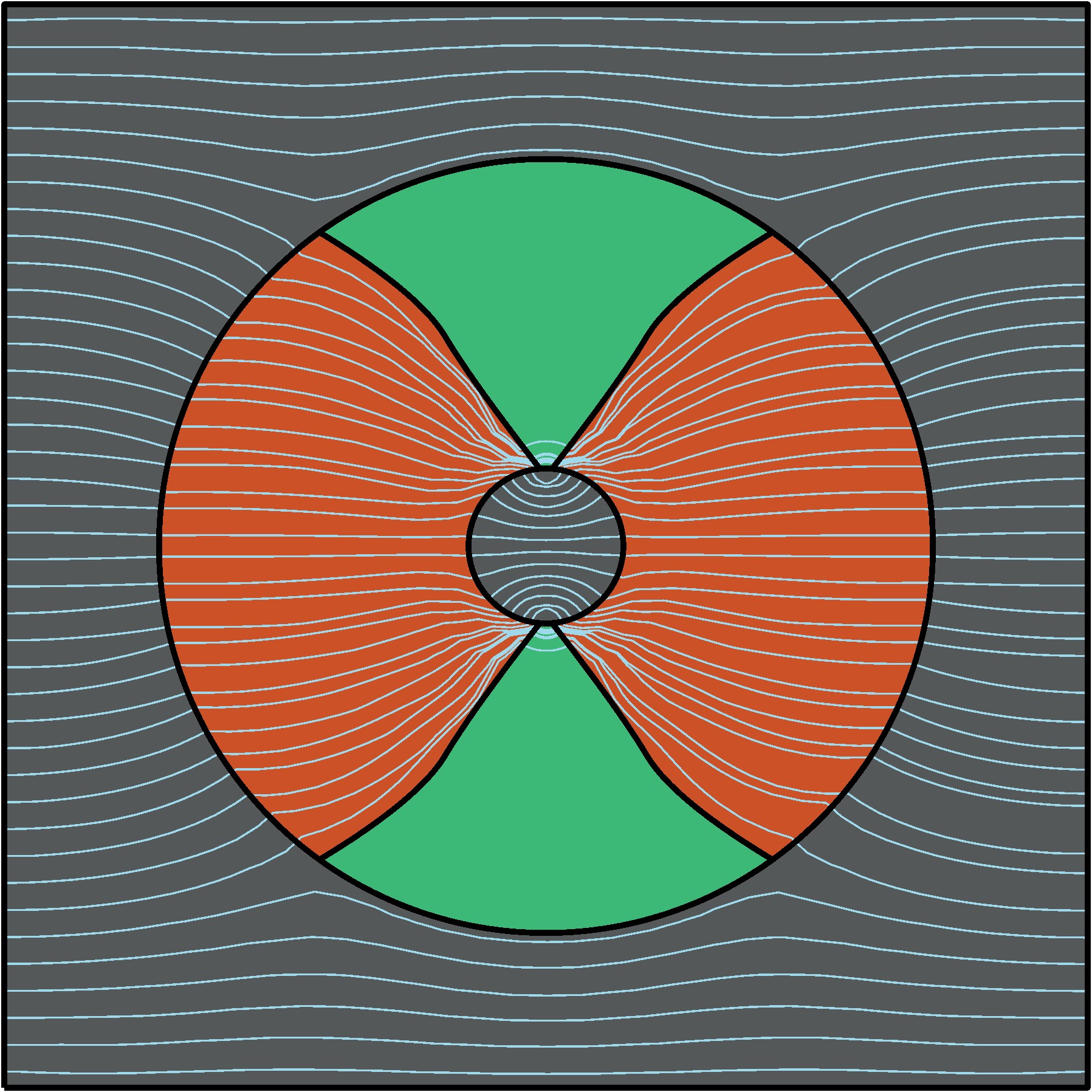}}
        \caption{\centering Type-A, $N_{\mathrm{sec}}=4$, $N_{\mathrm{var}}=4$}
        \label{figure:star_conv_plot_diff_p}
    \end{subfigure}    \begin{subfigure}[b]{0.23\textwidth}{\centering\includegraphics[width=1\textwidth]{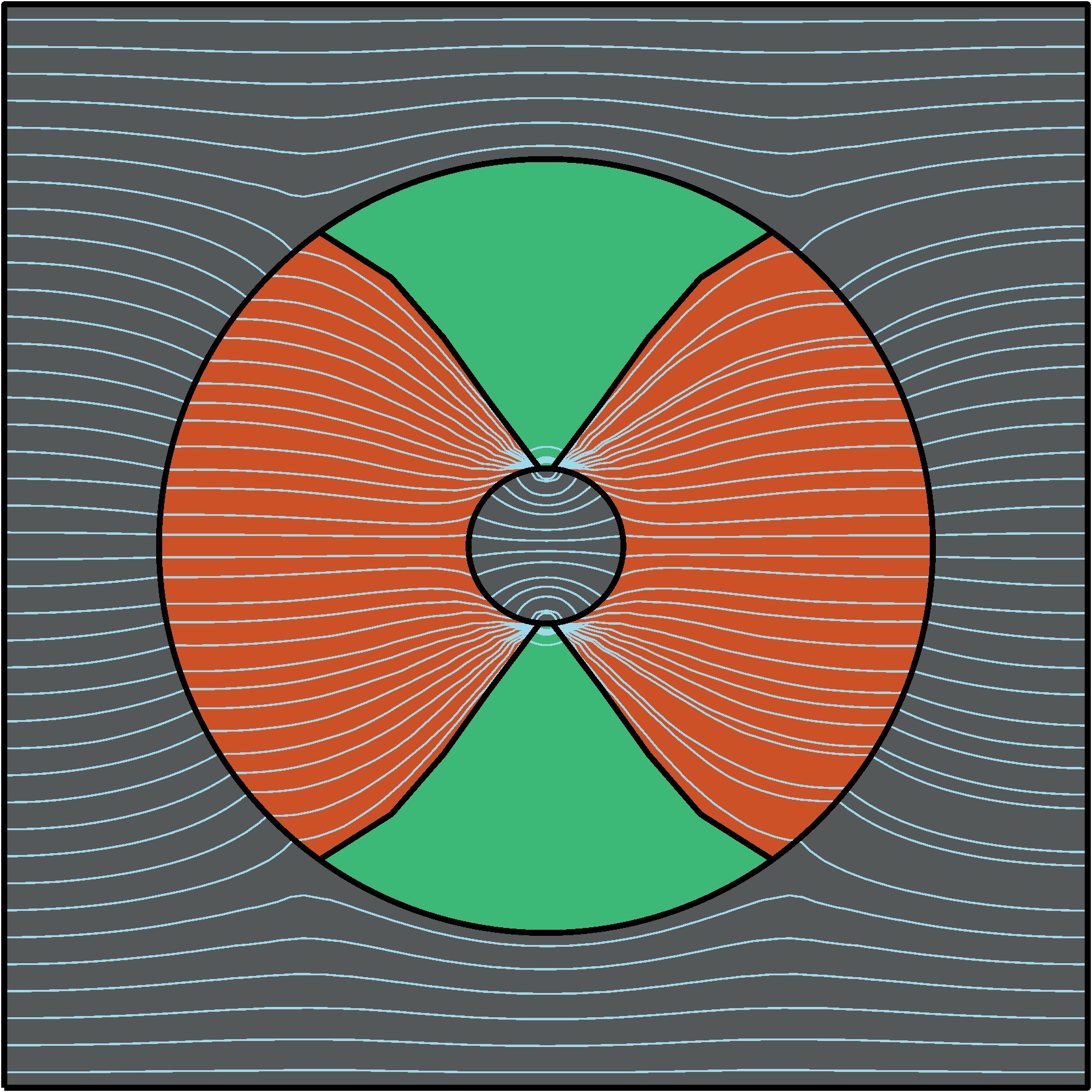}}
        \caption{\centering Type-A, $N_{\mathrm{sec}}=4$, $N_{\mathrm{var}}=5$}
        \label{figure:star_conv_plot_diff_p}
    \end{subfigure}    \begin{subfigure}[b]{0.23\textwidth}{\centering\includegraphics[width=1\textwidth]{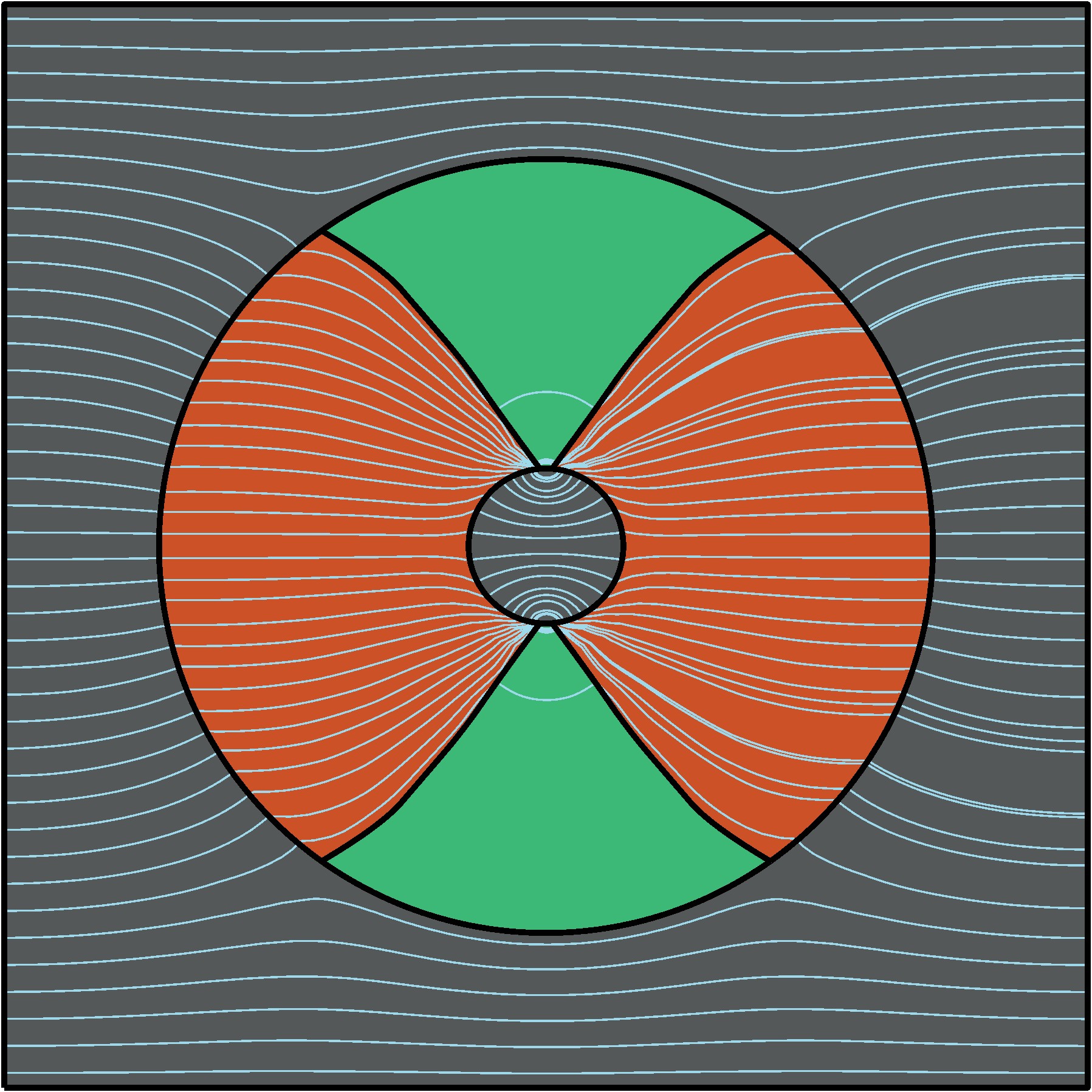}}
        \caption{\centering Type-A, $N_{\mathrm{sec}}=4$, $N_{\mathrm{var}}=6$}
        \label{figure:star_conv_plot_diff_p}
    \end{subfigure}\\
        \begin{subfigure}[b]{0.23\textwidth}{\centering\includegraphics[width=1\textwidth]{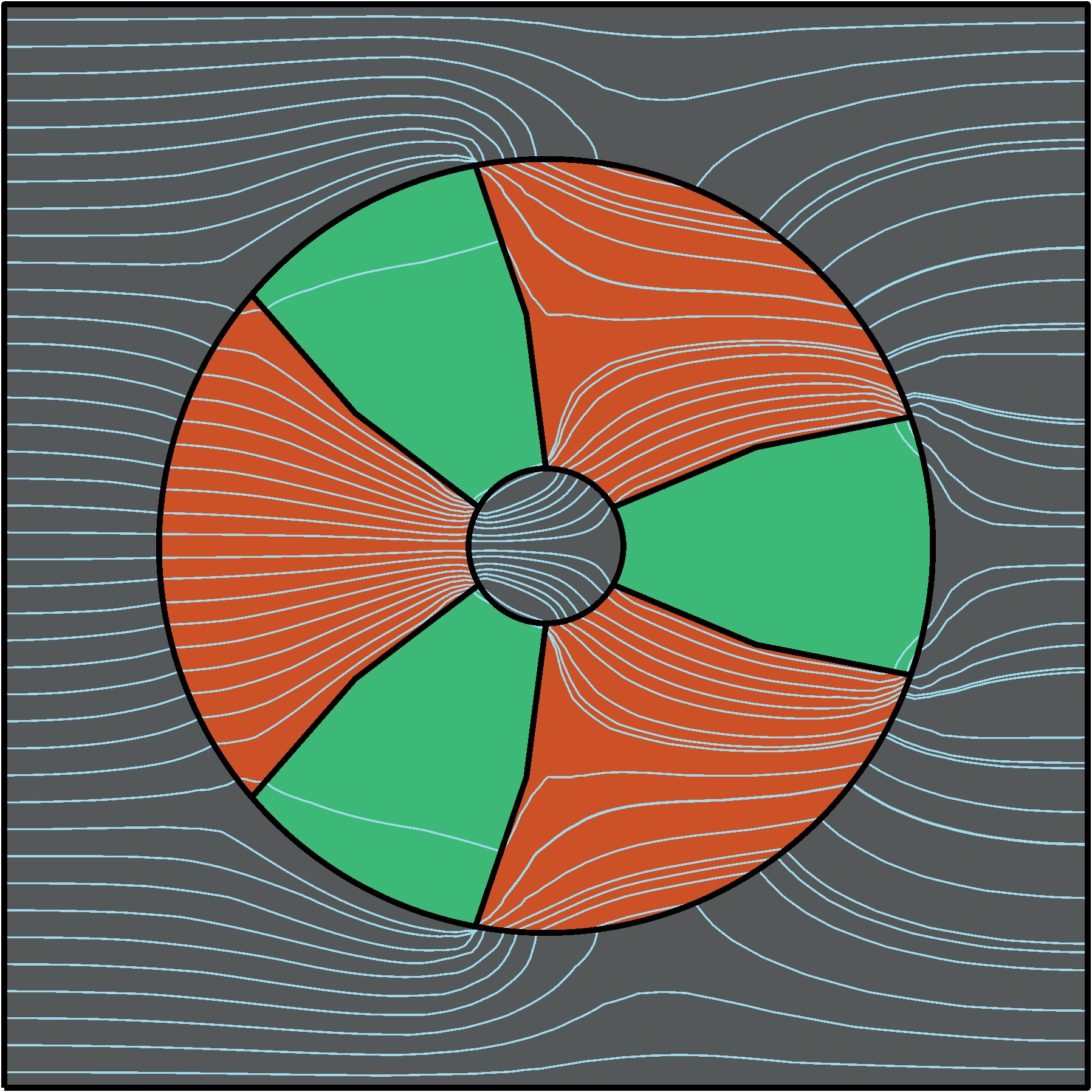}}
        \caption{\centering Type-B, $N_{\mathrm{sec}}=6$, $N_{\mathrm{var}}=3$}
        \label{figure:star_conv_plot_diff_p}
    \end{subfigure}
    \begin{subfigure}[b]{0.23\textwidth}{\centering\includegraphics[width=1\textwidth]{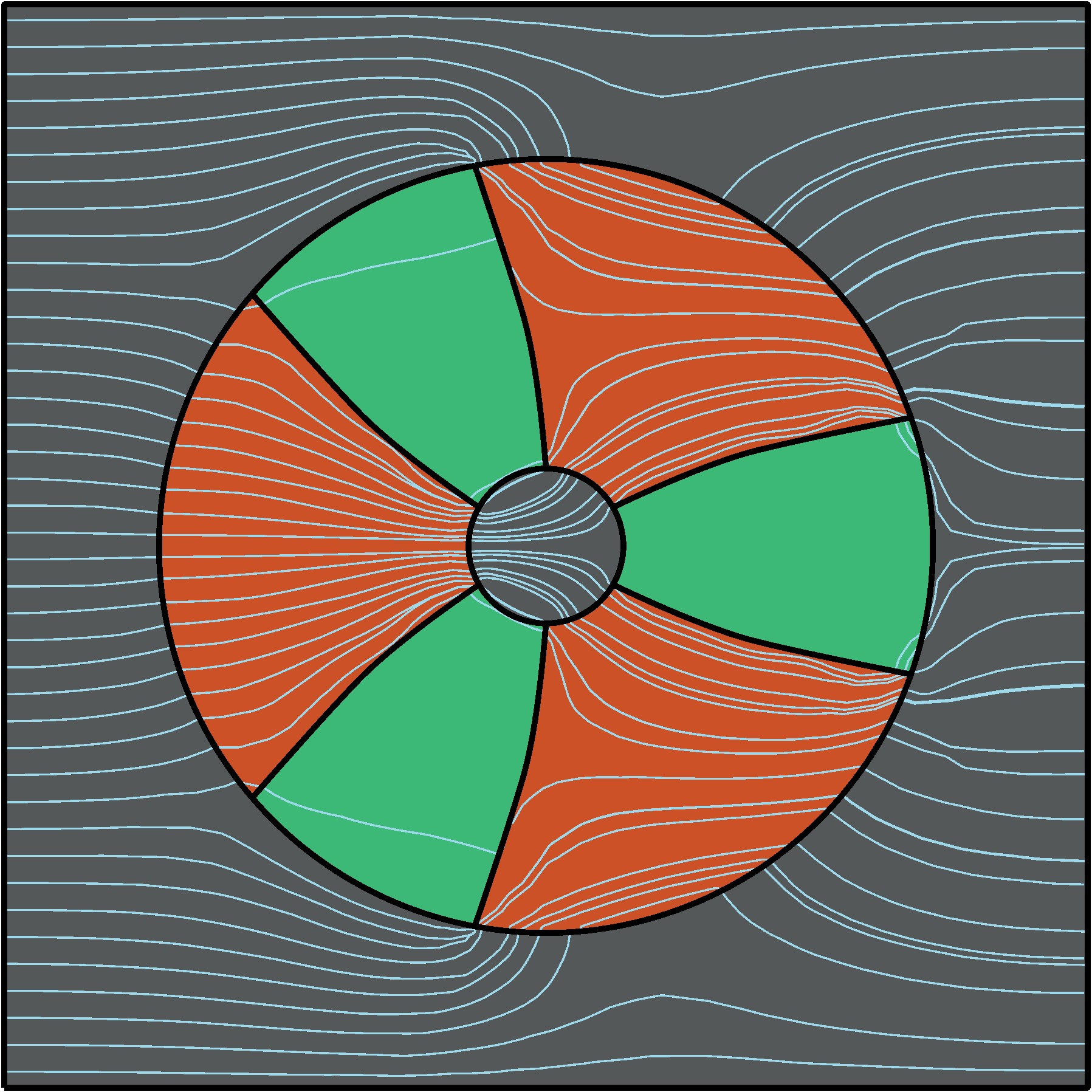}}
        \caption{\centering Type-B, $N_{\mathrm{sec}}=6$, $N_{\mathrm{var}}=4$}
        \label{figure:star_conv_plot_diff_p}
    \end{subfigure}    \begin{subfigure}[b]{0.225\textwidth}{\centering\includegraphics[width=1\textwidth]{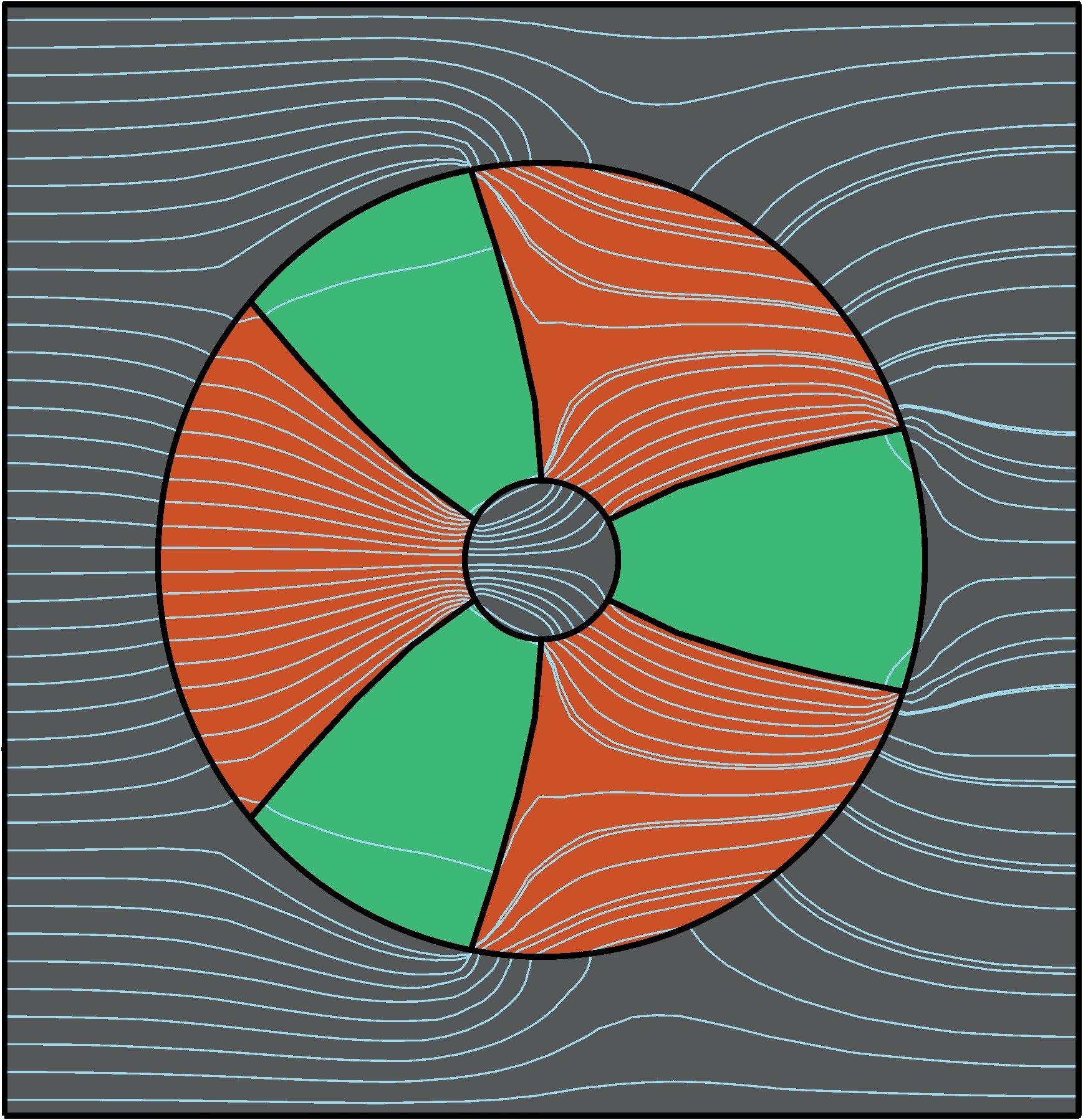}}
        \caption{\centering Type-B, $N_{\mathrm{sec}}=6$, $N_{\mathrm{var}}=5$}
        \label{figure:star_conv_plot_diff_p}
    \end{subfigure}    \begin{subfigure}[b]{0.23\textwidth}{\centering\includegraphics[width=1\textwidth]{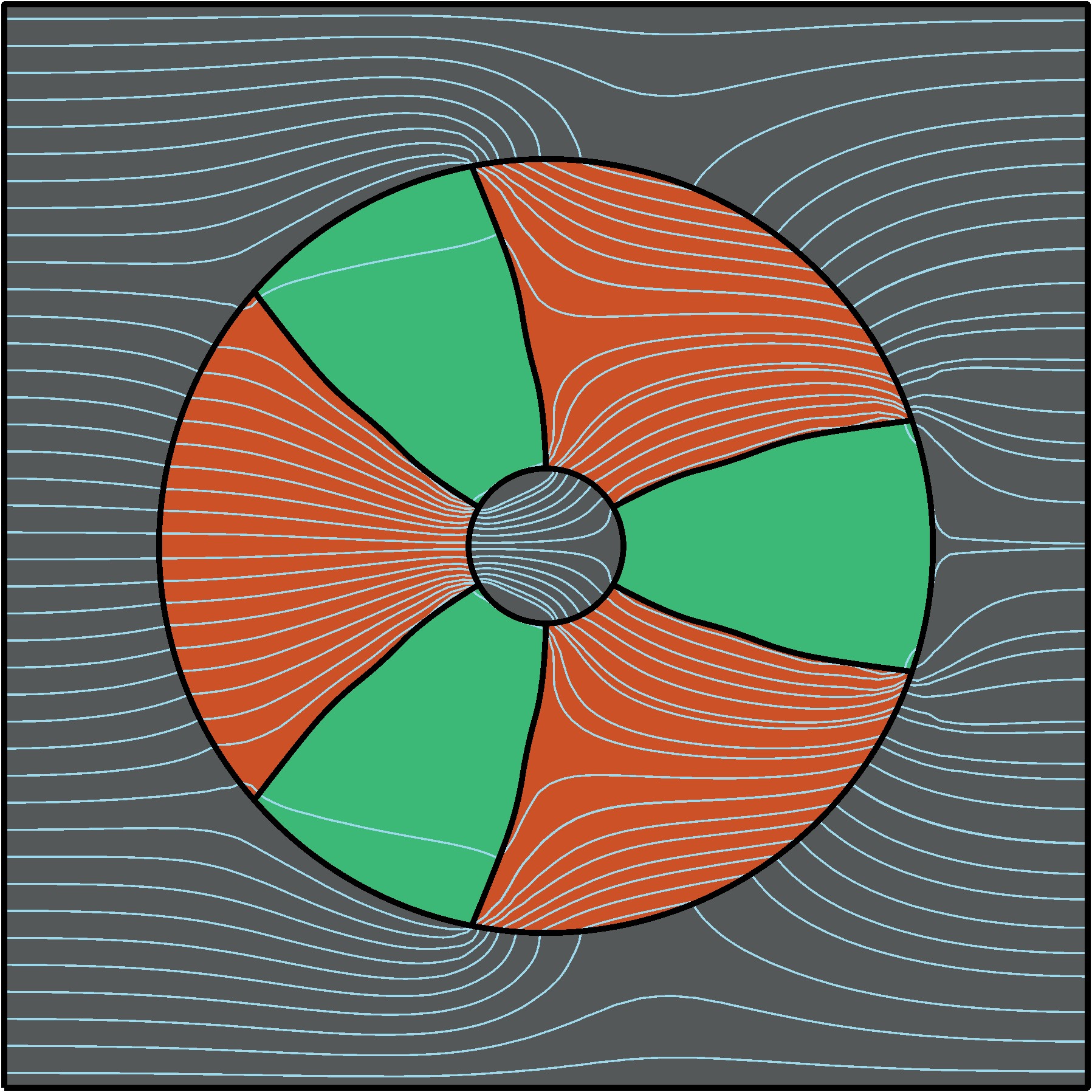}}
        \caption{\centering Type-B, $N_{\mathrm{sec}}=6$, $N_{\mathrm{var}}=6$}
        \label{figure:star_conv_plot_diff_p}
    \end{subfigure}\\
        \begin{subfigure}[b]{0.23\textwidth}{\centering\includegraphics[width=1\textwidth]{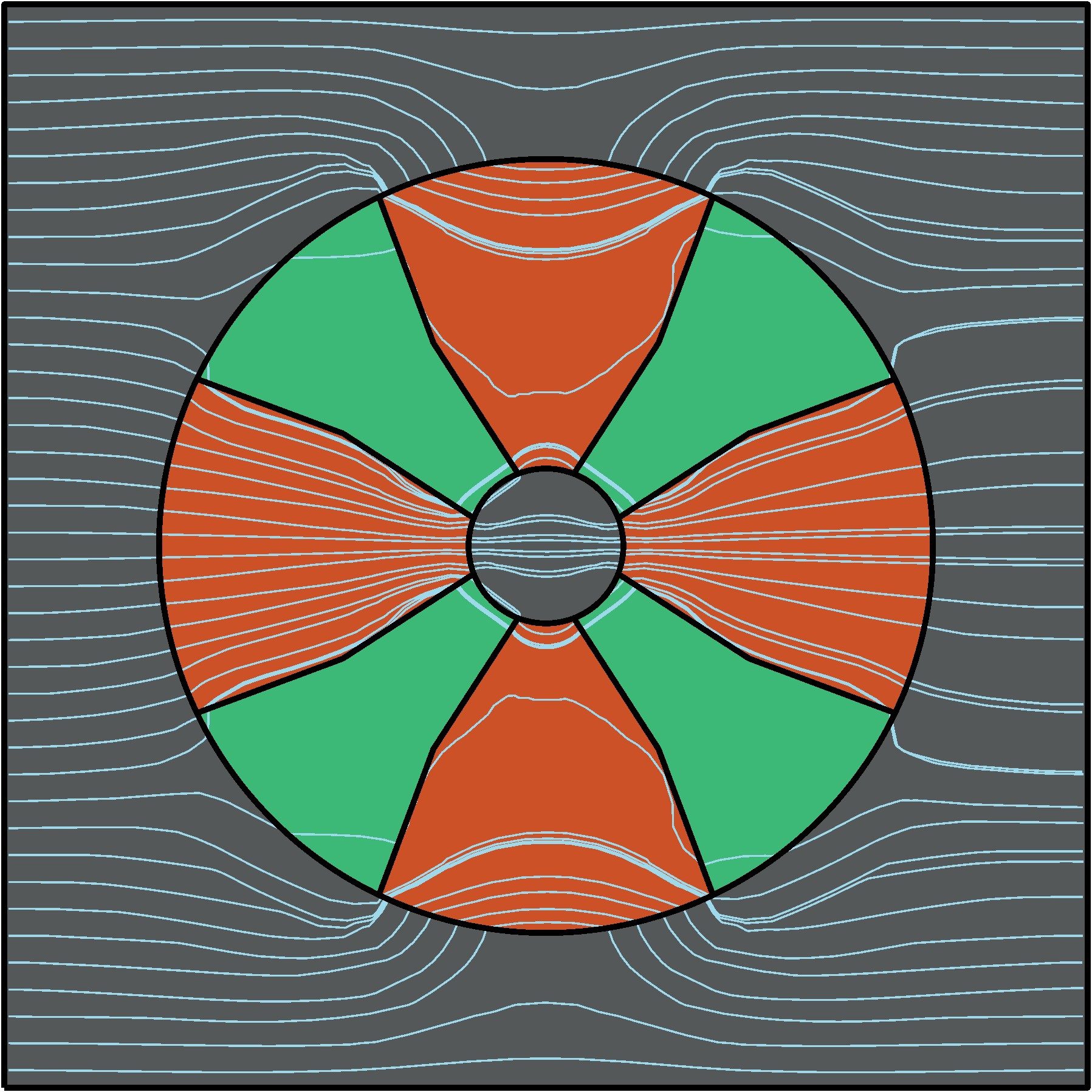}}
        \caption{\centering Type-C, $N_{\mathrm{sec}}=8$, $N_{\mathrm{var}}=3$}
        \label{figure:star_conv_plot_diff_p}
    \end{subfigure}
    \begin{subfigure}[b]{0.23\textwidth}{\centering\includegraphics[width=1\textwidth]{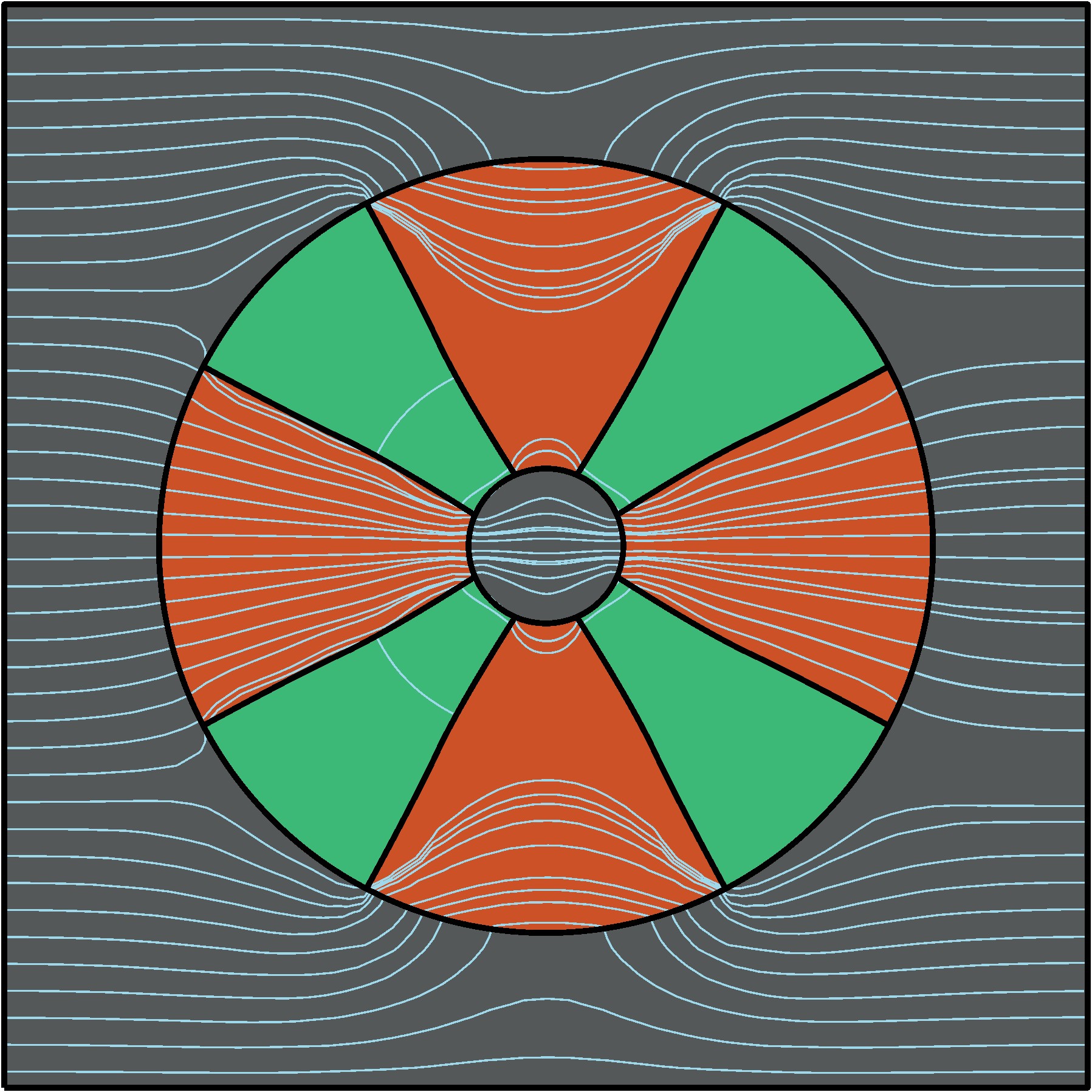}}
        \caption{\centering Type-C, $N_{\mathrm{sec}}=8$, $N_{\mathrm{var}}=4$}
        \label{figure:star_conv_plot_diff_p}
    \end{subfigure}    \begin{subfigure}[b]{0.23\textwidth}{\centering\includegraphics[width=1\textwidth]{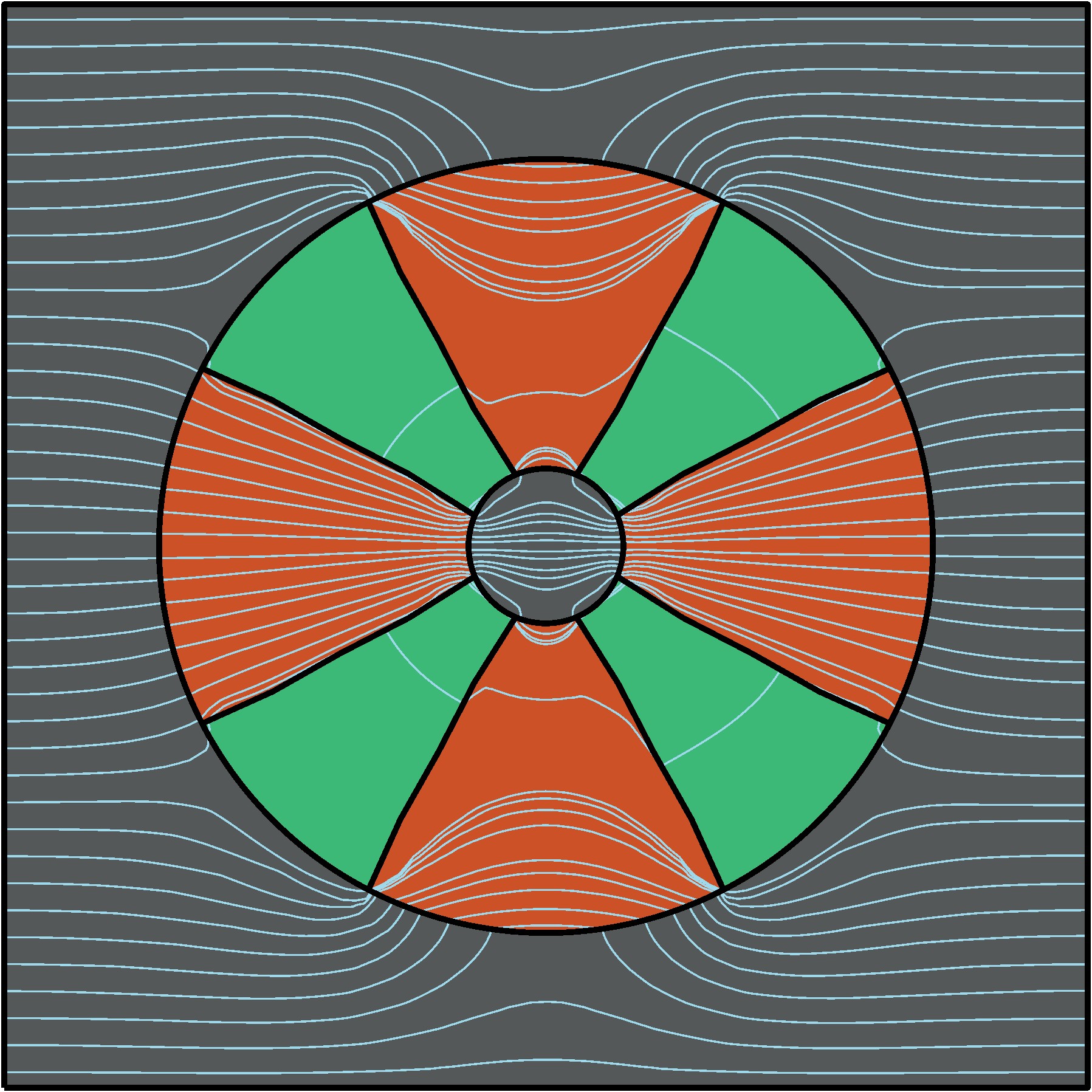}}
        \caption{\centering Type-C, $N_{\mathrm{sec}}=8$, $N_{\mathrm{var}}=5$}
        \label{figure:star_conv_plot_diff_p}
    \end{subfigure}    \begin{subfigure}[b]{0.23\textwidth}{\centering\includegraphics[width=1\textwidth]{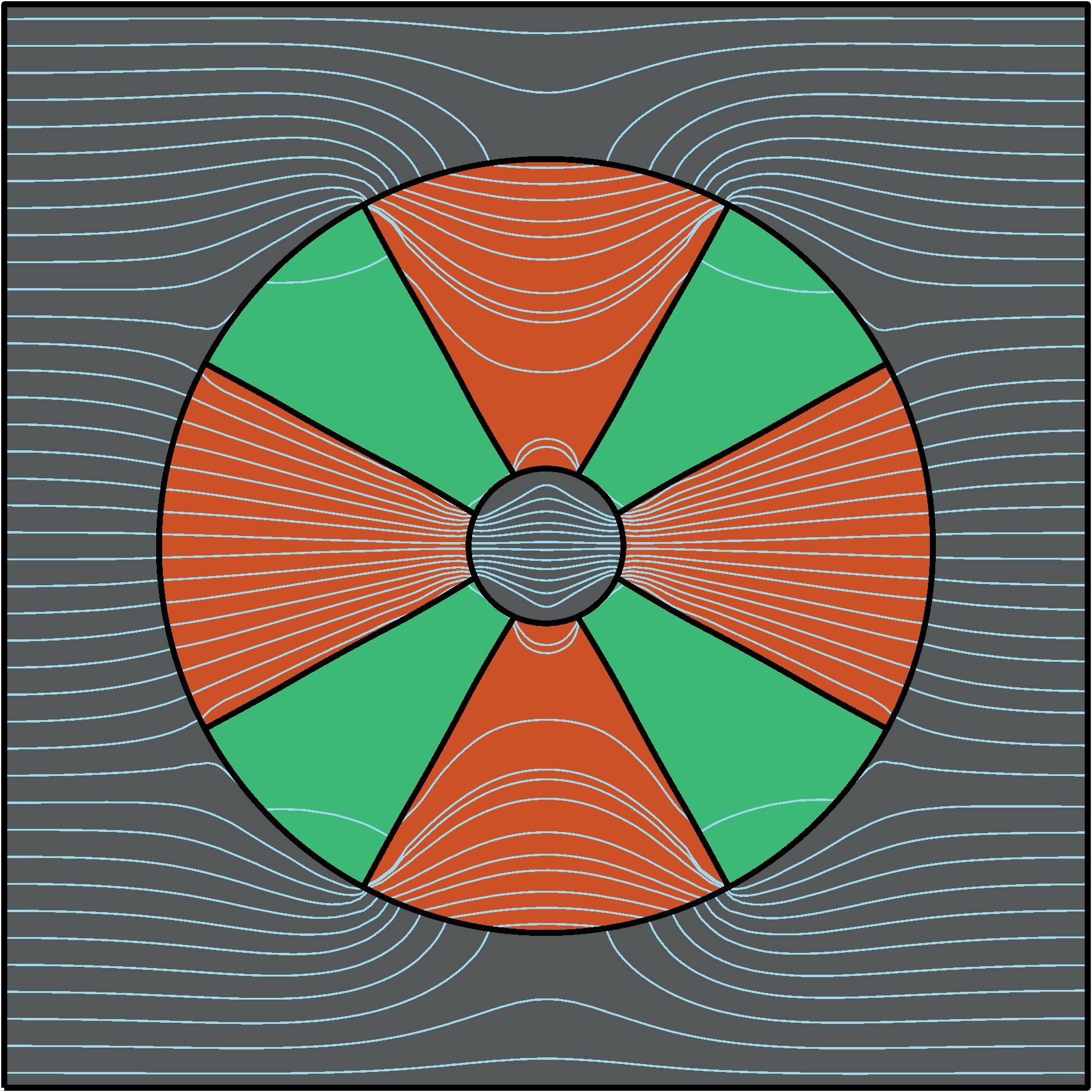}}
        \caption{\centering Type-C, $N_{\mathrm{sec}}=8$, $N_{\mathrm{var}}=6$}
        \label{figure:star_conv_plot_diff_p}
    \end{subfigure}\\
        \begin{subfigure}[b]{0.23\textwidth}{\centering\includegraphics[width=1\textwidth]{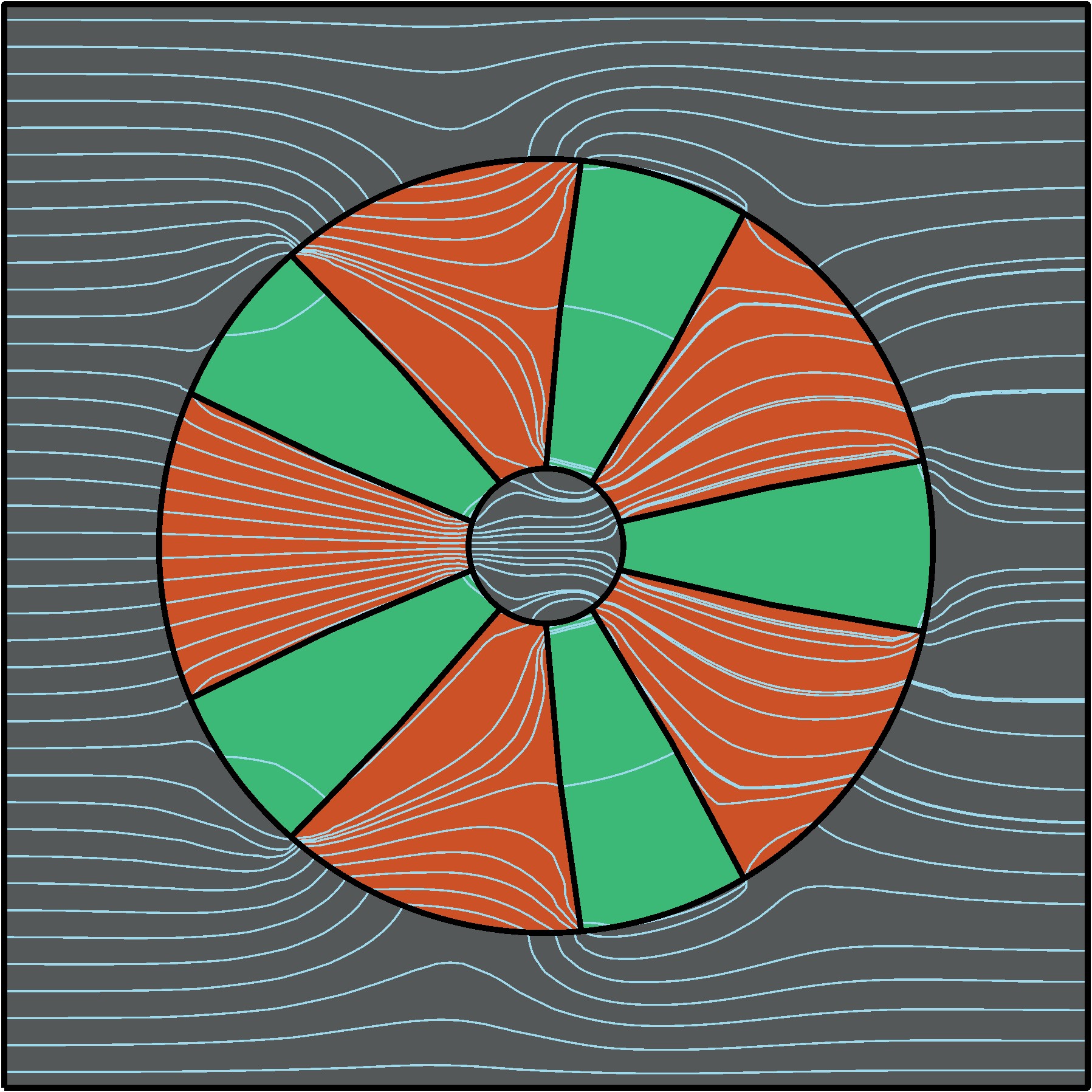}}
        \caption{\centering Type-D, $N_{\mathrm{sec}}=10$, $N_{\mathrm{var}}=3$}
        \label{figure:star_conv_plot_diff_p}
    \end{subfigure}
    \begin{subfigure}[b]{0.23\textwidth}{\centering\includegraphics[width=1\textwidth]{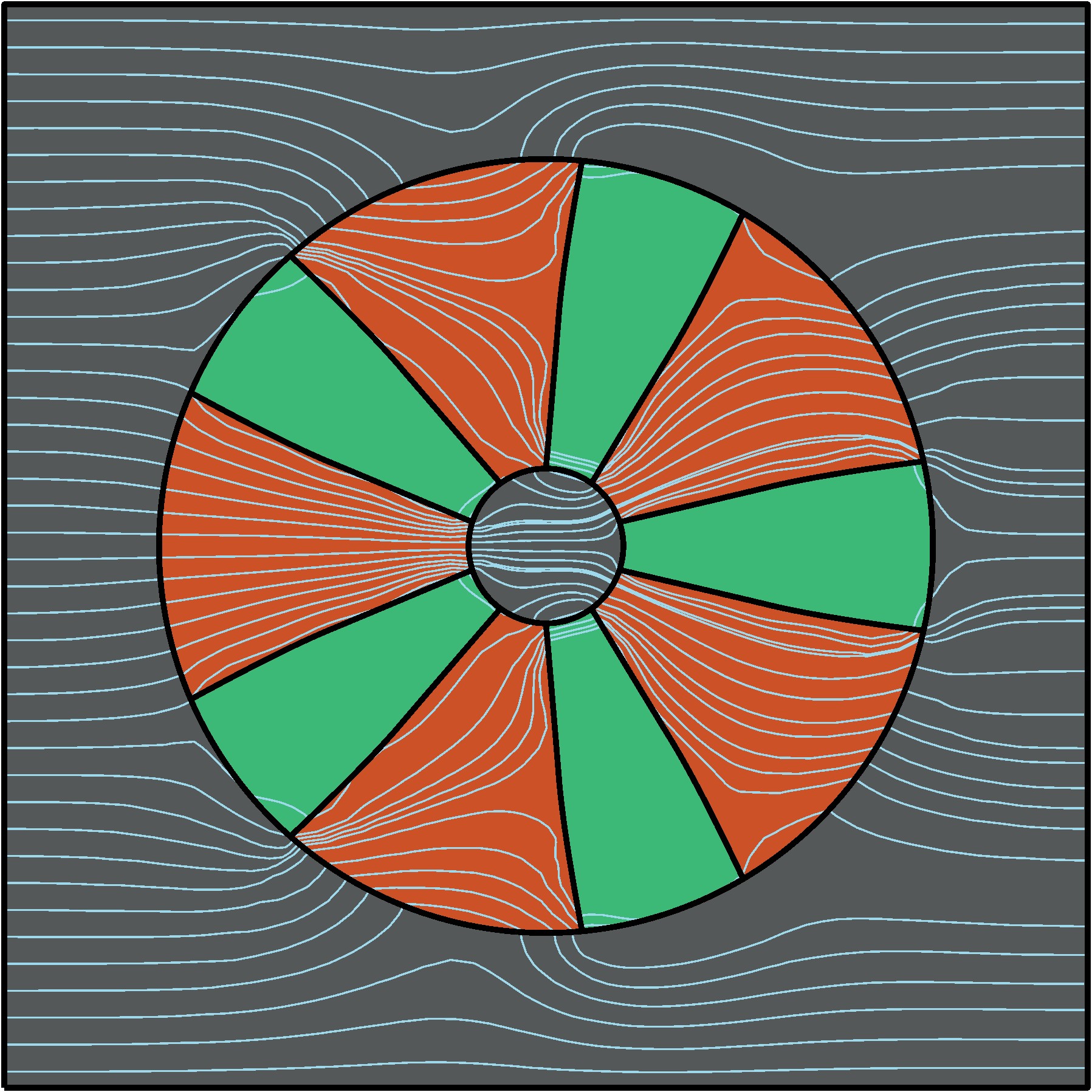}}
        \caption{\centering Type-D, $N_{\mathrm{sec}}=10$, $N_{\mathrm{var}}=4$}
        \label{figure:star_conv_plot_diff_p}
    \end{subfigure}    \begin{subfigure}[b]{0.23\textwidth}{\centering\includegraphics[width=1\textwidth]{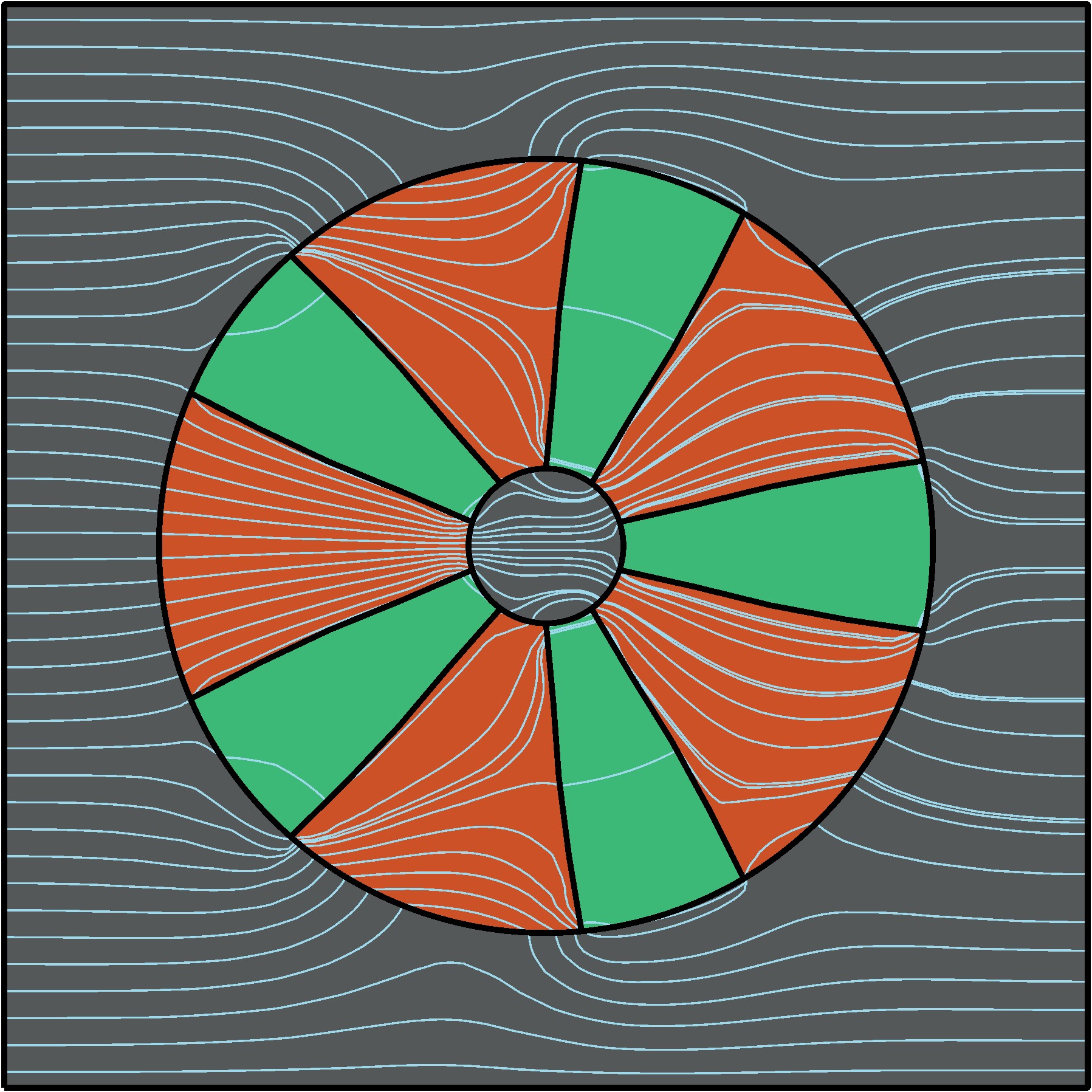}}
        \caption{\centering Type-D, $N_{\mathrm{sec}}=10$, $N_{\mathrm{var}}=5$}
        \label{figure:star_conv_plot_diff_p}
    \end{subfigure}    \begin{subfigure}[b]{0.23\textwidth}{\centering\includegraphics[width=1\textwidth]{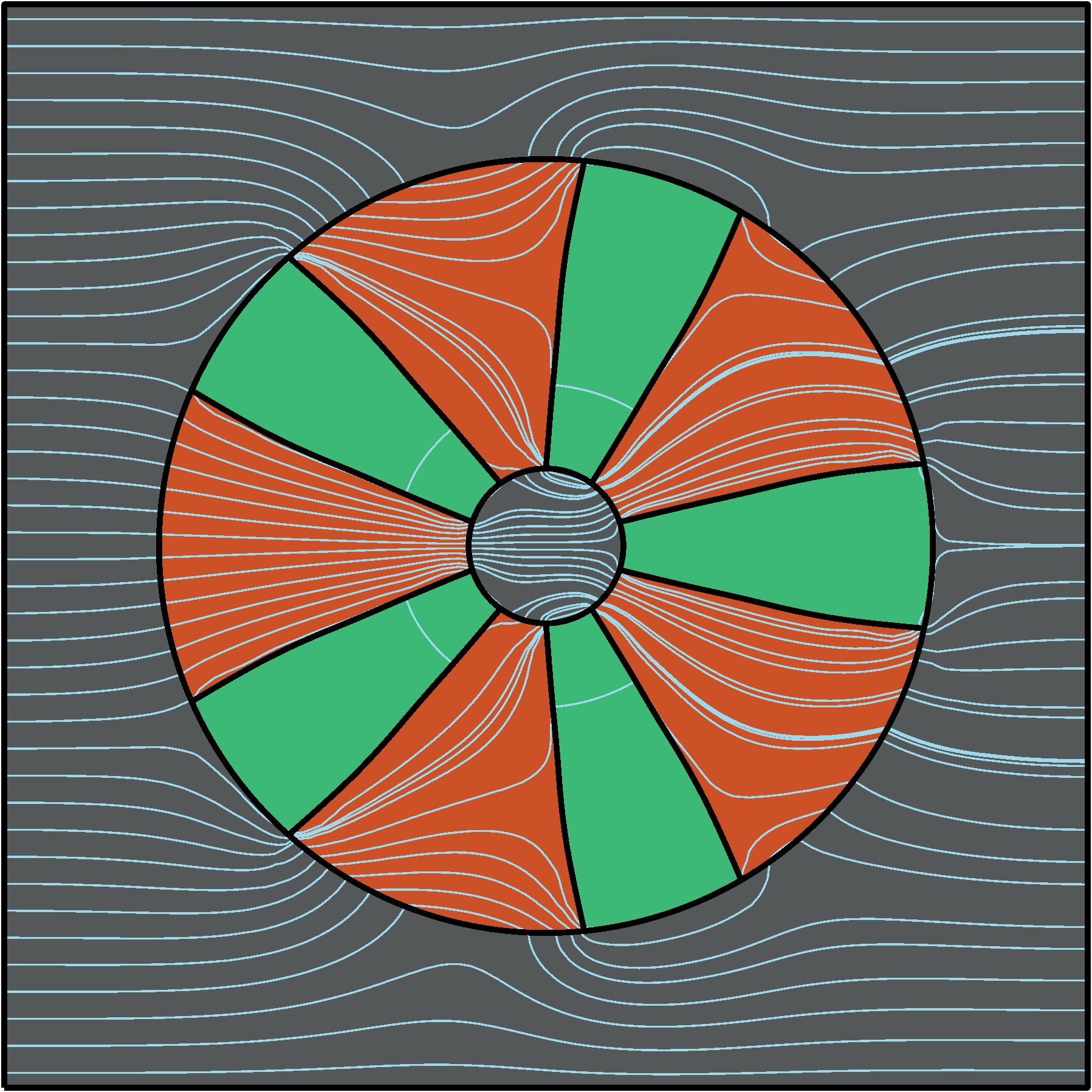}}
        \caption{\centering Type-D, $N_{\mathrm{sec}}=10$, $N_{\mathrm{var}}=6$}
        \label{figure:star_conv_plot_diff_p}
    \end{subfigure}
\caption{Flux flow in the optimized shape for different number of sectors and number of design variables. $N_{\mathrm{sec}}=$4 (Type-A), 6 (Type-B), 8 (Type-C), 10 (Type-D) and $N_{\mathrm{var}}=$3, 4, 5, 6. Type-A configuration concentrates more flux compared to other configurations. Increasing $N_{\mathrm{var}}$ only brush up the details of the optimized interfaces.}
 \label{fig:Cntr flux plot}
\end{figure} 

\par \fref{fig:Cntr obj funtion variation} and \tref{table:obj fun variation} show the concentrator function $\Psi_{\mathrm{flux}}$ variation with respect to number of sectors for $N_{\mathrm{var}}=3,4,5,6$. Here, we show $\Psi_{\mathrm{flux}}$-value for a geometry with straight lines as interfaces (i.e. geometry without any optimization, same as utilized in \cite{Chen2015}) as well as the optimized geometries obtained by three sample optimization runs. For each sample, PSO algorithm goes through and compares several solutions (completely different from another sample but in the design space) created by random numbers before eventually leading to an optimized shape at the end. We can consider each sample as a separate path from initial guess to final optimized solution. As can be seen from \fref{fig:Cntr obj funtion variation},  $f_{\mathrm{obj}}$ is almost constant for all three runs of optimization, which corroborates the reproducibility of the present optimization process and sufficiency of the exploited mesh for calculating $f_{\mathrm{obj}}$ accurately. \fref{fig:Cntr obj funtion variation} highlights the benefit achieved in $\Psi_{\mathrm{flux}}$-value with optimization. We can notice that Type-A configuration has bigger $\Psi_{\mathrm{flux}}$-value compared to all other configurations. In particular, $N_{\mathrm{sec}}=4$ performs better than any other number of sectors. For $N_{\mathrm{sec}}=4$, the un-optimized geometry concentrates around 3 times more flux than a base material flat plate, while the optimized geometry concentrates approximately 6 times more flux (refer \tref{table:obj fun variation}). 
\par Now, we analyse the effect of $N_{\mathrm{var}}$ on objective function. To understand the effect of $N_{\mathrm{var}}$, streamline plots are plotted for Type A to D configuration ($N_{\mathrm{sec}}=4,6,8,10$) in \fref{fig:Cntr flux plot}. From the figure, we can observe that the optimized shapes are more or less identical for different $N_{\mathrm{var}}$, and there is no significant gain by using larger $N_{\mathrm{var}}$. As large $N_{\mathrm{var}}$ is the indication of more degrees of freedom to manipulate the shape, increasing $N_{\mathrm{var}}$ only brushes up the details of the optimized interface.  Especially $N_{\mathrm{var}}=4,6$, which have $C^{1}$ continuity between interface NURBS elements, smoothens the interface between the sectors. Nevertheless, the improvement provided in the objective function value is very limited. 
\subsubsection{Shape optimization combined with conductivity optimization}
\label{sec:Shape optimization and k opt}
 \begin{figure}[htpb!]
    \centering
    \setlength\figureheight{1\textwidth}
    \setlength\figurewidth{1\textwidth}
    \begin{subfigure}[b]{0.48\textwidth}{\centering\includegraphics[width=1\textwidth]{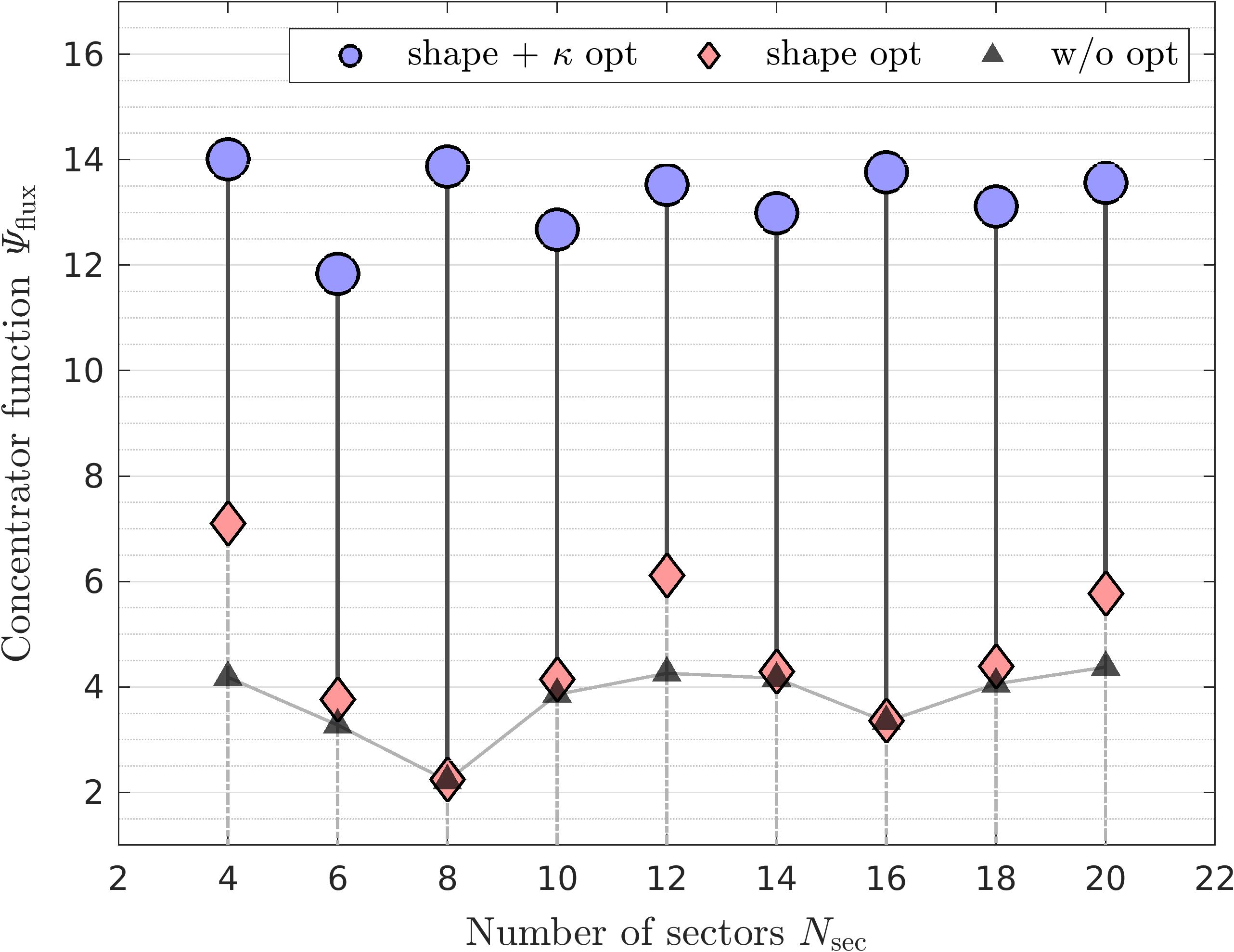}}
        \caption{$N_{\mathrm{var}}=3$}
        \label{figure:star_conv_plot_diff_p}
    \end{subfigure}
    \begin{subfigure}[b]{0.48\textwidth}{\centering\includegraphics[width=1\textwidth]{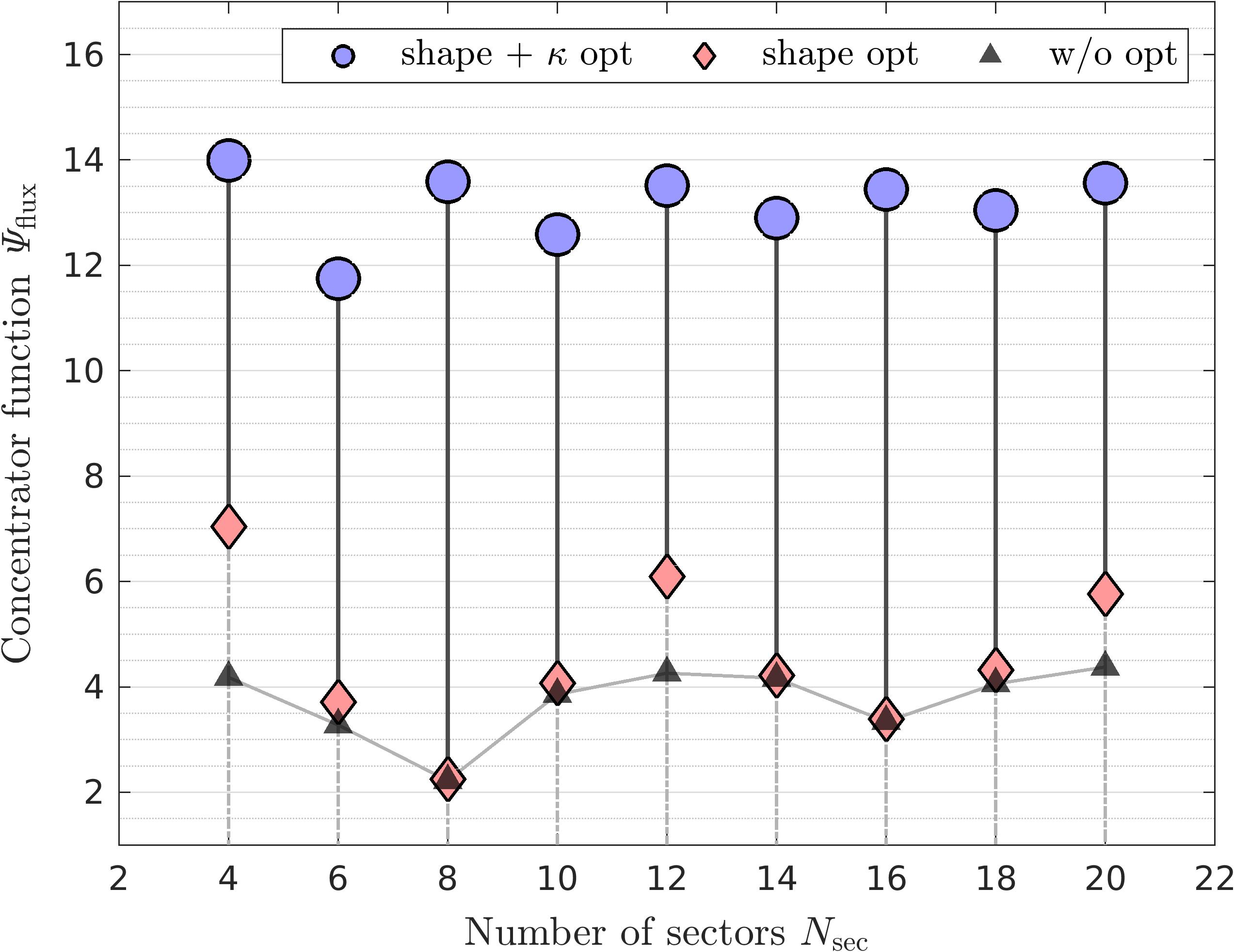}}
        \caption{$N_{\mathrm{var}}=4$}
        \label{figure:star_conv_plot_diff_p}
    \end{subfigure}\\
    \begin{subfigure}[b]{0.48\textwidth}{\centering\includegraphics[width=1\textwidth]{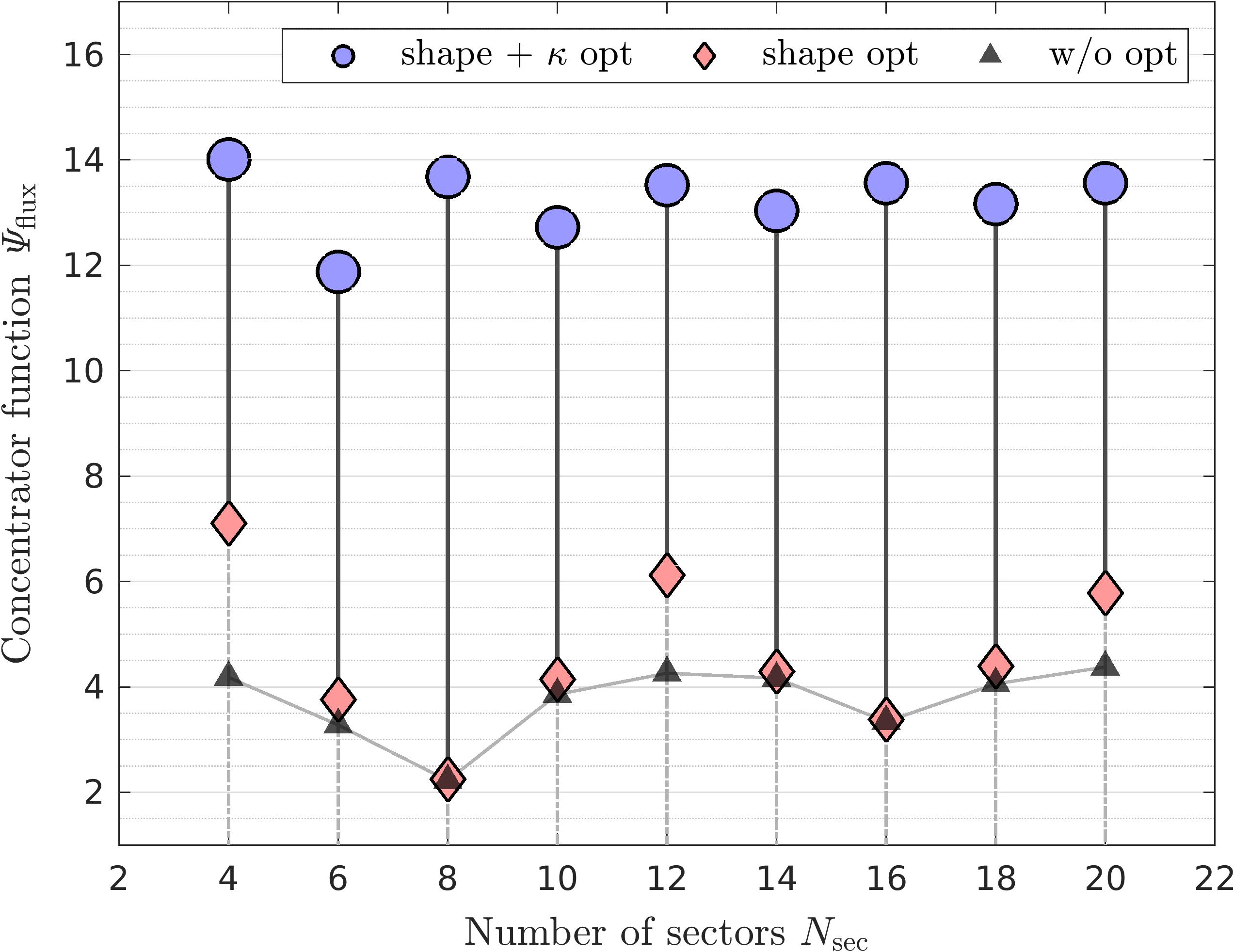}}
        \caption{$N_{\mathrm{var}}=5$}
        \label{figure:star_conv_plot_diff_p}
    \end{subfigure}
    \begin{subfigure}[b]{0.48\textwidth}{\centering\includegraphics[width=1\textwidth]{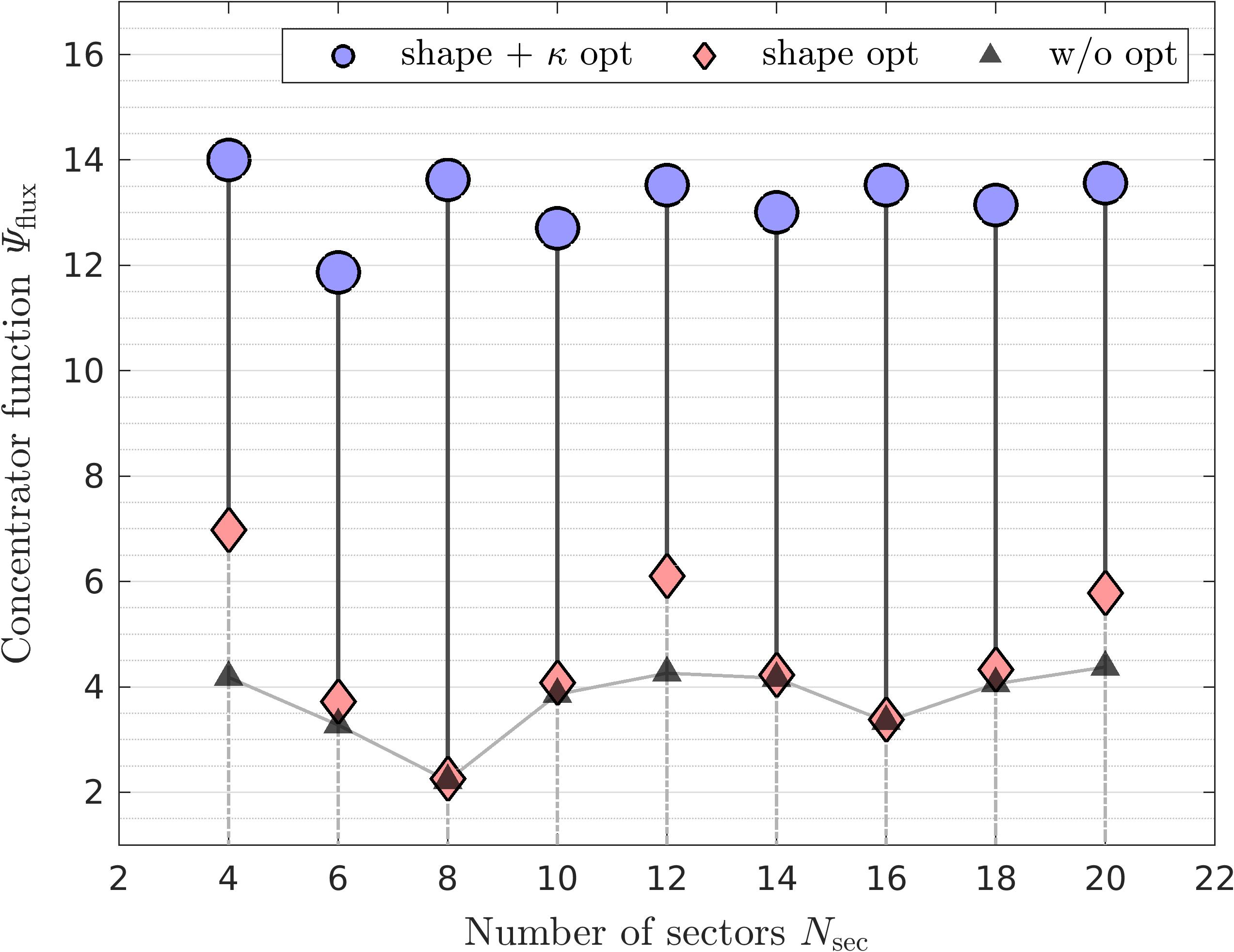}}
        \caption{$N_{\mathrm{var}}=6$}
        \label{figure:star_conv_plot_diff_p}
    \end{subfigure}
 \caption{Variation of concentrator function $\Psi_{\mathrm{flux}}$ of the optimized shape with respect to number of sectors for three case (I) without any optimization (II) only shape optimization, and (III) shape optimization combined with conductivity optimization. $N_{\mathrm{var}}=$3, 4, 5, 6. For case-III, The concentrated flux is approximately 10 to 13 times as large as for a base material plate, 2 to 3 times as large as for case-II \& 3 to 6 times as large as for case-I.}
 \label{fig:Cntr obj funtion variation with var k}
\end{figure}

\begin{figure}[htpb!]
    \centering
    \setlength\figureheight{1\textwidth}
    \setlength\figurewidth{1\textwidth}
    \begin{subfigure}[b]{0.22\textwidth}{\centering\includegraphics[width=1\textwidth]{Figures_chen2015case/chen2015casev5c_nSec2_nDvar6_fluxPlot_sample1.jpg}}
        \caption{\centering Type-A, $N_{\mathrm{sec}}=4$, shape opt}
        \label{fig:Cntr flux plot with var k a}
    \end{subfigure}
    \begin{subfigure}[b]{0.22\textwidth}{\centering\includegraphics[width=1\textwidth]{Figures_chen2015case/chen2015casev5c_nSec3_nDvar6_fluxPlot_sample1.jpg}}
        \caption{\centering Type-B, $N_{\mathrm{sec}}=6$, shape opt}
        \label{fig:Cntr flux plot with var k b}
    \end{subfigure}
    \begin{subfigure}[b]{0.22\textwidth}{\centering\includegraphics[width=1\textwidth]{Figures_chen2015case/chen2015casev5c_nSec4_nDvar6_fluxPlot_sample1.jpg}}
        \caption{\centering Type-C, $N_{\mathrm{sec}}=8$, shape opt}
        \label{fig:Cntr flux plot with var k c}
    \end{subfigure}
    \begin{subfigure}[b]{0.22\textwidth}{\centering\includegraphics[width=1\textwidth]{Figures_chen2015case/chen2015casev5c_nSec5_nDvar6_fluxPlot_sample1.jpg}}
        \caption{\centering Type-D, $N_{\mathrm{sec}}=10$, shape opt}
        \label{fig:Cntr flux plot with var k d}
    \end{subfigure}\\
    \begin{subfigure}[t]{0.22\textwidth}{\centering\includegraphics[width=1\textwidth]{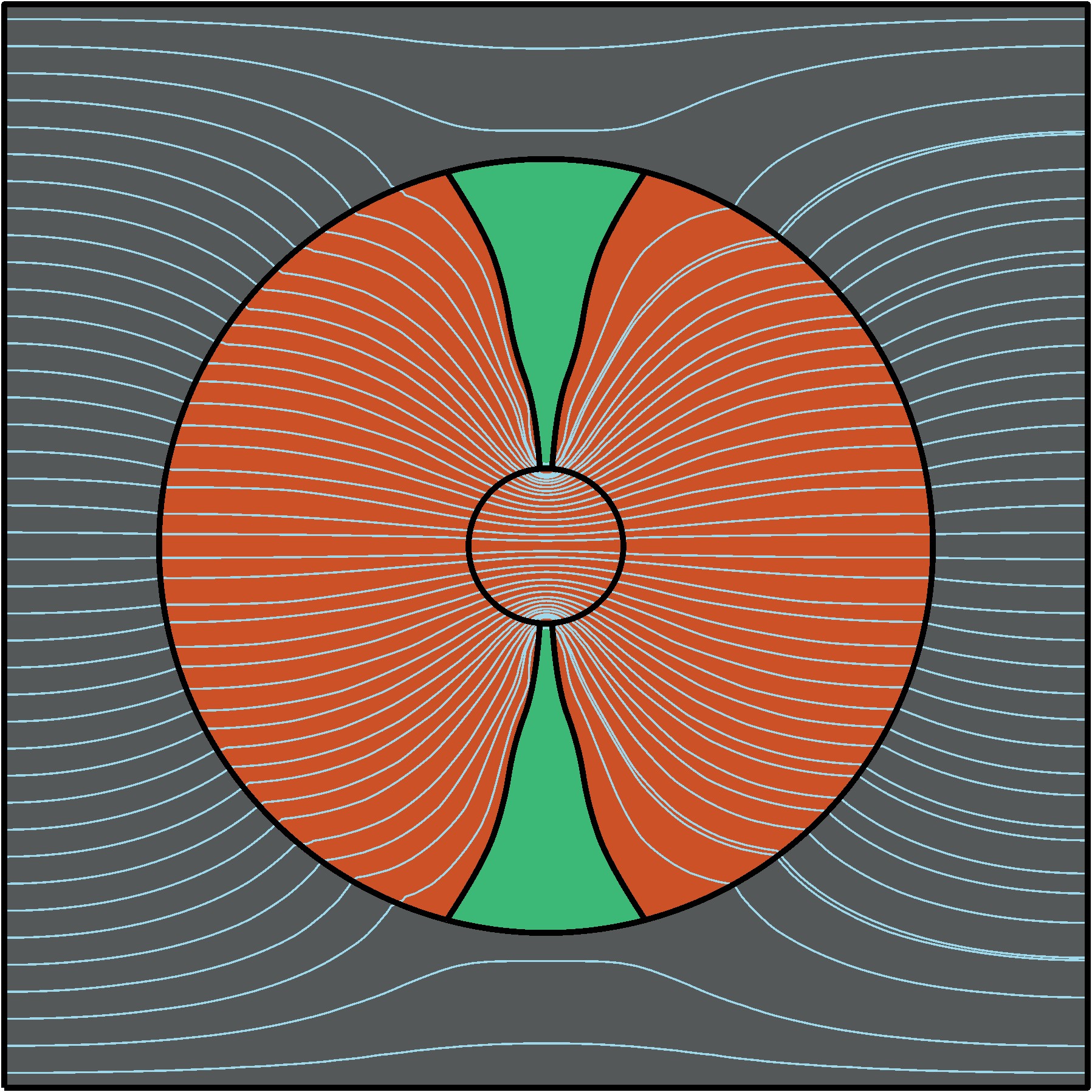}}
        \caption{\centering Type-A, $N_{\mathrm{sec}}=4$, shape + $\kappa$ opt}
        \label{fig:Cntr flux plot with var k e}
    \end{subfigure}
    \begin{subfigure}[t]{0.215\textwidth}{\centering\includegraphics[width=1\textwidth]{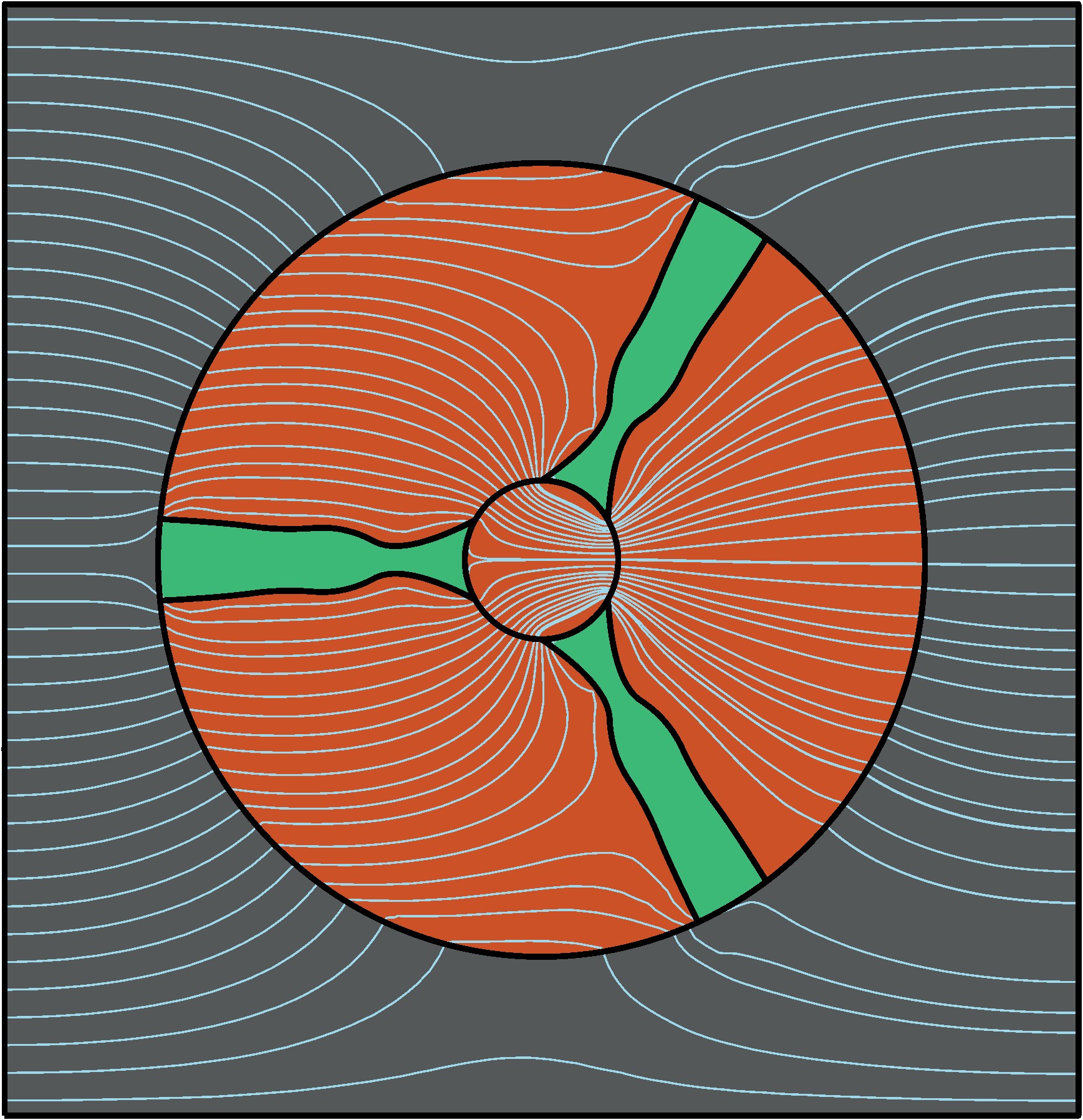}}
        \caption{\centering Type-B, $N_{\mathrm{sec}}=6$, shape + $\kappa$ opt}
        \label{fig:Cntr flux plot with var k f}
    \end{subfigure}
    \begin{subfigure}[t]{0.215\textwidth}{\centering\includegraphics[width=1\textwidth]{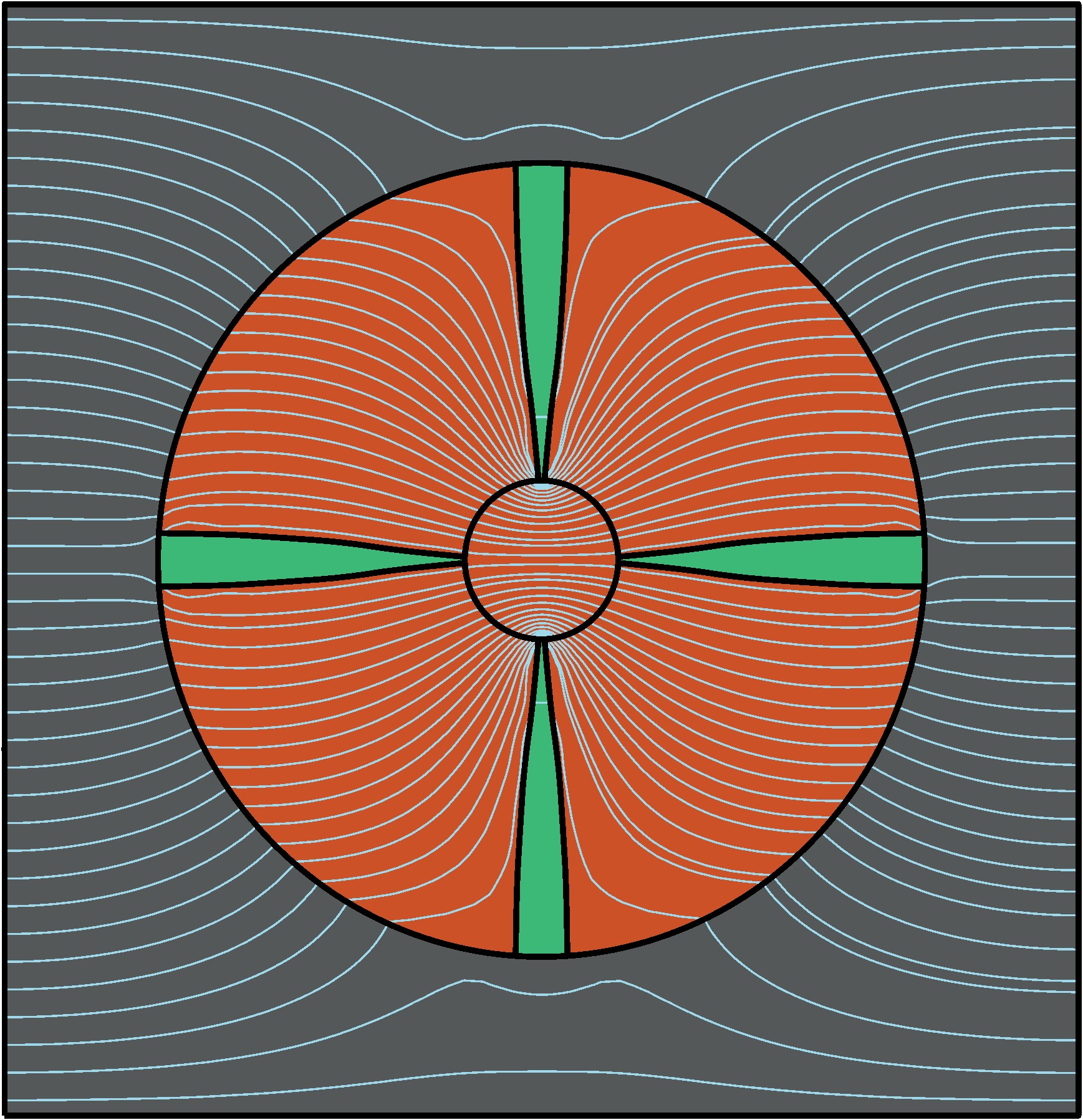}}
        \caption{\centering Type-C, $N_{\mathrm{sec}}=8$, shape + $\kappa$ opt}
        \label{fig:Cntr flux plot with var k g}
    \end{subfigure}
    \begin{subfigure}[t]{0.22\textwidth}{\centering\includegraphics[width=1\textwidth]{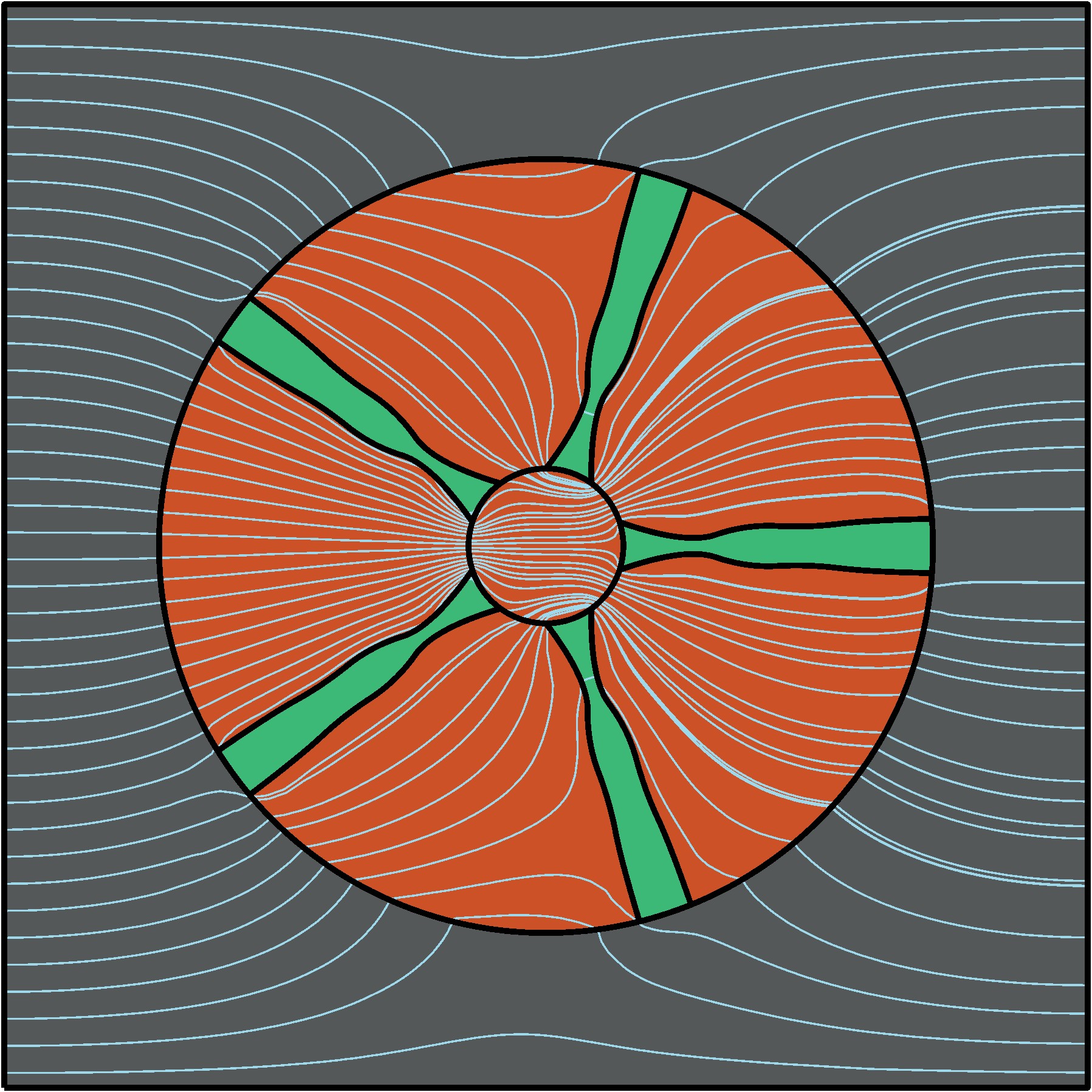}}
        \caption{\centering Type-D, $N_{\mathrm{sec}}=10$, shape + $\kappa$ opt}
        \label{fig:Cntr flux plot with var k h}
    \end{subfigure}
    \\
        \begin{subfigure}[b]{0.22\textwidth}{\centering\includegraphics[width=1\textwidth]{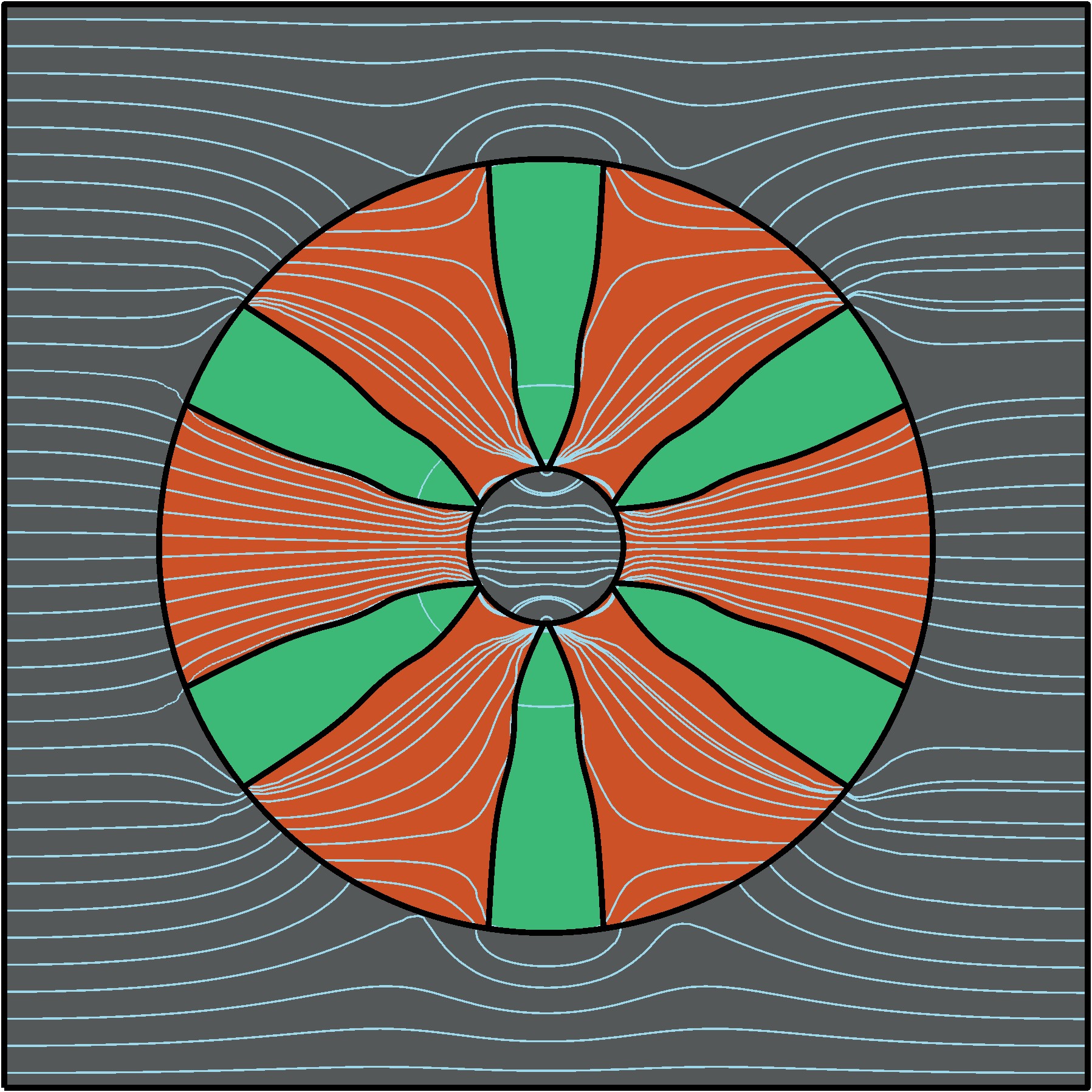}}
        \caption{\centering Type-A, $N_{\mathrm{sec}}=12$, shape opt}
        \label{fig:Cntr flux plot with var k i}
    \end{subfigure}
    \begin{subfigure}[b]{0.22\textwidth}{\centering\includegraphics[width=1\textwidth]{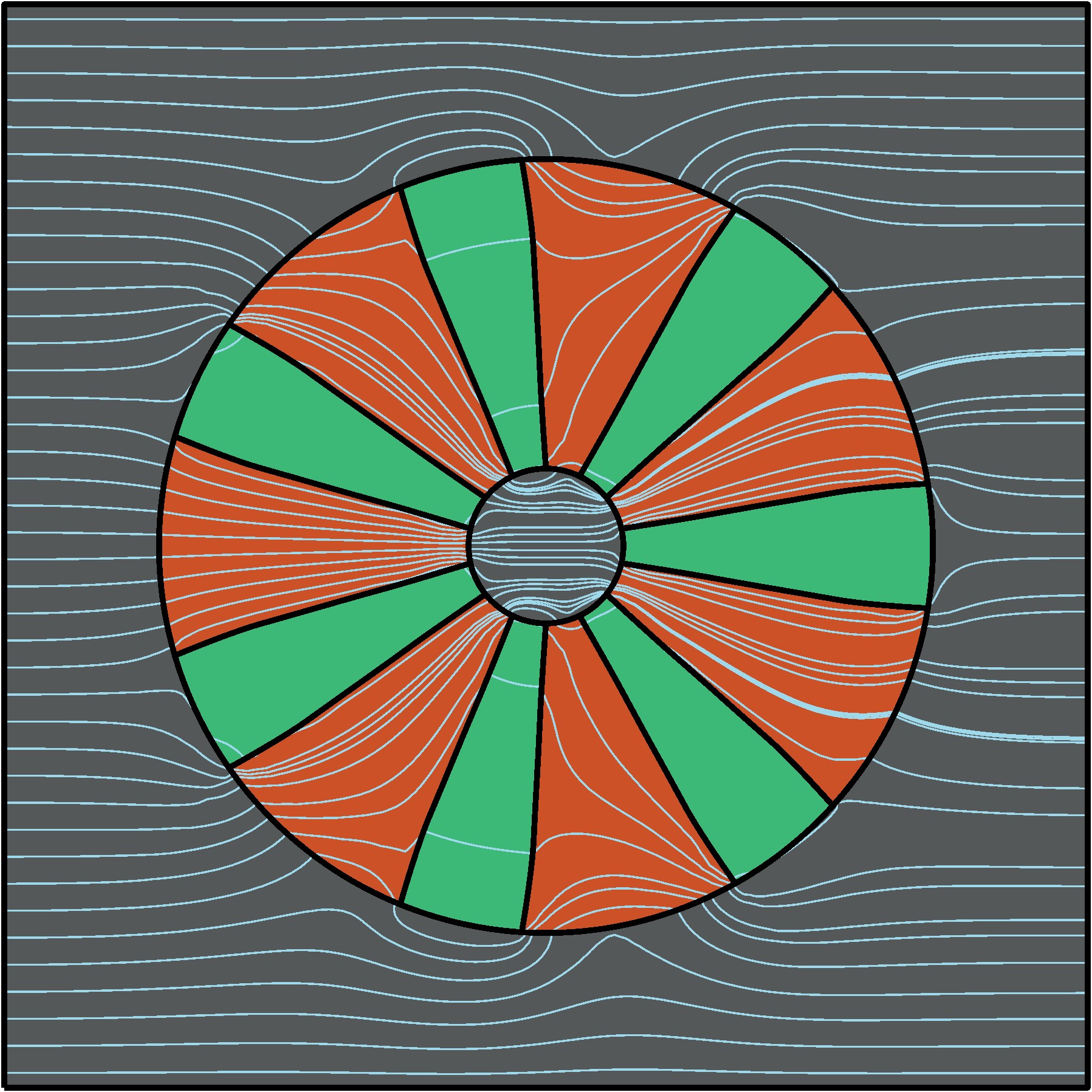}}
        \caption{\centering Type-B, $N_{\mathrm{sec}}=14$, shape opt}
        \label{fig:Cntr flux plot with var k j}
    \end{subfigure}
    \begin{subfigure}[b]{0.22\textwidth}{\centering\includegraphics[width=1\textwidth]{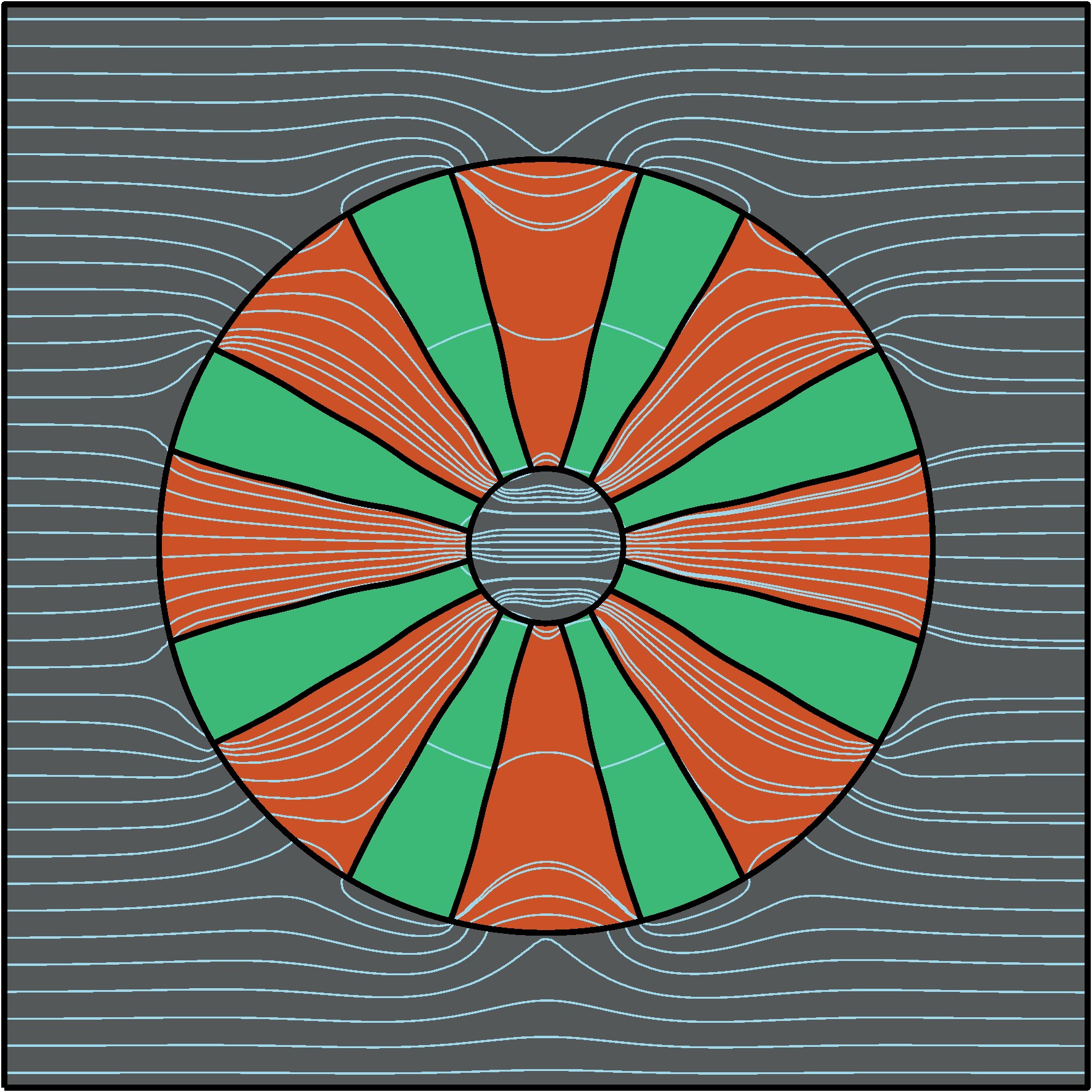}}
        \caption{\centering Type-C, $N_{\mathrm{sec}}=16$, shape opt}
        \label{fig:Cntr flux plot with var k k}
    \end{subfigure}
    \begin{subfigure}[b]{0.22\textwidth}{\centering\includegraphics[width=1\textwidth]{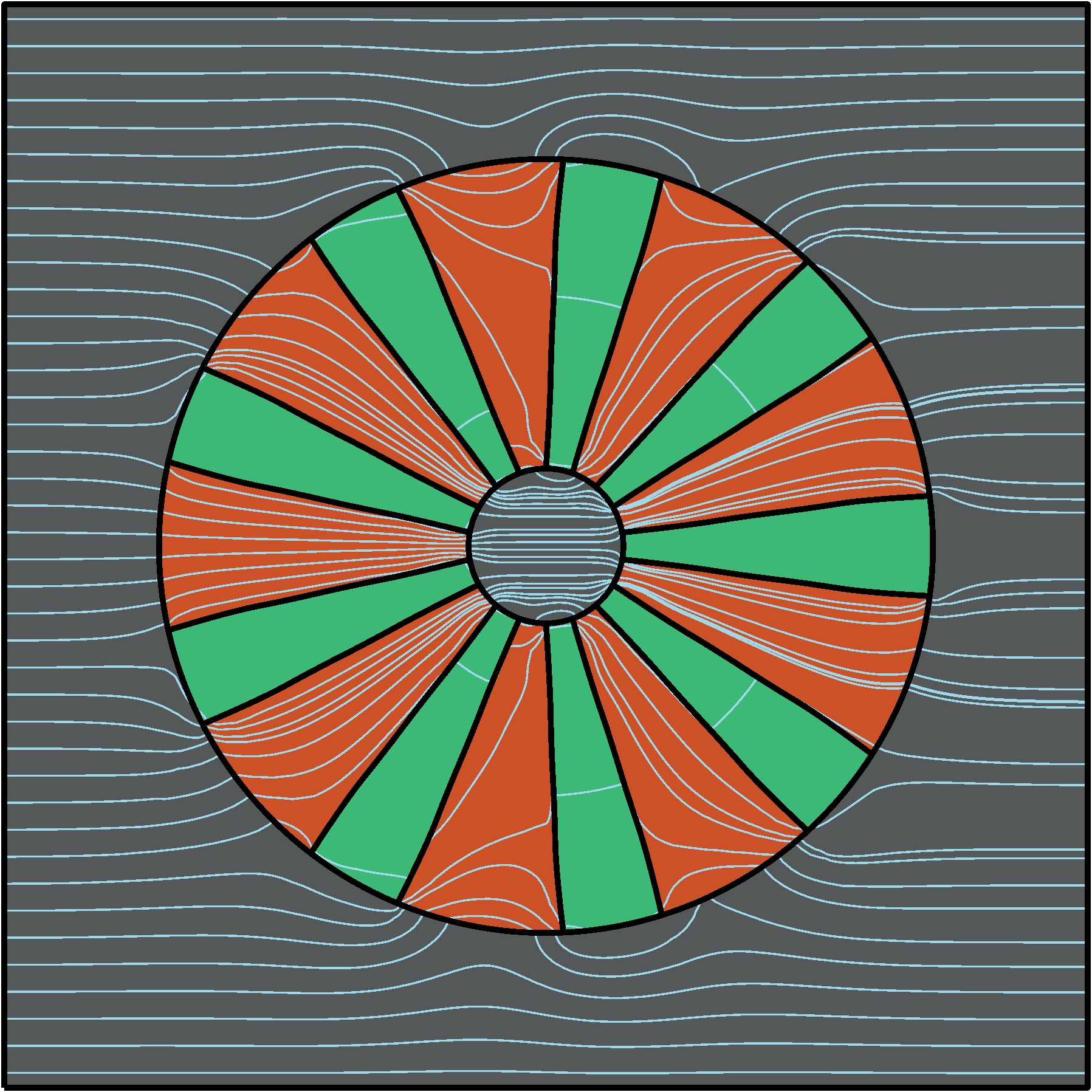}}
        \caption{\centering Type-D, $N_{\mathrm{sec}}=18$, shape opt}
        \label{fig:Cntr flux plot with var k l}
    \end{subfigure}\\
    \begin{subfigure}[b]{0.22\textwidth}{\centering\includegraphics[width=1\textwidth]{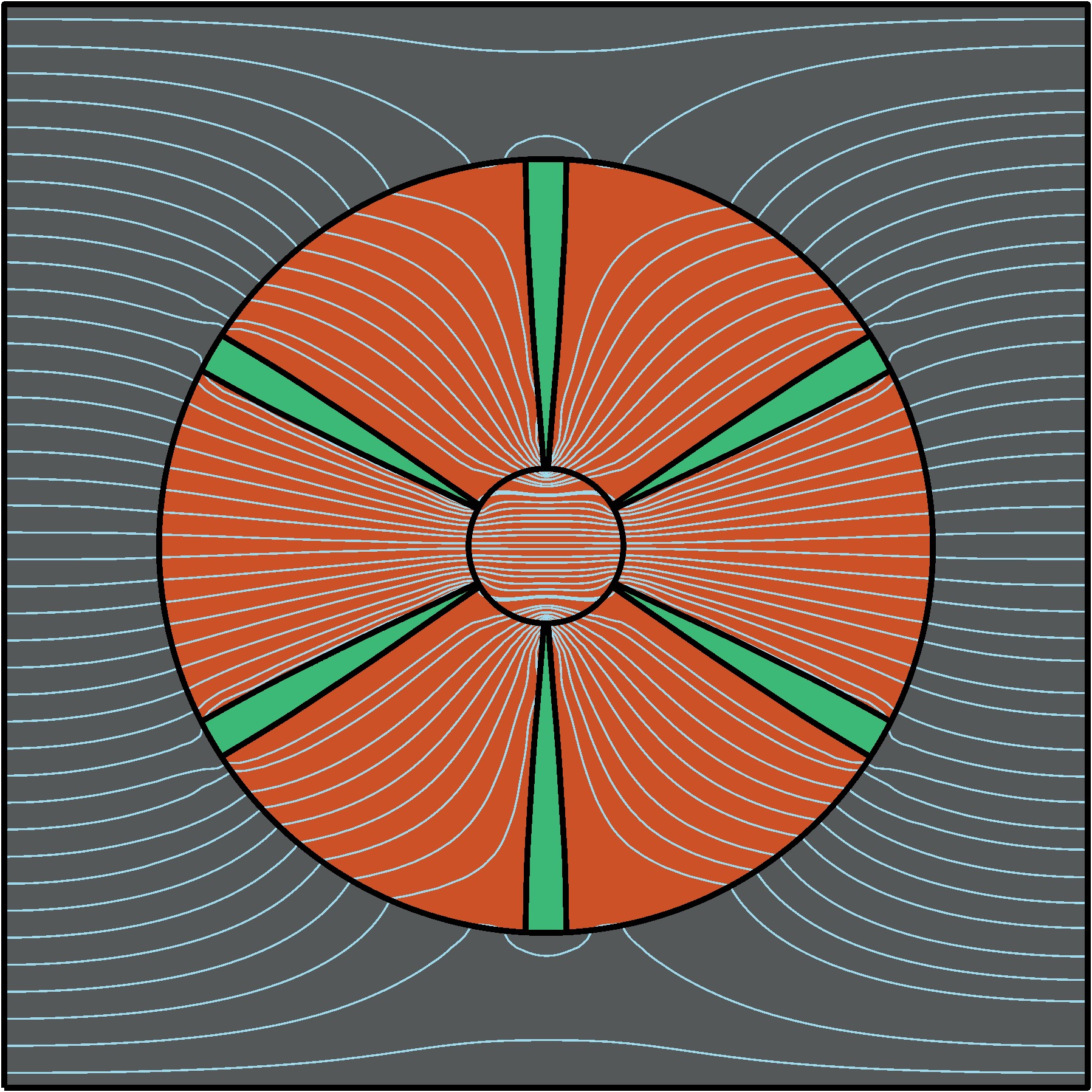}}
        \caption{\centering Type-A, $N_{\mathrm{sec}}=12$, shape + $\kappa$ opt}
        \label{fig:Cntr flux plot with var k m}
    \end{subfigure}
    \begin{subfigure}[b]{0.215\textwidth}{\centering\includegraphics[width=1\textwidth]{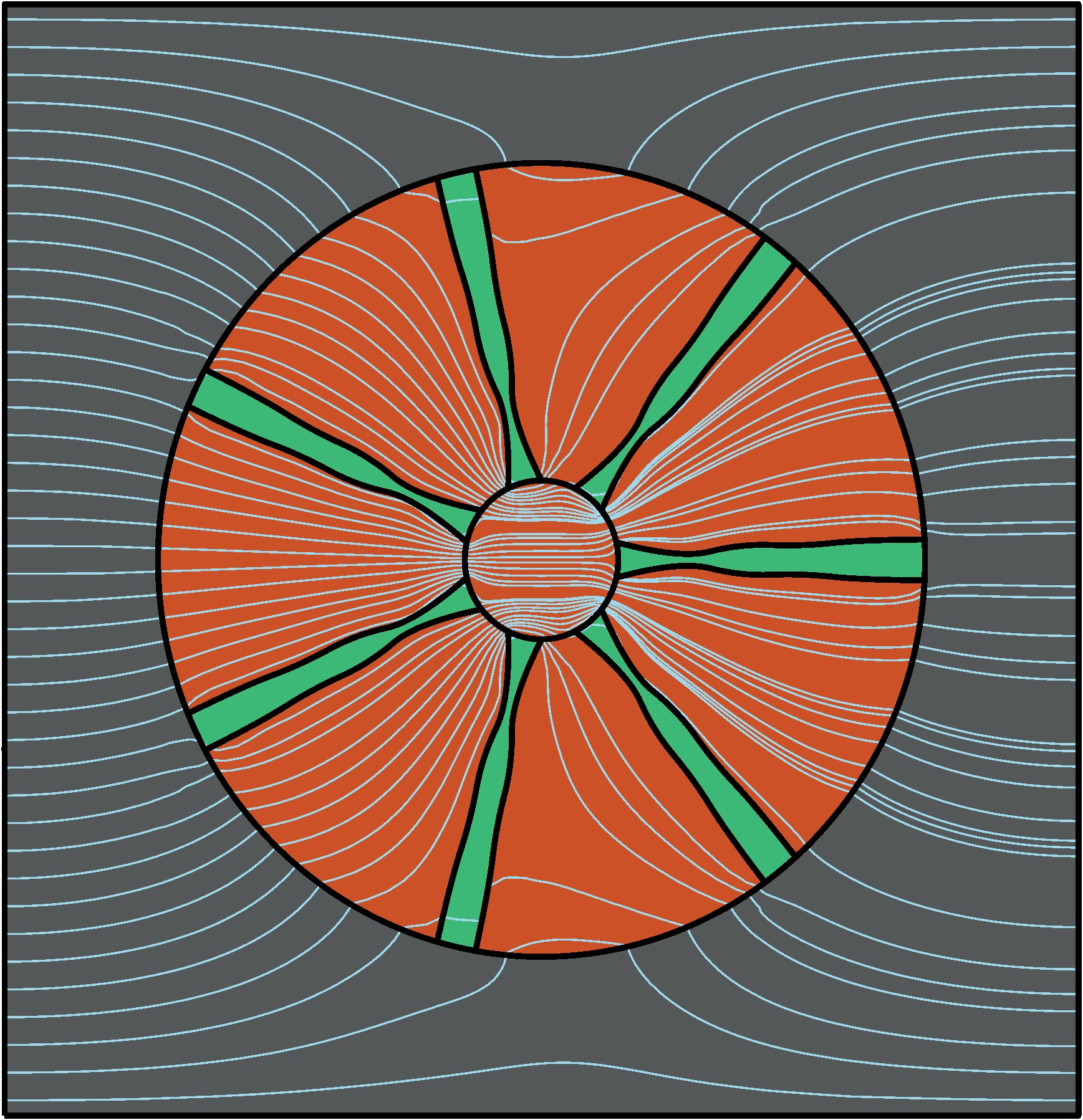}}
        \caption{\centering Type-B, $N_{\mathrm{sec}}=14$, shape + $\kappa$ opt}
        \label{fig:Cntr flux plot with var k n}
    \end{subfigure}
    \begin{subfigure}[b]{0.215\textwidth}{\centering\includegraphics[width=1\textwidth]{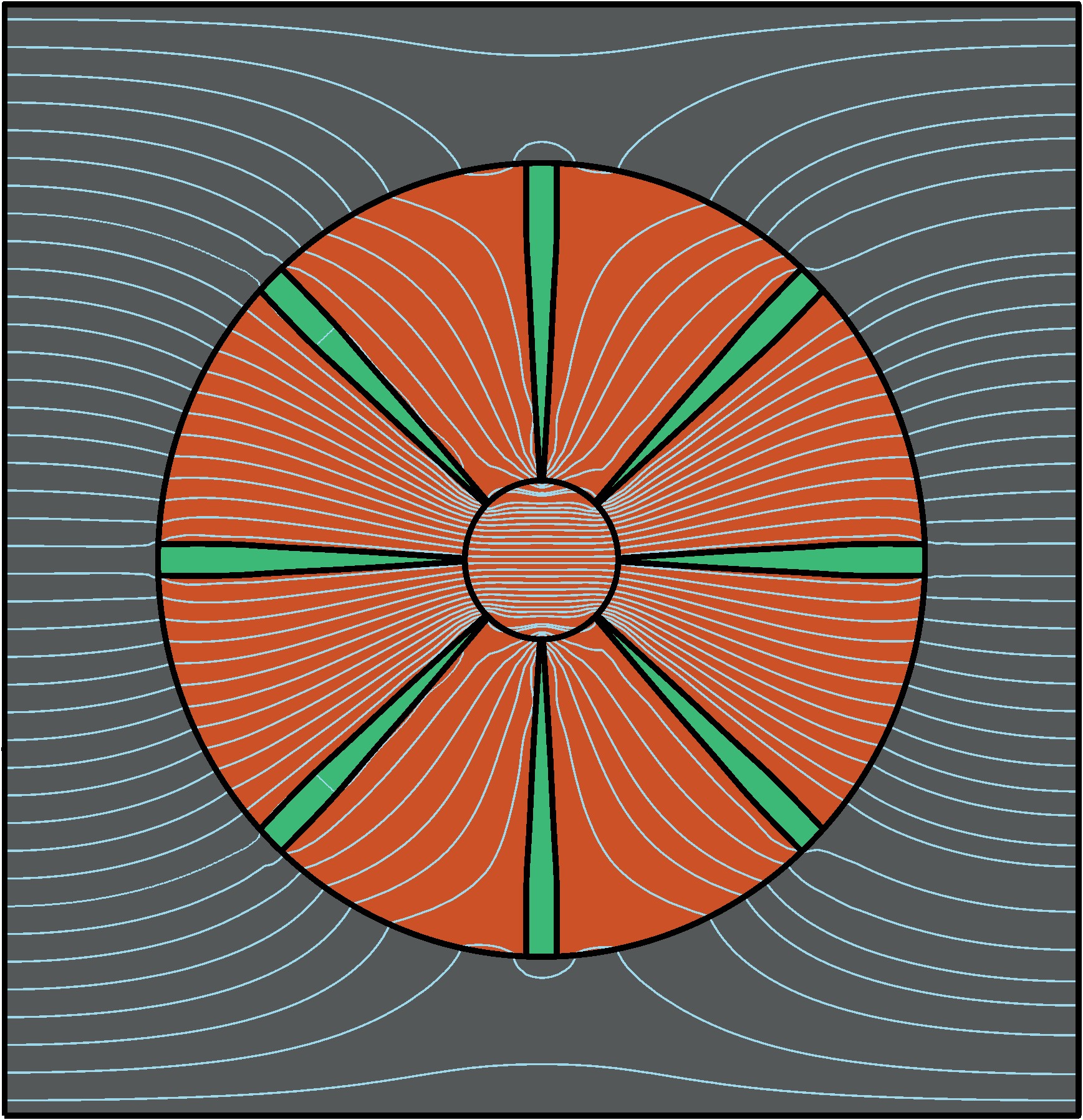}}
        \caption{\centering Type-C, $N_{\mathrm{sec}}=16$, shape + $\kappa$ opt}
        \label{fig:Cntr flux plot with var k o}
    \end{subfigure}
    \begin{subfigure}[b]{0.22\textwidth}{\centering\includegraphics[width=1\textwidth]{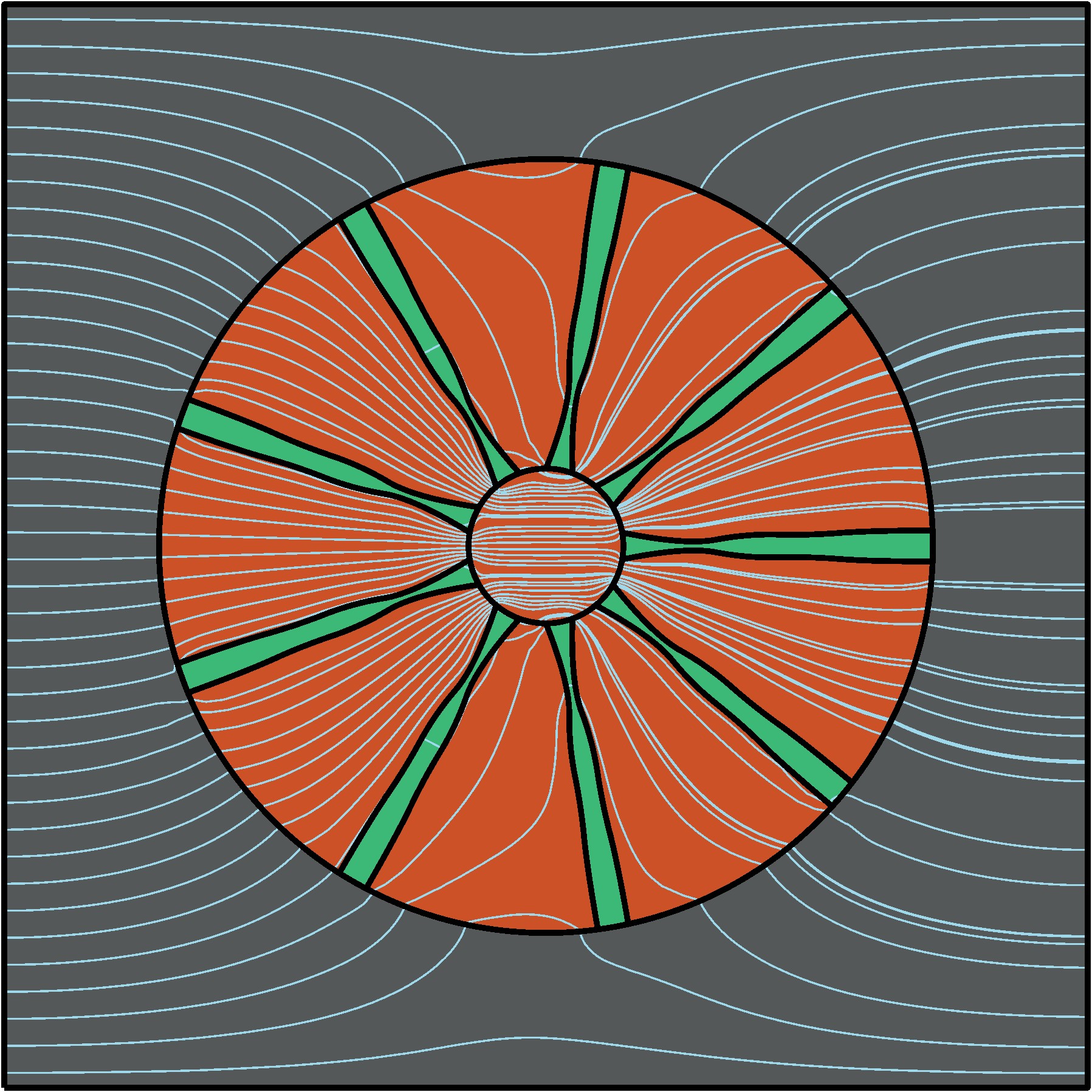}}
        \caption{\centering Type-D, $N_{\mathrm{sec}}=18$, shape + $\kappa$ opt}
        \label{fig:Cntr flux plot with var k p}
    \end{subfigure}\\
    \begin{subfigure}[b]{0.5\textwidth}{\centering\includegraphics[width=1\textwidth]{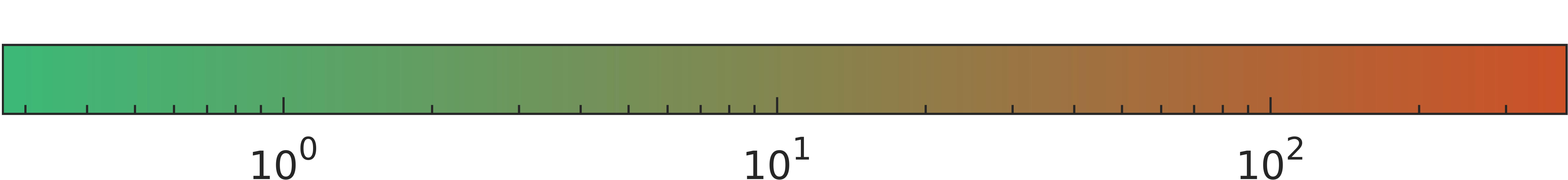}}
    \end{subfigure}
\caption{Flux flow in the optimized shape for different number of sectors for two cases (II) only shape optimization, and (III) shape + conductivity optimization. $N_{\mathrm{sec}}=$4, 12 (Type-A), 6, 14 (Type-B), 8, 16 (Type-C), 10, 20 (Type-D) and $N_{\mathrm{var}}=$6. Base material is shown in grey color. Copper in the core and PDMS sectors on $y$-axis (inline with $\Gamma_{\mathrm{flux}}$) provide benefits in concentrating more flux.}
 \label{fig:Cntr flux plot with var k}
\end{figure} 

\begin{table}[]
\renewcommand{\arraystretch}{1.06} 
\footnotesize
\caption{Objective function $f_{\mathrm{obj}}$ and concentrator function $\Psi_{\mathrm{flux}}$ values for case-I: without any optimization, case-II: shape optimization, and case-III: shape optimization combined with conductivity optimization. $N_{\mathrm{sec}}=$ 4, 6, 8, 10, 12, 14, 16, 18, 20. $N_{\mathrm{var}}=$ 3, 4, 5, 6.}
\begin{center}
\begin{tabular}{|m{4.4em}|m{4.2em}|m{5.3em}|m{2.4em}|m{5.3em}|m{2.4em}|m{5.3em}|m{2.8em}|} 
\hline
\rule{0pt}{3pt}
\multirow{2}{4.4em}{Number of sectors} & \multirow{2}{4.2em}{Number of design variables} & \multicolumn{2}{c|}{w/o opt}                           & \multicolumn{2}{c|}{Shape opt}                         & \multicolumn{2}{c|}{Shape + $\kappa$ opt}              \rule{0pt}{3pt}\\  \cline{3-8} 
  \rule{0pt}{15pt}                                 &                                             & \multicolumn{1}{c|}{$f_{\mathrm{obj}}$} & $\Psi_{\mathrm{flux}}$         & \multicolumn{1}{c|}{$f_{\mathrm{obj}}$} & $\Psi_{\mathrm{flux}}$         & \multicolumn{1}{c|}{$f_{\mathrm{obj}}$} & $\Psi_{\mathrm{flux}}$   \\ \hline
\footnotesize
\multirow{4}{*}{$N_{\mathrm{sec}}=4$} & $N_{\mathrm{var}}$=3 &  \multirow{4}{*}{2.391$\times 10^{-1}$} & \multirow{4}{*}{4.182} & 1.407$\times 10^{-1}$ & 7.105 & 7.138$\times 10^{-2}$ & 14.010\\
&$N_{\mathrm{var}}$=4 & &  & 1.421$\times 10^{-1}$ & 7.036 & 7.148$\times 10^{-2}$ & 13.991\\
&$N_{\mathrm{var}}$=5 & &  & 1.433$\times 10^{-1}$ & 6.977 & 7.143$\times 10^{-2}$ & 14.001\\
&$N_{\mathrm{var}}$=6 & &  & 1.407$\times 10^{-1}$ & 7.106 & 7.136$\times 10^{-2}$ & 14.013 \\
\hline                                              \multirow{4}{*}{$N_{\mathrm{sec}}=6$}&$N_{\mathrm{var}}$=3 & \multirow{4}{*}{3.051$\times 10^{-1}$} & \multirow{4}{*}{3.277} & 2.660$\times 10^{-1}$ & 3.760 & 8.448$\times 10^{-2}$ & 11.838 \\
&$N_{\mathrm{var}}$=4 & &  & 2.698$\times 10^{-1}$ & 3.706 & 8.512$\times 10^{-2}$ & 11.748 \\
&$N_{\mathrm{var}}$=5 & &  & 2.688$\times 10^{-1}$ & 3.721 & 8.422$\times 10^{-2}$ & 11.874 \\
&$N_{\mathrm{var}}$=6 & &  & 2.660$\times 10^{-1}$ & 3.760 & 8.420$\times 10^{-2}$ & 11.876 \\
\hline                                                \multirow{4}{*}{$N_{\mathrm{sec}}=8$}&                      
$N_{\mathrm{var}}$=3 & \multirow{4}{*}{4.503$\times 10^{-1}$} & \multirow{4}{*}{2.221} & 4.448$\times 10^{-1}$ & 2.248 & 7.207$\times 10^{-2}$ & 13.875 \\
&$N_{\mathrm{var}}$=4 & &  & 4.444$\times 10^{-1}$ & 2.250 & 7.357$\times 10^{-2}$ & 13.593 \\
&$N_{\mathrm{var}}$=5 & &  & 4.437$\times 10^{-1}$ & 2.254 & 7.337$\times 10^{-2}$ & 13.629 \\
&$N_{\mathrm{var}}$=6 & &  & 4.443$\times 10^{-1}$ & 2.251 & 7.307$\times 10^{-2}$ & 13.685 \\
\hline                                                \multirow{4}{*}{$N_{\mathrm{sec}}=10$}&                       
$N_{\mathrm{var}}$=3 & \multirow{4}{*}{2.592$\times 10^{-1}$} & \multirow{4}{*}{3.858} & 2.414$\times 10^{-1}$ & 4.143 & 7.887$\times 10^{-2}$ & 12.679 \\
&$N_{\mathrm{var}}$=4 & &  & 2.458$\times 10^{-1}$ & 4.068 & 7.946$\times 10^{-2}$ & 12.585 \\
&$N_{\mathrm{var}}$=5 & &  & 2.451$\times 10^{-1}$ & 4.080 & 7.871$\times 10^{-2}$ & 12.704 \\
&$N_{\mathrm{var}}$=6 & &  & 2.414$\times 10^{-1}$ & 4.143 & 7.858$\times 10^{-2}$ & 12.726 \\
\hline                                                \multirow{4}{*}{$N_{\mathrm{sec}}=12$}&                       
$N_{\mathrm{var}}$=3 & \multirow{4}{*}{2.348$\times 10^{-1}$} & \multirow{4}{*}{4.260} & 1.636$\times 10^{-1}$ & 6.113 & 7.393$\times 10^{-2}$ & 13.526 \\
&$N_{\mathrm{var}}$=4 & &  & 1.642$\times 10^{-1}$ & 6.091 & 7.397$\times 10^{-2}$ & 13.519 \\
&$N_{\mathrm{var}}$=5 & &  & 1.639$\times 10^{-1}$ & 6.102 & 7.395$\times 10^{-2}$ & 13.522 \\
&$N_{\mathrm{var}}$=6 & &  & 1.634$\times 10^{-1}$ & 6.122 & 7.393$\times 10^{-2}$ & 13.526 \\
\hline 
\multirow{4}{*}{$N_{\mathrm{sec}}=14$}&  
$N_{\mathrm{var}}$=3 & \multirow{4}{*}{2.401$\times 10^{-1}$} & \multirow{4}{*}{4.165} & 2.331$\times 10^{-1}$ & 4.290 & 7.698$\times 10^{-2}$ & 12.990 \\
&$N_{\mathrm{var}}$=4 & &  & 2.371$\times 10^{-1}$ & 4.217 & 7.751$\times 10^{-2}$ & 12.902 \\
&$N_{\mathrm{var}}$=5 & &  & 2.364$\times 10^{-1}$ & 4.229 & 7.688$\times 10^{-2}$ & 13.008 \\
&$N_{\mathrm{var}}$=6 & &  & 2.331$\times 10^{-1}$ & 4.291 & 7.669$\times 10^{-2}$ & 13.039 \\
\hline                                                 \multirow{4}{*}{$N_{\mathrm{sec}}=16$}&                       
$N_{\mathrm{var}}$=3 & \multirow{4}{*}{2.994$\times 10^{-1}$} & \multirow{4}{*}{3.340} & 2.979$\times 10^{-1}$ & 3.357 & 7.265$\times 10^{-2}$ & 13.764 \\
&$N_{\mathrm{var}}$=4 & &  & 2.950$\times 10^{-1}$ & 3.390 & 7.437$\times 10^{-2}$ & 13.446 \\
&$N_{\mathrm{var}}$=5& &  & 2.957$\times 10^{-1}$ & 3.382 & 7.392$\times 10^{-2}$ & 13.528 \\
&$N_{\mathrm{var}}$=6 & &  & 2.960$\times 10^{-1}$ & 3.379 & 7.372$\times 10^{-2}$ & 13.565 \\
\hline                                                \multirow{4}{*}{$N_{\mathrm{sec}}=18$}&                         
$N_{\mathrm{var}}$=3 & \multirow{4}{*}{2.465$\times 10^{-1}$} & \multirow{4}{*}{4.057} & 2.279$\times 10^{-1}$ & 4.388 & 7.621$\times 10^{-2}$ & 13.121 \\
&$N_{\mathrm{var}}$=4 & &  & 2.316$\times 10^{-1}$ & 4.318 & 7.663$\times 10^{-2}$ & 13.049 \\
&$N_{\mathrm{var}}$=5 & &  & 2.313$\times 10^{-1}$ & 4.323 & 7.603$\times 10^{-2}$ & 13.152 \\
&$N_{\mathrm{var}}$=6 & &  & 2.279$\times 10^{-1}$ & 4.389 & 7.593$\times 10^{-2}$ & 13.171 \\
\hline                                                 \multirow{4}{*}{$N_{\mathrm{sec}}=20$}&                        
$N_{\mathrm{var}}$=3 & \multirow{4}{*}{2.285$\times 10^{-1}$} & \multirow{4}{*}{4.376} & 1.733$\times 10^{-1}$ & 5.772 & 7.371$\times 10^{-2}$ & 13.566 \\
&$N_{\mathrm{var}}$=4 & &  & 1.736$\times 10^{-1}$ & 5.760 & 7.375$\times 10^{-2}$ & 13.559 \\
&$N_{\mathrm{var}}$=5 & &  & 1.731$\times 10^{-1}$ & 5.777 & 7.373$\times 10^{-2}$ & 13.563 \\
&$N_{\mathrm{var}}$=6 & &  & 1.729$\times 10^{-1}$ & 5.785 & 7.371$\times 10^{-2}$ & 13.566 \\
\hline                                                                 
\end{tabular} 
\label{table:obj fun variation}                          \end{center}                          
\end{table}  
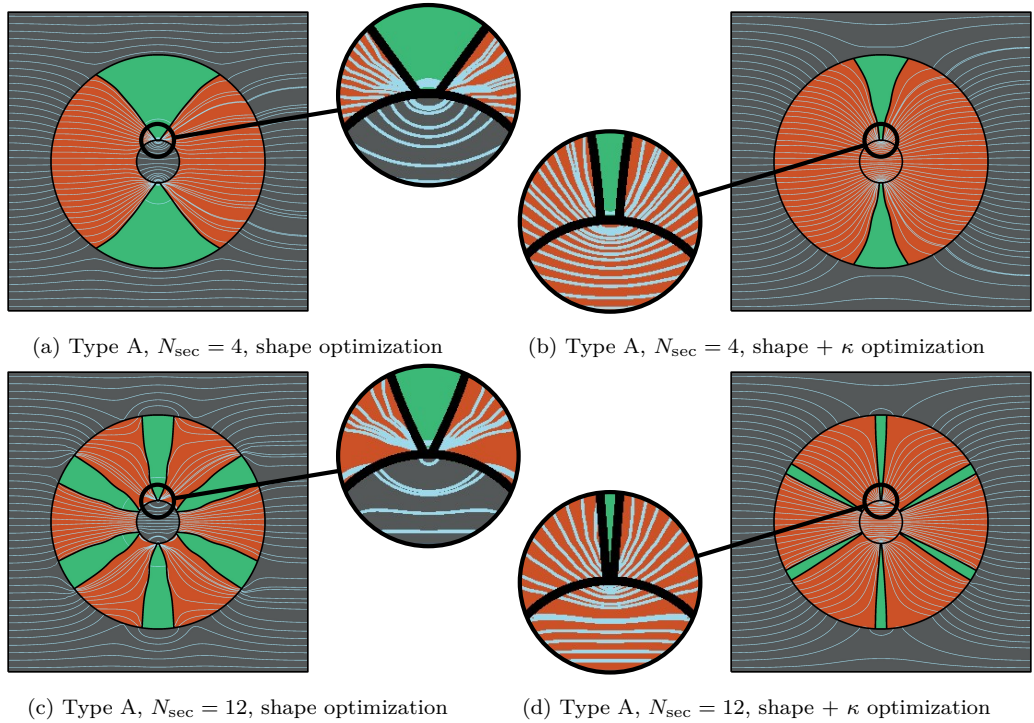
\begin{figure}[htbp!]
\centering
\begin{subfigure}[b]{0.42\textwidth}
\scalebox{0.8}{\begin{tikzpicture}[spy using outlines={circle,black,
    ,magnification=5.5,every spy on node/.append style={ultra thick},size=3cm, connect spies}]
\node {\pgfimage[interpolate=true,height=5cm]{Figures_chen2015case/chen2015casev5c_nSec2_nDvar6_fluxPlot_sample1.jpg}};
\spy[every spy on node/.append style={line width=2pt},spy connection path={
            \draw[line width=2pt] (tikzspyonnode) -- (tikzspyinnode);
        }]  on (0,0.35) in node [black,left,line width=2pt] at (6,1.1);
\end{tikzpicture}}
        \caption{Type A, $N_{\mathrm{sec}}=4$, shape optimization}
    \end{subfigure}
    \quad
 \begin{subfigure}[b]{0.42\textwidth}
\scalebox{0.8}{\begin{tikzpicture}[spy using outlines={circle,black,
    ,magnification=5.5,every spy on node/.append style={ultra thick},size=3cm, connect spies}]
\node {\pgfimage[interpolate=true,height=5cm]{Figures_chen2015case/chen2015casev5cKctrVar_nSec2_nDvar6_fluxPlot_sample1.jpg}};
\spy[every spy on node/.append style={line width=2pt},spy connection path={
            \draw[line width=2pt] (tikzspyonnode) -- (tikzspyinnode);
        }]  on (0,0.35) in node [black,right,line width=2pt] at (-6,-1);
\end{tikzpicture}}
        \caption{Type A, $N_{\mathrm{sec}}=4$, shape + $\kappa$ optimization}
    \end{subfigure}
\\
\begin{subfigure}[b]{0.42\textwidth}
\scalebox{0.8}{\begin{tikzpicture}[spy using outlines={circle,black,
    ,magnification=5.5,every spy on node/.append style={ultra thick},size=3cm, connect spies}]
\node {\pgfimage[interpolate=true,height=5cm]{Figures_chen2015case/chen2015casev5c_nSec6_nDvar6_fluxPlot_sample1.jpg}};
\spy[every spy on node/.append style={line width=2pt},spy connection path={
            \draw[line width=2pt] (tikzspyonnode) -- (tikzspyinnode);
        }]  on (0,0.35) in node [black,left,line width=2pt] at (6,1.1);
\end{tikzpicture}}
        \caption{Type A, $N_{\mathrm{sec}}=12$, shape optimization}
    \end{subfigure}
    \quad
 \begin{subfigure}[b]{0.42\textwidth}
\scalebox{0.8}{\begin{tikzpicture}[spy using outlines={circle,black,
    ,magnification=5.5,every spy on node/.append style={ultra thick},size=3cm, connect spies}]
\node {\pgfimage[interpolate=true,height=5cm]{Figures_chen2015case/chen2015casev5cKctrVar_nSec6_nDvar6_fluxPlot_sample1.jpg}};
\spy[every spy on node/.append style={line width=2pt},spy connection path={
            \draw[line width=2pt] (tikzspyonnode) -- (tikzspyinnode);
        }]  on (0,0.35) in node [black,right,line width=2pt] at (-6,-1);
\end{tikzpicture}}
        \caption{Type A, $N_{\mathrm{sec}}=12$, shape + $\kappa$ optimization}
    \end{subfigure}
\caption{Detailed view of flux concentration in the optimized shape at the junction for Type-A configuration for two cases (II) shape optimization and (III) shape optimization combined with conductivity optimization $N_{\mathrm{sec}}$=4, 12.}
 \label{fig:Cntr flux plot zoom view}
\end{figure} 
\par Next, we conducted an observation by allowing the thermal conductivity to vary with shape optimization. As our optimization uses non-gradient based algorithm, it is easy to implement the conductivity as a design variable apart from regular design variables. In this case, the materials in the concentrator and core region are allowed to change their conductivity patch-wise. The upper and lower limits for the conductivity box constraints are provided as $\kappa_{\rm copper}=398$ W/mK and $\kappa_{\rm PDMS}=0.27$ W/mK, respectively.  
\par The results are presented in \fref{fig:Cntr obj funtion variation with var k} and \tref{table:obj fun variation}. Here three different cases are compared: (I) without any optimization, (II) only shape optimization, and (III) shape optimization combined with conductivity optimization. It is observed that for each $N_{\mathrm{sec}}$ and $N_{\mathrm{var}}$ value, case-III provides the best resultss. The concentrated flux at $\Gamma_{\mathrm{flux}}$ is approximately 10 to 13 times as large as for a base material plate and 2 to 3 times as large as for case-II \& 3 to 6 times as large as for case-I. 
\par In order to understand the material distribution and corresponding optimized shape, the streamline plots for case-II and case-III are compared in \fref{fig:Cntr flux plot with var k}. In the figure, the color of a patch represents its thermal conductivity obtained by optimization. The color scale is defined over a logarithmic scale, where the orange color denotes a material with higher conductivity and green denotes a material with lower conductivity. However, the base material is separately kept in grey color, independent of the color scale.
\par It is evident from \fref{fig:Cntr flux plot with var k} that, for all configurations, the core region reaches the upper limit of conductivity range. Subsequently, the imposition of continuity conditions at interfaces as well as the symmetry constraints allow the copper sectors to cover a large periphery along outer radius. This large opening works as a gate to attract more flux, before eventually driving the same flux to the core region. Therefore, the large value of $\Psi_{\mathrm{flux}}$ for case III compared to case II is justified. Thus, in order to maximize the performance of the concentrator, the material with the highest conductivity should be associated to the inner core.
\par Apart from that, if we focus on the sector-wise material distribution in the concentrator region, it remains the same as before for Type-A configurations (\frefs{fig:Cntr flux plot with var k a}, \ref{fig:Cntr flux plot with var k e}, \ref{fig:Cntr flux plot with var k i}, \ref{fig:Cntr flux plot with var k m}). On the other hand, for Type-C configuration (\frefs{fig:Cntr flux plot with var k c}, \ref{fig:Cntr flux plot with var k g}, \ref{fig:Cntr flux plot with var k k}, \ref{fig:Cntr flux plot with var k o}), the materials in the concentrator region are exchanged. In other words, the sectors inline with $\Gamma_{\mathrm{flux}}$ should be of the lowest conductivity. Therefore, for Type-B (\frefs{fig:Cntr flux plot with var k b}, \ref{fig:Cntr flux plot with var k f}, \ref{fig:Cntr flux plot with var k j}, \ref{fig:Cntr flux plot with var k n}) and Type-D configurations (\frefs{fig:Cntr flux plot with var k d}, \ref{fig:Cntr flux plot with var k h}, \ref{fig:Cntr flux plot with var k l}, \ref{fig:Cntr flux plot with var k p})(where it is not possible to arrange the lowest conductivity along the sectors inline with $\Gamma_{\mathrm{flux}}$) for some instances the sectors swap materials, for some instances they do not.  
\par In case-II, one more point to note is that the flux concentration at the very small interface created between the sector with lower conductivity and inner core for Type-A configuration as shown in \fref{fig:Cntr flux plot zoom view}. The fundamental reason behind it is when the flux flows from a higher $\kappa$ patch to a lower $\kappa$ patch, the flux diverges before crossing the interface due to the normal flux continuity condition. The same effect occurs when the flux flows from the copper sector to the inner core which has comparatively low $\kappa$ nickel steel material). Eventually the diverged flux gets accumulated at the junction of three materials (shown in the detailed view in \fref{fig:Cntr flux plot zoom view}) before entering in the inner core. As this flux concentration can cause numerical burden, we have provided a limit on the minimum size of the interface by box constraints. However in case-III, the inner core takes the highest possible $\kappa$-value and avoids the flux concentration. The smooth flow of flux from the copper sector to the inner core is evident in \fref{fig:Cntr flux plot zoom view}. 

\subsubsection{Application of inclined flux on optimized shape}
\label{sec:inclined flux}

 \begin{figure}[htpb!]
    \centering
    \begin{subfigure}[t]{0.46\textwidth}{\centering\includegraphics[width=1\textwidth]{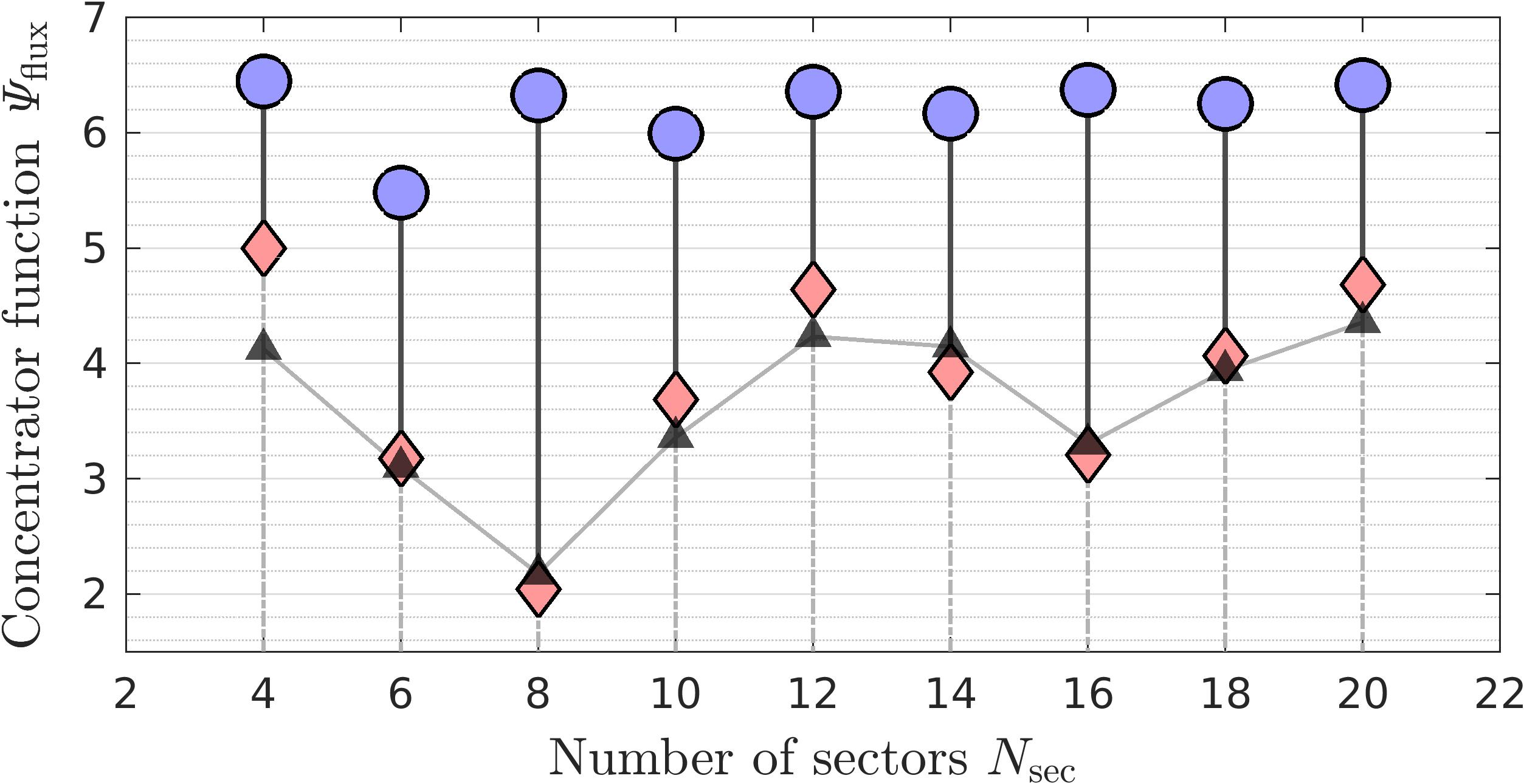}}
        \caption{$\phi=-3\pi/8$}
        \label{figure:star_conv_plot_diff_p}
    \end{subfigure}
    \begin{subfigure}[t]{0.46\textwidth}{\centering\includegraphics[width=1\textwidth]{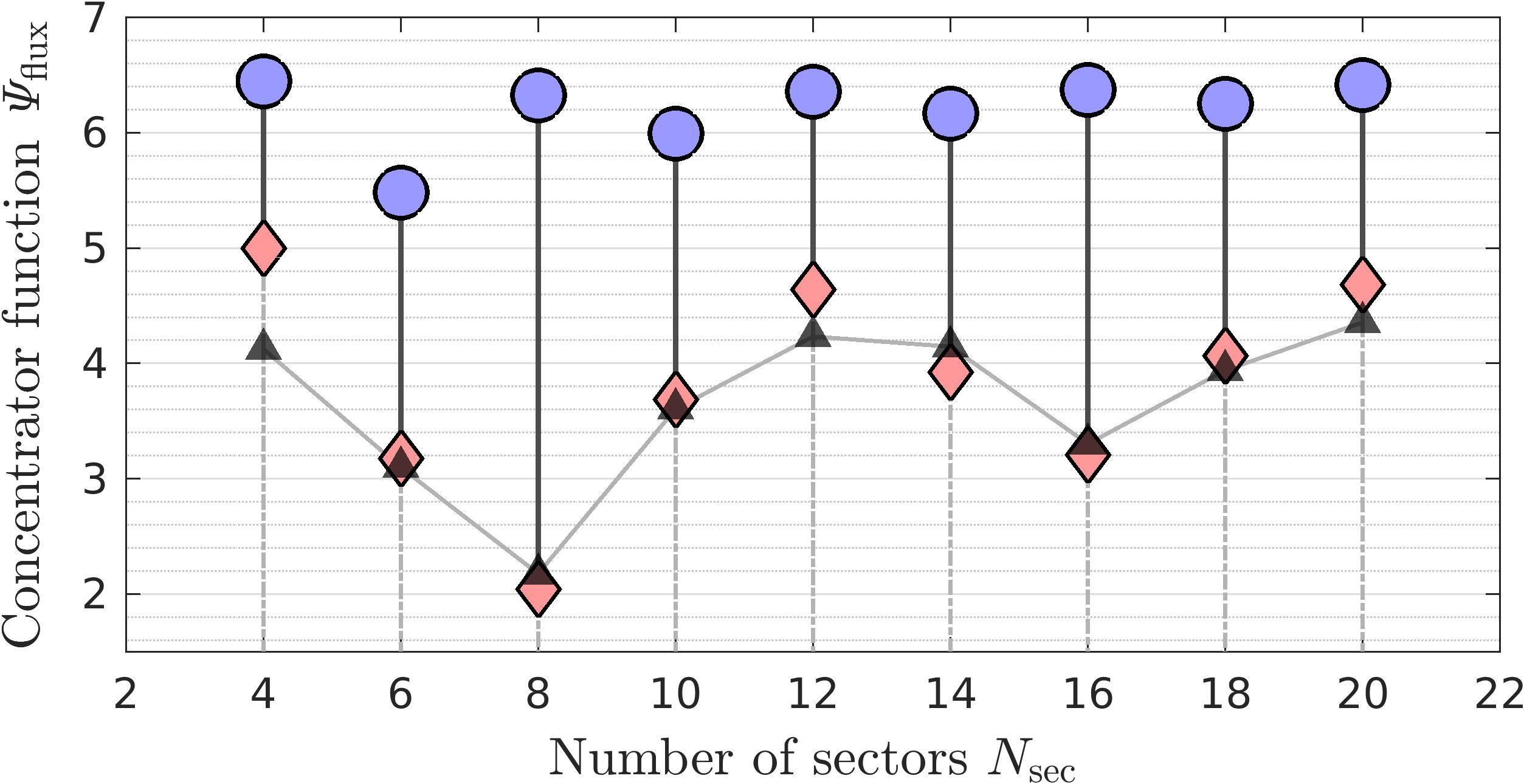}}
        \caption{$\phi=-\pi/4$}
        \label{figure:star_conv_plot_diff_p}
    \end{subfigure}\\
    \begin{subfigure}[t]{0.46\textwidth}{\centering\includegraphics[width=1\textwidth]{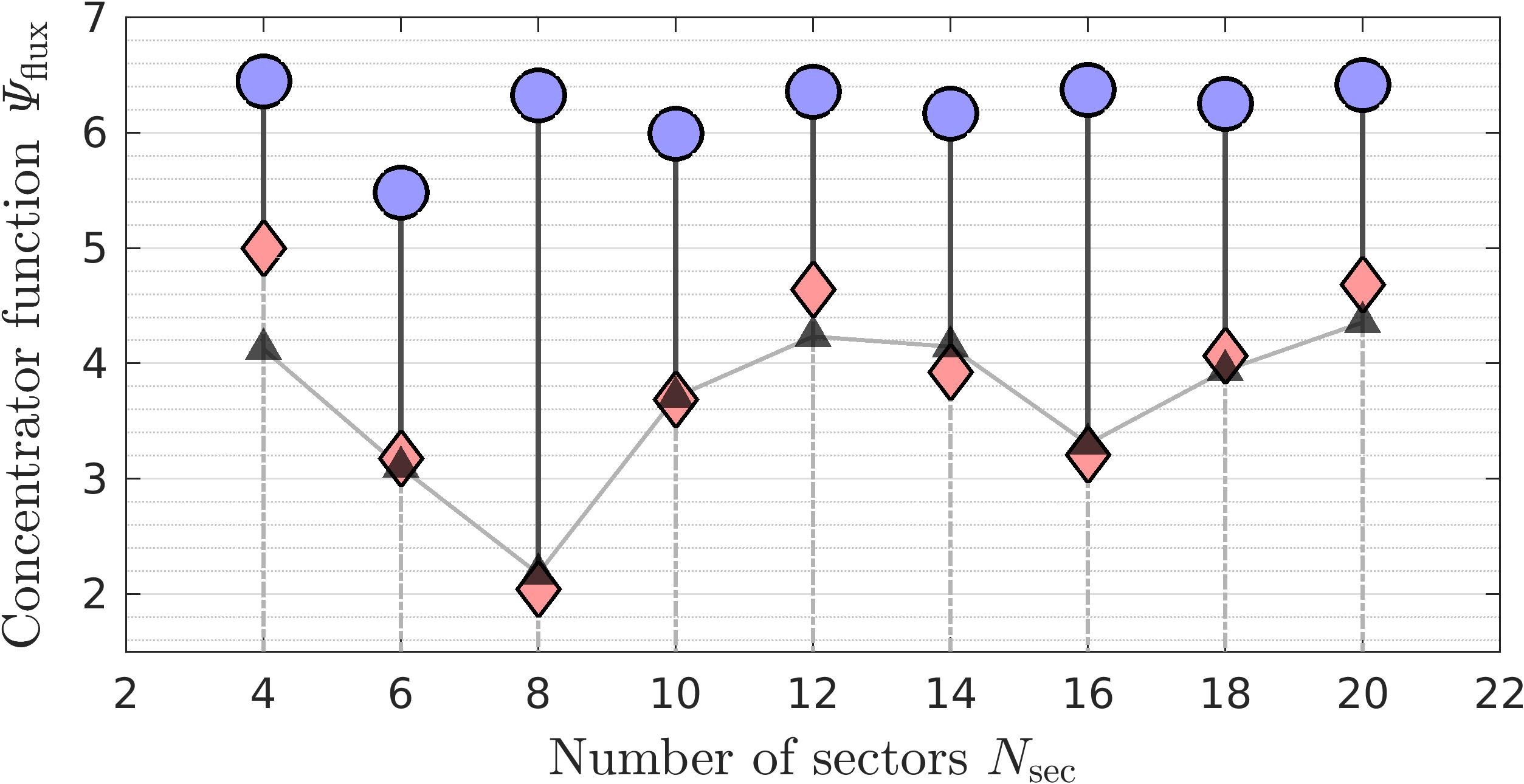}}
        \caption{$\phi=-\pi/8$}
        \label{figure:star_conv_plot_diff_p}
    \end{subfigure}
    \begin{subfigure}[t]{0.46\textwidth}{\centering\includegraphics[width=1\textwidth]{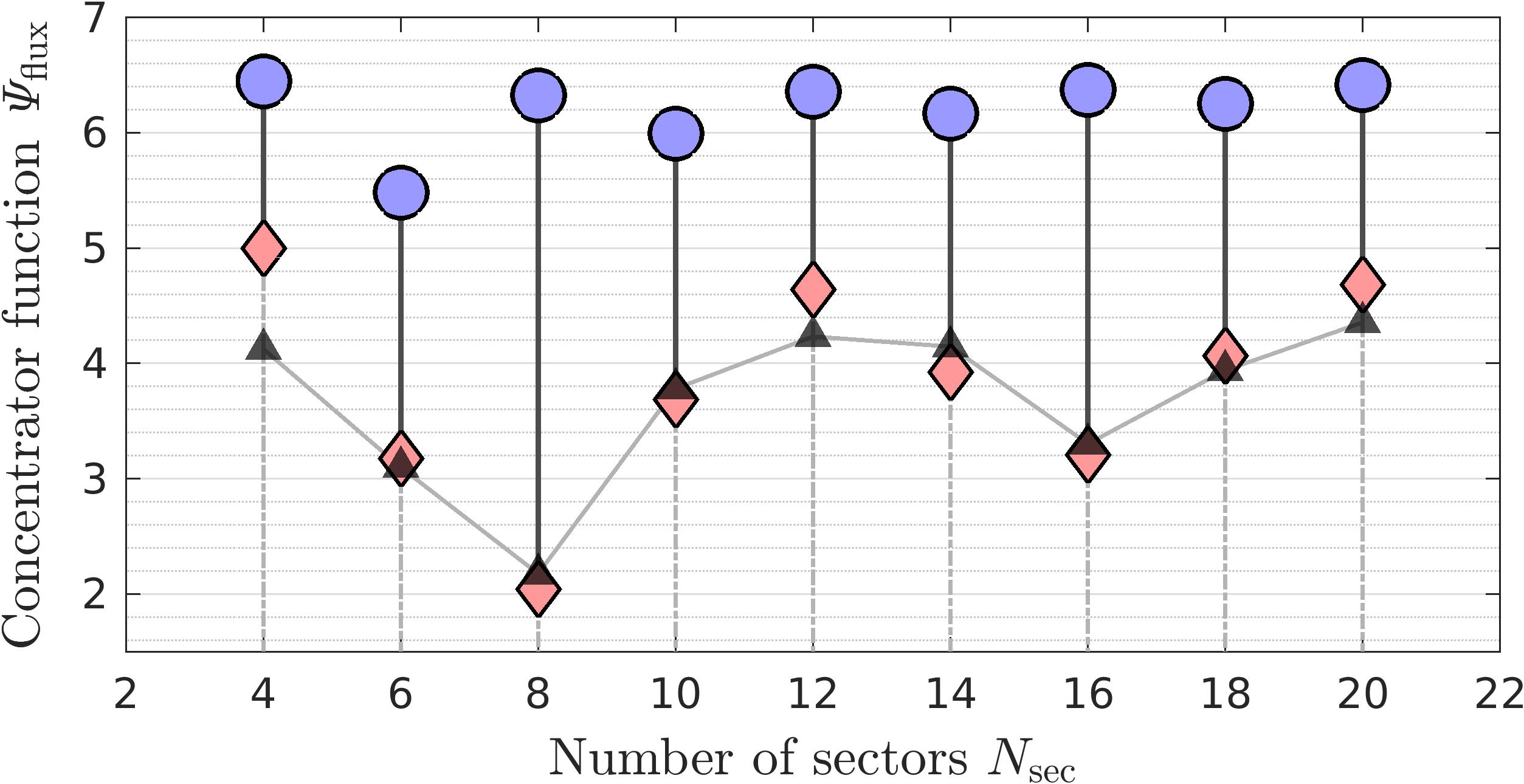}}
        \caption{$\phi=0$}
        \label{figure:star_conv_plot_diff_p}
    \end{subfigure}\\
    \begin{subfigure}[t]{0.46\textwidth}{\centering\includegraphics[width=1\textwidth]{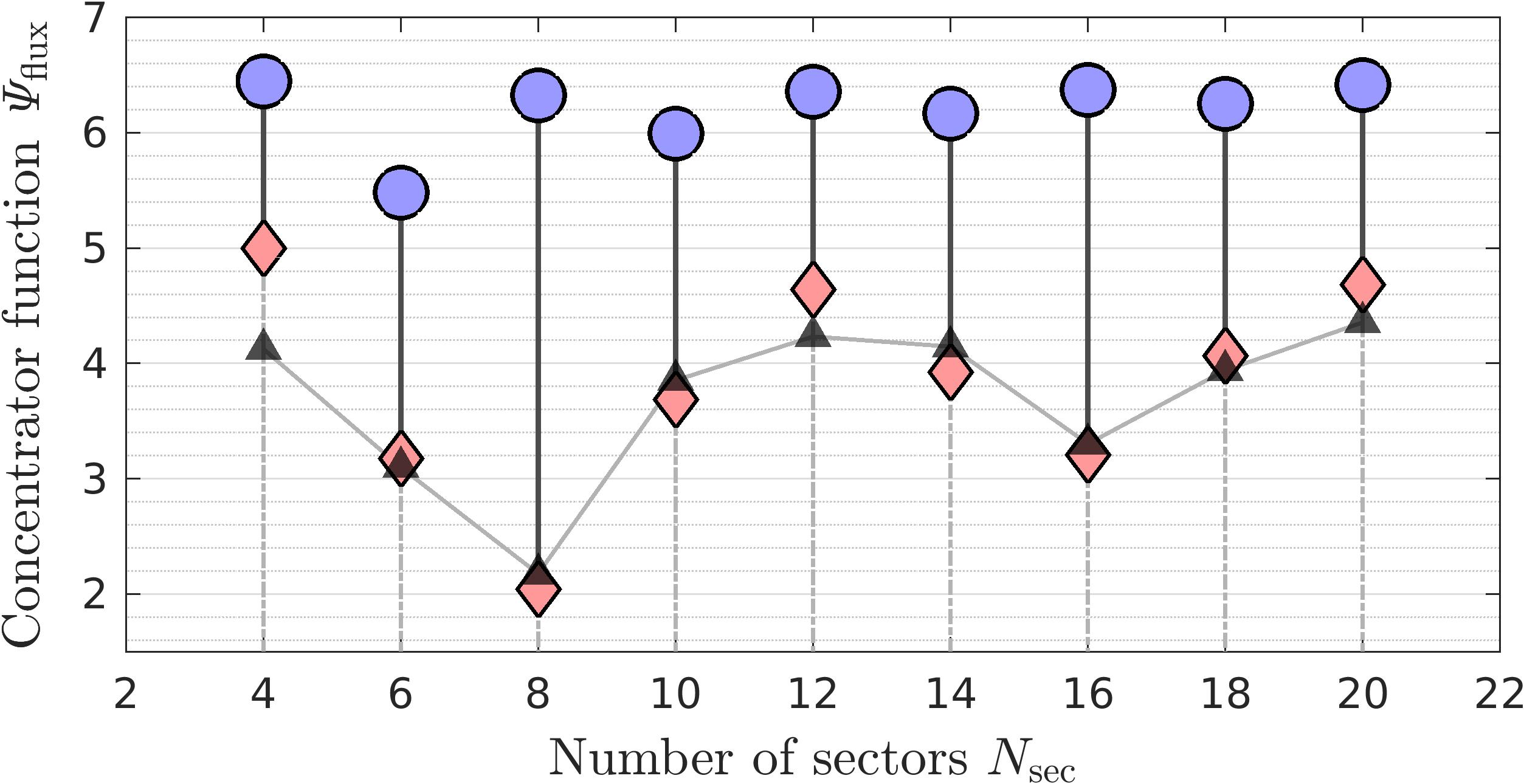}}
        \caption{$\phi=\pi/8$}
        \label{figure:star_conv_plot_diff_p}
    \end{subfigure}
    \begin{subfigure}[t]{0.46\textwidth}{\centering\includegraphics[width=1\textwidth]{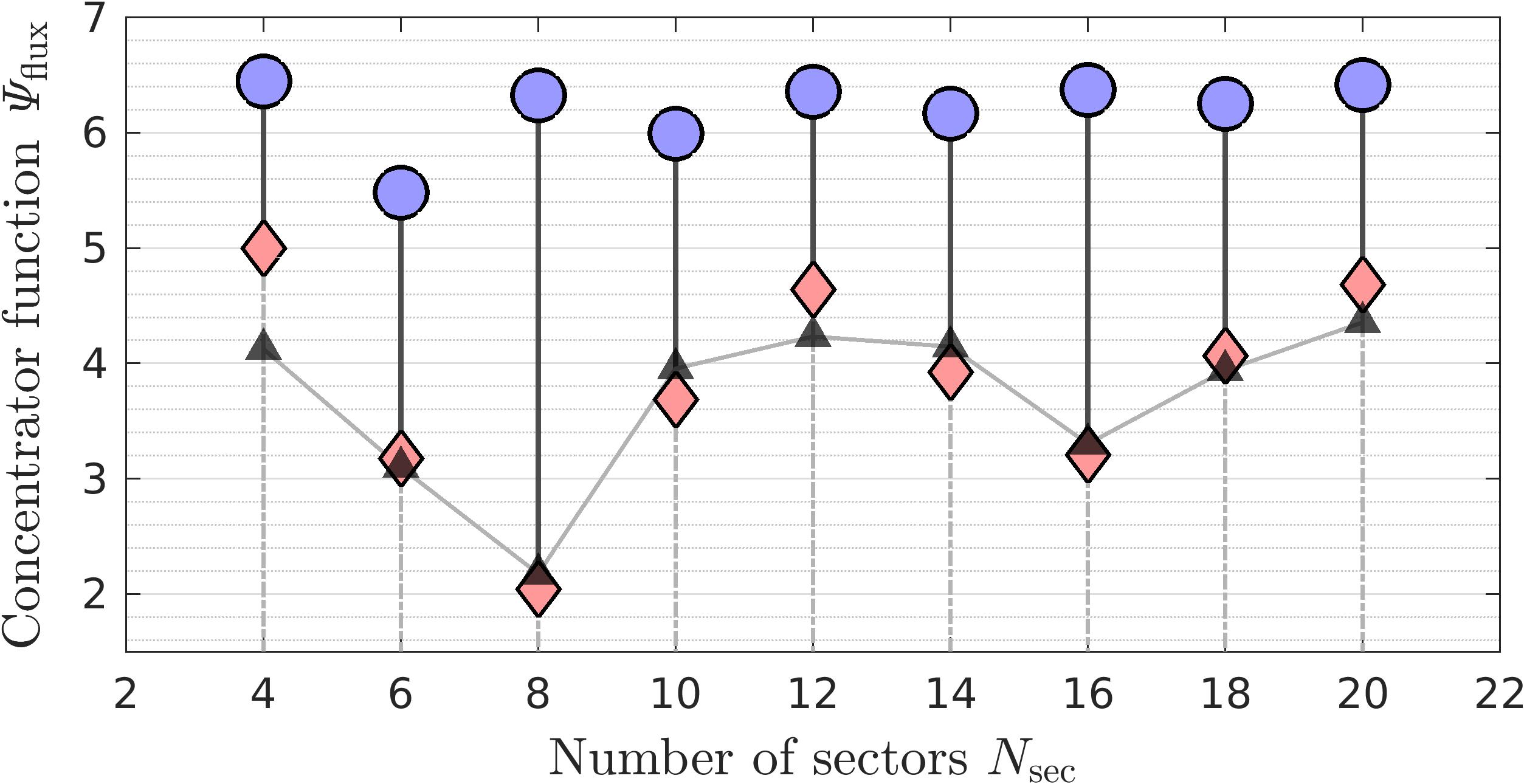}}
        \caption{$\phi=\pi/4$}
        \label{figure:star_conv_plot_diff_p}
    \end{subfigure}\\
    \begin{subfigure}[t]{0.46\textwidth}{\includegraphics[width=1\textwidth,valign=t]{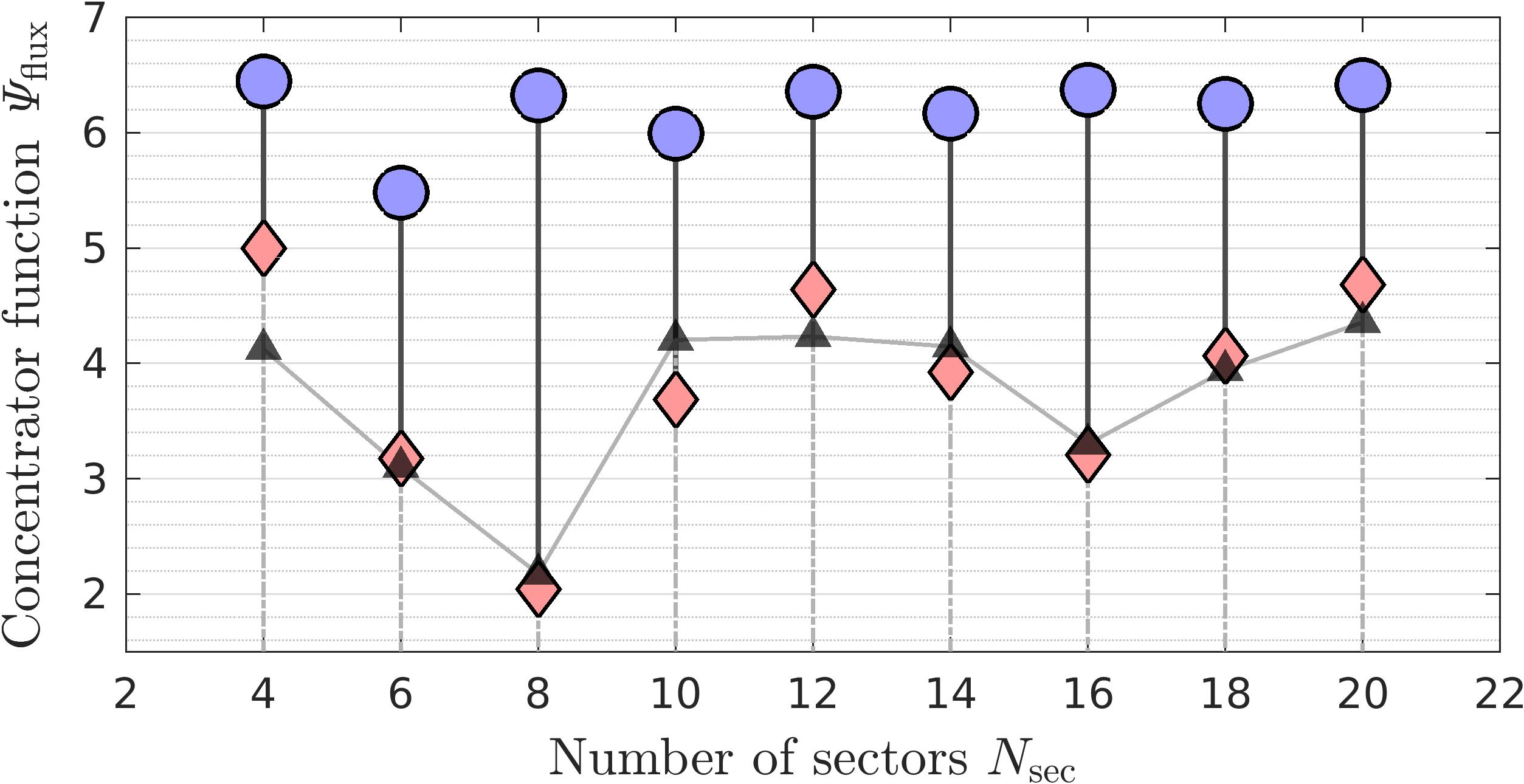}}
        \caption{$\phi=3\pi/8$}
        \label{figure:star_conv_plot_diff_p}
    \end{subfigure}
    \begin{subfigure}[t]{0.2\textwidth}{\vspace{0.1cm}\includegraphics[width=1\textwidth,valign=t]{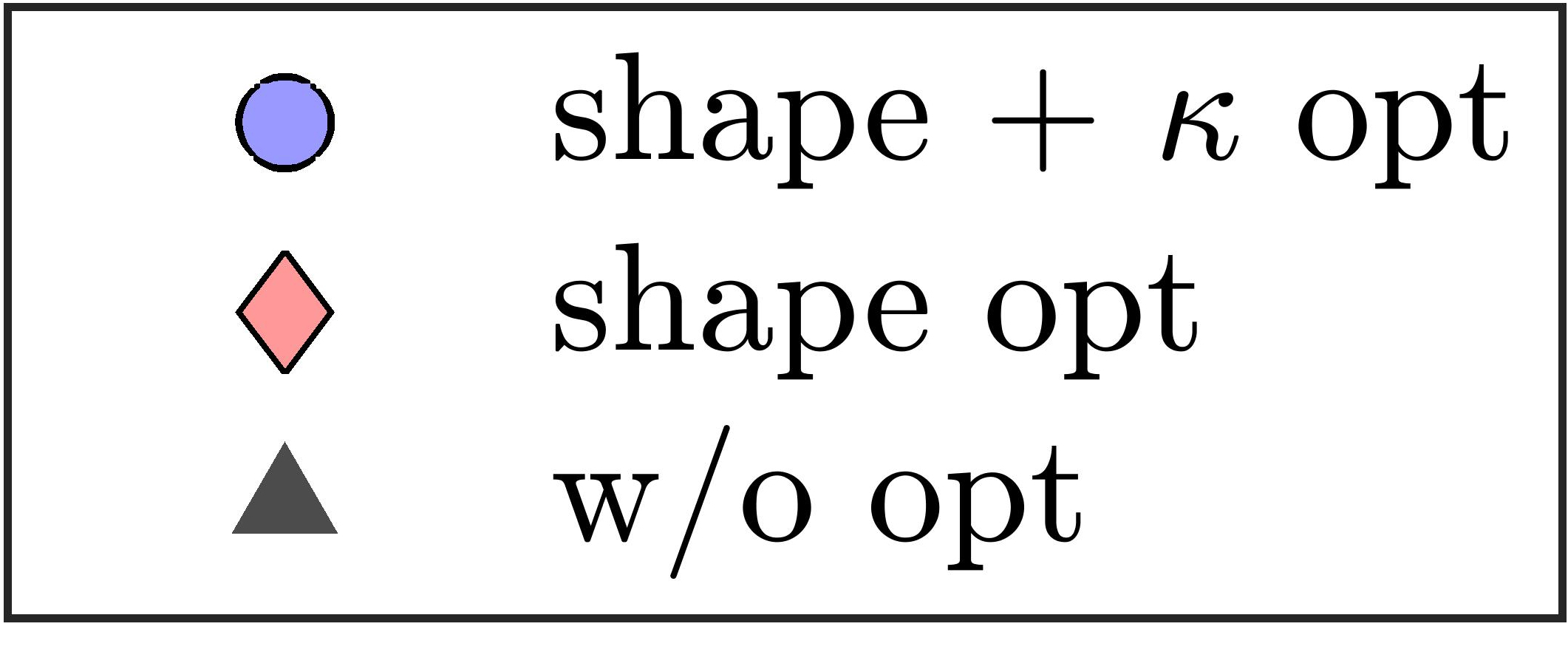}}
    \end{subfigure}
 \caption{Variation of concentrator function $\Psi_{\mathrm{flux}}$ of the optimized shape with respect to number of sectors for constant flux applied at angle $\phi$ = $-3\pi/8$, $-\pi/4$, $-\pi/8$, $0$, $\pi/8$, $\pi/4$ and $3\pi/8$ for three case (I) without any optimization (II) only shape optimization, and (III) shape optimization combined with conductivity optimization. $N_{\mathrm{var}}=$4. The optimized geometries for case-III perform better than case-I and II.}
 \label{fig:Cntr obj function variation with var k inc}
\end{figure}
In this study, we test the optimized geometries from case-I, II and III under application of constant flux at different directions. The constant flux of magnitude $Q_{N}= 7.143\times 10^{3}$ W is applied as the Neumann boundary conditions on all four sides of the plate. Also, the upper left corner is fixed at 300 K to make the boundary value problem well-posed. We consider seven different values for the angle $\phi$ of the applied flux with the $x$-axis ($\phi$=$-3\pi/8$, $-\pi/4$, $-\pi/8$, $0$, $\pi/8$, $\pi/4$ and $3\pi/8$). Since, as shown earlier, there is no apparent variation of $\Psi_{\mathrm{flux}}$ with respect to $N_{\mathrm{var}}$, we only consider one value $N_{\mathrm{var}}$, $N_{\mathrm{var}}$=4. The results are presented in \fref{fig:Cntr obj function variation with var k inc}. 
 \par From \fref{fig:Cntr obj function variation with var k inc}, it is evident that the optimized geometries achieved for case-III, which is combined shape and conductivity optimization, perform better than case-I and II. On the other hand, we do not observe any obvious superiority between case-I and II optimized geometries. Since the geometries for case-II are optimized for the specific boundary conditions, they will not necessarily perform well for other cases. However, the results for case-III are understandable considering relatively wider opening for copper sectors to attract more flux towards the core. One more thing to note for case-III geometries is that the effect of configuration (A/B/C/D) of sectors on $\Psi_{\mathrm{flux}}$ diminishes for larger $N_{\mathrm{sec}}$, and all configurations perform equally well. This comes from the fact that for larger $N_{\mathrm{sec}}$, almost the whole outer and inner peripheries are covered by copper and a very small part remains for PDMS. Nevertheless their thin structure, PDMS sectors are needed inline with $\Gamma_{\mathrm{flux}}$ to provide a barrier to the flux in $\Omega_{\mathrm{design}}$, and guide it towards $\Omega_{\mathrm{in}}$ and eventually through $\Gamma_{\mathrm{flux}}$.

\section{Thermal cloak-concentrator}
\label{sec:Thermal cloaked-concentrator}
\subsection{Problem definition}
\label{sec:Cloak+Cntr Problem definition}
\par
\begin{figure}[htbp!]
    \centering
    \begin{subfigure}[b]{0.45\textwidth}{\centering\includegraphics[width=1\textwidth]{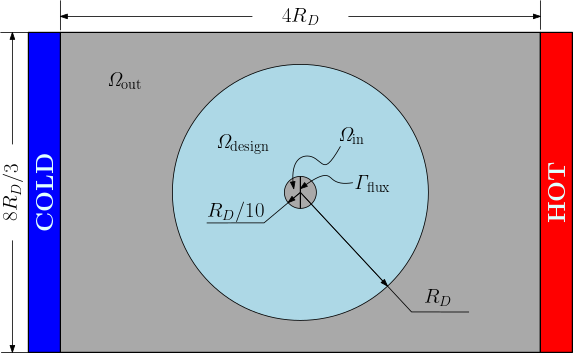}}
             \caption{\centering A metamaterial based cloak-concentrator in the base material plate}
             \label{fig:cloak + concentrator domain a}
    \end{subfigure}
    \begin{subfigure}[b]{0.43\textwidth}{\centering\includegraphics[width=1\textwidth]{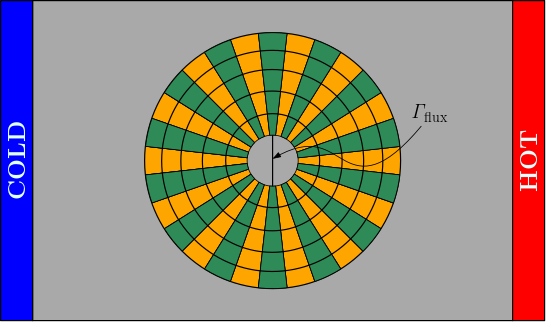}}
             \caption{\centering An annulus ring-sector shaped heat manipulator in the base material plate}
             \label{fig:cloak + concentrator domain b}
    \end{subfigure}
   \caption{Schematic design of (a) A thermal cloak-concentrator embedded in the base material plate to increase the flux concentration in  $\Omega_{\mathrm{in}}$, $\Omega_{\mathrm{design}}$ is the area of the cloak-concentrator, where we want to optimize the shape, $\Omega_{\mathrm{out}}$ is the outside area of remaining base material, $\Omega=\Omega_{\mathrm{in}} \cup \Omega_{\mathrm{design}}\cup \Omega_{\mathrm{out}}$; For our case, we assume $R_D$=10. The base material is iron. (b) An embedded matermaterial-based thermal cloak-concentrator made of block type of structure of copper and PDMS.}
\label{fig:cloak + concentrator domain}
\end{figure}

 In this example, we optimize the thermal cloak-concentrator for simultaneously concentrating flux and cloaking the inner core. The example is referred from~\cite{Fujii2020} and the problem details are kept unchanged to compare the results in the later stage. However, we apply different tools from the tools applied in~\cite{Fujii2020}. Instead of topology optimization with covariance matrix adaptation evolution strategy (CMA-ES), we utilize shape optimization with PSO algorithm. Additionally, we exploit isogeometric analysis (IGA) instead of Lagrange finite element method (FEM) to solve the boundary value problem. The schematics of the problem is similar to the last example as shown in \fref{fig:Concentrator problem schematics}, however the dimensions are different. The dimensions of the current geometry are shown in \fref{fig:cloak + concentrator domain a}. All dimensions are given relative to the outer radius of the heat manipulator, $R_D$. Here, we take $R_D$=10. We consider iron ($\kappa_{\rm iron}=67 $ W/mK) as the base material and copper \& PDMS (with slightly different values of conductivity from the last example) as the materials for thermal cloak-concentrator. The conductivity of copper and PDMS is taken as $\kappa_{\rm copper}=386$~W/mK and $\kappa_{\rm PDMS}=0.15$~W/mK respectively. The left boundary works as a sink with $0^{0}$C constant temperature, and the right boundary works as a source with $1^{0}$C. 

\subsection{Objective function}
\label{sec: Cloak+Cntr Objective function}
\par The objective of a thermal cloak is to avoid the disturbance caused by an obstacle/inner region, and mimic the temperature distribution in the outer base material as if there was no obstacle/inner region. Here, we will try to cloak the inner core, where we are also concentrating the flux, by a single heat manipulator. The combined objective function is defined as,
\begin{equation}
    f_{\mathrm{obj}}=\underbrace{\left\vert\dfrac{1}{\Psi_{\mathrm{flux}}}\right\vert^4}_{f_{\mathrm{obj,flux}}} +  \underbrace{\left\vert\Psi_{\mathrm{cloak}}\right\vert}_{f_{\mathrm{obj,cloak}}},    
\end{equation}
where the first part is related to the concentrator's objective and second is related to the cloak's objective. The concentrator function $\Psi_{\mathrm{flux}}$ is same as defined in \eref{eq: value of psi flux},  while the cloak function $\Psi_{\mathrm{cloak}}$ is defined as below,
\begin{equation}
    \Psi_{\mathrm{cloak}}=\dfrac{1}{\widetilde{\Psi}_{\mathrm{cloak}}} \int_{\Omega_{\mathrm{out}}} \vert T - \overline{T} \vert^2~d\Omega,
    \end{equation}
with $\widetilde{\Psi}_{\mathrm{cloak}}$ be the normalisation value given as,
\begin{equation}
    \widetilde{\Psi}_{\mathrm{cloak}}= \int_{\Omega_{\mathrm{out}}} \vert \widetilde{T} - \overline{T} \vert^2~d\Omega,
    \end{equation}
where $T$, $\overline{T}$ are same as described in \sref{sec:Cntr Objective function}. $\widetilde{T}$ is the temperature field when $\Omega_{\mathrm{design}}$ is filled with PDMS material.
\subsection{NURBS parameterization, design variables and constraints}
\label{sec:Cloak + cntr NURBS parameterization, design variables and constraints}
\par Most of the aspects of NURBS approximations, design variables and constraints are similar to the last example as mentioned \sref{sec:NURBS parameterization, design variables and constraints}. The geometry of the cloak-concentrator is taken as a structure made of several concentric annular rings connected at the periphery. Besides, each ring is made of sectors. Overall, the geometry of a cloak-concentrator is a structure made of sector-shaped blocks in annulus area as shown in \fref{fig:cloak + concentrator domain b}. Each block is considered as a different patch, and continuity conditions at interfaces are imposed by Nitsche's method. The design variables, constraints and symmetry conditions are kept unchanged from the last example. However, all the radius between the inner and outer radius are considered as the extra design variables in addition to the usual design variable.
 
\subsection{Results and discussion}
\label{sec: Cloak+Cntr Results and discussion}
\subsubsection{Conductivity optimization}
\label{sec: Cloak+Cntr Conductivity optimization}
\par At first, we need an initial topology for the cloak-concentrator to start the optimization. Since the new features cannot be created in the geometry during shape optimization, it is crucial to take an appropriate topology that can ensure sufficient scope for shape optimization. To get the initial topology, we run solely conductivity optimization. For the optimization, we impose $x$-axis and $y$-axis symmetry. The results of the conductivity optimization will exhibit suitable conductivity values for all blocks, which will assist us to define initial material distribution in cloak-concentrator.  We consider two initial geometries: one with $N_{\mathrm{ring}}=3$ \& $N_{\mathrm{sec}}=12$, another with $N_{\mathrm{ring}}=5$ \& $N_{\mathrm{sec}}=20$. \begin{figure}[htbp!]
    \centering
    \begin{subfigure}[b]{0.24\textwidth}{\centering\includegraphics[width=1\textwidth,height=0.0355\textheight]{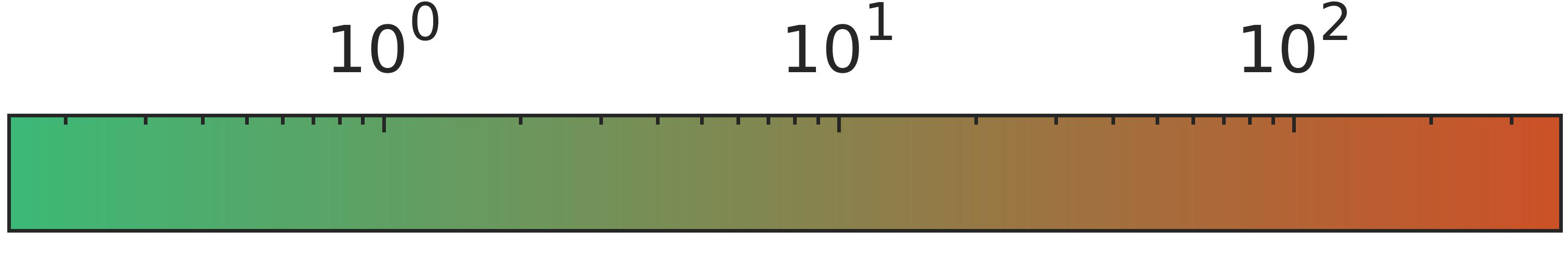}}
    \end{subfigure}
    \begin{subfigure}[b]{0.362\textwidth}{\centering\includegraphics[width=1\textwidth]{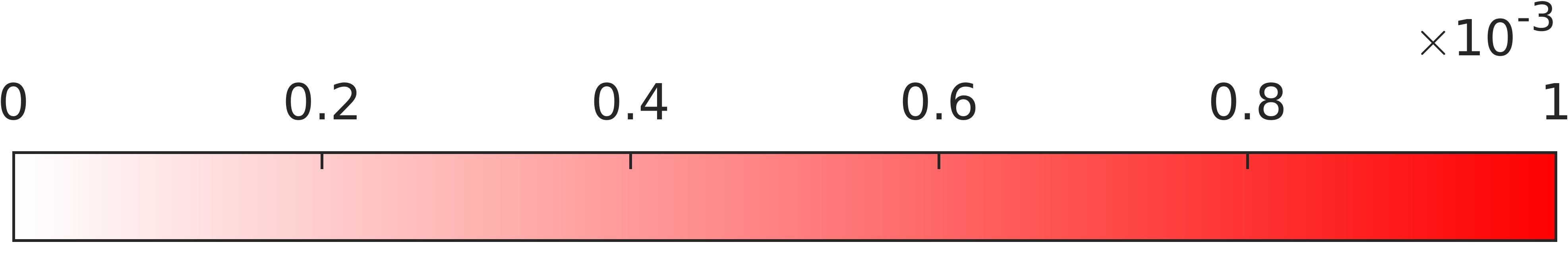}}
    \end{subfigure}
    \begin{subfigure}[b]{0.362\textwidth}{\centering\includegraphics[width=1\textwidth]{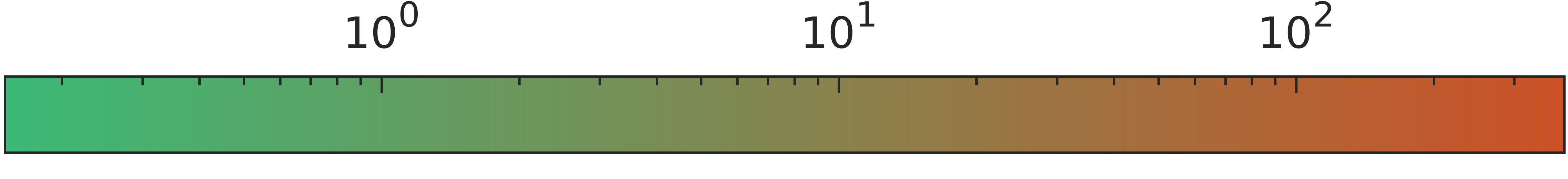}}
    \end{subfigure}\\
    \begin{subfigure}[b]{0.24\textwidth}{\centering\includegraphics[width=1\textwidth]{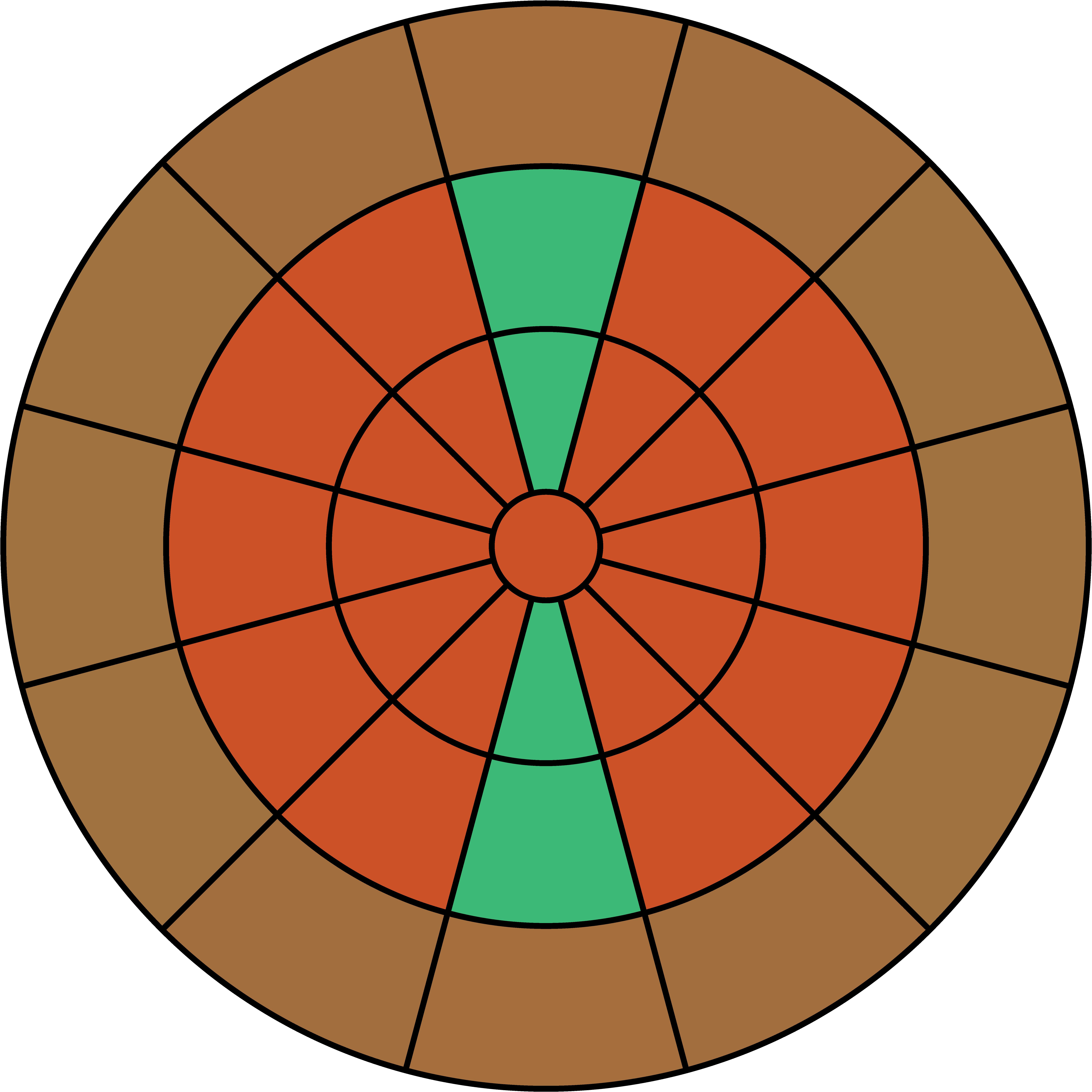}}
             \caption{\centering Material distribution for $N_{\mathrm{ring}}$=3, $N_{\mathrm{sec}}$=12}
             \label{fig:Schematic meta concentrator}
    \end{subfigure}
    \begin{subfigure}[b]{0.36\textwidth}{\centering\includegraphics[width=1\textwidth]{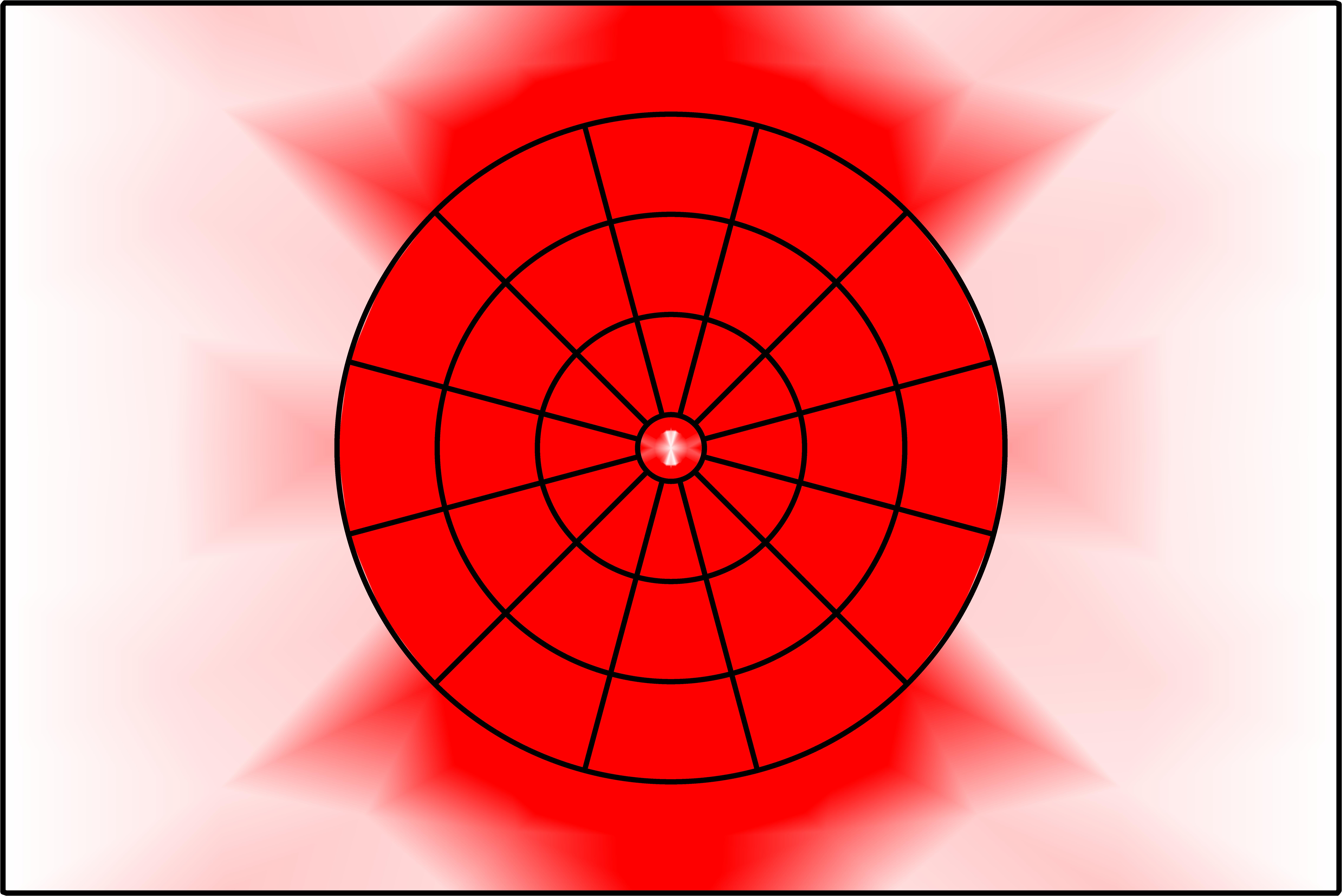}}
             \caption{\centering$|T-\overline{T}|$ for $N_{\mathrm{ring}}$=3, $N_{\mathrm{sec}}$=12, $\Psi_{\mathrm{cloak}}=1.425\times 10^{-5}$}
             \label{fig:Schematic meta concentrator}
    \end{subfigure}
    \begin{subfigure}[b]{0.36\textwidth}{\centering\includegraphics[width=1\textwidth]{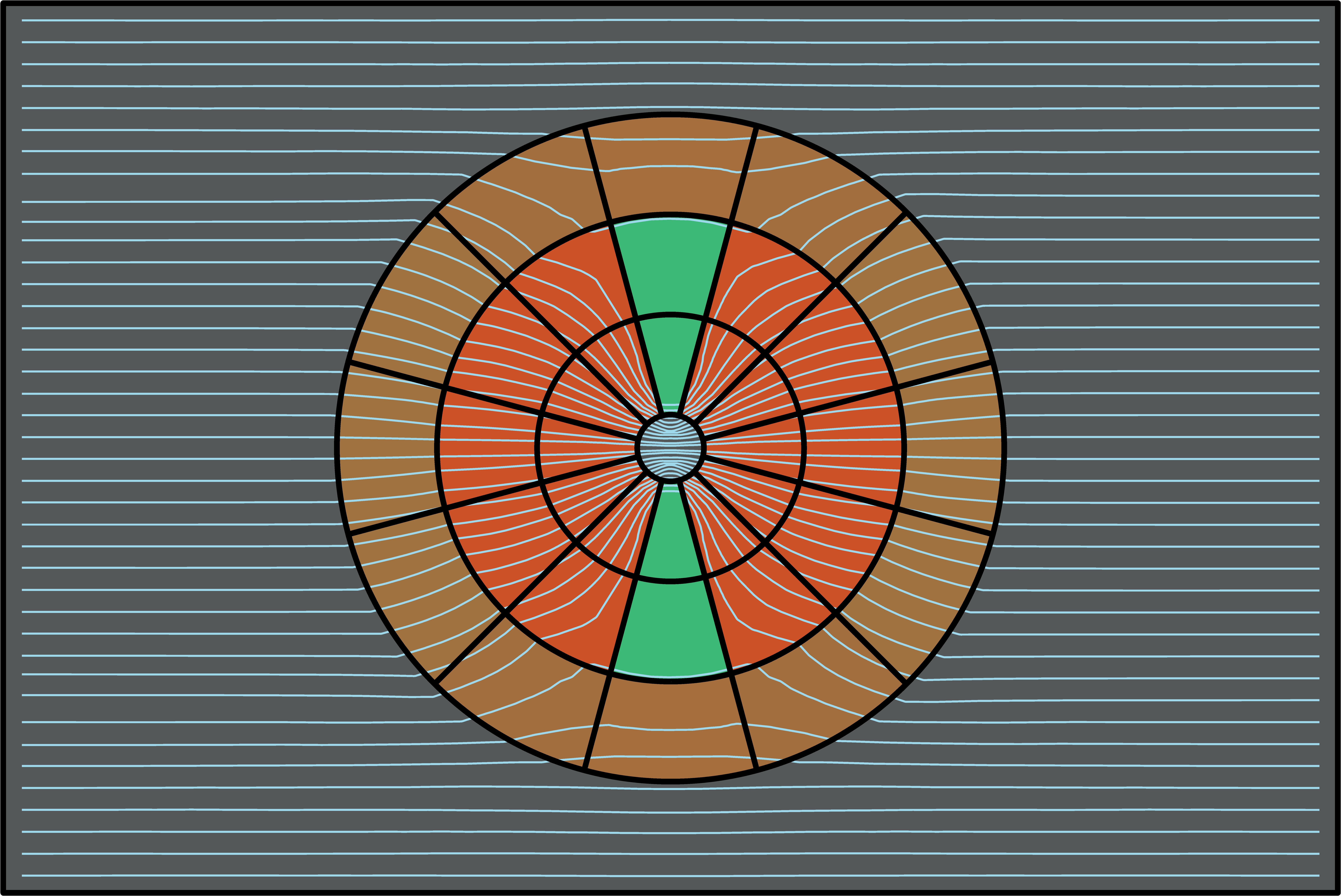}}
             \caption{\centering Flux flow for $N_{\mathrm{ring}}$=3, $N_{\mathrm{sec}}$=12, $\Psi_{\mathrm{flux}}=7.587$}
             \label{fig:Schematic meta concentrator}
    \end{subfigure}\\
    \begin{subfigure}[b]{0.24\textwidth}{\centering\includegraphics[width=1\textwidth]{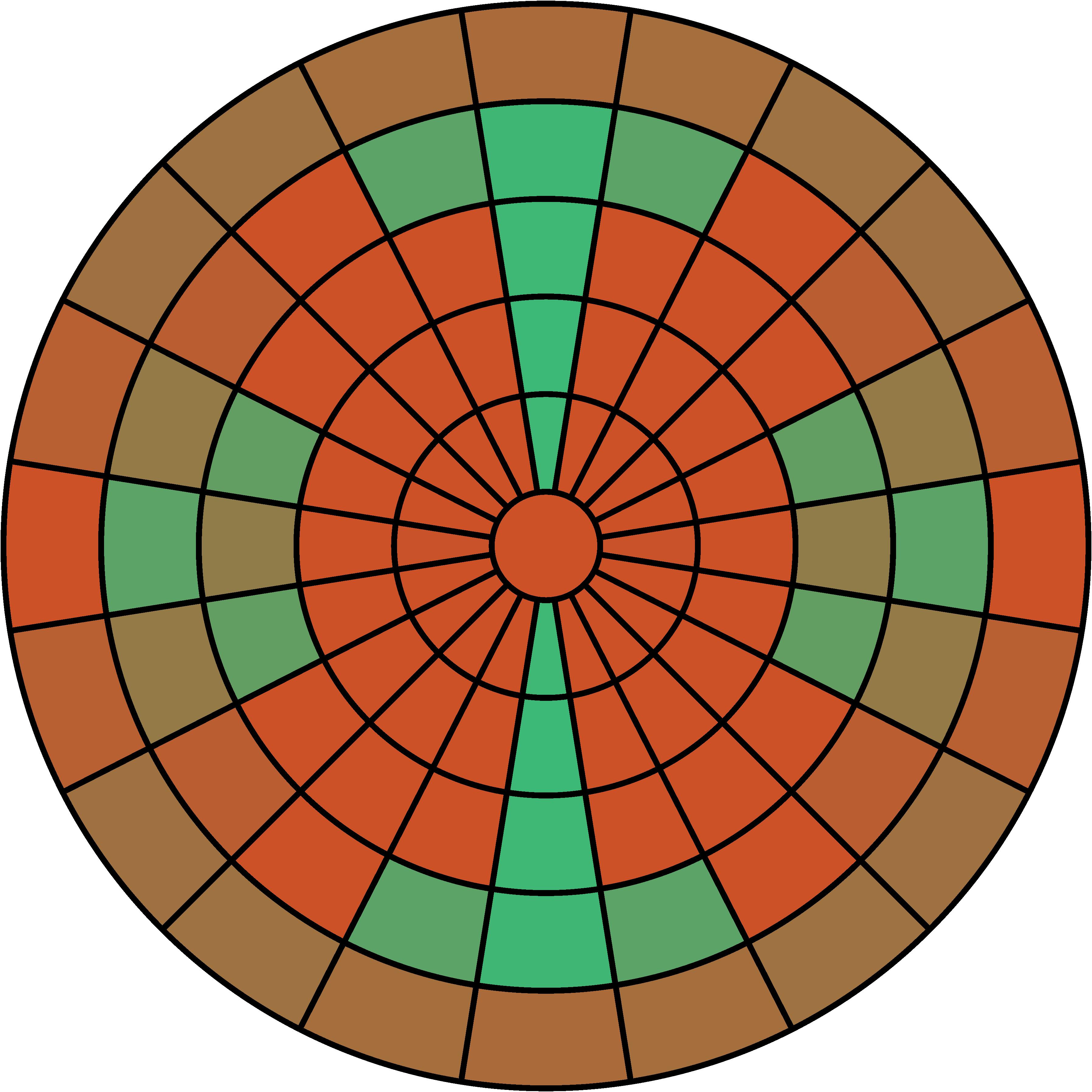}}
             \caption{\centering Material distribution for $N_{\mathrm{ring}}$=5, $N_{\mathrm{sec}}$=20}
             \label{fig:Schematic meta concentrator}
    \end{subfigure}
    \begin{subfigure}[b]{0.36\textwidth}{\centering\includegraphics[width=1\textwidth]{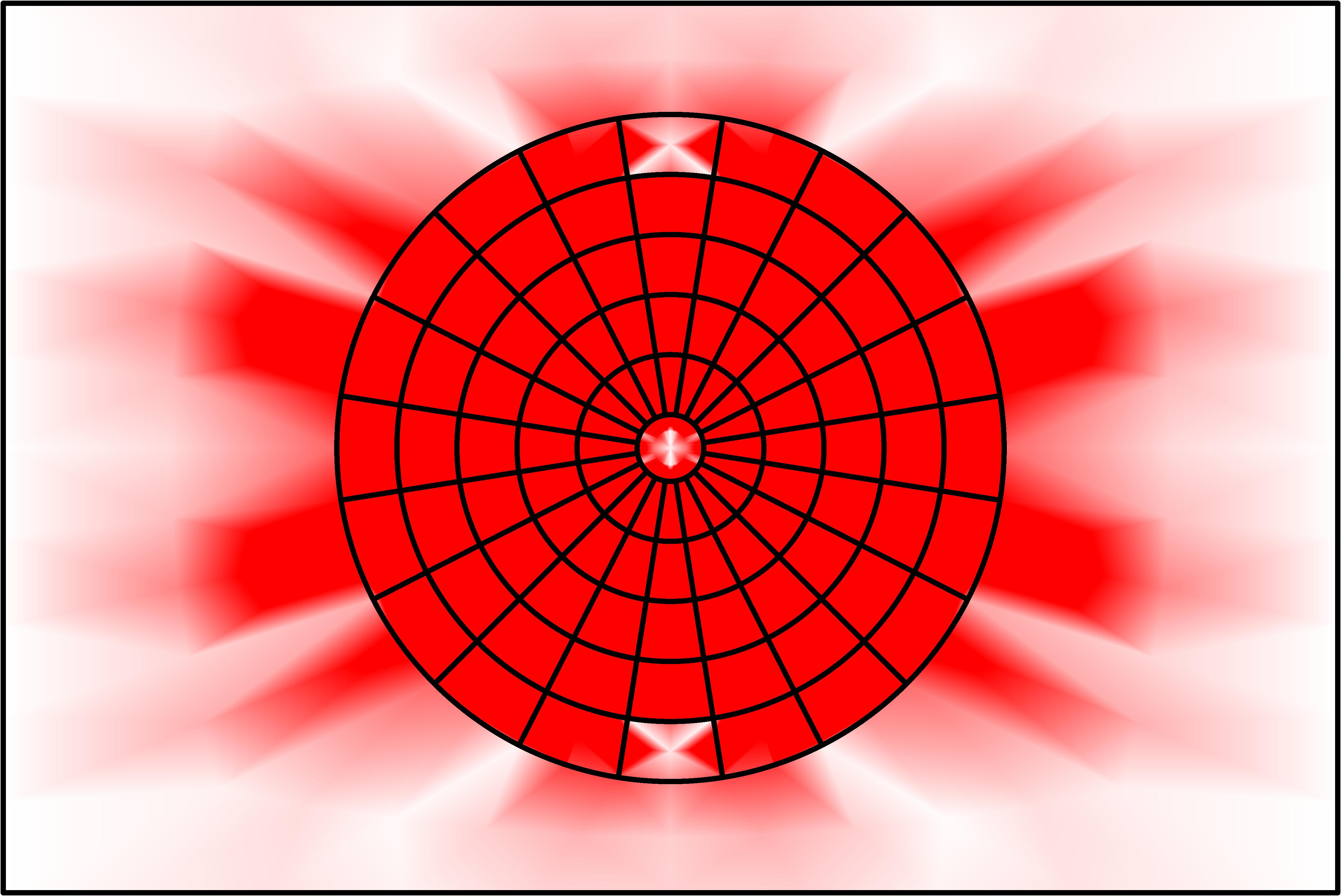}}
             \caption{\centering$|T-\overline{T}|$ for $N_{\mathrm{ring}}$=5, $N_{\mathrm{sec}}$=20,  $\Psi_{\mathrm{cloak}}=7.904\times 10^{-6}$}
             \label{fig:Schematic meta concentrator}
    \end{subfigure}
    \begin{subfigure}[b]{0.36\textwidth}{\centering\includegraphics[width=1\textwidth]{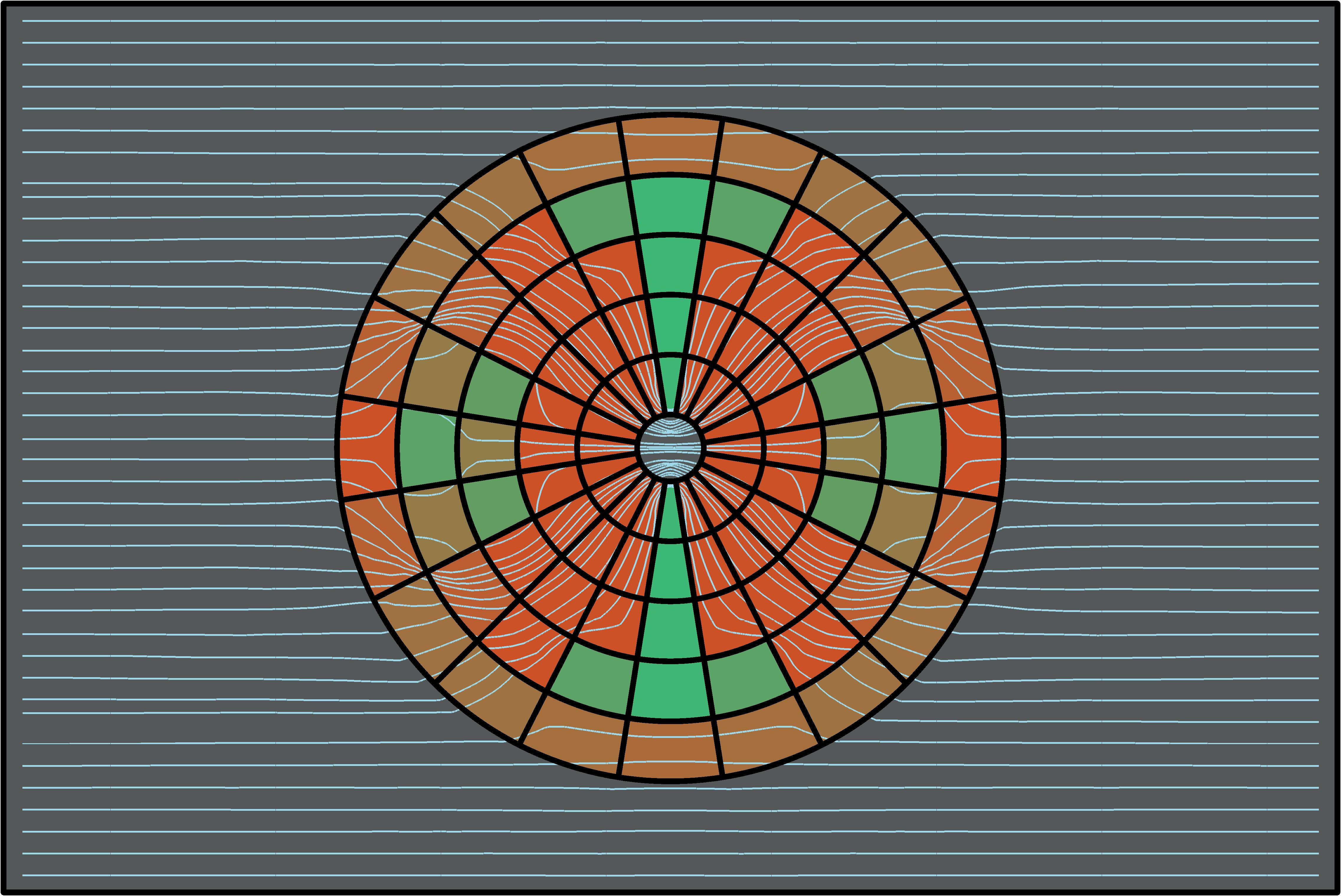}}
             \caption{\centering Flux flow for $N_{\mathrm{ring}}$=5, $N_{\mathrm{sec}}$=20, $\Psi_{\mathrm{flux}}=8.500$}
             \label{fig:Schematic meta concentrator}
    \end{subfigure}
   \caption{optimized block-wise $\kappa$-distribution of thermal cloak-concentrator; temperature disturbance and flux concentration after adding thermal cloak-concentrator in the base material plate. (a,b,c) $N_{\mathrm{ring}}$=3, $N_{\mathrm{sec}}$=12 $f_{\mathrm{obj}}=3.161\times10^{-4}$ (d,e,f) $N_{\mathrm{ring}}$=5, $N_{\mathrm{sec}}$=20, $f_{\mathrm{obj}}=1.995\times10^{-4}$. The optimized conductivity distribution primarily creates fan-type of shapes of low-conductivity material above and below inner core.}
\label{fig:cloak + concentrator MatDistr k opt}
\end{figure}
\begin{figure}[htbp!]
    \centering
    \begin{subfigure}[t]{0.46\textwidth}{\begin{center}\includegraphics[width=1\textwidth]{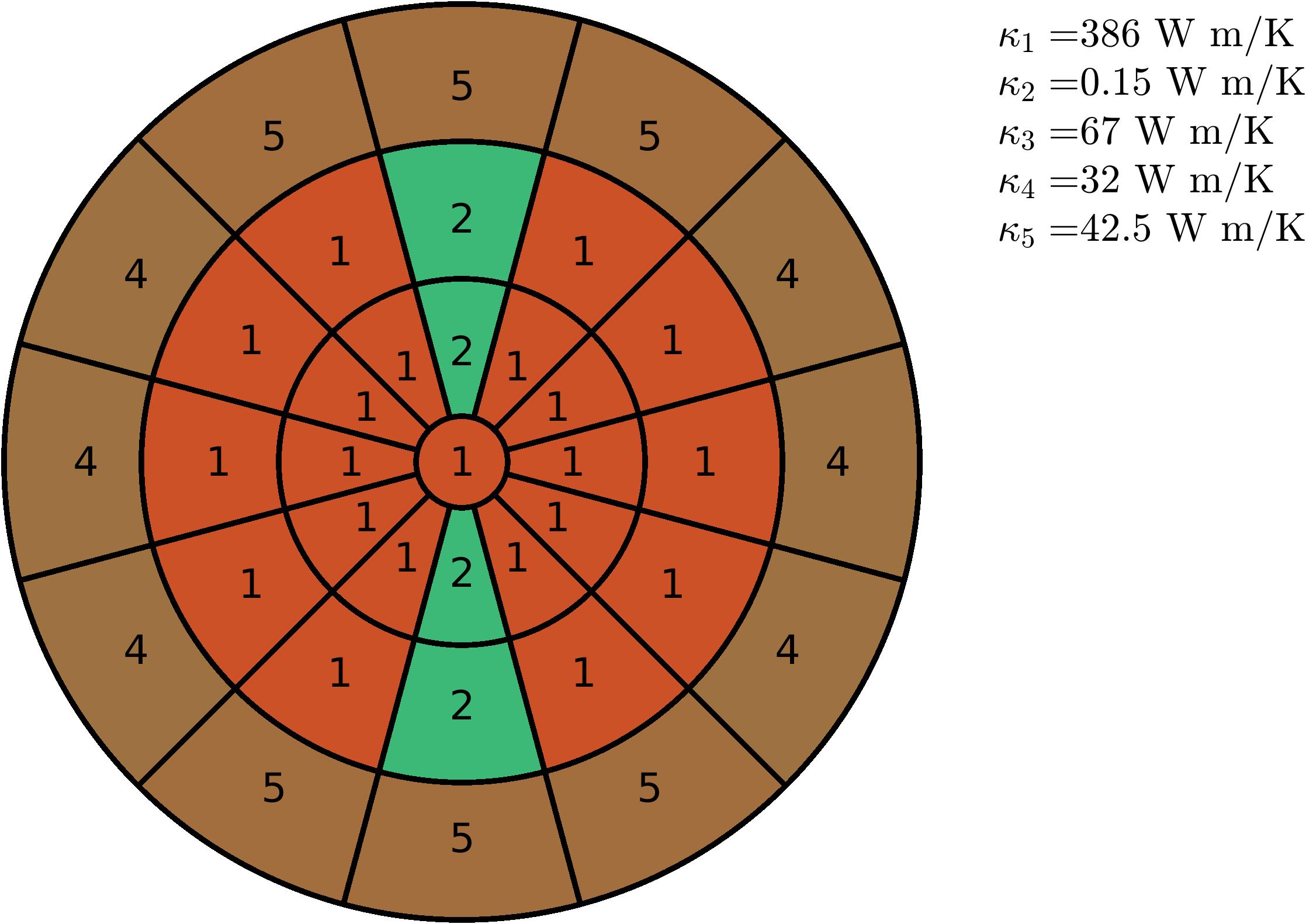}\end{center}}
             \caption{$N_{\mathrm{ring}}$=3, $N_{\mathrm{sec}}$=12. Conductivity values for all materials: $f_{\mathrm{obj}}=4.632\times 10^{-4}$ with $\Psi_{\mathrm{flux}}=7.688$ and $\Psi_{\mathrm{claok}}=1.769\times 10^{-4}$}
             \label{fig:Schematic meta concentrator}
    \end{subfigure}\quad
    \begin{subfigure}[t]{0.48\textwidth}{\begin{center}\includegraphics[width=1\textwidth]{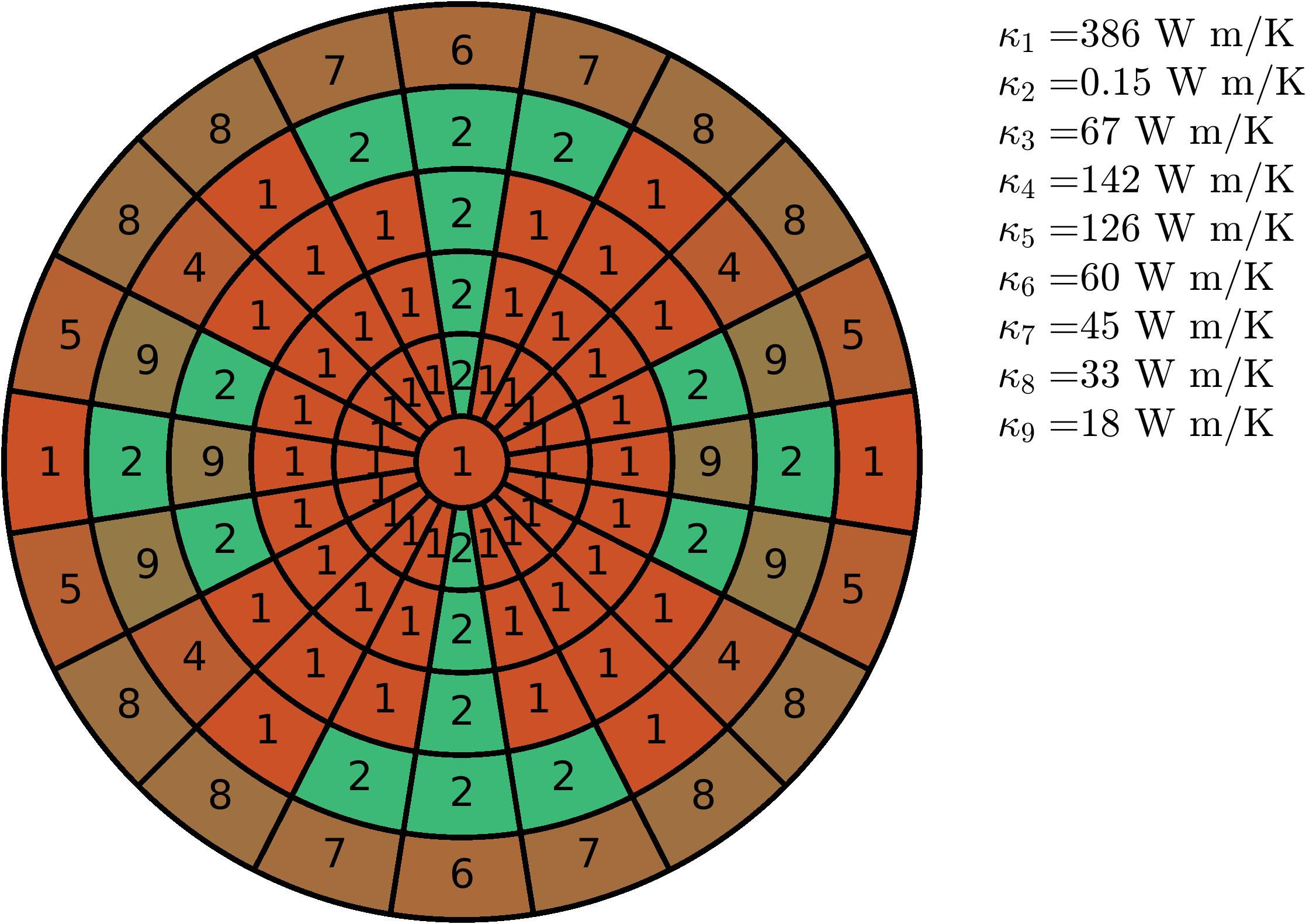}\end{center}}
             \caption{$N_{\mathrm{ring}}$=5, $N_{\mathrm{sec}}$=20. Conductivity values for all materials: $f_{\mathrm{obj}}=5.367\times 10^{-4}$ with $\Psi_{\mathrm{flux}}=8.212$ and $\Psi_{\mathrm{claok}}=3.168\times 10^{-4}$}
             \label{fig:Schematic meta concentrator}
    \end{subfigure}
\caption{Block-wise initial material distribution for shape optimization. To avoid large number of materials involved into the geometry, the blocks with close conductivity values (from conductivity optimization) are taken as a single material with a round-off conductivity value.}
\label{fig:cloak + concentrator MatNum k opt}
\end{figure}
\par The results obtained from conductivity optimization are shown in \fref{fig:cloak + concentrator MatDistr k opt}. \fref{fig:cloak + concentrator MatDistr k opt} shows the block-wise conductivity distribution, temperature disturbance $T-\overline{T}$ and flux flow. For the geometry with $N_{\mathrm{ring}}=3$ \& $N_{\mathrm{sec}}=12$, the optimization produces $f_{\mathrm{obj}}=3.161\times 10^{-4}$ with $\Psi_{\mathrm{flux}}=7.587$ and $\Psi_{\mathrm{cloak}}=1.425\times 10^{-5}$. Similarly for the geometry with $N_{\mathrm{ring}}=5$ \& $N_{\mathrm{sec}}=20$, $f_{\mathrm{obj}}=1.995\times 10^{-4}$ with $\Psi_{\mathrm{flux}}=8.500$ and $\Psi_{\mathrm{cloak}}=7.904\times 10^{-6}$. The optimized conductivity distribution primarily creates fan-type of shapes of low-conductivity material above and below inner core, similar to the optimized topology mentioned in~\cite{Fujii2020}. In addition to that, similar to the dam structure (a structure made of low $\kappa$ material obstructing the flux as it enters in cloak-concentrator along the $x$-axis) observed in~\cite{Fujii2020}, we also get low-conductivity material distribution in the direction of the incoming flux for $N_{\mathrm{ring}}=5$ \& $N_{\mathrm{sec}}=20$ geometry. As mentioned in~\cite{Fujii2020}, narrow structures of high $\kappa$ material appear between fan shape and base material in the optimal geometries. These narrow structures diminish the temperature disturbance produced by fan shapes. In our material distribution, the same cloaking effect is provided by the blocks with conductivity value close to the base material. After analysing the block-wise conductivity distribution, it is observed that some of the blocks have very close conductivity values. Now, to avoid increasing the number of materials involved in the geometry, the blocks with close conductivity values are taken as a single material with a round-off conductivity value. The updated material distribution and corresponding conductivity values are shown in \fref{fig:cloak + concentrator MatNum k opt}. However, the new conductivity values, which are round-off values of optimized values, changes $f_{\mathrm{obj}}$, $\Psi_{\mathrm{flux}}$ and $\Psi_{\mathrm{cloak}}$ to some extent. The altered values are shown in the caption. 
Lastly, the updated material distribution of block-type of geometry will be used as the initial topology for the shape optimization problem in the next stage. 
\begin{figure}[htbp!]
    \centering
    \begin{subfigure}[b]{0.24\textwidth}{\centering\includegraphics[width=1\textwidth,height=0.0355\textheight]{Figures_fujii2020case/fujii2020casev3cdMatblockCondCore_MatDistr_colorbar1.jpg}}
    \end{subfigure}
    \begin{subfigure}[b]{0.36\textwidth}{\centering\includegraphics[width=1\textwidth]{Figures_fujii2020case/fujii2020casev3cdMatblockCondCore_TempDiffPlot_colorbar1.jpg}}
    \end{subfigure}
    \begin{subfigure}[b]{0.36\textwidth}{\centering\includegraphics[width=1\textwidth]{Figures_fujii2020case/fujii2020casev3cdMatblockCondCore_FluxPlot_colorbar1.jpg}}
    \end{subfigure}\\
    \begin{subfigure}[t]{0.24\textwidth}{\centering
        \includegraphics[width=1\textwidth]{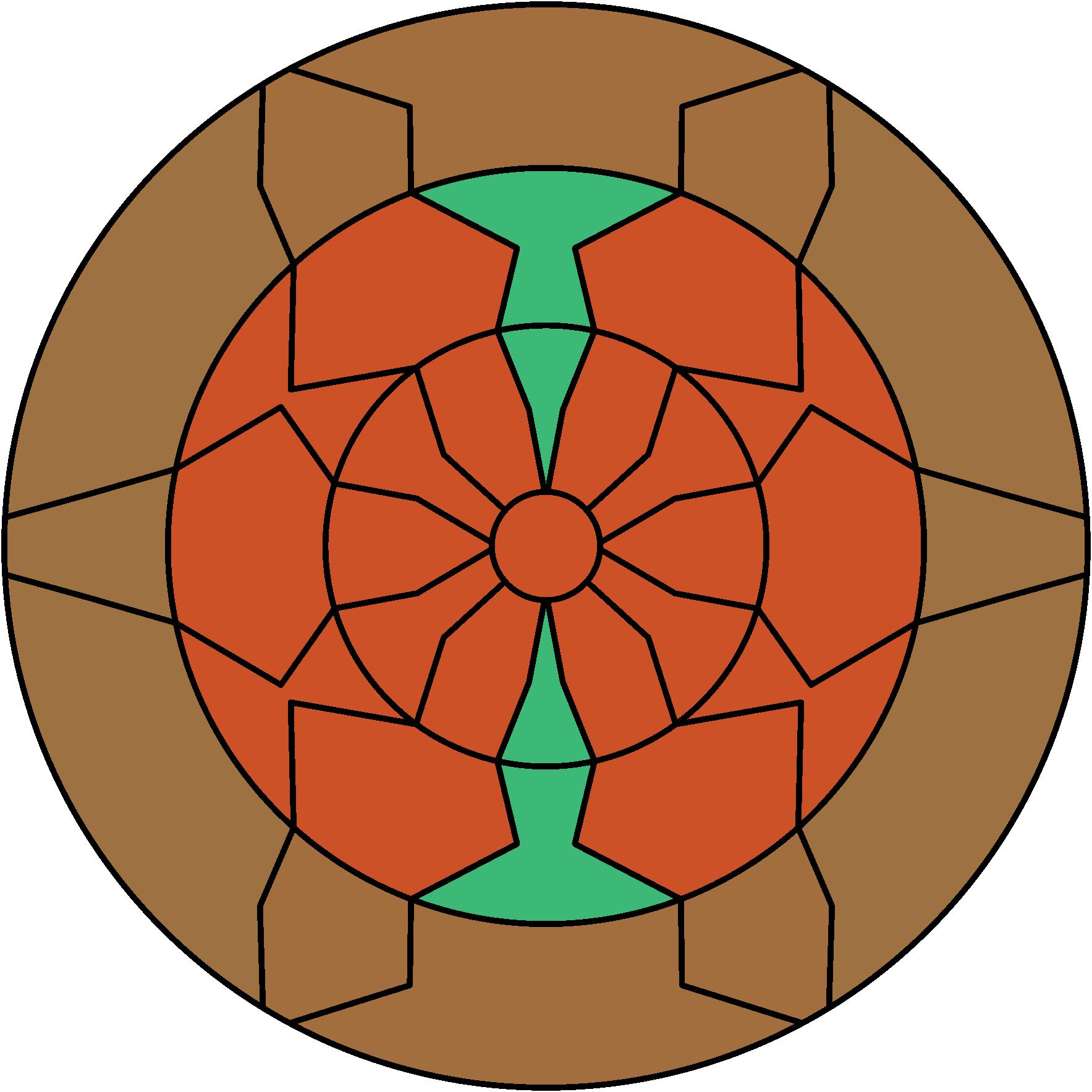}}
             \caption{\centering  Optimized shape for $N_{\mathrm{var}}$=23}
             \label{fig:Schematic meta concentrator}
    \end{subfigure}
    \begin{subfigure}[t]{0.36\textwidth}{\centering\includegraphics[width=1\textwidth]{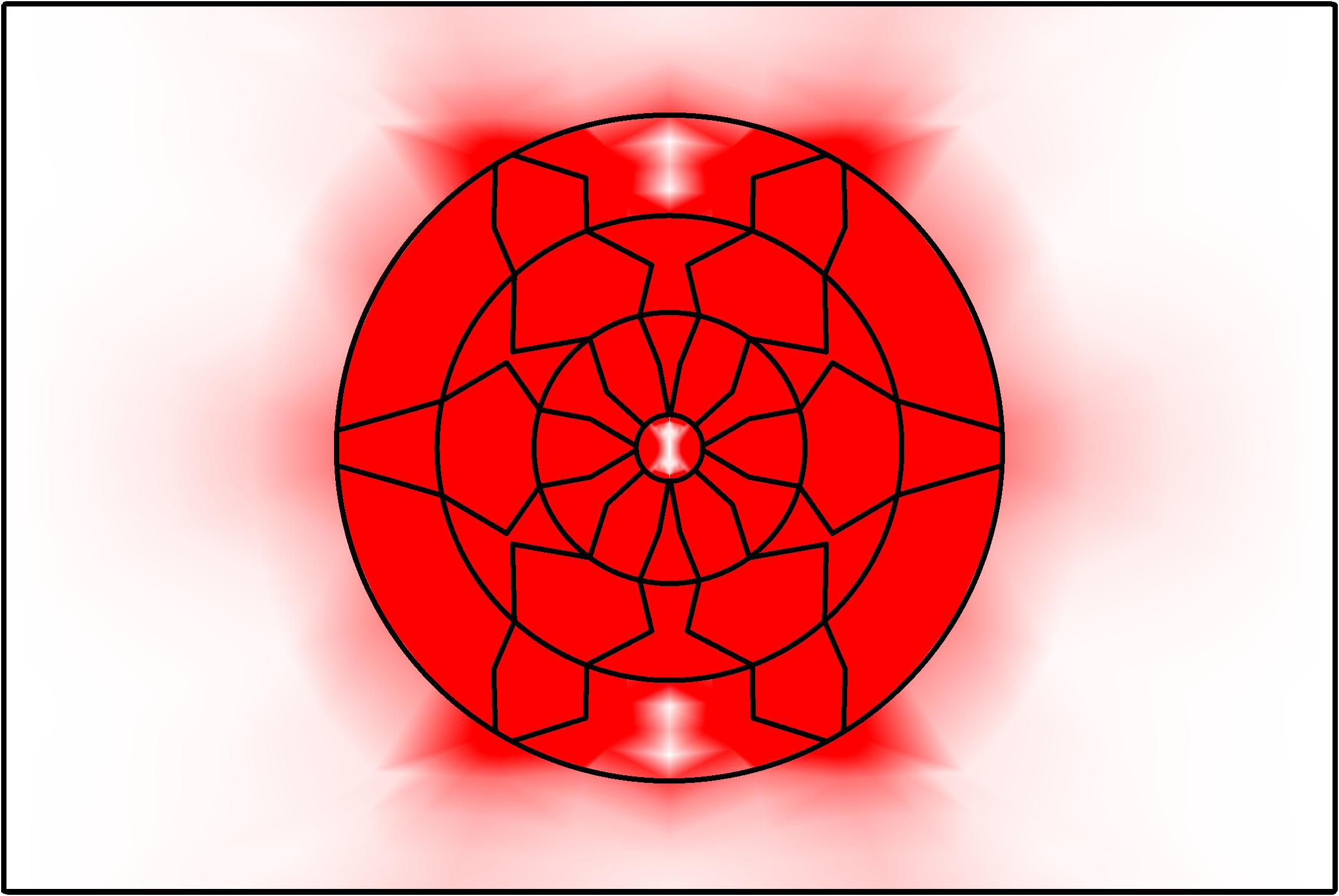}}
             \caption{\centering  $|T-\overline{T}|$ for $N_{\mathrm{var}}$=23, $\Psi_{\mathrm{cloak}}=4.749\times 10^{-6}$}
             \label{fig:Schematic meta concentrator}
    \end{subfigure}
    \begin{subfigure}[t]{0.36\textwidth}{\centering\includegraphics[width=1\textwidth]{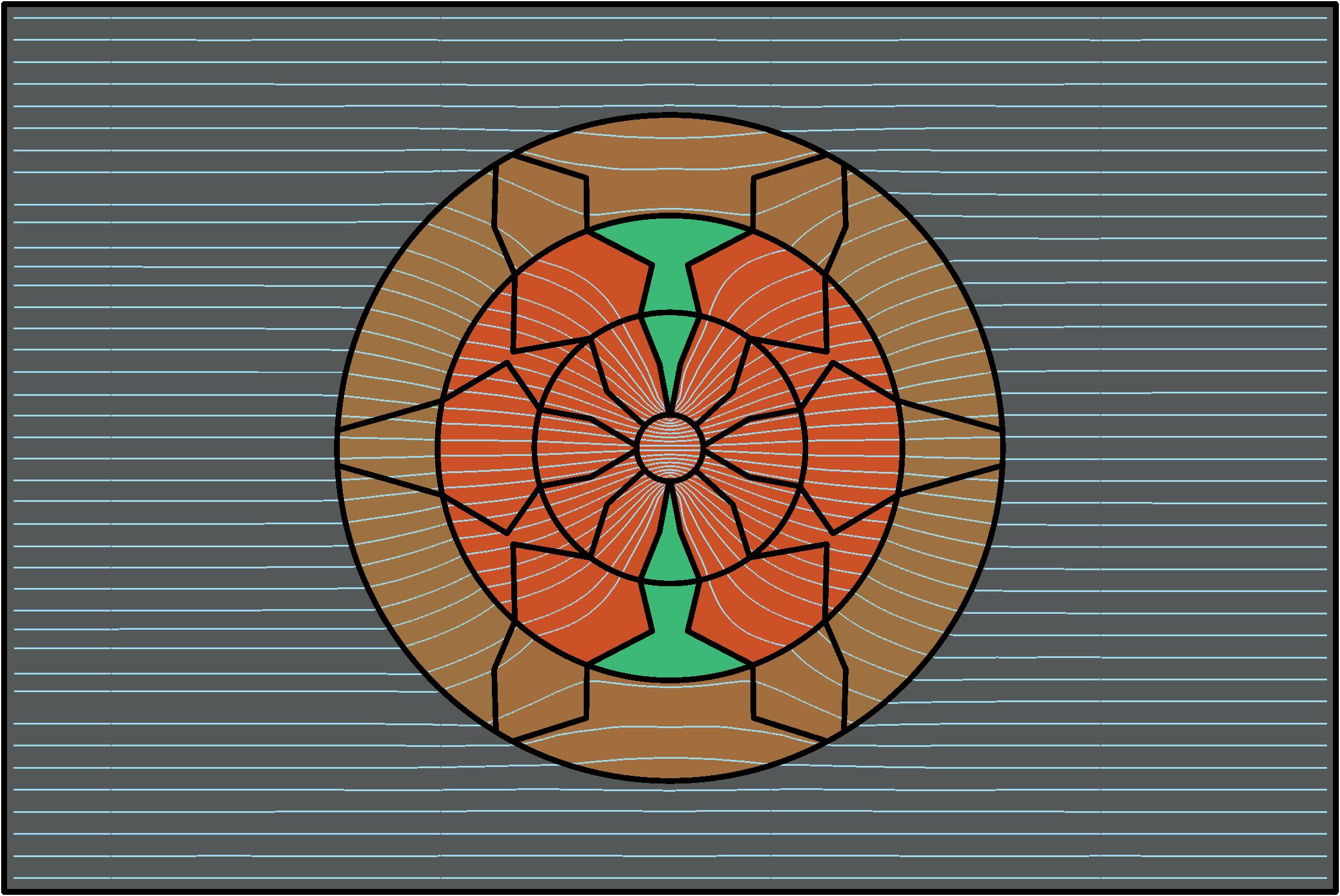}}
             \caption{\centering Flux flow for $N_{\mathrm{var}}$=23,\quad $\Psi_{\mathrm{flux}}=8.077$.}
             \label{fig:Schematic meta concentrator}
    \end{subfigure}\\
    \begin{subfigure}[t]{0.24\textwidth}{\centering\includegraphics[width=1\textwidth]{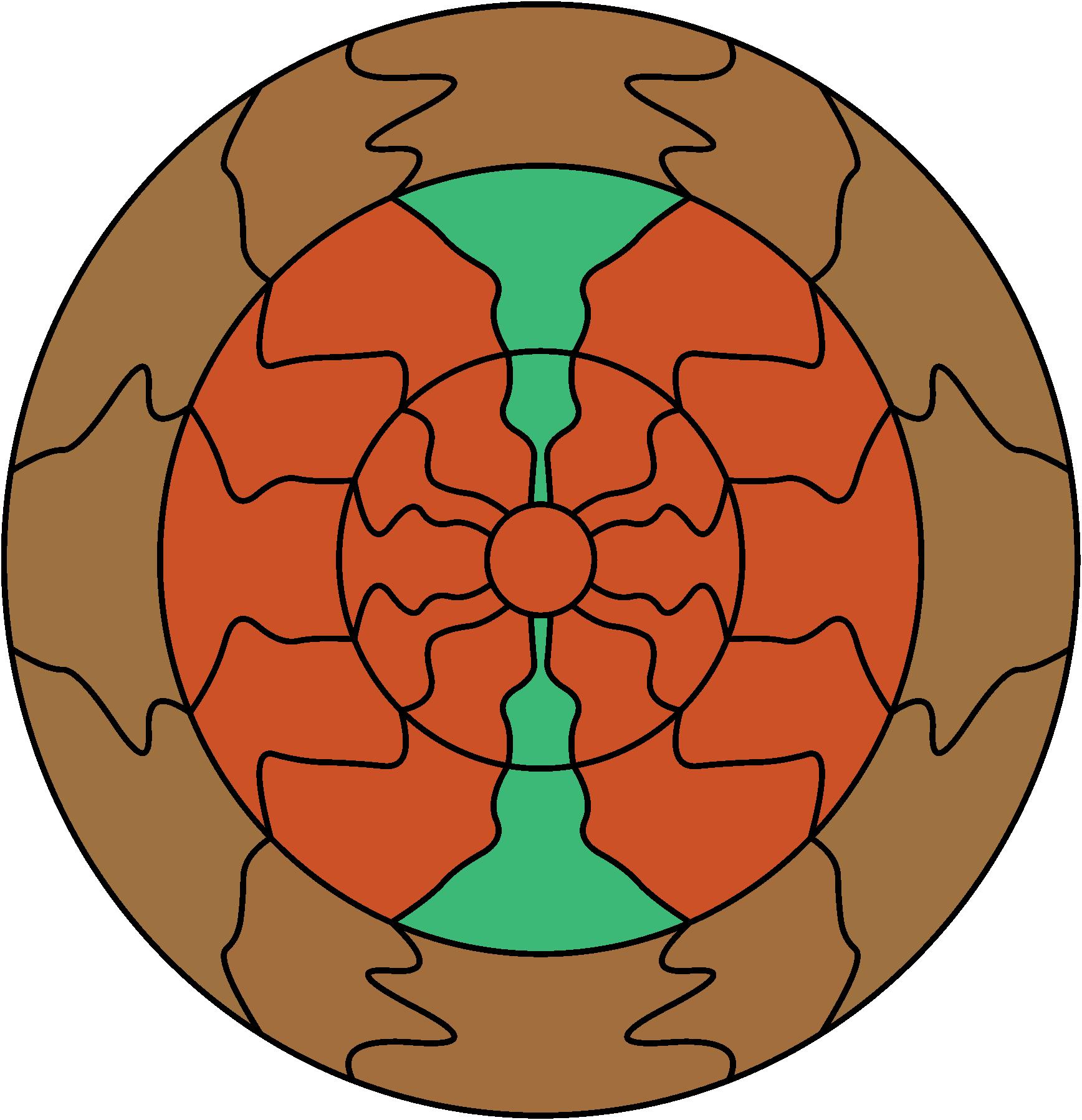}}
             \caption{\centering Optimized shape for $N_{\mathrm{var}}$=50}
             \label{fig:Schematic meta concentrator}
    \end{subfigure}
    \begin{subfigure}[t]{0.36\textwidth}{\centering\includegraphics[width=1\textwidth]{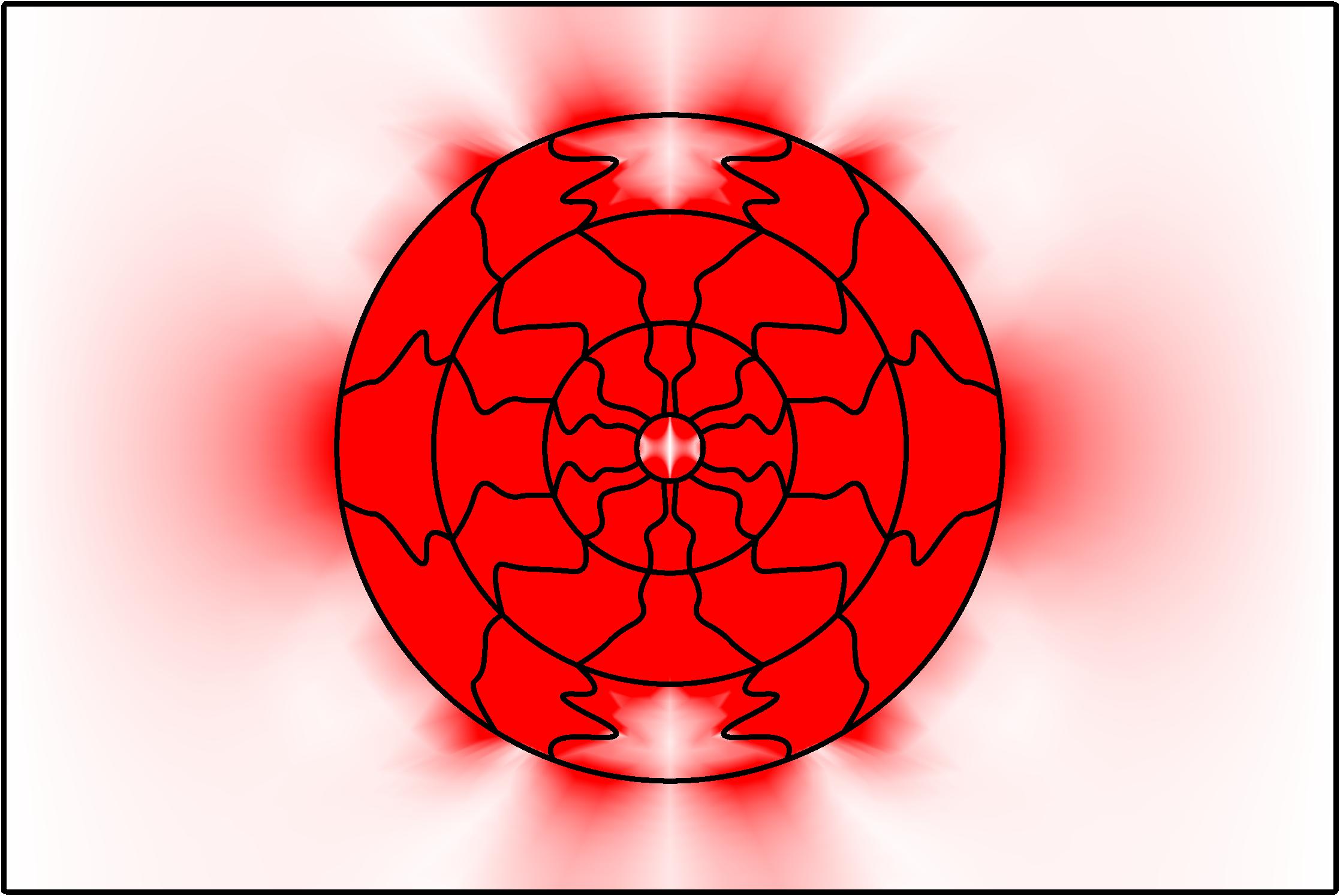}}
             \caption{\centering $|T-\overline{T}|$ for $N_{\mathrm{var}}$=50, $\Psi_{\mathrm{cloak}}=7.451\times 10^{-6}$}
             \label{fig:Schematic meta concentrator}
    \end{subfigure}
    \begin{subfigure}[t]{0.36\textwidth}{\centering\includegraphics[width=1\textwidth]{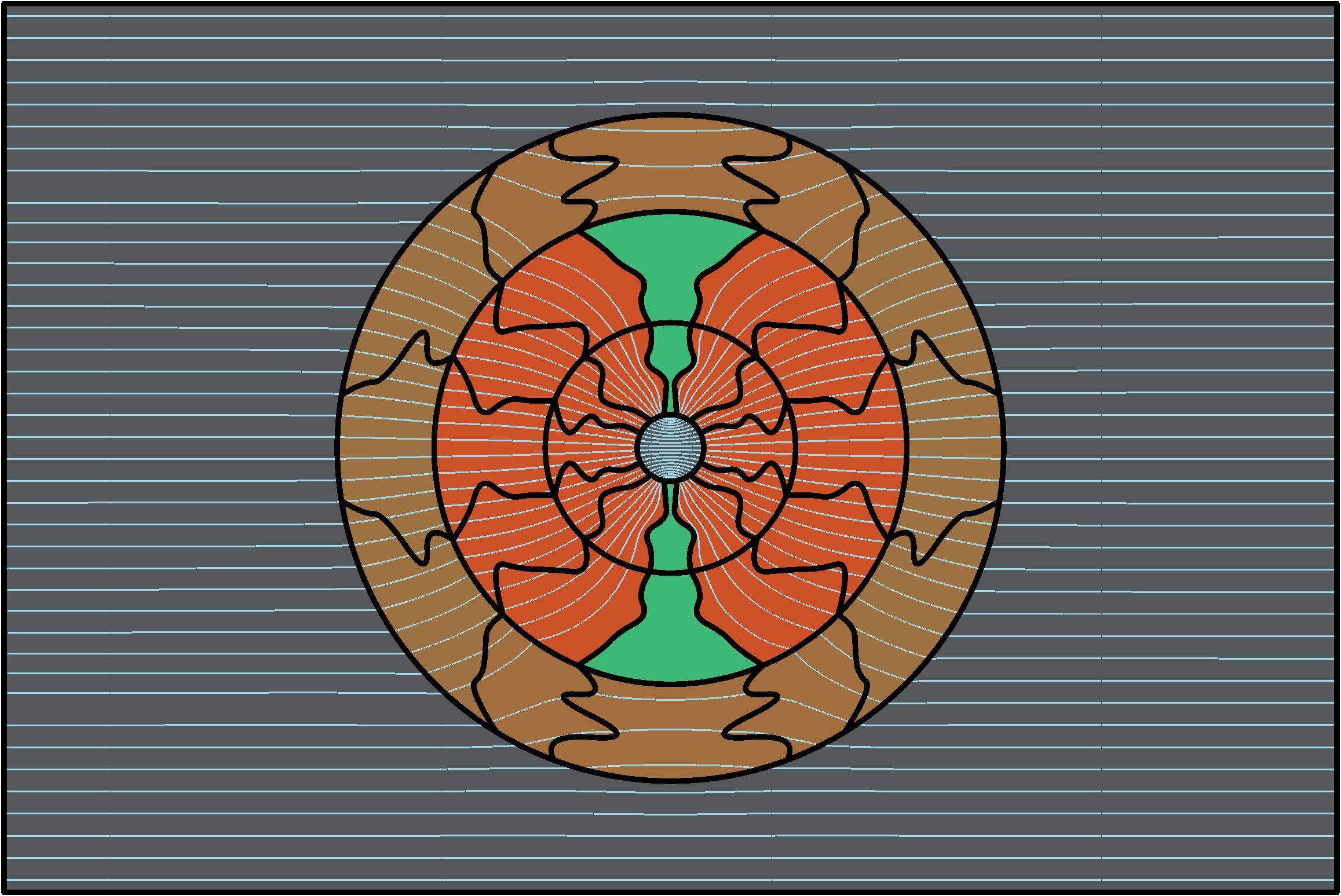}}
             \caption{\centering Flux flow for $N_{\mathrm{var}}$=50, \quad $\Psi_{\mathrm{flux}}=8.012$}
             \label{fig:Schematic meta concentrator}
    \end{subfigure}
   \caption{optimized shape of thermal cloak-concentrator; temperature disturbance ($T-\overline{T}$) and flux flow after adding thermal cloak-concentrator in the base material plate for $N_{\mathrm{ring}}$=3, $N_{\mathrm{sec}}$=12 (a,b,c) $N_{\mathrm{var}}=23$, $f_{\mathrm{obj}}=2.397\times10^{-4}$ (d,e,f) $N_{\mathrm{var}}=50$, $f_{\mathrm{obj}}=2.501\times10^{-4}$. For both cases, the concentrate flux is around 8 times as large as for the flat plate. The normalized temperature difference in the outer region is in the order of $10^{-6}$.}
\label{fig:cloak + concentrator objfn 1}
\end{figure}

\begin{figure}[htbp!]
    \centering
    \begin{subfigure}[b]{0.24\textwidth}{\centering\includegraphics[width=1\textwidth,height=0.0355\textheight]{Figures_fujii2020case/fujii2020casev3cdMatblockCondCore_MatDistr_colorbar1.jpg}}
    \end{subfigure}
    \begin{subfigure}[b]{0.36\textwidth}{\centering\includegraphics[width=1\textwidth]{Figures_fujii2020case/fujii2020casev3cdMatblockCondCore_TempDiffPlot_colorbar1.jpg}}
    \end{subfigure}
    \begin{subfigure}[b]{0.36\textwidth}{\centering\includegraphics[width=1\textwidth]{Figures_fujii2020case/fujii2020casev3cdMatblockCondCore_FluxPlot_colorbar1.jpg}}
    \end{subfigure}\\
     \begin{subfigure}[b]{0.24\textwidth}{\centering\includegraphics[width=1\textwidth]{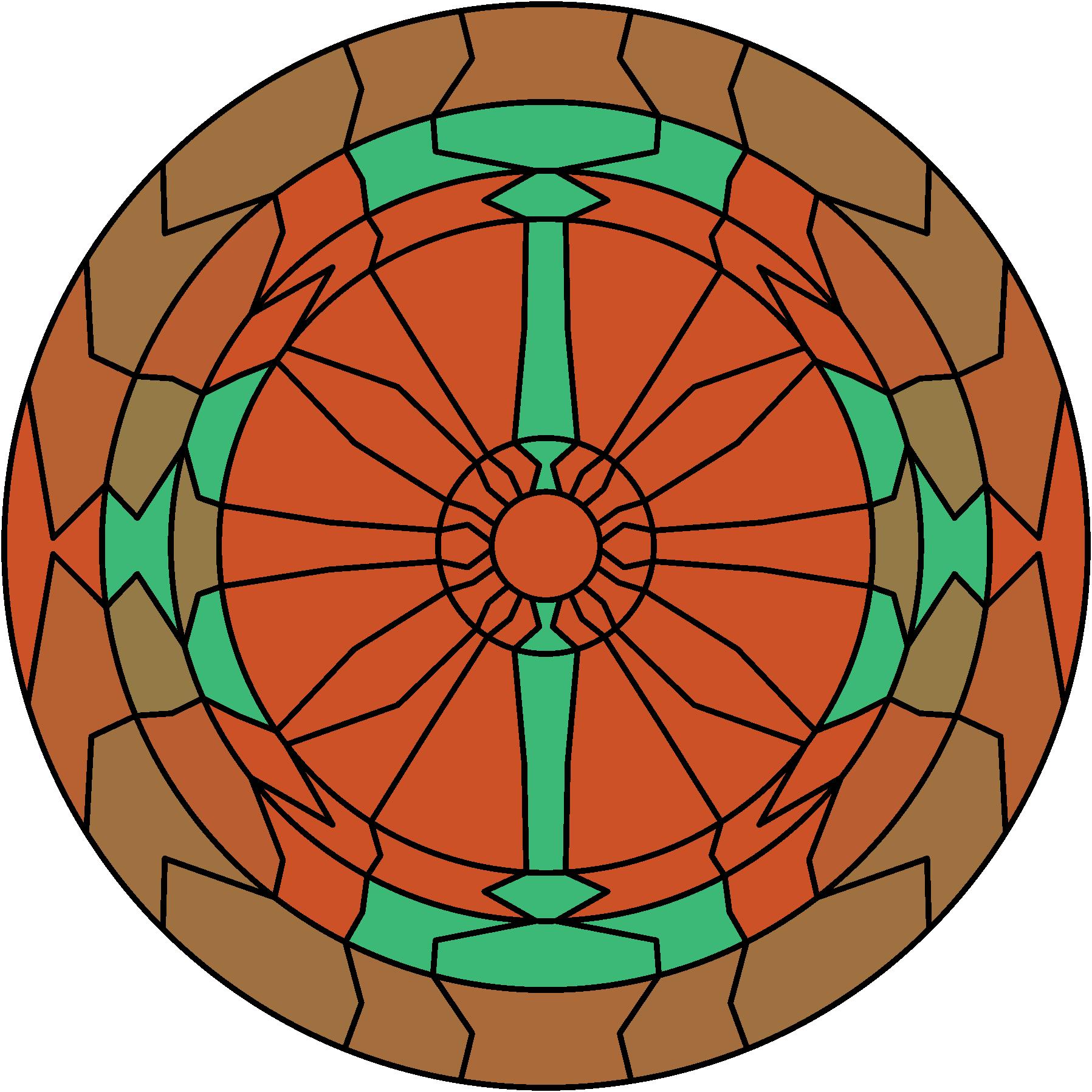}}
             \caption{\centering Opt. shape $N_{\mathrm{ring}}$=5, $N_{\mathrm{sec}}$=20, $N_{\mathrm{var}}$=59}
             \label{fig:Schematic meta concentrator}
    \end{subfigure}
    \begin{subfigure}[b]{0.36\textwidth}{\centering\includegraphics[width=1\textwidth]{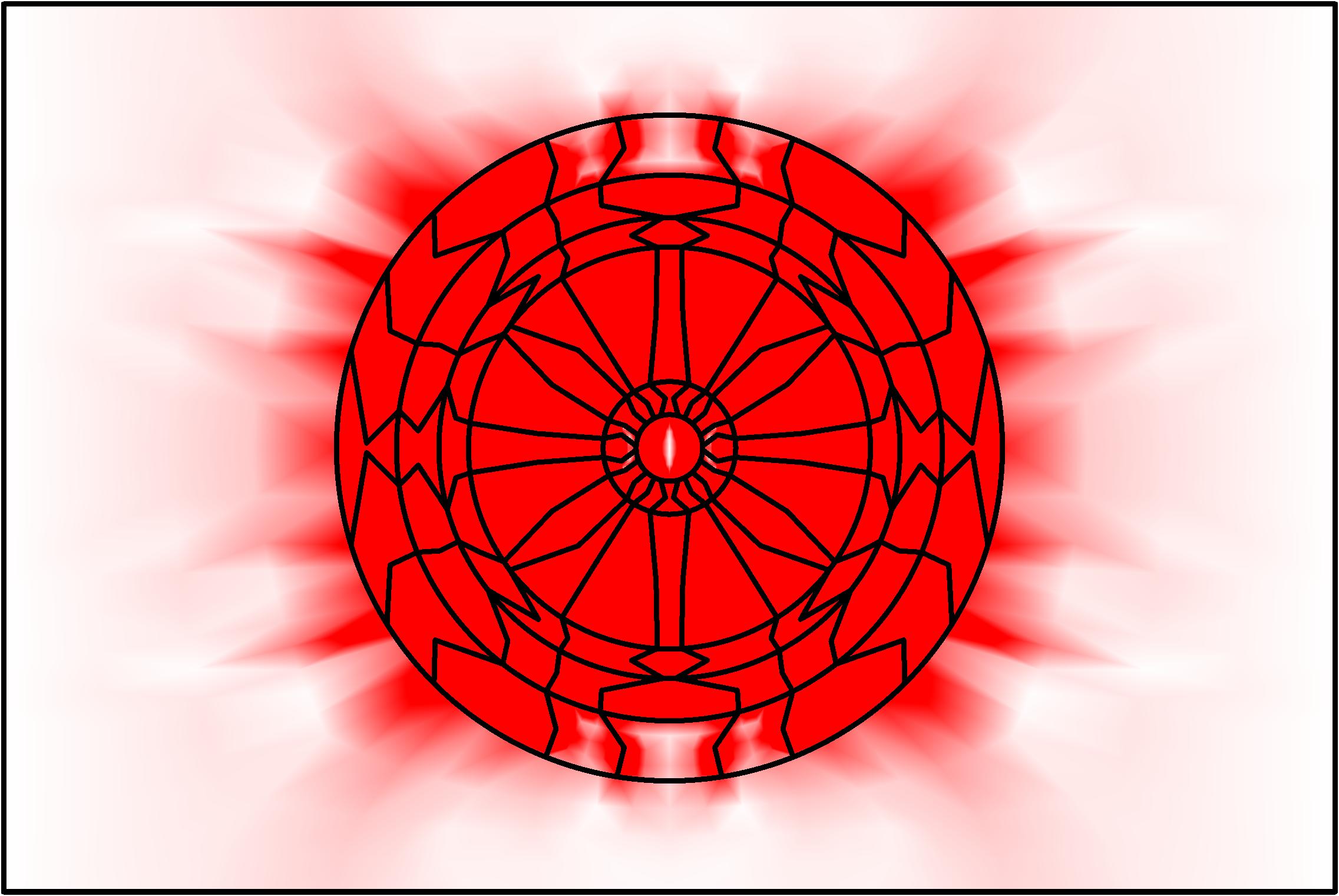}}
             \caption{\centering $|T-\overline{T}|$ for $N_{\mathrm{ring}}$=5, $N_{\mathrm{sec}}$=20, $N_{\mathrm{var}}$=59, $\Psi_{\mathrm{cloak}}=7.363\times 10^{-6}$}
             \label{fig:Schematic meta concentrator}
    \end{subfigure}
    \begin{subfigure}[b]{0.36\textwidth}{\centering\includegraphics[width=1\textwidth]{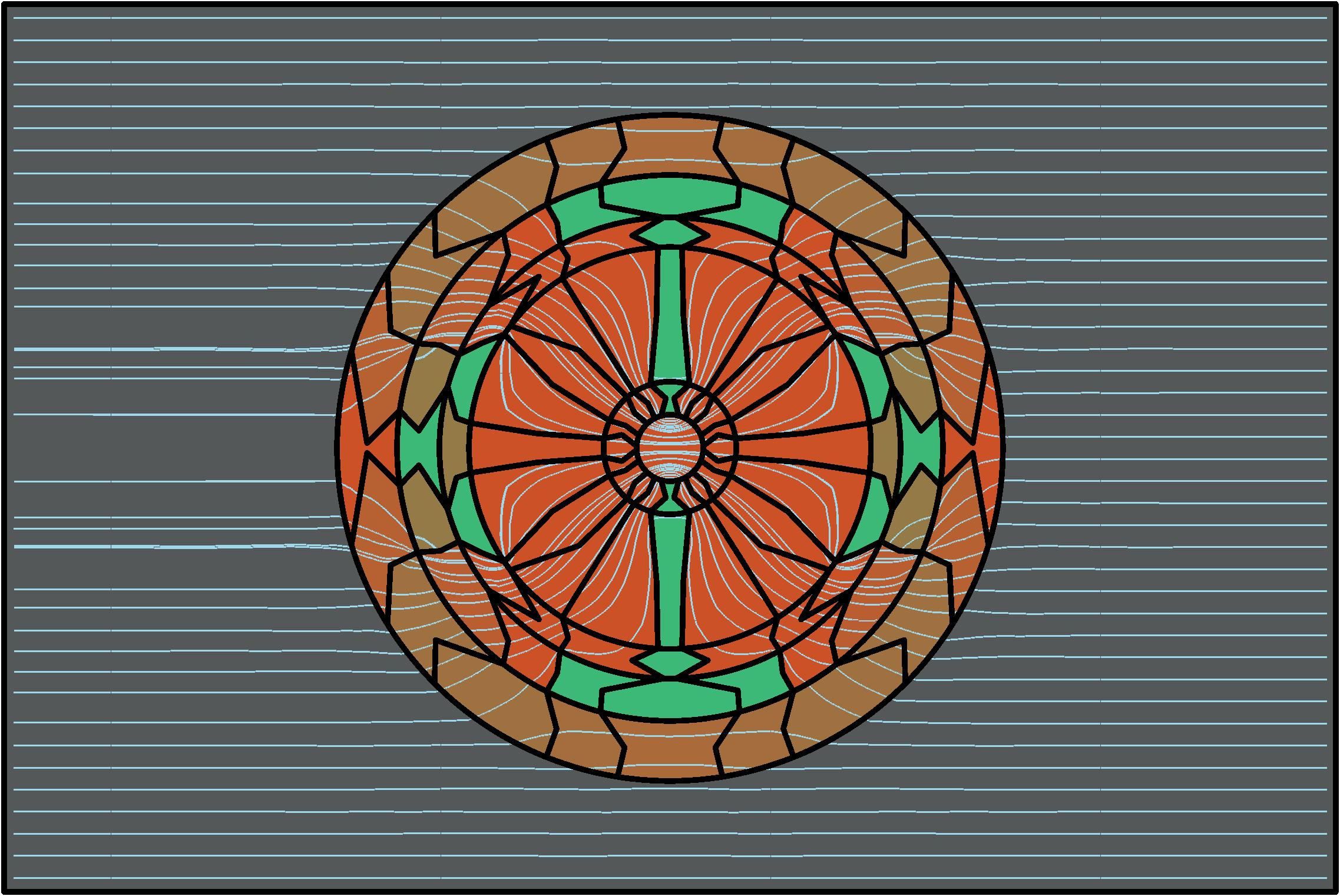}}
             \caption{\centering Flux flow for $N_{\mathrm{ring}}$=5, $N_{\mathrm{sec}}$=20, $N_{\mathrm{var}}$=59, $\Psi_{\mathrm{flux}}=8.316$}
             \label{fig:Schematic meta concentrator}
    \end{subfigure}\\
     \begin{subfigure}[b]{0.24\textwidth}{\centering\includegraphics[width=1\textwidth]{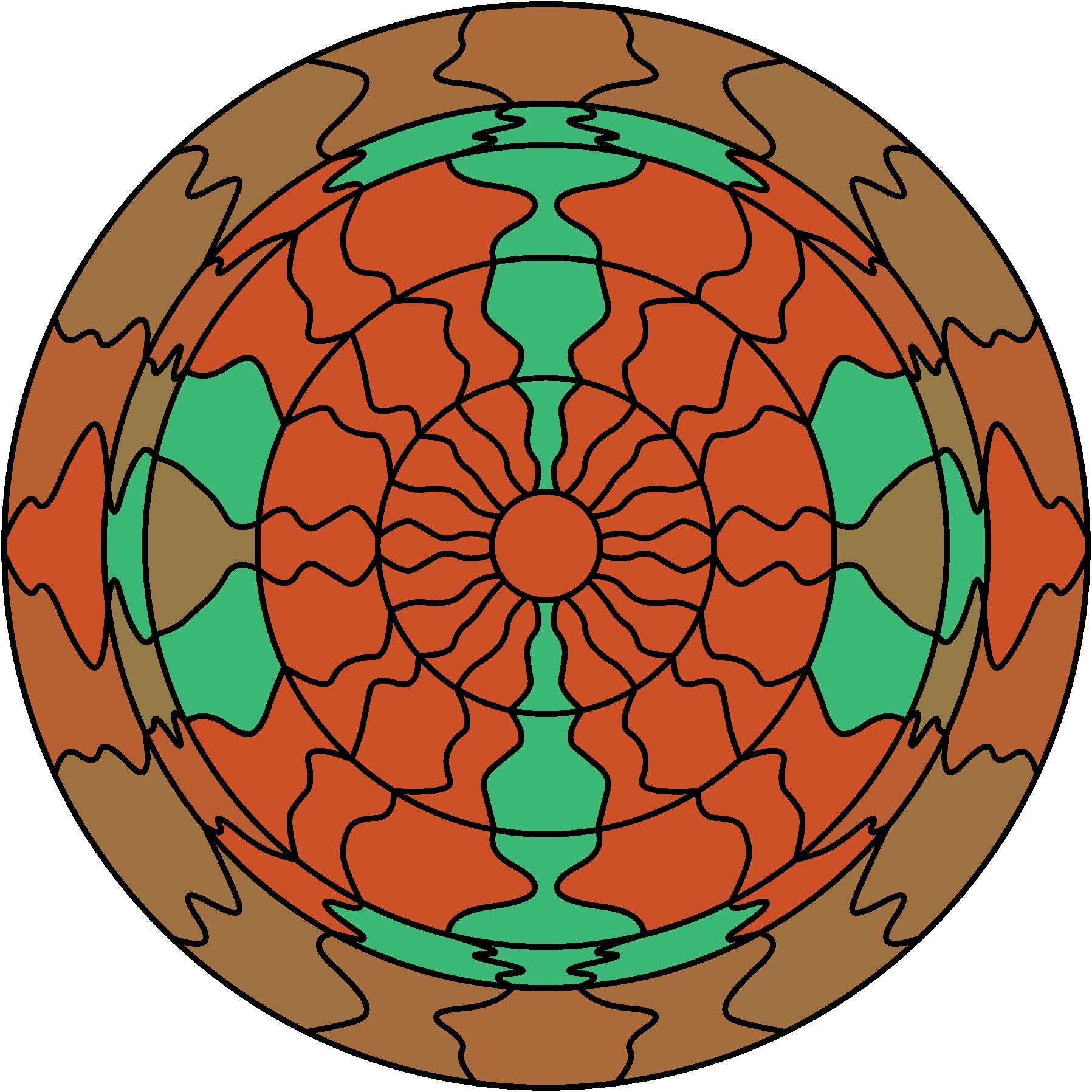}}
             \caption{\centering Opt. shape $N_{\mathrm{ring}}$=5, $N_{\mathrm{sec}}$=20, $N_{\mathrm{var}}$=132}
             \label{fig:Schematic meta concentrator}
    \end{subfigure}
    \begin{subfigure}[b]{0.36\textwidth}{\centering\includegraphics[width=1\textwidth]{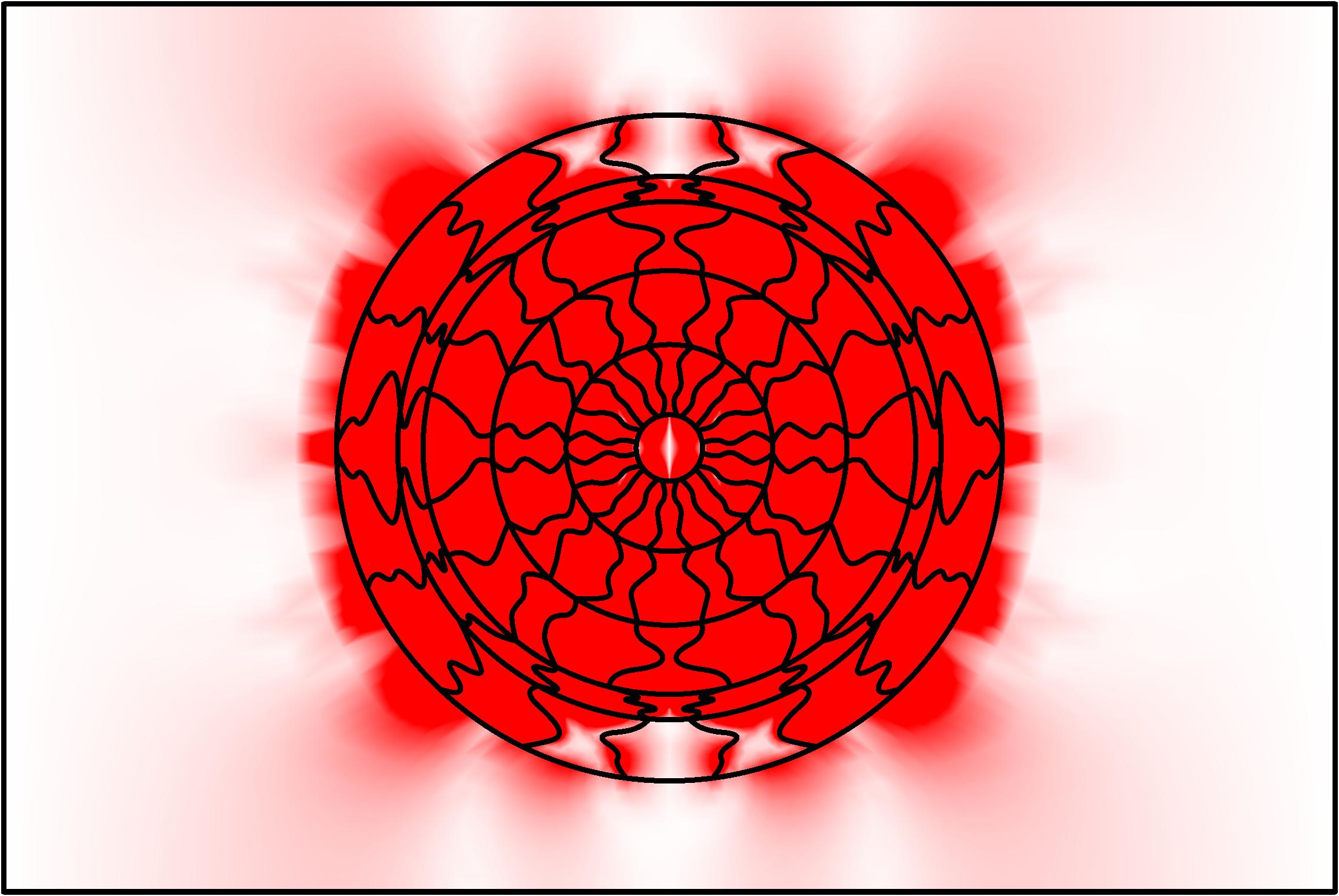}}
             \caption{\centering $|T-\overline{T}|$ for $N_{\mathrm{ring}}$=5, $N_{\mathrm{sec}}$=20, $N_{\mathrm{var}}$=132, $\Psi_{\mathrm{cloak}}=1.232\times 10^{-5}$}
             \label{fig:Schematic meta concentrator}
    \end{subfigure}
    \begin{subfigure}[b]{0.36\textwidth}{\centering\includegraphics[width=1\textwidth]{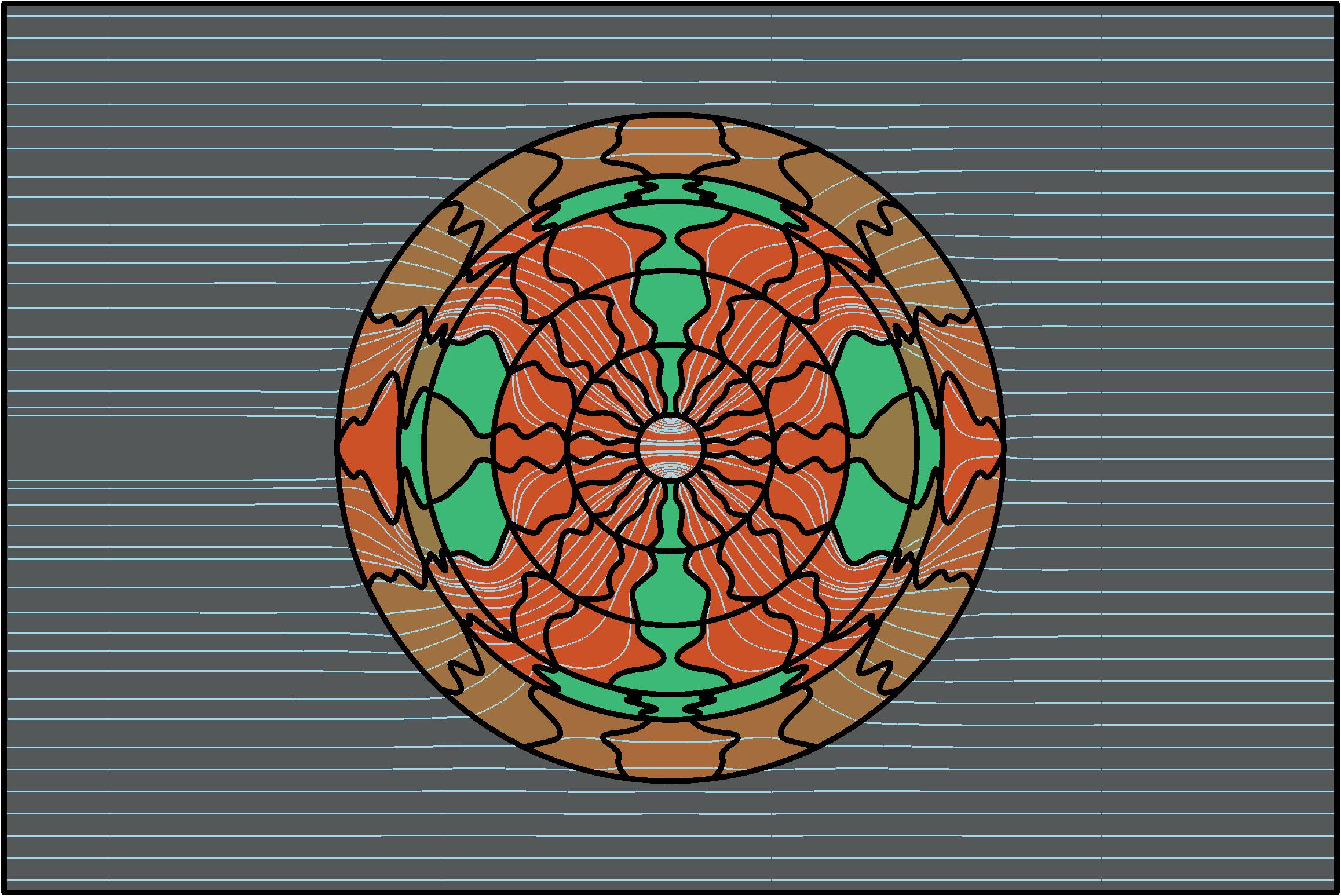}}
             \caption{\centering Flux flow for $N_{\mathrm{ring}}$=5, $N_{\mathrm{sec}}$=20, $N_{\mathrm{var}}$=132, $\Psi_{\mathrm{flux}}=8.318$}
             \label{fig:Schematic meta concentrator}
    \end{subfigure}
   \caption{optimized shape of thermal cloak-concentrator; temperature disturbance ($T-\overline{T}$) and flux flow after adding thermal cloak-concentrator in the base material plate for $N_{\mathrm{ring}}$=5, $N_{\mathrm{sec}}$=20 (a,b,c) $N_{\mathrm{var}}=59$, $f_{\mathrm{obj}}=2.165\times10^{-4}$ (d,e,f) $N_{\mathrm{var}}=132$, $f_{\mathrm{obj}}=2.212\times10^{-4}$. For both cases, the concentrate flux is around 8 times as large as for the flat plate. The normalized temperature difference in the outer region is in the order of $10^{-5}$-$10^{-6}$.}
\label{fig:cloak + concentrator objfn 2}
\end{figure}
\newcolumntype{L}[1]{>{\raggedright\let\newline\\\arraybackslash\hspace{0pt}}m{#1}}
\newcolumntype{C}[1]{>{\centering\let\newline\\\arraybackslash\hspace{0pt}}m{#1}}
\newcolumntype{R}[1]{>{\raggedleft\let\newline\\\arraybackslash\hspace{0pt}}m{#1}}
\begin{table}   
\caption{Comparison between literature, conductivity optimization and shape optimization results for thermal cloak concentrator.}  
\renewcommand{\arraystretch}{1.4}                 
\centering                                            \small              
\begin{tabular}{|C{10em}|C{4.4em}|C{5.3em}|C{2.4em}|C{5.3em}|C{2.4em}|}   
\hline 
\multirow{2}{10em}{\centering Optimization method} & \multirow{2}{4.4em}{\centering Number of design variables} &  \multicolumn{2}{|c|}{$N_{\mathrm{ring}}=3$, $N_{\mathrm{sec}}=12$}& \multicolumn{2}{|c|}{$N_{\mathrm{ring}}=5$, $N_{\mathrm{sec}}=20$}\\[2ex]
\cline{3-6}
 & & $\Psi_{\mathrm{cloak}}$ & $\Psi_{\mathrm{flux}}$  & $\Psi_{\mathrm{cloak}}$ & $\Psi_{\mathrm{flux}}$\\[1ex]
\hline  
Literature results~\cite{Fujii2020} & 2850 & 7.64$\times 10^{-5}$ & 7.16 &7.64$\times 10^{-5}$ & 7.16 \\
\hline  
Conductivity optimization & & 1.425$\times 10^{-5}$& 7.587 &7.904$\times 10^{-6}$ & 8.500\\
\hline  
Initial geometry for shape optimization & & 1.769$\times 10^{-4}$& 7.688 &3.168$\times 10^{-4}$ & 8.212 \\
\hline  
\multirow{4}{10em}{\centering Shape optimization}   & 23 &4.749$\times 10^{-6}$ & 8.077 & & \\
\cline{2-6}
& 50 &7.451$\times 10^{-6}$ & 8.012 & & \\
\cline{2-6}
& 59 &  & &7.363$\times 10^{-6}$ & 8.316\\
\cline{2-6}
&132 &  & &1.232$\times 10^{-5}$ & 8.318 \\
\hline                                                                                                              
\end{tabular}                                   
                                                   
\label{table:Cloak + cntr objfn comparison}                                                   
\end{table} 
\subsubsection{Shape optimization}
\label{sec: Cloak+Cntr Shape optimization}
\par We now perform shape optimization for the geometries shown in \fref{fig:cloak + concentrator MatNum k opt}. The results of the shape optimization are presented in \frefs{fig:cloak + concentrator objfn 1}-\ref{fig:cloak + concentrator objfn 2}. For each geometry we consider two different values of $N_{\mathrm{var}}$. The optimized shapes for all cases generate negligible disturbance in the temperature profile ($T-\overline{T}$) as shown in \frefs{fig:cloak + concentrator objfn 1}-\ref{fig:cloak + concentrator objfn 2}. Also, we can observe that the material distribution in the inner rings guide the flux towards the inner core, and the material distribution in the outermost ring suppresses all the temperature disturbance created by the inner region. Thus, we can say that the shape and conductivity values of the patches on the outermost ring are very critical to the cloaking performance. In \tref{table:Cloak + cntr objfn comparison}, results from literature, conductivity optimization, and shape optimization are compared. From the comparison, we can say that, the shape optimization process can produce slightly better results than the topology optimization results from literature~\cite{Fujii2020} with much fewer design variables. However, this advantage also comes with the complexity of using more than two materials. We also use high conductivity copper in the inner core, instead of base material iron as in ~\cite{Fujii2020}.
However, the increment $N_{\mathrm{var}}$ does not provide any specific advantages. 
\par For the first type of geometry with $N_{\mathrm{ring}}=3$ \& $N_{\mathrm{sec}}=12$, $N_{\mathrm{var}}=23$ and  $N_{\mathrm{var}}=50$, while for the second type of geometry with $N_{\mathrm{ring}}=5$ \& $N_{\mathrm{sec}}=20$, $N_{\mathrm{var}}=59$ and  $N_{\mathrm{var}}=132$. The optimized shape effectively reduces the disturbance caused by inner core as well as concentrate the flux. For all four cases, the concentrate flux is around 7 times as large as for the flat plate. The normalized temperature difference in the outer region is very small (mostly in the order of $10^{-6}$). The exact values of $\Psi_{\mathrm{flux}}$ and $\Psi_{\mathrm{cloak}}$ are presented in \frefs{fig:cloak + concentrator objfn 1}-\ref{fig:cloak + concentrator objfn 2}. The optimized shape results are  $10^{-1}$ order better (both in $\Psi_{\mathrm{flux}}$ and $\Psi_{\mathrm{cloak}}$) than the results from literature.






\section{Conclusions}
\label{sec:Conclusions}
\par In the present article, we investigated the shape optimization method for metam-\\aterial-based heat manipulators. The approach works as a tool to find better geometries with good manufacturability and higher efficiency. The proposed method utilized a gradient-free Particle Swarm Optimization (PSO) algorithm. The geometry and solution fields are approximated using NURBS basis functions. It enables easy control of shape and provides good geometrical accuracy without meshing, remeshing, or any special post-processing approaches. Also, Nitsche’s method is used to impose interface continuity conditions. We presented examples of a thermal concentrator and a thermal cloak-concentrator to demonstrate the efficiency of the proposed method.
\par For the thermal concentrator, we analyzed sector-type geometries made of two materials. We studied the effect of geometric parameters such as the number of sectors, the number of design variables, the imposed symmetry conditions, the conductivity and spatial distribution of member materials, and the direction of incident flux on the optimized shape and performance of the thermal concentrator. The shape optimization indicates that,
\begin{itemize}
    \item With applied symmetry across $x$-axis and center axes of sectors, the optimized geometries can accumulate heat flux 2 to 7 times as large as for un-optimized geometries with straight sector edges. 
    \item Type A  ($N_{\mathrm{sec}}=8\ell-4$,~$\ell=1,2,...$) configurations perform better than other types of configurations, as the low $\kappa$ sectors for Type A configurations lie inline with $\Gamma_{\mathrm{flux}}$ (across which the concentrated flux is measured) in $\Omega_{\mathrm{design}}$. The low $\kappa$ material obstructs the flux flow circumventing the core and guides it towards the core, and eventually, through $\Gamma_{\mathrm{flux}}$. 
    \item By associating the high $\kappa$ material to the core, the openings (along the outer radius) for the high $\kappa$ sectors increase significantly. That provides a wider area to draw more flux towards the core. Also, the heat flux transfer from $\Omega_{\mathrm{design}}$ to $\Omega_{\mathrm{in}}$ becomes smoother. 
    \item The flux concentration capacity of the concentrator can be improved up to 2 to 3 times by placing high $\kappa$ material in the core and low $\kappa$ material inline with $\Gamma_{\mathrm{flux}}$ in $\Omega_{\mathrm{design}}$.
\end{itemize}
\par For the thermal cloak-concentrator, we analyzed the sector-shaped block-type geometries. We studied the effect of geometric parameters such as the number of sectors and the number of rings, the conductivity and spatial distribution of member materials. The shape optimization indicates that,
\begin{itemize}
    \item With the optimized geometries,  the thermal cloak-concentrator can collect 7 to 8 times as much flux as the un-optimized geometries. It also cloaks the core by diminishing the normalized temperature disturbance up to a negligible value of order $10^{-5}$ -$10^{-6}$. 
    \item With more member materials, the results in terms of the objective function are around $10^{-1}$ order better than the literature results (which are based on topology optimization) with fewer design variables.
\end{itemize}
\par The proposed method is applicable to any heat manipulator including thermal cloaks, active cloaks, heat inverters, heat illusions, and so on. Since it uses a gradient-free optimization algorithm, it can also handle non-differentiable problems. However, the limitation of the method is that it can not generate a new topology. Given that, the initial topology is crucial for the optimization. Furthermore, it can be extended to 3D applications, which is the subject of future studies.

\section*{Acknowledgements}
St\'ephane P.A. Bordas acknowledges the funding from the European Union’s Horizon 2020 research and innovation programme under grant agreement No 811099 TWINNING Project DRIVEN for the University of Luxembourg.
\bibliographystyle{model1-num-names}
\bibliography{thermal_metamaterial.bib}

\end{document}